\documentclass[a4paper,oneside,12pt,openright]{book}

\usepackage[utf8]{inputenc}
\usepackage[english]{babel}
\usepackage{bold-extra}
\usepackage{subfig}
\usepackage{booktabs}
\usepackage[T1]{fontenc}
\usepackage [a4paper,left=3cm,bottom=3cm,right=3cm,top=3cm]{geometry}
\usepackage[swapnames]{frontespizio}
\usepackage{mathtools}
\usepackage{longtable}
\usepackage{amsmath,amssymb}
\usepackage{bbold}
\usepackage{hyperref}
\usepackage[version=4]{mhchem}
\usepackage[output-decimal-marker={,}]{siunitx}
\usepackage{graphicx,setspace}
\usepackage{amsmath, amsfonts, amssymb, amsthm,mathrsfs,wrapfig,sidecap,eucal,amssymb,braket,appendix}
\usepackage{fancyhdr}
\usepackage[font=small,format=hang,labelfont={sf,bf}]{caption}
\usepackage{bm}
\usepackage{mdframed}
\usepackage{hyperref}
\usepackage{tikz}
\usepackage{hyperref}
\usepackage{float}
\usepackage{braket, mhchem, amscd}
\usepackage{listings, xcolor}

\newcommand{\be}{\begin{equation}}
\newcommand{\ee}{\end{equation}}
\newcommand{\bea}{\begin{eqnarray}}
\newcommand{\eea}{\end{eqnarray}}
\newcommand{\brr}{\begin{array}}
\newcommand{\err}{\end{array}}
\newcommand{\bit}{\begin{itemize}}\newcommand{\eit}{\end{itemize}}
\newcommand{\ben}{\begin{enumerate}}\newcommand{\een}{\end{enumerate}}

\newcommand{\ba}{\begin{array}}
\newcommand{\ea}{\end{array}}

\def\lf{\left}

\def\non{\nonumber}

\def\ri{\right}

\def\si{\sigma}
\def\Om{\Omega}
\newcommand{\mlab}[1]{\label{#1}}

\def\1{{_{1}}}\def\2{{_{2}}}
\def\bp{{\bf {p}}}\def\bk{{\bf {k}}}\def\bx{{\bf {x}}}\def\ba{{\bf {a}}}\def\bw{{\bf {w}}}

\def\noHe0{:\;\!\!\;\!\!:H_e(0):\;\!\!\;\!\!:}
\def\noHm0{:\;\!\!\;\!\!:H_\mu(0):\;\!\!\;\!\!:}

\def\lf{\left}

\def\non{\nonumber}

\def\ri{\right}

\def\si{\sigma}
\def\Om{\Omega}

\def\1{{_{1}}}\def\2{{_{2}}}

\def\bogu{\rho^{\bk}_{12}}
\def\bogv{\lambda^{\bk}_{12}}

\def\bogubst{{\rho}^{\bk\, *}_{12}}
\def\bogub{{\rho}^{\bk}_{12}}
\def\bogvb{{\lambda}^{\bk}_{12}}

\def\ebogo{e^{i\left(\omega_{\bk,1}+\omega_{\bk,2}\right) t}}

\def\bogocoeffAlpha{{\cal A}_{(\Om,\Om'),\,\vec{k}}^{(\si,\si')}}

\def\bogocoeffBeta{{\cal B}_{(\Om,\Om'),\,\vec{k}}^{(\si,\si')}}

\def\bogocoeffBetater{{\cal B}_{(\Om,\Om''),\,\vec{k}}^{(\si,\si'')}}

\def\bogocoeffAlphater{{\cal A}_{(\Om,\Om''),\,\vec{k}}^{(\si,\si'')}}

\def\bogocoeffAlphaquintst{{\cal A}_{(\Om',\Om''),\,\vec{k}'}^{(\si',\si'')\,*}}
\def\bogocoeffBetaquintst{{\cal B}_{(\Om',\Om''),\,\vec{k}'}^{(\si',\si'')\,*}}
\def\Betasiomst{{\cal B}_{(\Om,\Om''),\,\vec{k}}^{(\si,\si')\,*}}
\def\Betasiom{{\cal B}_{(\Om',\Om''),\,\vec{k}'}^{(\si,\si')}}

\def\densityBB{N_{\mathcal{BB}}}

\def\densityAA{N_{\mathcal{AA}}}
\def\densityAB{N_{\mathcal{AB}}}
\def\densityBA{N_{\mathcal{BA}}}
\def\uuu{U}

\def\lu{u}
\def\lv{v}
\def\lnu{\nu}

\newcommand{\lp}{\ell_{\rm p}}
\newcommand{\mpl}{m_{\rm p}}
\newcommand{\gn}{G_{\rm N}}

\renewcommand{\d}{\mbox{${\rm d}$}}
\newcommand{\Ep}{\mathcal{E}_{\rm p}}
\newcommand{\EX}{{$\eta$-$\xi$ }}

\newcommand{\etad}{\eta_{_\delta}}
\newcommand{\xid}{\xi_{_\delta}}
\newcommand{\I}{{_I}}
\newcommand{\II}{{_{I\!I}}}
\newcommand{\III}{{_{I\!I\!I}}}
\newcommand{\IV}{{_{I\!V}}}
\newcommand{\IId}{{_{I\!I_{\!\delta}}}}

\newcommand{\IQ}{{_I,_{I\!I}}}
\newcommand{\R}{{\Bbb R}}
\newcommand{\Vk}{{\bk}}
\newcommand{\Vp}{{\bf p}}
\newcommand{\Vx}{{\bx}}

\begin {document}

\begin{titlepage}
   \begin{minipage}{0.46\textwidth}
   \includegraphics[width=0.41\textwidth]{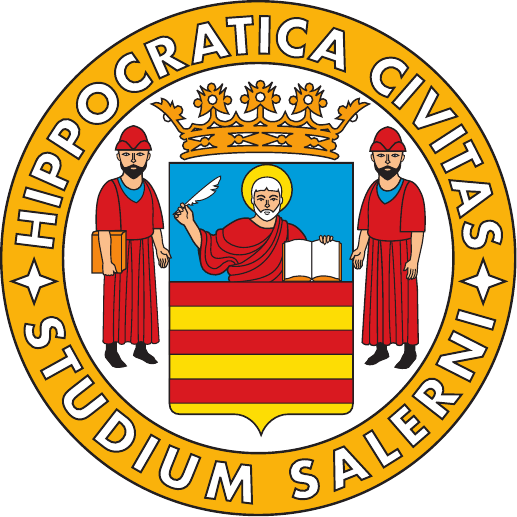}
    \end{minipage}
    \hspace{1.5cm}
    \begin{minipage}{0.46\textwidth}
    \includegraphics[width=1\textwidth]{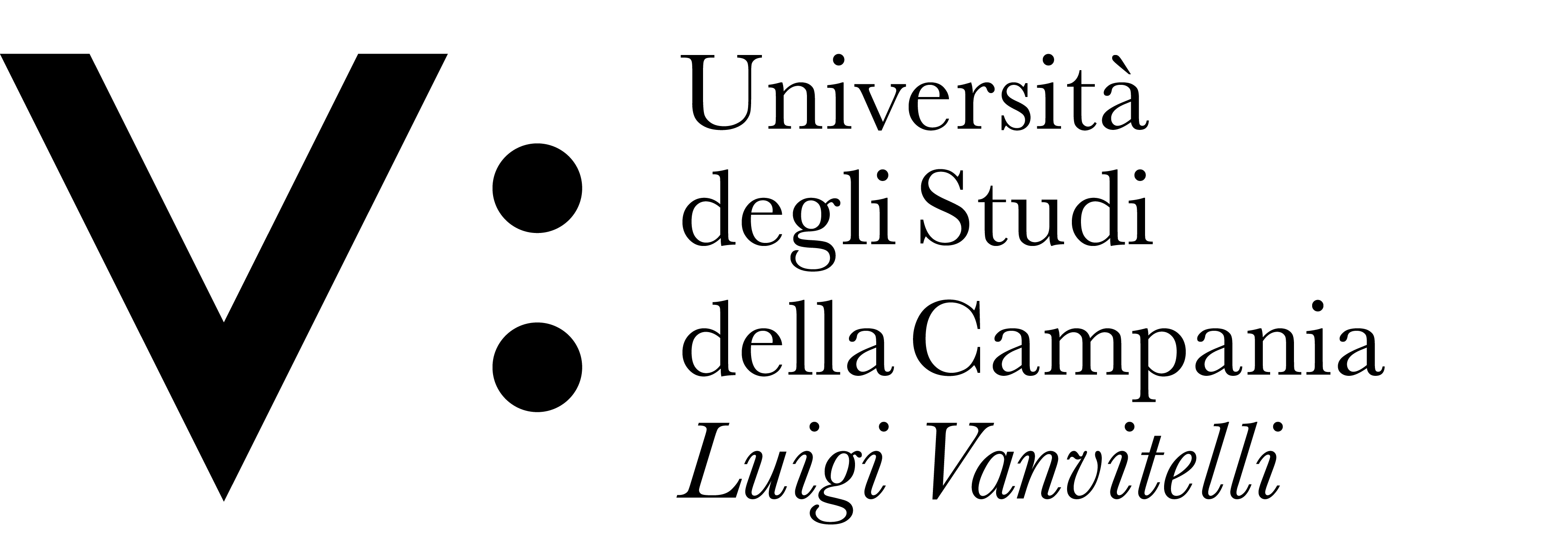}
    \end{minipage}
   \vspace{-0.018\textwidth}
    \begin{center}

    \vspace*{0.3cm}
        \large {\textbf {Universit\`a degli Studi di Salerno}} \\
        \vspace{0.1cm}Dipartimento di Fisica ``E.R. Caianiello'' \\
        \vspace{0.6cm} \large {\bf Universit\`a degli Studi della Campania ``Luigi Vanvitelli''}\\ 
       \vspace{0.1cm} Dipartimento di Matematica e Fisica \\
            \vspace*{0.8cm}
               \emph{A dissertation submitted in partial fulfillment of the\\ requirements 
               for the degree of Doctor of Philosophy}\\
               \vspace{0.2cm}
               \emph{in}\\
               \vspace{0.2cm}
               \bf{Physics} (FIS/02)\\
                 \vspace{0.08cm}
           \rule{\textwidth} {0.002\textwidth} 
        \Large  Non-thermal aspects of Unruh effect\vspace{-2mm}
        \rule{\textwidth} {0.002\textwidth} 
                  \end{center}
                  
                    \vspace{0.9cm}

%
%
%

  \begin{minipage}{0.46\textwidth}
        \hspace{-4mm} {\bf Author} \\
        
        \vspace{-5mm}
           \hspace{-4mm} Giuseppe Gaetano Luciano \\
        
          \end{minipage}

           \hspace{8.7cm}
          \begin{minipage}{0.46\textwidth}
          
          \vspace{-17mm}
          
        \hspace{-3.98mm}{\bf PhD School Coordinator} \\
        
        \vspace{-5.3mm}
 \hspace{-5.2mm} Ch.mo Prof. Cataldo Godano \\
        
          \end{minipage}
          
 \vspace{1cm}
 
    \begin{minipage}{2\textwidth}
  \vspace{0.05cm}
 \hspace{8.27cm}  {\bf Advisor}\\
   
   \vspace{-0.5cm}
   \hspace{8.29cm} Ch.mo Prof. Massimo Blasone\\
   
   \end{minipage}
   
    
      
    
    
    
\vspace{3cm}
\centering
{\bf PhD Program in ``Matematica, Fisica e Applicazioni'' XXXI Ciclo}

   \end{titlepage}

\makeatletter
\newcommand{\partialslash}{\mathpalette\@partialslash}
\newcommand{\@partialslash}[1]{%
  \ooalign{\raise.2ex\hbox{$#1\mkern1mu/$}\cr$#1\partial$\cr}}
\makeatother

\newpage\null\thispagestyle{empty}\newpage

${}$
\vspace{4 cm}
\begin{flushright}
\emph{In loving memory of my mother}
\end{flushright}

\newpage\null\thispagestyle{empty}\newpage

\frontmatter

\chapter*{Abstract}
The search for a consistent and empirically established relation among
general relativity, quantum theory and thermodynamics is the most challenging 
task of theoretical physics since 
the discovery of the Hawking effect.
The emergence of a temperature in 
spacetimes endowed with event horizon(s) has unveiled
the existence of a fertile but largely uncharted territory, in 
which general covariance, gravity, thermal
and quantum effects are intimately connected. 
Even though black holes would be 
the best arena to explore such an interplay, 
observational evidences for their existence 
are still lacking, thus suggesting to address the issue  in  
less exotic contexts. 
In this sense, a tantalizing framework  
is provided by the quantum field theory (QFT) in curved spacetime. 
Specifically, 
the \emph{Unruh effect}, along with its distinctive thermal 
features arising from the Rindler horizon structure, 
represents the first important step towards unifying the ``quantum''  and ``gravity''
worlds via the equivalence principle.  
Furthermore, it turns out to be an essential ingredient for 
the general covariance of QFT, as recently witnessed by the study
of the decay of accelerated particles 
in both the laboratory and comoving frames.
Not least, the possibility to extend it to interacting field theory and 
to a broad class of other spacetimes, 
makes this prediction one of the most firmly rooted results in QFT. Hence, 
waiting for a solid and completely
successful theory of quantum gravity, a scrupulous investigation 
of Unruh effect, and, in particular, of
any deviations of Unruh spectrum from a strictly thermal behavior, 
may offer a promising window to new physics 
in the current limbo between general relativity and quantum framework.

\medskip
Starting from the outlined picture, in this Thesis we look at
the connection between geometric properties of spacetimes
and ensuing thermal quantum phenomena from a non-traditional 
perspective, based on the analysis of
``perturbative'' effects that undermine the standard scenario. 
As a test bench for our analysis, we consider
the Unruh radiation detected by an eternally, linearly accelerated 
observer in the inertial vacuum.
After a brief discussion on the derivation of the Unruh effect in QFT and the status of experimental tests, we examine to what extent the characteristic Planck spectrum of Unruh radiation gets spoilt: $i)$ for \emph{mixed fields}, and specifically 
for neutrinos, which are 
among the fundamental constituents of the Standard Model,
$ii)$ in the presence  of a \emph{minimal fundamental length} arising from gravity in Planck regime. 
On the one hand, it is shown that the Unruh distribution loses 
its thermal identity
when taking into account flavor mixing,  
due to the interplay between the Bogoliubov transformation
associated to the Rindler horizon and the one
responsible for mixing in QFT. Implications of such a result are 
tackled in the context of the inverse $\beta$-decay, also in light
of the recent debate on a possible violation of the general covariance of QFT induced by mixing.
On the other hand, we focus on the effects
triggered by 
deformations of the Heisenberg uncertainty principle at Planck scale, 
exploring the possibility to constrain the characteristic parameter of the Generalized
Uncertainty Principle (GUP)
via the Unruh effect. 
The question is addressed 
of whether these seemingly unrelated frameworks
have some roots in common or, in other words, if their features 
can be understood in a deeper way so that they appear to be merged.
Along this line, we lay the foundations to provide a unifying perspective of these effects, which still relies on a purely geometric interpretation of their origin.
As future prospects, we finally study our results
in connection with  neutrino physics beyond the Standard Model, experiments on Planck-scale effects on the analogue Hawking-Unruh radiation and
entanglement properties in accelerated frames.

\newpage\null\thispagestyle{empty}\newpage

\chapter*{Acknowledgements}
Let me be a bit less rigorous and academic just for a while \dots there 
are many people I would like to thank in this thesis work. 
Some of them gave me a direct support, some other indirect, 
but not less important. First of all, I would like to express my sincere 
gratitude to all members of the Theoretical Physics research group at 
University of Salerno. The senior staff: my supervisor, Prof. Massimo Blasone, 
my (next to leading order) supervisor, Prof. Gaetano Lambiase, and the
``top of the pyramid'' (even if formally retired), Prof. Giuseppe Vitiello, for the 
long sessions of discussions (and food) we had together and the 
continuous improvement they provided to my work; my colleagues PhD 
students: Luciano ``octopus broth'', Luca S. and Luca B, for constructive debates
mixed with funny moments.
Outside the research group, I also want to cite Carmine Napoli for brotherly and technical
support, and the other colleagues
I shared my office with. 
The amount of  hours spent having fun with them is by far larger than 
the time devoted to rigorous scientific activities.

A section of these acknowledgments cannot but be reserved to my two 
thesis referees, Prof. Petr Jizba (whom I also thank for his great willingness during my PhD 
study abroad in Prague) and Prof. Alex Eduardo de Bernardini.
While reviewing the manuscript, they provided me great suggestions, 
drawing my attention on a number of occasional
(and not) mistakes. I am also indebted to Dr. Fabio Scardigli,  Prof. Achim Kempf and Dr. Fabio Dell'Anno
for providing me with food for thought and very insightful comments.

Of course, in this long list, I could not forget my brotherly friends, 
an unexpectedly (but lovely) ``special'' person, Edda, and my family, that always 
supported me along these years. Maybe they did not help me writing a 
paper or solving an integral, but I do not think that, without them, 
I would have been able to be where now I am.

\newpage\null\thispagestyle{empty}\newpage

\tableofcontents

\newpage\null\thispagestyle{empty}\newpage

\chapter*{Conventions and abbreviations}
\addcontentsline{toc}{chapter}{Conventions}
\begin{flushright}
\emph{``I'm a voracious reader,\\ and I like to explore all sorts of writing \\without prejudice and without paying any attention \\to labels, conventions or silly critical fads.''}\\[1mm]
- Carlos Ruiz Zafon -
\\[6mm]
\end{flushright}
\Large{\textbf{Units and metric}}\vspace{5mm}
\normalsize
${}$\\
Throughout the manuscript, we set
\begin{equation}
\hbar\,=\,c\,=\,k_B\,=\,G\,=\,1\,,
\label{units}
\end{equation}
unless explicitly stated otherwise. Furthermore, we work in $1+3$-dimensions, using the 
conventional timelike signature for the metric:
\begin{equation}
(+\,-\,-\,-)\,,
\end{equation}
except for Chapter~\ref{Unified formalism for Thermal Quantum Field Theories: a geometric viewpoint}.

The following tensor index notation is used for the 
Riemann and Ricci curvature tensors:
\begin{eqnarray}
{R^{\rho}}_{\sigma\mu\nu}&=&\partial_{\mu}\Gamma^{\rho}_{\nu\sigma}
\,-\,\partial_{\nu}\Gamma^{\rho}_{\mu\sigma}
\,+\,\Gamma^{\rho}_{\mu\lambda}\Gamma^{\lambda}_{\nu\sigma}
\,-\,\Gamma^{\rho}_{\nu\lambda}\Gamma^{\lambda}_{\mu\sigma}\,\\[2mm]
R_{\mu\nu}&=&{R^{\rho}}_{\sigma\rho\nu}\,
\end{eqnarray}
where the Christoffel symbols are usually defined in terms of the 
metric tensor $g_{\mu\nu}$ as
\be
\Gamma^{\rho}_{\nu\sigma}\,=\,\frac{1}{2}\,g^{\rho\lambda}
\left(\partial_{\sigma}g_{\lambda\nu}\,+\,\partial_{\nu}g_{\lambda\sigma}
\,-\,\partial_{\lambda}g_{\nu\sigma}\right)\,.
\ee
Formulae can be changed from our notation to the opposite 
spacelike convention $(-\,+\,+\,+)$ by  reversing the signs of 
$g_{\mu\nu}$, $\square\equiv g^{\mu\nu}\nabla_\mu\nabla_\nu$, 
${R^{\rho}}_{\sigma\mu\nu}$, $R_{\mu\nu}$, ${T_{\mu}}^{\nu}$, 
but leaving $R_{\rho\sigma\mu\nu}$, ${R_{\mu}}^\nu$, $R$ and 
$T_{\mu\nu}$ unchanged.
\\[8mm]
${}$\\
\Large{\textbf{Special characters and abbreviations}\\[5mm]
\normalsize{The following special characters and abbreviations are used throughout: 
\vspace{-6mm}
\begin{longtable}{lp{0.35\columnwidth}}
\caption*{}
\label{tab:longtable} \\
\toprule
\qquad\qquad\qquad\quad\quad Character & \qquad Meaning \\
\midrule
\endfirsthead
\multicolumn{2}{l}{\footnotesize\itshape
\qquad\qquad\qquad\qquad\qquad\qquad\quad\quad\qquad\quad Continua dalla pagina precedente}  \\
\toprule
\endhead
\bottomrule
\multicolumn{2}{l}{\footnotesize\itshape Continued on next page} \\
\endfoot
\bottomrule
\multicolumn{2}{r}{\footnotesize\itshape} \\
\endlastfoot
\hspace{33.4mm}* &\hspace{7mm} complex conjugate\\ 
\hspace{32.1mm} $\dagger$ o h.c.&\quad\, \,  Hermitian conjugate\\
\hspace{32mm} ${}^{-}$ & \quad\,\,\,\,\,\,Dirac adjoint\\
\hspace{32mm} $\frac{\partial}{\partial x^{\mu}}$ o $\partial_{\mu}$ &\quad \, \, partial derivative\\
\hspace{33.2mm}$\nabla_{\mu}$ &\quad\, \, covariant derivative\\
\hspace{31.7mm} Re\,(Im)&\quad \,\,\,\,\,\,real (imaginary) part\\
\hspace{33.2mm}tr& \qquad trace\\
\hspace{33.2mm}$\ln$& \qquad natural logarithm\\
\hspace{33.2mm}$\log$& \qquad  decimal logarithm\\
\hspace{33.1mm}$\Gamma$&\qquad Gamma function\\
\hspace{33.1mm}$K_{\alpha}(x)$&\qquad K-Bessel function of \\
&\hspace{8.5mm}order $\alpha$ and argument\\ 
&\hspace{8.5mm}$x$\\
\hspace{33.3mm}$[A,B]$&\qquad $AB-BA$\\
\hspace{33.3mm}$\{A,B\}$&\qquad $AB+BA$\\
\hspace{33.6mm}$\simeq$ &\qquad approximately equal to\\
\hspace{33.6mm}$\sim$ & \qquad order of magnitude of\\
\hspace{33.6mm}$\equiv$ & \qquad {defined to be equal to}\\
\hspace{33.6mm}$\varpropto$ & \qquad {proportional to}\\
\hspace{33.6mm}$::$ & \qquad normal ordering\\
\end{longtable}

By convention, greek letters are used for $4$-dimensional spacetime indices 
running from $0$ to $3$, while latin letters are reserved 
for $3$-dimensional spatial indices running from $1$ to $3$.

\newpage\null\thispagestyle{empty}\newpage

\chapter{Introduction}
\begin{flushright}
\emph{``The World is Made of Events, \\not Things.''}\\[1mm]
- Carlo Rovelli -\\[6mm]
\end{flushright}
The connection among the notion of time, quantum theory,
thermodynamics and general relativity is at the heart
of a number of debated issues.
Among these, a still vibrant subject of investigation is to understand how
the physically evident time flow arises from the
``timelessness'' of the hypothetically fundamental
Quantum Field Theory (QFT)\footnote{See, for example, Refs.~\cite{Rovelli:1990jm} for a detailed discussion 
on the ``issue of time'' in quantum gravity.}. On the other hand, no less suggestive 
are the lack of a statistical theory of the gravitational field~\cite{Rovelli:1993ys}
and the elusive thermal features of quantum theory in spacetimes endowed with event horizon(s),
the traces of which are recognizable in such pivotal
 phenomena as the Hawking black hole
radiation~\cite{Hawking:1974sw} and the Unruh effect~\cite{Unruh:1976db}.
It is generally accepted that these arguments may suggest the existence of a 
not yet fully explored territory, in which geometry of spacetime, thermal and quantum effects 
are closely related. In the absence of a well-established 
theory of quantum gravity, the QFT in curved backgrounds
provides the most consistent 
test bench to investigate this puzzling scenario to date.

Within such a framework, one of the pioneering attempts to merge 
thermal effects and intrinsic
properties of non-trivial backgrounds 
was performed in Ref.~\cite{Israel:1976},
where the ``vacuum'' perceived 
by stationary observers outside a black hole 
(and also by uniformly accelerated observers in Minkowski spacetime) 
was interpreted in terms of the finite-temperature
field theory of Takahashi and Umezawa (the so-called Thermo Field Dynamics (TFD) 
formalism~\cite{UM0}). Specifically, it was noted
that the Hartle-Hawking vacuum in the black hole theory
and the thermal vacuum in TFD have  the same formal
expression, although they have different physical meanings: 
whilst the former is a thermal state for a static observer
in proximity of a black hole,  the latter exhibits analogous properties
for an inertial observer in Minkowski spacetime.
Along this line, in Refs.~\cite{GUI90} the question
was addressed of whether a direct relationship between 
geometric features and thermo fields could be found via the introduction
of a suitable background: efforts in this sense
culminated in the construction of the \EX spacetime, a flat manifold
with a complex horizon structure where the (zero temperature)
vacuum for quantum fields corresponds to the thermal state for a static observer
in Minkowski spacetime. Furthermore, 
both the imaginary- and real-time approaches 
to thermal field theory can be naturally connected within this framework, 
which thus provides 
the proper background  to develop a unified formalism
for quantum field theories at finite temperature.

Remarkably, endeavors to translate 
inherent characteristics of physical systems
in terms of geometric properties of spacetime were also put forward 
in various other contexts: in the standard QFT, for instance, it was found
that \emph{flavor mixing} relations hide a Bogoliubov transformation
responsible for the unitary inequivalence of flavor and mass
representations and their related vacuum structures~\cite{Blasone:1995zc,bosonmix}.
Recently, the same analysis was carried out for an eternally, linearly accelerated (Rindler) 
observer in flat background~\cite{Blasone:2017nbf,Blasone:2018byx}. 
In that case, it was emphasized that the Bogoliubov transformation arising
from the superposition of fields with different masses
and the one
associated with the geometric (horizon) structure of Rindler spacetime 
combine \emph{symmetrically} in the calculation of the Unruh distribution, 
suggesting a possible geometric interpretation for the origin of mixing too. 
Similarly, in Refs.~\cite{Nirouei,Nirouei:2011zz,Scardaus} the geometric imprints of 
the existence of a \emph{minimal fundamental length} on the Hawking and Unruh radiations 
were addressed in the context of the 
Generalized Uncertainty Principle (GUP).
The above background picture could also help to shed new light on such relevant
issues as the cosmological holography~\cite{Lloyd} or $\alpha$-vacua in AdS/CFT correspondence~\cite{Chamblin}.

\smallskip
Besides formal aspects, it is worth noting that a subtle 
common thread running through the aforementioned effects
is the possibility to spoil the thermal identity
of Hawking and Unruh radiations via the appearance 
of non-trivial corrections in the particle spectrum.
In light of the central  r\^ole played by these phenomena 
in several areas of QFT (including interacting field theory and quantum theory
in a large class of non-trivial spacetimes), the investigation of
this scenario and, in particular, of any deviations of the Hawking and Unruh spectra
from a purely thermal profile, could provide a pragmatic way to 
probe new physics in the limbo between
gravitational and quantum ``worlds''. In the context of flavor mixing in QFT, 
for instance, it was shown that the Unruh distribution does  
acquire a subdominant (non-thermal) contribution 
depending on the mixing angle and the squared mass difference.
This was proved for both bosons~\cite{Blasone:2017nbf} and fermions~\cite{Blasone:2018byx},
and, specifically, for neutrinos, which are among the elementary particles in the Standard Model (SM). Apart from
highlighting a non-universal character  
of the Unruh effect even in such an ordinary framework as the SM, the obtain result
may be a starting point
to fix new bounds on the characteristic parameters of neutrino mixing.
A somewhat exotic behavior for the Hawking and Unruh radiations was also
derived within the non-canonical 
QFT, where non-thermal effects were found to be tied in with 
geometric deformations of the Heisenberg uncertainty 
relation descending from quantum gravity contexts
(see, for example, Refs.~\cite{Nirouei:2011zz,Scardaus} 
and therein). The analysis of these aspects at both theoretical
level (via gedanken experiments on 
the radiation emitted by 
large black holes~\cite{MM} or
the formation of micro 
black holes~\cite{FS}) and phenomenological level
(through the study of the effects induced by the Generalized Uncertainty Principle in analogue gravity experiments)
can clarify to what extent
Planck-scale effects do undermine the bases of QFT. 

To give a more comprehensive overview of the current state-of-the-art, 
we mention that the idea of looking at non-thermal signatures of the
Hawking radiation has been arousing growing interest during the
last two decades in a plethora of other quantum
contexts, ranging from particle production by spherical
bodies collapsing into extremal
Reissner-Nordstr\"om black holes~\cite{Liberati:2000sq}, to the emission by  
Schwarzschild Anti-de Sitter~\cite{Hossain:2013qda} and
Kerr-Newman~\cite{Han:2009zzb} sources. In the same way, 
classical effects responsible
for non-thermal corrections in the infrared regime are provided by
grey-body factors, adiabaticity 
or phase space constraints~\cite{Visser:2014ypa}.
On the other hand,  for what concerns the Unruh radiation, distortions of the spectrum
have been derived in the evaluation of the Casimir-Polder force
between two relativistic uniformly accelerated atoms~\cite{Marino:2014rfa}, 
in the polymer (loop) 
quantization applied to the calculation of
the two-point function along
Rindler trajectories~\cite{Hossainviolat},
and in the analysis of the response function of a detector moving with 
non-uniform (and/or only temporarily switched on)
acceleration~\cite{Ahmadzadegan:2018bqz}. 

Taking stock of the arguments discussed so far,  it follows
that a wide-ranging investigation of non-thermal features of the 
Hawking and Unruh effects
is far from being trivial, as it involves disparate 
contexts not yet fully understood.
Furthermore,  
it may have valuable implications 
in a variety of challenging domains, including
theoretical tests of gravity theories and
hypothetical  violations 
of the equivalence principle within the framework of QFT on curved background. 
In connection with the latter aspect, in particular, it is interesting to note that, 
in spite of some accidental similarities appearing in the respective derivations, 
the Hawking and Unruh effects cannot be regarded as two sides of a unique phenomenon, 
as they do not proceed from the
same underlying mechanism (pair creation
close to an \emph{observer independent} geometric horizon
in the case of the Hawking radiation, versus an \emph{observer/coordinate 
dependent} horizon for the Unruh effect~\cite{Birrell}). In light of this, 
the following questions may naturally 
arise: what if accelerated observers detected a non-thermal vacuum
spectrum, whereas stationary observers outside a black hole did not (or vice versa)?
Could this finding be interpreted as a failure of the equivalence principle?
Positive answers on this matter should not be greeted with
skepticism; footprints of violations of the equivalence principle  
were already shown up in the context of the Hawking-Unruh effect.
In Ref.~\cite{Singleton:2011vh}, specifically, 
it was found that the temperature measured by a 
detector at rest in the background
of a Schwarzschild black hole is \emph{higher} than the 
one recorded by a uniformly accelerating device in Minkowski spacetime.
Similarly, non-trivial differences between the two phenomena were 
derived in Ref.~\cite{stait-gardner} from studies about pair production by lasers in vacuum.
In what follows, however, we shall not take specific care of these
issues, reserving them for forthcoming discussions.

\medskip
It is in the picture described above that our work is set. 
We begin by reviewing the most  common thermal features of quantum theory in non-trivial
backgrounds, focusing on
the Hawking and Unruh radiations. 
By pursuing a solid research-line about the decay of accelerated particles~\cite{Ahluwalia:2016wmf,Blasone:2018czm,BlasonePOS,Cozzella:2018qew}, 
we also explain why the Unruh effect 
is \emph{mandatory} to maintain the general covariance of 
QFT,  in spite of the skepticism of
part of the community~\cite{Unruh effect no, Oriti:1999xq, Narz}. 
After setting the stage,  
we move on to the study of a cluster
of higher-order effects that undermine the
standard well-known findings. Broadly speaking, such an analysis is intended to be a ``stress test'' of QFT, 
namely a tool to investigate whether unrelated well-established branches of 
the theory (as event horizon thermodynamics, flavor mixing and quantum contexts
with a minimal length) still provide a consistent framework when patched together.
On the basis of the obtained results, we
finally explore the possibility to trace these exotic behaviors 
back to some common root, offering preliminary hints
for the analysis of currently debated topics.

\smallskip
The Thesis is structured as 
follows\footnote{The work is to a large extent 
built on our recent papers~\cite{Blasone:2017nbf,Blasone:2018byx,Scardaus,Blasone:2018czm,noepl,JizLuc}, 
but some results are based on still unpublished ideas.}:
\emph{Chapter 1} contains a pedagogical explanation 
of the Unruh effect within the canonical QFT framework 
(a brief mention to the ``twin'' calculation of the
Hawking radiation is also made). 
To this aim, different complementary derivations are illustrated: besides
the original Bogoliubov transformation approach~\cite{Unruh:1976db},
an unusual evaluation based on the time-dependent Doppler
shift~\cite{Alsing:2004ig} and a more standard analysis involving the
response function of a uniformly accelerated detector~\cite{Birrell} are proposed.
We also introduce the less known Letaw-Pfautsch-Bell-Leinaas effect~\cite{Letaw1981,Bell:1982qr}, 
namely, the  circular counterpart 
of the Unruh effect. Finally, we speculate on the possibility to seek
for experimental evidences of the  Unruh radiation, clarifying the persisting state of confusion
and the frequent concerns raised in literature about its real existence. 

\emph{Chapter 2} is basically from the papers~\cite{Blasone:2017nbf,Blasone:2018byx,noepl},  
with a little bit more details and revisions. In this context,  
the topic of the Unruh effect for mixed fields is addressed.
As a consequence of the interplay between the Bogoliubov transformation 
hiding in field mixing and the one arising from the geometric structure of Rindler background,
the Unruh vacuum distribution gets non-trivially modified, 
resulting in the sum of the standard Planckian spectrum, plus non-thermal corrections depending upon
the mixing parameters (a detailed review of the QFT treatment of flavor mixing 
can be found in Appendix~\ref{QFT of fm}). Based on such an outcome, 
we give a preliminary interpretation of
the origin of mixing in terms of geometric properties of the spacetime.

\emph{Chapter 3} is built upon Refs.~\cite{Blasone:2018czm,BlasonePOS}.
In connection with the issue of vacuum effects for mixed fields, in this Chapter
we provide a theoretical ``check'' of Unruh's discovery in the context
of the decay of accelerated protons with mixed neutrinos. 
This analysis fits in with a well-established line of research, which extends
from the original works by Matsas and Vanzella~\cite{Matsas:1999jx,Matsas:1999jx2,Matsas:1999jx3}
on the necessity of the Unruh effect for maintaining the general covariance of QFT in the absence 
of mixing, to the recent debate about possible internal inconsistencies 
when mixed fields are
involved~\cite{Ahluwalia:2016wmf, Blasone:2018czm,BlasonePOS,Cozzella:2018qew}.
The possibility that in this case
non-thermal like corrections to the Unruh bath become mandatory 
to keep the QFT generally covariant
(at least within 
a certain approximation) is finally investigated. 
In Appendix~\ref{Cozzella}
we also comment on some findings of 
Ref.~\cite{Cozzella:2018qew}, 
where the same topic, although with different assumptions, is analyzed.

Relying on Ref.~\cite{Scardaus}, in \emph{Chapter 4}
we examine to what extent  the Unruh distribution is
affected by Planck-scale corrections
in the guise of geometric deformations of the uncertainty relation. 
Using a generalized commutator quadratic in the momentum~\cite{Nirouei:2011zz,Scardaus}, 
we show that the characteristic thermal profile of the Unruh radiation gets 
non-trivially spoilt: for small deviations from 
the canonical QFT, 
however, the resulting spectrum can be rearranged so that
it recovers its standard behavior, but with a shifted temperature depending on the
deforming parameter. Besides a genuinely theoretical interest 
within the frameworks of  
black hole physics and string theory (where deformed uncertainty relations were 
originally addressed), 
the outlined dependence may open 
new perspective avenues for laboratory
tests of GUP effects, also in light of the growing number of analogue 
gravity experiments that have been 
carrying out in the last years~\cite{silke,Iorio:2017vtw}.

\emph{Chapter 5} is devoted to study the QFT formalisms 
at finite temperature and density~\cite{JizLuc}, such as the Path Ordered Method, 
the Closed Time Path formalism,
the Matsubara approach and the Thermo Field Dynamics.
Searching for a unified perspective on all these seemingly unrelated
techniques, we introduce the so-called \EX spacetime~\cite{GUI90},
a flat manifold with complexified $S^1\times R^3$ topology, whose thermal
features are discussed in connection with the properties of 
curved backgrounds 
endowed with event horizon(s). Specifically, we analyze 
the possibility to enrich the framework 
originally built on by Gui~\cite{GUI90} and the ensuing definition of a one-to-one map
between \emph{vacuum} Green's functions
in  \EX spacetime and Mills representation of 
\emph{thermal} Green's functions in Minkowski spacetime. 
We also emphasize how this enlarged scenario could be useful
to extend the connection between \EX spacetime and Thermal Quantum Field Theories
to out-of-thermal-equilibrium systems.

\emph{Chapter 6} contains closing remarks and an outlook at future prospects. 
Although the whole framework is conceptually simple, specific computations are sometime lengthy
and, for the reader convenience, we confine mathematical technicalities to the final Appendices.

\newpage\null\thispagestyle{empty}\newpage

\mainmatter
\chapter{The emerging part of an iceberg called quantum vacuum: the Unruh effect}
\label{The fascinating puzzle of quantum vacuum: the Hawking-Unruh effect}
\begin{flushright}
\emph{``Maybe the universe \\is a vacuum fluctuation.''}\\[1mm]
- Edward~P.~Tryon -\\[6mm]
\end{flushright}

\label{the tip}
Vacuum is one of the most exquisite and puzzling concepts 
in Quantum Field Theory (QFT). Notwithstanding
the name, it exhibits 
a variety of properties that underpin distinctive, though 
usually exceedingly small, physical
effects. To get an idea of how rich the vacuum structure is, 
the hypothesis by Tryon~\cite{Tryon:1973xi} about the genesis of Universe comes in handy: 
``\emph{maybe the Universe itself is a vacuum fluctuation}'', that is to say
``\emph{the Universe began as a single particle arising from an absolute 
vacuum}'' in a manner similar to ``\emph{how virtual particles come into existence and 
then fall back into non-existence... It's just possible that there might 
have been absolutely nothing out of which came a particle so potent 
that it could blossom into the entire Universe''.}

If we had a ``quantum'' magnifying glass, we
would observe that vacuum is not empty at all; actually, it would appear 
as a turmoil of (virtual)  particles continuously popping in and out of existence.
One of the most striking manifestations of these ephemeral
objects is the Casimir effect~\cite{Casimir:1948dh}, 
the relevance of which has been growing in the last years 
in a broad class of domains, ranging
from quantum computing~\cite{Benenti} to biology~\cite{Brado}.
This phenomenon arises from the alteration by metal boundaries of the zero 
point electromagnetic energy between them.  In the same way
as metallic plates can disturb the electromagnetic vacuum, the
curvature of the spacetime should affect, in principle, all vacua, due to the coupling 
of gravity with all (massive) fields. In this context, the Hawking effect
provides the most eloquent example of the 
fundamental r\^{o}le played by the quantum vacuum in regime of extreme gravitation~\cite{Hawking:1974sw}.

The discovery that black holes can evaporate emitting a thermal
radiation has led to a profound connection between the 
properties of  spacetime and laws of QFT and
thermodynamics, offering precious hints as to what 
we should expect from a complete theory of quantum gravity. 
Being related to such esoteric objects as black holes, however, 
Hawking radiation never came under the spotlight of experimental physics.

\begin{figure}[t]
\centering
\resizebox{8.5cm}{!}{\includegraphics{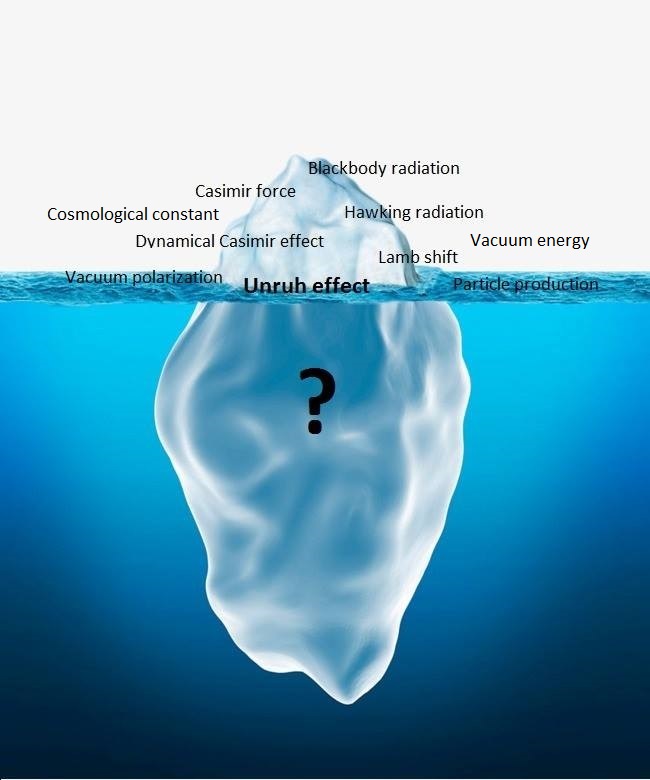}}
\caption{\small{The iceberg-like nature of quantum vacuum: 
a creative representation of known and (almost) unknown vacuum effects and properties.
}}
\label{figure:iceberg}
\end{figure}

In 1976, a then rookie W.~Unruh, while working on various 
aspects of black hole evaporation, unveiled one of the most 
impressive results of QFT: from the point of view of a uniformly 
accelerated detector, empty space contains a gas of particles 
at a temperature proportional to its proper acceleration 
$\alpha$~\cite{Unruh:1976db}. Roughly speaking, it may be said
that an accelerating observer in Minkowski vacuum would feel a 
warm wind of particles at $T\sim \alpha$, whereas inertial observer 
would be frozen at $0\, \mathrm{K}$ (see Fig.~\ref{pictunruh} for a pictorial
representation of the Unruh effect).

According to the equivalence principle, such a prediction can be 
regarded, at least locally, as the inertial non-gravitational counterpart of the Hawking radiation, confirming 
Fulling's previous achievements that the particle content of QFT is observer-dependent even in the case of flat spacetime~\cite{Fulling:1972md}. In spite of this, however, some skepticism on the real existence of the Unruh effect has been expressing during the years~\cite{Unruh effect no, Oriti:1999xq, Narz}.
The rather frequent concerns raised in literature have thus motivated the search for phenomenological evidences that could definitively solve the matter\footnote{As a matter of fact, there is no shortage of physicists who claim that the Unruh effect has already been observed~\cite{Smolyaninov}!}.  In this sense, the pioneers were Bell and Leinaas~\cite{Bell:1982qr}, who tried to interpret the depolarization of 
electrons in a storage ring through the Unruh effect, using spin as a sensitive
thermometer. An alternative approach was subsequently proposed by M\"{u}ller~\cite{Muller:1997rt}
in the context of the decay of accelerated
particles. In Refs.~\cite{Matsas:1999jx,Matsas:1999jx2,Matsas:1999jx3,Suzuki:2002xg}, in particular, with reference to the inverse 
$\beta$-decay, it was shown that 
the Unruh thermal bath is indeed mandatory to ensure that the decay rates
of accelerated protons in the inertial and comoving frames coincide, also when considering 
flavor mixing for emitted neutrinos~\cite{Blasone:2018czm,Cozzella:2018qew} (see Chapter~\ref{The necessity 
of the Unruh effect in QFT: the proton} for a complete treatment of 
this issue).  As remarked in Ref.~\cite{Matsas:1999jx2}, however, searching for experimental evidences of the 
Unruh radiation in this context is unfeasible, due to the relatively small
accelerations achievable on Earth 
(with typical accelerations of the LHC, for example, the proton
lifetime would be of the order of $10^{3\times10^8}\,\mathrm{yr}$, a time out of reach even for the
long-lived physicist!).  More viable, on the other hand, was the strategy 
suggested by Higuchi \emph{et al.} in Refs.~\cite{Higuchi:1992we}, where it was
highlighted that the emission of a bremsstrahlung photon from an accelerated charge
as described by an inertial observer can be interpreted in the comoving frame
as either the emission or the absorption of zero-energy Rindler photon in the Unruh
thermal bath. 
Recently, further attempts to measure out signatures of the Unruh effect 
have been  pursued by investigating the behavior of
accelerated atoms in micro-wave cavities~\cite{Scully:2003zz, Belyanin}, the backreaction radiation emitted by electrons in ultra-intense lasers~\cite{Chen:1998kp,Schutzhold:2006gj}, the thermal spectra  in high-energy collisions~\cite{Barshay:1977hc,Becattini:1995if, Becattini:1997rv,Kharzeev,Becattini:2001fg}, the Berry phase~\cite{MartinMartinez:2010sg} or some particular  generalization~\cite{Capolupo:2015rha}, up to the latest efforts within the framework of classical electrodynamics~\cite{Cozzella:2017ckb} and through the detection of  antiparticles in the Unruh radiation~\cite{Rosabal:2018hkx}. 
Proposals for experimental confirmations have also been put forward considering in-the-lab analogues of the Unruh effect~\cite{Unruh:1980cg}, or dealing with those experiments that successfully detected the Sokolov-Ternov effect~\cite{Akhmedov:2007xu}.
\begin{figure}[t]
\centering
\resizebox{15.3cm}{!}{\includegraphics{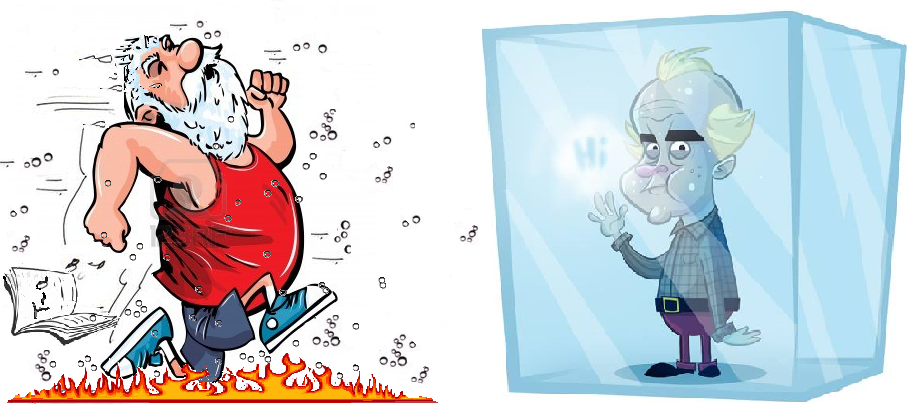}}
\caption{\small{A pictorial representation of the Unruh effect: Unruh (on the left)
is feeling warm in ``his'' bath due to the wind of particles at $T\sim\alpha$; by contrast, the inertial observer (on the right)
is frozen at $T=0\, \mathrm{K}$.
}}
\label{pictunruh}
\end{figure}

Unfortunately, all of these endeavors have been frustrated so far, thus leaving
the issue of the real existence of the Unruh effect still open. In view of this, 
we believe  the persisting state of confusion cannot but be interpreted in light 
of the ambiguity revolving around this phenomenon at the ontological level. 
In this sense, we propose one of the three following scenarios as possible way out 
of the controversy:
\begin{itemize}
\item the Unruh effect really \textbf{exists} and it is \textbf{observable}. The lack of experimental confirmations would stem from the fact that measuring the Unruh effect is a daunting task (to observe a temperature of $1\,\mathrm{K}$, an acceleration of $10^{20}\,\mathrm{m/s^2}$ would indeed be necessary); the creative proposals that have been continuously devising may provide new insights in the foreseeable future.

\item the Unruh effect is theoretically feasible, but, in practice, it is \textbf{not observable}, since it requires 
asymptotically Rindler boundary conditions that are  unattainable in any physical  situation~\cite{Narz} (such an interpretation, however, seems now to have run its course, as widely discussed in Refs.~\cite{Fulling:2004xy,Crispino:2007eb});

\item the question whether or not the Unruh effect is real/observable is physically unfounded (and, 
indeed, it is matter of debate among philosophers~\cite{Earman:2011zz}). All we can say (and that is enough!) is that the Unruh effect is an alternative \textbf{description} of ordinary aspects of QFT  in flat background in terms of a coordinate chart that is ``adapted'' to  accelerated observers (see the next Section for more details). In other terms, it is not really a novel phenomenon, but rather an unavoidable consequence of looking at known phenomena from a different perspective. In this sense, the claim that ``an accelerated observer immersed in the inertial vacuum will experience a thermal bath with a temperature proportional to the magnitude of his acceleration'' is nothing more than the evidence that ``an observer comoving with a rotating frame will undergo a centrifugal force depending on his radial acceleration'' (we remark that  the parallelism between the quantum Unruh effect and the classical centrifugal force was first used in Ref.~\cite{Matsas:1999jx3}).  Thus, the Unruh effect is absolutely essential for constructing a fully-fledged covariant QFT, even in the presence of interacting fields~\cite{Bisognano:1975ih}, and, as such, it does not require any further verification beyond those of QFT itself~\cite{Matsas:1999jx3, Cozzella:2017ckb}.
\end{itemize}

In line with a series of previous works~\cite{Matsas:2002ep}, we lean towards the last hypothesis, 
although we do not exclude \emph{a priori} the possibility that the Unruh radiation does indeed ``live'' 
among us. In what follows, we try to argue our position investigating the
question from both standard and novel perspectives. Before proceeding further, however,
we consider it appropriate to trace a brief \emph{historical excursus} on the Unruh effect, 
starting from its original derivation in QFT, going through some of the alternative approaches developed in literature, and finally concluding with a brief mention on its rotational analog
(the so-called Letaw-Pfautsch-Bell-Leinaas effect) and the ongoing controversies concerning its existence.

\section{QFT in Rindler spacetime}
\label{tuet}
To make the analysis as transparent as possible, consider the $1+1$-dimensional Minkowski spacetime with metric\footnote{A more general $1+3$-dimensional derivation
of the Unruh effect will be given in the next Chapter.}
\be
ds^2\,=\,dt^2\,-\,dx^2\,.
\label{twodimMin}
\ee
Upon the coordinate transformation
\begin{eqnarray}
\label{rindcoordrind}
\left\{\begin{array}{rcl}
t&=&\frac{1}{a}\,e^{a\xi}\sinh\,a\eta\,\\[2mm]
x&=&\frac{1}{a}\,e^{a\xi}\cosh\,a\eta\,
\end{array}\right.\,,
\label{rc}
\end{eqnarray}
with $-\infty<\eta,\xi<+\infty$, 
and $a$ positive constant, the metric Eq.~\eqref{twodimMin} becomes
\be
\label{metrica}
ds^2\,=\,e^{2\hspace{0.3mm}a\hspace{0.3mm}\xi}\,(d\eta^2\,-\,d\xi^2)\,.
\ee
The coordinates $(\eta$, $\xi)$ in Eq.~\eqref{rindcoordrind} are known as \emph{Rindler coordinates}~\cite{Rindler:1966zz}: they cover only the right (Rindler) wedge of Minkowski spacetime, namely $R_+=\{x|x>|t|\}$ (see Fig.~\ref{figure:RindlerNONTHERMAL}).
The left sector $R_-=\{x|x<-|t|\}$ can be obtained by reflecting the first
in the $t$- and then the $x$-axis. It is a trivial matter to verify that all curves of constant
$\eta$ are straight lines in the $(x,t)$ plane coming from the origin, while curves of constant position $\xi$ are hyperbolae,
\be
\label{trajeje}
x^2\,-\,t^2\,=\,\frac{1}{a^2}\,e^{2a\hspace{0.3mm}\xi}\,=\,\mathrm{const}\,.
\ee

\begin{figure}[t]
\centering
\resizebox{9.5cm}{!}{\includegraphics{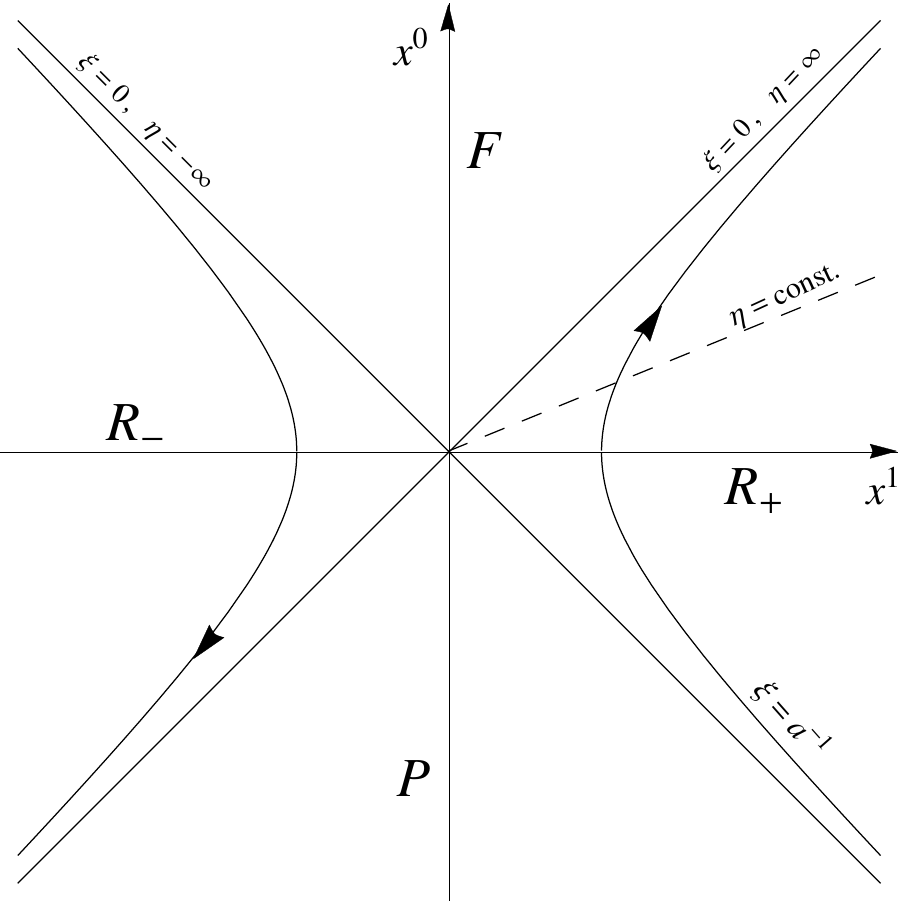}}
\caption{\small{The proper coordinate system of a uniformly accelerated observer in Minkowski spacetime.The hyperbola represents the world line of an observer with proper acceleration $\alpha$.
}}
\label{figure:RindlerNONTHERMAL}
\end{figure}

As it will be discussed in Section~\ref{Unruh} of the next Chapter, 
these represent the world lines of uniformly accelerated (Rindler)
observers with null asymptotes $x=\pm\,t$ (or $\eta=\pm\infty$) and
proper acceleration 
\be
\label{prac}
\alpha\,=\,a\hspace{0.3mm}e^{-a\hspace{0.3mm}\xi}\,.
\ee
The proper time of these observers is given by
\be
\label{prtime}
\tau\,=\,e^{a\hspace{0.2mm}\xi}\hspace{0.3mm}\eta\,.
\ee
We now comment on the non-trivial causal structure of 
the Rindler wedges: since uniformly accelerated observers
approach, but never cross, the asymptote $x=t$ ($x=-t$), 
this acts as future (past) event horizon. An accelerated observer 
in $R_+$ is thus causally disjoint 
from one in $R_-$, as events in such regions can only be 
connected by a somewhere spacelike line. For completeness, 
note also that events in the remaining future ($F$) and
past ($P$) quadrants can be connected to both $R_+$ and 
$R_-$ by null rays.

Aware of these caveats, let us quantize the Klein-Gordon
scalar field $\phi$ in both  Minkowski and Rindler supports.
To avoid unnecessary technicalities, in the next we consider the simplest treatment of
a real massless field; the extension to the massive case 
will be touched on in the next Chapter.

In $1+1$-dimensions, the Klein-Gordon wave equation in Minkowski
coordinates reads (i.e., Eq.~\eqref{eqn:esplicitKleinGordon} of Appendix~\ref{plane wave}
with $n=2$ and $m=0$)
\be\label{twodimMinspacetime}
\square\hspace{0.2mm}\phi
\,\equiv\,
\left(\frac{\partial^2}{\partial t^2}\,-\,\frac{\partial}{\partial x^2}\right)\hspace{-0.5mm}\phi
\,=\,0\,.
\ee
Orthonormal solutions are given by standard plane waves,
\be
\label{mo}
U_k(t,x)\,=\,{(4\pi\omega_k)}^{-\frac{1}{2}}\,e^{i(kx-\omega_kt)}\,,
\ee
which are of positive frequency $\omega_k=|k|$ with respect to the 
timelike Killing vector $\partial_t$. 

The question now arises as to how the above formalism gets modified
in the Rindler background.  To this aim, let us observe that, in the
regions $R_+$ and $R_-$, the metric Eq.~(\ref{metrica}) is conformal
to the Minkowski one, since it reduces to $d\eta^2-d\xi^2=0$ 
under the conformal transformation $g_{\mu\nu}\rightarrow e^{-2a\hspace{0.2mm}\xi}\,g_{\mu\nu}$. Furthermore, because the wave equation (\ref{twodimMinspacetime}) is 
conformally invariant, it can be written in Rindler coordinates as~\cite{Birrell}
\be
e^{2a\hspace{0.2mm}\xi}\,\square\hspace{0.2mm}\phi
\,=\,\left(\frac{\partial^2}{\partial \eta^2}\,-\,\frac{\partial}{\partial \xi^2}\right)\hspace{-0.5mm}\phi
\,=\,0\,,
\ee
which has orthonormal solutions of the same functional form as Eq.~\eqref{mo}, i.e.
\be
u_k(\eta,\xi)\,=\,{(4\pi\omega_k)}^{-\frac{1}{2}}\,e^{i(k\xi\,\pm\,\omega_k\eta)}.
\ee
The positive (negative) sign refers to the left (right) Rindler wedge;
roughly speaking, the sign reversal is due to the fact that the boost
Killing vector $\partial_\eta$ is future oriented in $R_+$, while it is
past oriented in $R_-$.

Now, consider the following sets:
\begin{equation}
\label{1}
{}^{R_{+}}u_k\,=\,\left\{\begin{array}{rcl}
&&\hspace{-8mm}{(4\pi\omega_k)}^{-\frac{1}{2}}\,e^{i(k\xi\,-\,\omega_k\eta)}\,,\quad \mathrm{in}\,\, R_{+} \\[2mm]
&&\hspace{-8mm}0\,,\hspace{32.3mm}\quad\mathrm{in}\,\, R_{-}
\end{array}\right.\,,
\ee
\begin{equation}
\label{2}
{}^{R_{-}}u_k\,=\,\left\{\begin{array}{rcl}
&&\hspace{-8mm}0\,,\hspace{32.3mm}
\quad \mathrm{in}\,\, R_{+} \\[2mm]
&&\hspace{-8mm}{(4\pi\omega_k)}^{-\frac{1}{2}}\,e^{i(k\xi\,-\,\omega_k\eta)}\,,\quad\mathrm{in}\,\, R_{-}
\end{array}\right.\,.
\ee
They are complete in the Rindler wedges $R_+$ and $R_-$, 
respectively, but not in the whole of Minkowski spacetime. By 
joining them, however, we obtain a set that is so complete.
Additionally, they can be analytically continued into the future 
and past regions (provided that $a$ becomes imaginary), 
since the lines are constant time $\eta$ global Cauchy surfaces~\cite{Boulware:1974dm}.
Thus, they represent a proper basis for  quantizing the field
in Minkowski spacetime, in the same way as the plane wave basis Eq.~\eqref{mo}.

Let us then expand the $\phi$ field in either set
\begin{equation}
\phi\,=\,\int d{k}\, \Big\{a_{{k}}\, \uuu_{{k}}\,+\, {a_{k}}^\dagger\, \uuu_{k}^{\hspace{0.3mm}*} \Big\},
\label{eqn:expansMINKO}
\end{equation}
or
\be
\label{eqn:expansRINDO}
\phi\,=\,\int d{k}\, \Big\{b^{(1)}_{{k}}\, {}^{R_{-}}u_k\,+\,b^{(1)\,\dagger}_{{k}}\, {}^{R_{-}}u^*_k\,+\,b^{(2)}_{{k}}\, {}^{R_{+}}u_k\,+\,b^{(2)\,\dagger}_{{k}}\, {}^{R_{+}}u^*_k\Big\}\,,
\ee
with both $a$ and $b$ satisfying canonical commutators,
\begin{eqnarray}
[a_k, a^\dagger_{k'}]&=&\delta(k-k'),\\[2mm]
[b_k^{(i)}, b^{(j)\,\dagger}_{k'}]&=&\delta_{ij}\,\delta(k-k'),
\end{eqnarray}
and all other commutators vanishing. 
 
The expansions Eqs.(\ref{eqn:expansMINKO})-(\ref{eqn:expansRINDO}) 
naturally lead to two alternative definitions of vacuum state, 
\begin{eqnarray}
\label{mnvac}
&a_k|0\rangle_{\mathrm{M}}\,=\,0\,,\qquad \forall k\,,&\\[2mm]
\label{rnvac}
&b^{(1)}_{{k}}|0\rangle_{\mathrm{R}}\,=\,b^{(2)}_{{k}}|0\rangle_{\mathrm{R}}\,=\,0\,,\qquad \forall k\,.&
\end{eqnarray}
The former is the vacuum for Minkowski (inertial) observers, since
it is defined so that there are no positive frequency quanta
with respect to the inertial time $t$; similarly, $|0\rangle_{\mathrm{R}}$ represents the 
vacuum for Rindler observers. There are several ways to prove that 
these two states are not equivalent~\cite{Unruh:1976db,Gibbons:1976pt,Lee:1985rp,Cande}:
following the original lines of thought~\cite{Unruh:1976db}, 
here we make use of the Bogoliubov transformation approach,
showing that  a non-trivial condensate structure of ``non-inertial'' particles is induced into
the vacuum $|0\rangle_{\mathrm{M}}$. 

In this connection, let us inspect
the analyticity properties of both sets in Eq.~(\ref{mo}) and Eqs.~(\ref{1})-(\ref{2}):
because of the sign reversal in the exponent at the crossover point between
$R_+$ and $R_-$ (i.e., the origin of the spacetime), the functions ${}^{R_{+}}u_k$ do not
go over smoothly to ${}^{R_{-}}u_k$ passing from $R_+$ to $R_-$ (and vice-versa). As a result,
${}^{R_{+}}u_k$ and ${}^{R_{-}}u_k$ are non-analytic at this point. 
By contrast, the plane waves $U_k$ are analytic in the whole
of Minkowski spacetime, and such a property remains true for any linear
superposition of these modes with positive frequency. It thus arises that the
functions Eqs.~(\ref{1})-(\ref{2}) cannot be a pure combination of
positive frequency plane waves, but they must be a mixture of these modes 
with both  positive 
and negative frequencies. Since the definition of vacuum excitations 
is intimately related to the one of positive frequency modes (with respect
to a given timelike Killing vector), we conclude that the 
vacua Eqs.~(\ref{mnvac}), (\ref{rnvac}) 
are not equivalent, namely the vacuum associated with one set
of modes contains particles with respect to the other. This is precisely
the same ambiguity in the definition of particle concept arising in the QFT
in curved spacetime~\cite{Birrell}. 

Although the modes ${}^{R_{\pm}}u_k$ are not a good choice in the sense described
above, one can check that the two normalized combinations~\cite{Unruh:1976db}
\begin{eqnarray}
\tilde u_k^{(1)}&=&{\left[2\sinh(\pi\omega_k/a)\right]}^{-1/2}\left[e^{\frac{\pi\omega_k}{2a}}\hspace{0.3mm}{}^{R_{+}}u_k\,+\,e^{-\frac{\pi\omega_k}{2a}}\hspace{0.3mm}{}^{R_{-}}u^*_{-k}\right],\\[2mm]
\tilde u_k^{(2)}&=&{\left[2\sinh(\pi\omega_k/a)\right]}^{-1/2}\left[e^{-\frac{\pi\omega_k}{2a}}\hspace{0.3mm}{}^{R_{+}}u^*_{-k}\,+\,e^{\frac{\pi\omega_k}{2a}}\hspace{0.3mm}{}^{R_{-}}u_{k}\right],
\end{eqnarray}
share the same analyticity properties, and thus the same vacuum  
$|0\rangle_{\mathrm{M}}$, of plane waves 
(for a proof of this, see, for example, Ref.~\cite{Birrell}).
To derive the particle content of Minkowski vacuum, 
let us then expand the field in terms of these new combinations:
\be
\phi\,=\,\int d{k}\, \Big\{d^{(1)}_{{k}}\,\tilde u_k^{(1)}\,+\,d^{(1)\,\dagger}_{{k}}\, \tilde u^{(1)\,*}_k\,+\,d^{(2)}_{{k}}\, \tilde u^{(2)}_k\,+\,d^{(2)\,\dagger}_{{k}}\,\tilde u^{(2)\,*}_k\Big\}\,,
\label{expcomb}
\ee
with
\be
\label{vacuumd}
d^{(1)}_{{k}}|0\rangle_{\mathrm{M}}\,=\,d^{(2)}_{{k}}|0\rangle_{\mathrm{M}}\,=\,0\,,\qquad \forall k\,,
\ee
according to our previous discussion. The Bogoliubov transformation
relating the $d$- and $b$- operators can be derived by equating the
expansion Eqs.~\eqref{eqn:expansRINDO}, \eqref{expcomb} on 
a spacelike surface at constant time $\eta$ and multiplying both sides first
by ${}^{R_{-}}u_k$ and then by ${}^{R_{+}}u_k$. Exploiting the
orthonormality property of  ${}^{R_{\pm}}u_k$, a straightforward
calculation leads to
\begin{eqnarray}
\label{bogounobogo}
b_k^{(1)}&=&{\left[2\sinh(\pi\omega_k/a)\right]}^{-1/2}\left[e^{\frac{\pi\omega_k}{2a}}d_k^{(2)}\,+\,e^{-\frac{\pi\omega_k}{2a}}\hspace{0.3mm}d^{(1)\,\dagger}_{-k}\right],\\[2mm]
\label{bogoduebogo}
b_k^{(2)}&=&{\left[2\sinh(\pi\omega_k/a)\right]}^{-1/2}\left[e^{\frac{\pi\omega_k}{2a}}d_k^{(1)}\,+\,e^{-\frac{\pi\omega_k}{2a}}\hspace{0.3mm}d^{(2)\,\dagger}_{-k}\right].
\end{eqnarray}
Hence, if the field is in the Minkowski vacuum $|0\rangle_{\mathrm{M}}$, the
spectrum of particles in mode $k$ detected by the
Rindler observer will be
\be
_{\mathrm{M}}\langle 0|\hspace{0.3mm}b_k^{(1,2)\,\dagger}\,b_k^{(1,2)}|0\rangle_{\mathrm{M}}
\,=\,
\frac{1}{e^{\frac{2\pi\omega}{a}}-1}\,,
\label{Usp}
\ee
that is the Bose-Einstein thermal distribution for radiation at 
temperature $T_0\,=\,a/2\pi$. Using the metric Eq.~\eqref{metrica} 
and the definition of proper acceleration, Eq.~\eqref{prac}, the temperature 
$T_{\rm{U}}$ as measured by the Rindler observer will be given by the
Tolman relation~\cite{Landau}
\be
\label{billTemperature}
T_{\rm U}\,=\,(g_{00})^{-\frac{1}{2}}\,T_0\,=\,\frac{\alpha}{2\pi}\,.
\ee
Therefore, from the point of view of a uniformly accelerated observer, 
the inertial vacuum appears as a thermal bath of particles
with temperature proportional to the proper acceleration\footnote{To 
address the Unruh effect in the presence
of flavor mixing, in the next Chapter we will exhibit 
an alternative derivation of the spectrum Eq.~(\ref{Usp}) for a \emph{massive field}, 
based on the diagonalization
of the Lorentz boost generator.}. 
This is, in short, the essence of the main result obtained by Unruh in
Ref.~\cite{Unruh:1976db}, and the temperature Eq.~\eqref{billTemperature} is usually 
referred to as the Unruh (or Fulling-Davies-Unruh~\cite{Fulling:1972md,davies}) temperature.

\section{Detector model}
\label{Detmod}
One might suspect that the foregoing result is merely a mathematical
curiosity, but physically nonsense, since the particle concept 
does not have universal significance in the absence of a 
globally timelike Killing vector. In this connection, however, 
Unruh pointed out that the response of a localized particle detector 
must be determined by the dependence of quantum fields on the detector's 
proper time, not the time of a global coordinate system~\cite{Unruh:1976db}. 
To argue this, he devised a detector model (later developed by
DeWitt~\cite{DeWitt}), showing that a uniformly accelerated device in the
Minkowski vacuum does indeed respond as though it were static, but immersed in a thermal bath 
with temperature $T=T_{\rm U}$.

In order to illustrate this, consider an idealized point-like detector 
with internal energy levels $E>E_0$, where $E_0$ is the energy of the
ground state, and coupled with a massless  
scalar field $\phi$ via a monopole interaction. Denoting by $x(\tau)$
the world line of the detector, the field-detector interaction is described by the Lagrangian
\be
\label{deL}
\mathcal{L}(\tau)=c\hspace{0.5mm}m(\tau)\hspace{0.3mm}\phi[x(\tau)]\,,
\ee
where $\tau$ is the detector's proper time,  $c$ a small coupling constant and 
$m$ the monopole moment operator.

Now, suppose the field $\phi$ is initially in the Minkowski vacuum $|0\rangle_{\mathrm{M}}$
given in Eq.~(\ref{mnvac}). Due to the interaction, however, both the field and the
detector will not remain in general in their ground states. Assuming the detector (field) 
undergoes the transition $E_0\rightarrow \widetilde E$ ($|0\rangle_{\mathrm{M}}\rightarrow|\psi\rangle$),
for sufficiently small couplings the amplitude will be given by first order perturbation 
theory as~\cite{Birrell}
\be
\label{transitiondetec}
\mathcal{A}_\psi(\widetilde E)\,=\,i\,\langle\widetilde{E}, \psi|
\int_{-\infty}^{\infty}d\tau\,\mathcal{L}(\tau)\,|0_M,E_0\rangle\,,
\ee
which can be factorized by writing the time-evolution of
$m(\tau)$ explicitly, i.e., $m(\tau)=e^{i\,H_0\tau}\,m(0)\,e^{-i\,H_0\tau}$, 
with $H_0|E\rangle=E|E\rangle$. A straightforward calculation then leads to
\be
\mathcal{A}_\psi(\widetilde E)
\,=\,i\hspace{0.5mm}c\,\langle \widetilde E|m(0)|E_0\rangle\int_{-\infty}^{\infty}d\tau
e^{i\left(\widetilde E-E_0\right)\tau}\langle\psi|\phi[x(\tau)]|0\rangle_{\mathrm{M}}\,.
\ee
The transition probability to \emph{all} possible excited states 
will be obtained by squaring
the modulus of $\mathcal{A}_\psi$, and summing over the
complete set of detector energy levels $E$ and field states $|\psi\rangle$. This gives
\be
\label{probaba}
\mathcal{P}\,=\,c^2 \sum_{E} {|\langle E|m(0)|E_0\rangle|}^2\,\mathcal{F}(E-E_0)\,,
\ee
where
\be
\label{resp}
\mathcal{F}(E)\,=\,\int_{-\infty}^{\infty}d\tau\int_{-\infty}^{\infty}d\tau' e^{-iE\left(\tau-\tau'\right)}
\hspace{0.3mm}D^{+}\hspace{-0.5mm}\left(x(\tau),x(\tau')\right),
\ee
with $D^{+}\hspace{-0.5mm}\left(x(\tau),x(\tau')\right)$ being
the positive frequency Wightman function,
\be
\label{pfwf}
D^{+}\hspace{-0.5mm}\left(x(\tau),x(\tau')\right)\,=\,{}_{\mathrm{M}}\langle0|\hspace{0.3mm}\phi\hspace{0.3mm}[x(\tau)]\hspace{0.3mm}\phi\hspace{0.3mm}[x(\tau')]\hspace{0.3mm} |0\rangle_{\mathrm{M}}\,.
\ee
The function $\mathcal{F}(E)$ in Eq.~(\ref{probaba}) is usually referred to as 
the \emph{detector response function}: it is clearly independent of
the details of the detector and, roughly speaking, represents
the bath of particles the device experiences during its motion. 
By contrast, the remaining term, which gives the \emph{selectivity}
of the detector to such a bath, is inherently related to the internal structure of the
device itself.

In situations where the detector is
in equilibrium with the field, i.e., when the whole system
is invariant under time translations in the comoving frame
($\tau\rightarrow\tau+\mathrm{const}$),
to avoid unpleasant divergences in 
the calculation of the response function, one typically considers a
coupling that switches off adiabatically as $\tau\rightarrow\pm\infty$, 
so that the integrals in Eq.~(\ref{resp}) are somehow regularized
(see, for example, Ref.~\cite{Birrell} for a more detailed discussion).
An alternative solution may be to deal with the excitation probability
per unit proper time, 
\be
\mathcal{P}/T\,=\,c^2 \sum_{E} {|\langle E|m(0)|E_0\rangle|}^2\,
\int_{-\infty}^{\infty}d(\Delta\tau)\,e^{-i(E-E_0)\Delta\tau}D^{+}(\Delta\tau)\,,
\label{traperunitpt}
\ee
rather than the probability Eq.~\eqref{probaba}, where $\Delta\tau=\tau-\tau'$
and $T$ is the total proper time. 

To finalize the calculation of the transition rate, 
details of the detector's trajectory are obviously needed.
Since we are interested in  the evaluation of the
response function of a uniformly accelerated 
detector, let us then consider the trajectory Eq.~\eqref{trajeje}, 
here rewritten as
\be
x\,=\,{(t^2+\alpha^{-2})}^{\frac{1}{2}}\,,
\ee
where $\alpha$ is the detector's proper acceleration (see Eq.~\eqref{prac}).
Exploiting the relation Eq.~\eqref{rc} between the coordinate time $t$ and the
detector's proper time $\tau$, the positive
Wightman function $D^{+}(\Delta\tau)$  takes the form
\be
D^{+}(\Delta\tau)\,\propto\,-\sum_{k=-\infty}^{\infty}{\left(\Delta\tau\,-\,2i\varepsilon\,+\,\frac{2\pi ik}{\alpha}\right)}^{-2}\,,
\ee
where the small imaginary part $i\varepsilon$, $\varepsilon>0$, arises from the evaluation of the Wightman function
on the appropriate integration contour~\cite{Birrell}.
Inserting into Eq.~\eqref{traperunitpt} and performing the 
Fourier transform, we finally obtain
\begin{equation}
\label{finaobt}
\mathcal{P}/T\,\propto\,c^2\sum_{E}\frac{\left(E-E_0\right){|\langle E|m(0)|E_0\rangle|}^2}{e^{\frac{2\pi\left(E\,-\,E_0\right)}{\alpha}}\,-\,1}\,.
\ee
The Planckian spectrum ${\left[e^{\frac{2\pi\left(E\,-\,E_0\right)}{\alpha}}\,-\,1\right]}^{-1}$
appearing in the transition probability thus shows that, at equilibrium, 
an accelerated detector in the vacuum $|0\rangle_{M}$ behaves as though
it were unaccelerated, but immersed in a thermal bath at the temperature 
$T\equiv T_{\mathrm{U}}=\frac{\alpha}{2\pi}$, in conformity with the result Eq.~\eqref{billTemperature}.

Undoubtedly, this marks a point in favor of the Unruh result against all the Unruh-skeptics.
Nevertheless, for those who are still rooted in the belief that 
the Unruh effect is merely a mathematical artifact, 
in the next Section we shall exhibit an alternative derivation of Eq.~\eqref{billTemperature} 
based on the time-dependent  \emph{Doppler effect}.

\section{The Unruh temperature via the time dependent Doppler effect}
Consider a  plane wave  massless mode of frequency $\omega_k$
and momentum $k$ parallel to the direction along
which the Rindler observer is accelerated (in our case, 
the $x$-direction), so that $\omega_k=k$~\cite{Alsing:2004ig}. In the reference frame comoving with the
observer, the frequency $\omega_k'$ transforms according to
\be
\label{Lortrans}
\omega'_k\,=\,\gamma\left[\omega_k\,-\,k\,v(\tau)\right]\,=\,\gamma\,\omega_k\left[1\,-\,v(\tau)\right]\,,
\ee
where $\gamma={\left[1-v(\tau)\right]}^{-\frac{1}{2}}$ is the Lorentz factor and $v(\tau)$ is the velocity of the accelerated observer as measured in the laboratory frame,
\be 
\label{vtau}
v(\tau)\,=\,\frac{\alpha\,t(\tau)}{\sqrt{1+\alpha^2\,t^2(\tau)}}\,.
\ee 
Using the first of Eqs.~\eqref{rc}, it follows that
\be
\label{velocityexpl}
v(\tau)\,=\,\tanh\alpha\tau\,,
\ee
leading to
\begin{equation}
\label{omtransf}
\omega'_k\,=\,\omega_k\,e^{-\alpha\tau}\,,
\end{equation}
(notice that, for $k$ antiparallel to the $x$-direction, 
one would simply obtain 
a change of sign in the exponent of the above equation).
For small values of $\alpha\tau$, 
one has
\be
\omega'_k\,\simeq\,\omega_k\hspace{0.4mm}(1\,-\,\alpha\tau)\,,
\ee 
which is the usual Doppler effect. Equation~\eqref{omtransf}
thus involves
a \emph{time-dependent} Doppler shift
for the Rindler observer. The time-dependent
phase is accordingly defined as
\be
\varphi(\tau)\,\equiv\,\int_{0}^{\tau}d\tau'\omega_k(\tau')\,=\,-\frac{\omega_k}{\alpha}\,e^{-\alpha\tau}\,.
\label{tdphaseshift}
\ee
Therefore, for a wave propagating along the $x$-direction, 
the power spectrum $S(\Omega)$ 
detected by the accelerated observer will be given by~\cite{Alsing:2004ig}
\be
\label{Ft}
S(\Omega)\,\propto\,{\bigg|\int_{-\infty}^{\infty}d\tau e^{i\Omega\tau}\,e^{i\varphi(\tau)}\bigg|}^2
\,=\,
\frac{2\pi}{\Omega\hspace{0.3mm}\alpha}\,\frac{1}{e^{\frac{2\pi\Omega}{\alpha}}\,-\,1}\,,
\ee
where we have used the formula~\cite{Grad}
\be
\label{intformula}
\int_{0}^{\infty}dx\,x^{\mu-1}e^{ib\,x}\,=\, \frac{\Gamma(\mu)}{b^\mu}\,e^{i\mu\pi/2}\,,
\ee
with $\Gamma$ being the Euler Gamma function.

Thus, because of  the time-dependent Doppler shift, 
the accelerated observer will detect a Planckian frequency spectrum, which is
indicative of a  Bose-Einstein distribution with $T=\alpha/2\pi$.

Strictly speaking,  Eq.~\eqref{Ft} has been derived
for a single plane wave frequency $\omega_k$; as shown in Ref.~\cite{Alsing:2004ig}, however, 
the result can be easily extended to the case of a field.
This confirms (once again) the validity of the Unruh effect, 
a difficult pill to swallow for Unruh's opponents!

\section{Unruh effect for interacting theories: a brief sketch}
On top of that, the Unruh effect has  also been derived within the 
framework of interacting field theories. A rigorous
treatment of such an extension requires notions
of axiomatic field theory, and thus goes beyond the scope of the present work.
In the following, we simply describe the essentials, redirecting the reader to the
references cited therein for a more comprehensive  analysis.

To begin with, let us briefly discuss the original works by
Bisognano and Wichmann~\cite{Bisognano:1975ih}, 
who proved that the Minkowski vacuum,
restricted to the right or
left Rindler wedges (Fig.~\ref{figure:RindlerNONTHERMAL}), behaves like a thermal 
Kubo-Martin-Schwinger (KMS) state when identifying the time-evolution operator
with a Lorentz boost, in line with 
Unruh's finding Eq.~\eqref{Usp}.
In this connection, it is worth recalling one of the possible, equivalent definitions
of the KMS condition: consider, for simplicity, a quantum
system with Hamiltonian ${H}$ and discrete (orthonormal) eigenstates 
$|n\rangle$ of energy $E_n$. As known, the thermal average  of
a generic operator ${A}$ in a state with inverse temperature 
$\beta=1/T$ can be expressed as (see the Thermo Field Dynamics (TFD)
formalism in Chapter~\ref{Unified formalism for Thermal Quantum Field Theories: a geometric viewpoint}
for more details)
\be
\label{medtherm}
\langle {A}\rangle_{\beta}\,=\,Z^{-1}(\beta)\sum_n e^{-\beta E_n}\,\langle n| {A}|n\rangle
\,=\, Z^{-1}(\beta)\, \mathrm{Tr}(e^{-\beta H}{A})\,,
\ee
where $Z(\beta)=\sum_n e^{-\beta E_n}=\mathrm{Tr}(e^{-\beta{H}})$ 
is the grand-canonical partition function and the trace is taken over the full
Hilbert space.

Note that the above thermal average can be realized as expectation value
on a pure state, provided that the degrees of 
freedom of the system are doubled.
In other terms, if $\mathcal{H}$
is the Hilbert space spanned by $|n\rangle$, we can construct
a pure state $|\beta\rangle$ in the ``doubled'' Hilbert space $\mathcal{H}\otimes\mathcal{H}$
such that
\be
\label{betabeta}
|\beta\rangle\,=\,Z^{-1/2}(\beta)\sum_n e^{-\beta E_n/2}\,|n,\tilde{n}\rangle\,,
\ee
and
\be
\label{easytocheck}
\langle {A}\rangle_{\beta}\,=\,\langle\beta|{A}^{(e)}|\beta\rangle\,,
\ee
where $\otimes$ denotes the tensor product, $|n,\tilde{n}\rangle=|n\rangle\otimes|\tilde{n}\rangle$, with
${H}|n\rangle=E_n|n\rangle$, ${\tilde{H}}|\tilde{n}\rangle=E_n|\tilde{n}\rangle$
and ${A}^{(e)}={I}\otimes{A}$, ${I}$ being
the identity operator. Of course, such a doubling is also reflected 
in the structure of the time-evolution operator, yielding
\be
\label{timevop}
e^{-i{H}^{(e)}\tau}\,=\,e^{i{H}^{(e)}\tau}\otimes e^{-i{\tilde{H}}^{(e)}\tau}\,.
\ee
Following Ref.~\cite{Crispino:2007eb}, we now define
an anti-unitary involution ${J}^{(e)}$ satisfying
\be
\label{auinvolution}
{J}^{(e)}\,=\,\alpha|n\rangle\otimes|m\rangle\,=\,\alpha^*|m\rangle\otimes|n\rangle\,,
\ee
where $\alpha\in\Bbb{C}$. Using this relation, one can prove that
${J}^{(e)}$ commutes with the time-evolution operator in 
Eq.~(\ref{timevop}). Furthermore, for any operator $A$ 
such that ${A}|n\rangle=\sum_{m}A_{mn}|m\rangle$, 
it follows that
\be
\label{KMS}
e^{-{H}^{(e)}\beta/2} {A}^{(e)}|\beta\rangle\,=\,{J}^{(e)}{{A}^{(e)\,\dagger}}|\beta\rangle\,,
\ee
where the state $|\beta\rangle$ must given (up to an irrelevant global phase factor) by Eq.~\eqref{betabeta}.

In algebraic field theory, Eq.~\eqref{KMS} provides the KMS thermal condition~\cite{Haag:1967sg}, and
the state $|\beta\rangle$ is a KMS state at inverse
temperature $\beta$. Thus, proving the Unruh effect within this framework
amounts to prove that the Minkowski vacuum restricted
to the right (left) Rindler wedge satisfies the KMS condition at 
$\beta=2\pi/\alpha$ with respect to the generator of Lorentz boosts, 
regarded as a generalized Hamiltonian in the Rindler coordinates.
Notice also that the rather artificial doubling of the degrees of freedom 
introduced above make intuitive sense in this context
as extension of the QFT from the right to the left wedge of Minkowski spacetime. 
The boost generator which allows for such an extension can be naturally
written in the form Eq.~\eqref{auinvolution}, for the corresponding
Killing vector field has opposite directions in the two wedges (see 
the discussion in Section~\ref{tuet}).

In $1+1$-dimensions, the involution $J^{(e)}$ is defined
through its action on the vacuum $|0\rangle_{\mathrm{M}}$
and on the ladder operators in Eq.~\eqref{eqn:expansRINDO}:
\begin{eqnarray}
&J^{(e)}|0\rangle_{\mathrm{M}}\,=\,|0\rangle_{\mathrm{M}}\,,&\\[2mm]
&J^{(e)}\hspace{0.2mm}b_k^{(1)}\hspace{0.2mm}J^{(e)}\,=\,b_k^{(2)}\,,&\\[2mm]
&J^{(e)}\hspace{0.2mm}b_k^{(1)\,\dagger}\hspace{0.2mm}J^{(e)}\,=\,b_k^{(2)\,\dagger}\,,&
\end{eqnarray}
with ${[J^{(e)}]}^2=I\otimes I$. Thus, in this context 
${J}^{(e)}$ is nothing more than the $PCT$ transformation $\phi(t,x)\rightarrow\phi(-t,-x)$
leading from $R_{+}$ to $R_{-}$. With these definitions, 
it is a trivial matter to check that the Bogoliubov transformations Eqs.~\eqref{bogounobogo}, \eqref{bogoduebogo} do indeed lead to the KMS condition Eq.~\eqref{KMS}.

We can now address the derivation of the Unruh effect
by Bisognano and Wichmann~\cite{Bisognano:1975ih} for a generic 
interacting scalar field
satisfying the Wightman axioms, the original formulation of which
can be found in Ref.~\cite{Wightman:1967jda}.
To this aim, let us observe that, denoting by $e^{i\hspace{0.2mm}K\hspace{0.05mm}a}$
the boost operator corresponding to the rotation 
\begin{eqnarray}
\label{eqa}
t\rightarrow t(a)&\equiv&t\cosh a\hspace{0.2mm}\alpha\,+\,x\sinh a\hspace{0.2mm}\alpha\,,\\[2mm]
x\rightarrow x(a)&\equiv&t\sinh a\hspace{0.2mm}\alpha\,+\,x\cosh a\hspace{0.2mm}\alpha\,,
\label{eqb}
\end{eqnarray}
in the imaginary 
Minkowski spacetime ($x, i\hspace{0.2mm}t)$, the transformation $(t,x)\rightarrow(-t,-x)$ can
be promptly obtained by setting $a=i\pi/\alpha$ (notice that, in our treatment, the r\^oles
of $a$ and $\alpha$ are reversed with respect to Ref.~\cite{Crispino:2007eb}). Starting from
these equations, Bisognano and Wichmann showed that the
states obtained by multiplying the vacuum $|0\rangle_{\mathrm{M}}$
by a finite number of operators of the form $\int d^2x\, f(x)\,\phi(x)$, 
with $f(x)$ being defined in the right Rindler wedge, 
belong to the domain of the operator $e^{-\beta K}$ for $0\le\beta\le\pi/\alpha$.
Roughly speaking, for the $1+1$-dimensional scalar field theory, 
this can be translated into the form
\be
\label{intotheform}
e^{-K\pi/\alpha}\,\phi\hspace{-0.5mm}\left(t^{(1)},x^{(1)}\right)\cdot\cdot\cdot\phi\hspace{-0.5mm}\left(t^{(n)},x^{(n)}\right)|0\rangle_{\mathrm{M}}
\,=\,\phi\hspace{-0.5mm}\left(-t^{(1)},-x^{(1)}\right)\cdot\cdot\cdot\phi\hspace{-0.5mm}\left(-t^{(n)},-x^{(n)}\right)|0\rangle_{\mathrm{M}}\,,
\ee
where the state $|0\rangle_{\mathrm{M}}$ is a
unique Poincar\'e invariant vacuum, which is assumed to exist,
if $(t^{(i)}, x^{(i)})$, $i=1,...,n$, are spatially separated points
in the right Rindler wedge of Minkowski spacetime. 
The above relation turns out to be crucial in showing the Unruh
effect, since it can be easily converted into the KMS condition Eq.~\eqref{KMS}
for $H^{(e)}=K$, $\beta=2\pi/\alpha$ and $J^{(e)}$ representing 
the $PCT$-transformation generator for operators $A^{(e)}$ acting
in the left wedge $R_{-}$. Thus, the whole demonstration boils down to prove 
Eq.~\eqref{intotheform}.

To make our analysis as transparent as possible, 
here we deal with the simplest case of
a free $1+1$-dimensional massless scalar field with $n=1$.
Exploiting the relations $K|0\rangle_{\mathrm{M}}=0$
and $a_k|0\rangle_{\mathrm{M}}=0$, with $a_k$ given in
Eq.~\eqref{eqn:expansMINKO}, it follows that
\begin{eqnarray}
\label{proof}
\non
\hspace{-5mm}e^{iKa}\phi({t,x})|0\rangle_{\mathrm{M}}&=&
\phi\left(t(a),x(a)\right)|0\rangle_{\mathrm{M}}\\[3mm]
&=&\hspace{-3mm}\int dk\,
{(4\pi\omega_k)}^{-\frac{1}{2}}\,e^{i(\omega_kx-kt)\sinh(\alpha a)}\,e^{i(\omega_kt-kx)\cosh(\alpha a)}a^\dagger_k|0\rangle_{\mathrm{M}},
\end{eqnarray}
where we have used the field expansion Eq.~\eqref{eqn:expansMINKO}
and the transformations Eqs.~\eqref{eqa} and~\eqref{eqb}, with
 $a$ being a real parameter. The above relation 
can also be analytically extended to imaginary values of $a$,
and in particular from $a=0$ to $a=i\pi/\alpha$,
if the point $(t,x)$ is in the right Rindler wedge. For such points,
we can then write
\be
e^{-K\pi/\alpha}\phi({t,x})|0\rangle_{\mathrm{M}}\,=\,
\phi\left(-t,-x\right)|0\rangle_{\mathrm{M}}\,,
\ee
which is exactly Eq.~\eqref{intotheform} with $n=1$. 
Finally, by inserting the field expansion Eq.~\eqref{eqn:expansRINDO}
in terms of Rindler modes, we infer the Bogoliubov transformations
relating Minkowski and Rindler ladder operators, respectively (see Section~\ref{tuet}),
and, thus, the Unruh effect Eq.~(\ref{Usp}).

\medskip
The Unruh effect can  also  be derived for interacting
theories with arbitrary potential $V(\phi)$ in the path
integral approach, for both scalars~\cite{Unruh1984} 
and spinors~\cite{Gibbons:1976pt,Unruh1984}.
In this context, it must be shown that
\be
\label{thcond}
{}_{\mathrm{M}}\langle 0|\, T\left[\phi(x)\phi(x')\right]|0\rangle_{\mathrm{M}}\,=\,
\frac{\mathrm{Tr}\left\{e^{-\beta K}\hspace{0.2mm}T\hspace{-0.2mm}\left[\phi(x)\phi(x')\right]\right\}}{\mathrm{Tr}\left(e^{-\beta K}\right)}\,,
\ee
where the trace is over all the states of the system, $K$
is the boost-operator defined above and $\beta=2\pi/\alpha$
(similar considerations apply to the case of a $n$-point
correlation function). To this aim, consider the Lagrangian density
for the massless scalar field in $1+1$-dimensions with potential $V(\phi)$,
\be
\label{lagdensscal}
\mathcal{L}\,=\,\frac{1}{2}\left(\frac{\partial^2\phi}{\partial t^2}\,-\,\frac{\partial^2\phi}{\partial x^2}\right)\,-\,V(\phi)\,.
\ee
Recasting the Rindler coordinates Eq.~\eqref{rindcoordrind} in the form
\be
\label{rindcoordcrisp}
t\,=\,\rho\sinh\alpha\tau,\qquad x\,=\,\rho\cosh\alpha\tau\,,
\ee
where $\rho=a^{-1}\,e^{a\xi}$ and $\alpha\tau=a\eta$, Eq.~\eqref{lagdensscal} 
becomes
\be
\label{lagnewfomrind}
\mathcal{L}\,=\,\alpha\rho\left[\frac{1}{2\alpha^2\rho^2}\,\frac{\partial^2\phi}{\partial \tau^2}\,-\,\frac{1}{2}\frac{\partial^2\phi}{\partial \rho^2}\right]\,-\,V(\phi)\,.
\ee
Let us now define the Euclidean action as
\be
\label{euclact}
S^R_E(\beta)\,\equiv\,-\int_{0}^{\beta}d\tau_e\int_{0}^{\infty}\d\rho\, \mathcal{L}_{\tau_e}\,,
\ee
where $\tau_e=i\tau$ is the Euclidean time 
and the field $\phi$ is $\beta$-periodic, i.e., $\phi(\tau_e+\beta,\rho)=\phi(\tau_e,\rho)$.
The condition Eq.~\eqref{thcond} can then be proved
by observing that both sides 
are obtained by analytically continuing  the same function $D_{2\pi/\alpha}\big((t_e, x_e), (t_e', x_e')\big)$
to imaginary values of time, where
\be
\label{Dbeta}
D_\beta\big((t_e, x_e), (t_e', x_e')\big)\,\equiv\,\frac{\int_{\phi(\tau_e=0)=\phi(\tau_e=\beta)}\left[D\phi\right]\phi(t_e, x_e)\,\phi(t_e', x_e')\,e^{-S^R_E(\beta)}}{\int_{\phi(\tau_e=0)=\phi(\tau_e=\beta)}\left[D\phi\right]\,e^{-S^R_E(\beta)}}\,,
\ee
with $t_e=\rho\sin \alpha\tau_e$ and $x_e=\rho\cos \alpha\tau_e$ (in particular, 
the r.h.s. of Eq.~\eqref{thcond} 
is obtained by the analytic continuation 
$\tau\rightarrow\tau=-i\tau_e$, while
the l.h.s. by  $t\rightarrow t=-it_e$).

In closing, we stress that the Bisognano-Wichmann result
has also been extended to the class of Schwarzschild and de Sitter
spacetimes~\cite{Sewell:1982zz}, showing in this way
the close connection between the Unruh and Hawking effects.

\section{Rotational analog of the Unruh effect}
In the framework of the whole discussion of this Chapter, the
other side of the coin is provided by a circularly moving 
observer.  As well known, rotational motion involves 
acceleration, albeit radially directed, which does not change the 
magnitude of the velocity. The question thus arises whether 
in this case there exists a corresponding Unruh effect and, accordingly,
an analog of the temperature Eq.~(\ref{billTemperature}).
Surprisingly, the state-of-the-art knowledge looks even more puzzling 
when dealing with such an issue; as explained in Ref.~\cite{Scholarpedia},
disputes mainly arise from the ambiguity in the notion of\hspace{1mm} ``circularly accelerated coordinate system'' and
in the definition of a globally timelike Killing vector for 
the rotating observer.
In this Section, going through the historical 
development of the debate, we try to shed
some light on the persisting controversies on the rotational analog
Unruh effect. For the following review,
we basically comply with Ref.~\cite{Crispino:2007eb}.

As a starting point, let us calculate the excitation rate of a
circularly moving Unruh-DeWitt detector coupled with a massless
scalar field, in analogy with what discussed 
in Section~\ref{Detmod}
for the linearly accelerated case. To this aim, consider 
a rotating device in the $xy$-plane at the radius $r=r_0$
with angular velocity
$\Omega\equiv\frac{d\theta}{dt}>0$. Here, $r$ and $\theta$ 
denote the usual polar coordinates.

Using Eqs.~\eqref{deL}, \eqref{transitiondetec}, 
the first order amplitude for the transition of the detector (field)
from the ground state $E_0$ ($|0\rangle_{\mathrm{M}}$) to an excited state $E$ ($|\psi\rangle$) 
read
\be
\mathcal{A}_\psi(E)\,=\,i\hspace{0.5mm}c\,\langle E|m(0)|E_0\rangle\int_{-\infty}^{\infty}d\tau
e^{i\left(\widetilde E-E_0\right)\tau}\langle\psi|\phi[x^{\mu}(\tau)]|0\rangle_{\mathrm{M}}\,,
\ee
yielding the transition probability per unit time, Eq.~\eqref{traperunitpt}.

Now, by writing the world line of the detector explicitly,
\be
\label{wle}
t\,=\,\gamma\tau,\quad x\,=\,r_0\hspace{0.2mm}\cos(\Omega\,t), \quad y=r_0\hspace{0.2mm}\sin(\Omega\,t)\,,
\ee
and setting $r_0\,\Omega<1$ so that the world line is timelike and the Lorenz factor
$\gamma={\left(1-r_0^2\,\Omega^2\right)}^{-\frac{1}{2}}$ is well-defined,
it is a trivial matter to check that~\cite{Crispino:2007eb}
\begin{eqnarray}
\label{trprobababia}
\non
\mathcal{P}/T&\propto&c^2 \sum_{E} {|\langle E|m(0)|E_0\rangle|}^2\,
\int_{-\infty}^{\infty}d(\Delta\tau)\,e^{-i(E-E_0)\Delta\tau}\\[2mm]
&&\times\,{\left[-\gamma^2{\left(\Delta\tau\,-\,i\varepsilon\right)}^2\,+\,4\,r_0^2\sin^2(\Omega\hspace{0.4mm}\gamma\hspace{0.4mm}\Delta\tau/2)\right]}^{-1}\,.
\end{eqnarray}
with $\varepsilon>0$. The excitation rate is thus \emph{non-vanishing}:
the above expression was numerically evaluated in Ref.~\cite{Letaw:1979wy}, 
and in Ref.~\cite{Bell:1982qr} it was first
used to interpret the depolarization 
of electrons in storage rings from
the point of view of comoving observers through the Unruh
effect. The latter work, together with Ref.~\cite{Bell:1986ir}, 
has played a key r\^ole in persuading a large part of physicists that the Unruh
thermal bath does indeed have real physical significance.

The question now arises as to how the transition probability
Eq.~\eqref{trprobababia} would appear for observers
corotating with the detector. As shown in Ref.~\cite{Crispino:2007eb},
however, some confusion occurs when trying to define the particle
concept within this framework, because
of the lack of a globally timelike Killing vector field (physically, 
the situation resembles the creation of particles by rotating black holes. 
The surface on which the Killing vector
 $K=\partial/\partial t+\Omega\,\partial/\partial \theta$
ceases to be timelike is indeed called \emph{ergosphere}).
If one does not take into 
account this fact and goes on defining naively the particle content
of the theory, the vacua for the inertial and rotating observer
would be identified, contradicting the previous result that
 detectors undergoing circular motion do have
a non-vanishing excitation rate. 
 In this case, therefore, quoting Ref.~\cite{Crispino:2007eb},
``\emph{either we have a suitable
way to extract the particle content from the theory or it may be
better not to introduce such a concept at all}''.

However, the situation becomes less puzzling when considering a 
detector  rotating with constant angular velocity $\Omega$
and enclosed in a limited region~\cite{levis,Daviesdetect}.
In that case, indeed, no ambiguity arises at all between 
the detector response function and its interpretation in terms
of the particle content defined by rotating observers.

To show this, consider a device inside a circle 
of radius $r$ such that
\be
\label{radius}
r\,=\,\rho\,<\,1/\Omega\,,
\ee
and impose the following Dirichlet boundary conditions on 
the massless scalar field:
\be
\phi\hspace{-0.2mm}\left(t, r=\rho, \theta\right)\,=\,0\,,
\ee 
(the discussion can be promptly extended to 
the case of a rotating detector confined inside a cylindrical surface, 
since the transverse extra-dimension does not affect the following calculation).
In this case, the positive frequency Wightman function Eq.~(\ref{pfwf}) takes
the form
\be
D^{+}(\Delta\tau)
\,=\,
\sum_{m=-\infty}^{\infty}\sum_{n=1}^{\infty}\,C_{mn}^2\,J^2_{m}(\alpha_{mn}\,r_0/\rho)\,e^{-i\left(\omega_{mn}-m\Omega\right)\gamma\Delta\tau}\,,
\label{pwowo}
\ee
where we have used the worldline of the rotating detector, Eq.~\eqref{wle}.
Here $m\in\Bbb{Z}$, $n\in\Bbb {N_{+}}$ are the parameters identifying
the positive frequency field modes with respect to inertial observers,
\be
\label{pfminert}
u_{mn}(t,r,\theta)\,=\,C_{mn}\,J_{m}(\alpha_{mn}r/\rho)\,e^{im\theta}\,e^{-i\omega_{mn}t}\,,
\ee
where $\alpha_{mn}$ is the $n$-th (non-vanishing) zero of the 
Bessel function of the first kind
$J_{m}$, $\omega_{mn}=\alpha_{mn}/\rho$, 
and  $C_{mn}$ is fixed
by the orthonormality condition of the modes 
$u_{mn}$.
A straightforward substitution of Eq.~\eqref{pwowo} into the 
definition Eq.~\eqref{traperunitpt} of transition rate per unit
proper time leads to
\begin{eqnarray}
\label{prdoro}
\non
\mathcal{P}/T&=&c^2 \sum_{E} {|\langle E|m(0)|E_0\rangle|}^2
\sum_{m=-\infty}^{\infty}\sum_{n=1}^{\infty}\,C_{mn}^2\,J^2_{m}(\alpha_{mn}\,r_0/\rho)\\[2mm]
&&\times\int_{-\infty}^{\infty}d(\Delta\tau)\,e^{-i\left[(E-E_0)\,+\,(\omega_{mn}-m\Omega)\hspace{0.2mm}\gamma\right]\Delta\tau}\,.
\end{eqnarray}
By integrating over $\Delta\tau$, we thus find that
the transition rate is proportional to
$\delta\big((E-E_0)\,+\,(\omega_{mn}-m\Omega)\hspace{0.2mm}\gamma\big)$.
Obviously, for $\Omega>0$, all modes with $m\le 0$ do not contribute to the sum in
Eq.~\eqref{prdoro}, being $E>E_0$. On the other hand, for $m>0$, 
the necessary condition to have a non-vanishing contribution
is that $m\Omega>\omega_{m1}=\alpha_{m1}/\rho$, 
where $\alpha_{m1}$ is the first non-trivial zero of the Bessel
function $J_{m}$. As shown in Ref.~\cite{Abram}, however,
such a condition can never be satisfied, since $\alpha_{mn}>m$,
which would imply $\Omega\hspace{0.2mm}\rho>1$, in contrast with
the original constraint Eq.~(\ref{radius}). Thus, unlike the previous case 
Eq.~\eqref{trprobababia}, the rotating
detector will exhibit a \emph{vanishing} response function when confined inside 
a boundary.

It is worthwhile to remark that the obtained result
has also an unambiguous interpretation in terms of
the particle content defined by the rotating observer.
According to what discussed above, indeed, it is now
possible to define a globally timelike Killing vector
for such an observer. For this purpose, let us recast the
positive frequency modes in
Eq.~\eqref{pfminert} into the form
\be
\label{pfminertnewform}
\widetilde{u}_{mn}(t,r,\theta')\,=\,C_{mn}\,J_{m}(\alpha_{mn}r/\rho)\,e^{im\theta'}\,e^{-i\widetilde{\omega}_{mn}t}\,,
\ee
where $\widetilde\omega_{mn}=\omega_{mn}-m\Omega>0$
for any  $m$ and $\theta'=\theta-\Omega\hspace{0.2mm}t$, 
with $t$ representing the proper time of a rotating 
observer with angular velocity $\Omega$ lying at $r = 0$.
Thus, these modes are of 
positive frequency even with respect to the corotating observer; hence, 
the Bogoliubov transformation connecting
the inertial and rotating modes, Eqs.~\eqref{pfminert} and~\eqref{pfminertnewform}, 
respectively, is trivial~\cite{Crispino:2007eb,Denardo:1978dj}, as no
mixing arises at all among positive- and negative-frequency solutions of the two sets.
By a calculation similar to that in Eq.~\eqref{Usp}, 
we can then conclude that the Minkowski vacuum is equivalent to
the vacuum defined by the confined rotating observer, in line with the vanishing
response rate found in Eq.~\eqref{prdoro}.

\medskip
In closing, we remark that the whole analysis of this Chapter can be
extended to arbitrary stationary accelerated frames~\cite{Letaw1981}, 
of which linearly and circularly accelerated ones are two extreme 
special cases. In general, there will appear both an event horizon and 
an ergosphere. Therefore, the ensuing temperature and  detector response 
will arise as combinations 
of the Unruh and Letaw-Pfautsch-Bell-Leinaas effects.

\section{Hawking radiation}
For the sake of comparison, in this Section we analyze
the Hawking effect for a massless scalar field in $1+1$-dimensional
spacetime\footnote{A more realistic $1+3$-dimensional analysis can be found in Ref.~\cite{Mukhanov}.}. Following Ref.~\cite{Mukhanov}, we show that calculations 
formally reproduce the ones leading to the Unruh result (see Section~\ref{tuet}).

Consider the gravitational field outside 
a non-rotating black hole of mass $M$ with null electric charge. In $1+3$-dimensions, 
this is described by the well-known Schwarzschild metric,
\be
\label{Scm}
ds^2\,=\,\left(1-\frac{2M}{r}\right)dt^2\,-\,\frac{dr^2}{1-\frac{2M}{r}}\,-\,r^2\,(d\theta^2+\sin^2\theta\,d\varphi^2).
\ee
To make calculations as simple as possible, let us focus
on a $1+1$-dimensional black hole, assuming the metric
to be equal to the time-radial part of the above expression. 
Outside the black hole (i.e., for $r>2M$), 
by introducing the ``tortoise coordinate'' $r^*(r)$ such that
\be
\label{tort}
dr^*\,=\,\frac{dr}{1\,-\,\frac{2M}{r}}\,\Longrightarrow\,r^*(r)\,=\,r\,-\,2M\,+\,2M\log\left(\frac{r}{2M}\,-\,1\right), 
\ee
with $-\infty<r^*<\infty$, the metric becomes
\be
\label{tortmet}
ds^2\,=\,\left(1\,-\,\frac{2M}{r(r^*)}\right)\left[dt^2-{dr^*}^2\right]\,.
\ee
This can be further manipulated by means of the tortoise
lightcone coordinates 
\be
\label{tortlight}
\tilde{u}\,\equiv\, t-r^*,\quad \tilde{v}\equiv t+r^*\,,
\ee 
obtaining
\be
\label{newscwmetric}
ds^2\,=\,\left(1\,-\,\frac{2M}{r(\tilde{u},\tilde{v})}\right)d\tilde{u}\,d\tilde{v}\,.
\ee
Both the Schwarzschild and lightcone coordinates introduced above 
are singular on the Schwarzschild horizon $r=2M$. Furthermore, 
the latter describe only the region external to the black hole, $r>2M$.
To cover the whole spacetime, let us then define the  
set of Kruskal-Szekeres lightcone coordinates as
\be
\label{Kruskal-Szekeres lightcone coordinates}
u\,=\,-4M\exp\left(-\frac{\tilde{u}}{4M}\right),\quad v\,=\,4M\exp\left(\frac{\tilde{v}}{4M}\right)\,,
\ee
which vary in the intervals $-\infty<u<0$ and $0<v<\infty$, respectively.
In terms of these coordinates, the metric Eq.~\eqref{newscwmetric} can be cast 
into the form
\be
\label{newnewmetric}
ds^2\,=\,\frac{2M}{r(u,v)}\exp\left(1\,-\,\frac{r(u,v)}{2M}\right)du\,dv\,,
\ee
which is regular at radius $r=2M$. Therefore, the singularity of the 
Schwarzschild metric at $r=2M$ is a mere \emph{coordinate singularity}.
The Kruskal-Szekeres coordinates can be analytically extended
to $u>0$ and $v<0$, where the metric Eq.~\eqref{newnewmetric}
is still well-defined. Thus, they span the entire Schwarzschild spacetime, providing
the proper set of coordinates we are looking for.

Using Eqs.~\eqref{tort} and \eqref{tortlight}, we can now 
derive the relations between the original Schwarzschild coordinates $(t,r)$
and the Kruskal-Szekeres lightcone coordinates $(u,v)$:
\begin{eqnarray}
\label{coordinatesa}
uv\,=\,-16M^2\left(\frac{r}{2M}\,-\,1\right)\exp\left(\frac{r}{2M}\,-\,1\right),\\[3mm]
\label{coordinatesb}
{\left(\frac{v}{u}\right)}^2\,=\,\exp\left(\frac{t}{M}\right).\hspace{21mm}
\end{eqnarray}
which are valid, via analytic continuation, for arbitrary values of 
$u$ and $v$.  From the first of these equations, it arises 
that the black hole horizon, $r=2M$, corresponds to $u=v=0$.
Therefore, the Schwarzschild spacetime has two horizons resolved only
in the Kruskal-Szekeres coordinates: the \emph{past horizon} for $v=0$ (corresponding
to $t\rightarrow-\infty$) and the \emph{future horizon} for $u=0$ (corresponding to $t\rightarrow\infty$).

In order to analyze the causal structure of 
the Schwarzschild spacetime, let us now
introduce the timelike and spacelike coordinates $T$
and $R$ according to
\be
\label{timspaccord}
u\,=\,T\,-\,R\,,\quad v\,=\,T\,+\,R\,.
\ee
The \emph{Kruskal diagram} for the 
Schwarzschild spacetime can be straightforwardly drawn
in the $(T,R)$ plane (see Ref.~\cite{Mukhanov} for more details).
Similarly to what discussed in Section~\ref{tuet}, null geodesics
$u=\mathrm{const}$ and $v=\mathrm{const}$ are
straight lines at $\pm\,\pi/4$ angles in this plane. 
From Eq.~\eqref{coordinatesa}, it follows 
that the hypersurfaces $r = \mathrm{const}$
are given by $uv=T^2-R^2=\mathrm{const}$.
For $r>2M$, one has $uv<0$ and the lines
$r=\mathrm{const}$ are timelike, while for $r<2M$, corresponding to
$uv>0$, they are spacelike. Thus, the 
Schwarzschild coordinate $r$ can be interpreted as the
standard spacelike radial coordinate only outside the horizon
$(r>2M)$.  Similarly, from Eq.~\eqref{coordinatesb}
the surfaces $t=\mathrm{const}$ are straight lines
in the $(T,R)$ plane. The Schwarzschild
coordinate $t$ is interpreted as the usual time for $r>2M$, 
but it becomes a spatial coordinate for $r<2M$. The hypersurface $r=0$
represents a physical singularity, being the curvature invariants
infinite on it.

\subsection{Field quantization and Hawking effect}
Consider a scalar field with the action
\be
\label{action}
S[\phi]\,=\,\frac{1}{2}\int g^{\mu\nu}\partial_\mu\phi\,\partial_\nu\phi\,\sqrt{-g}\,d^2x
\ee
in a $1+1$-dimensional spacetime. Exploiting the 
conformally invariance of the action, the solution of the
field equation can be written in both the lightcone
tortoise coordinates Eq.~\eqref{tortlight}, 
\be
\label{eqn:firstform}
\phi\,=\,\tilde A\left(\tilde{u}\right)\,+\,\tilde B(\tilde{v})\,,
\ee
and the lightcone Kruskal-Szekeres coordinates 
Eq.~\eqref{Kruskal-Szekeres lightcone coordinates},
\be
\label{eqn:secondform}
\phi\,=\, A\left({u}\right)\,+\,B({v})\,,
\ee
with $A$, $\tilde A$, $B$ and $\tilde B$ being suitable functions.

Similar to the scalar field quantization in the Rindler spacetime, 
the mode
\be
\label{rimov}
e^{-i\omega\tilde{u}}\,=\,e^{-i\omega(t-r^*)}
\ee
describes a right moving 
wave of positive frequency $\omega$ with respect to 
the time $t$, which propagates away from
the black hole.
Since the proper time of an observer at rest at infinity
coincides with $t$,\footnote{For $r\rightarrow\infty$, indeed, one has $ds^2\rightarrow d\tilde u\,d\tilde{v}=dt^2\,-\,{dr^*}^2$.}, he will associate the concept of particle 
with such a mode, expanding the field according to
\be
\label{firstfiexp}
\phi\,=\,\int_0^{\infty}\frac{d\Omega}{{(2\pi)}^{1/2}}\frac{1}{\sqrt{2\Omega}}\left[e^{-i\Omega\tilde u}\,b^-_\Omega\,+\,e^{i\Omega\tilde u}\,b^+_\Omega\right]\,+\,(\mathrm{left\,\,moving\,\,component})\,,
\ee
where $b^\pm_\Omega$ are the creation and annihilation operators
of a particle with frequency $\Omega$. The corresponding vacuum $|0\rangle_{\mathrm{B}}$
is defined as usual, 
\be
\label{Boulvac}
b^-_\Omega\,|0\rangle_{\mathrm{B}}\,=\,0\,,
\ee
and it is called \emph{Boulware vacuum}.

By pursuing the parallelism with the Rindler analysis,
let us recall that the tortoise coordinates cover only the exterior of the
Schwarzschild black hole. Thus, they play the same
r\^ole in  Schwarzschild spacetime as the Rindler coordinates
in Minkowski background. Moreover, the 
Boulware vacuum $|0\rangle_{\mathrm{B}}$ 
turn out to be similar to the Rindler vacuum defined in Eq.~\eqref{rnvac}.
On the other hand, the Kruskal-Szekeres coordinates 
are non-singular on the horizon and span
the entire Schwarzschild spacetime. In this sense, 
they are similar to the inertial
coordinates in Minkowski spacetime. In terms of 
the coordinates Eq.~\eqref{timspaccord}, the metric close to the horizon
becomes
\be
\label{metrinewfor}
ds^2\,\rightarrow\,du\,dv\,=\,dT^2\,-\,dR^2\,,
\ee
which means that particles detected by an observer
crossing the horizon are associated
with positive-frequency modes with respect to the time $T$.
The Kruskal ``vacuum'' $|0\rangle_{\mathrm{K}}$ for such an observer
can be defined by expanding the field operator in terms
of the Kruskal-Szekeres lightcone coordinates, 
\be
\label{fexpaKSco}
\phi\,=\,\int_0^{\infty}\frac{d\omega}{{(2\pi)}^{1/2}}\frac{1}{\sqrt{2\omega}}\left[e^{-i\omega u}\,a^-_\omega\,+\,e^{i\omega u}\,a^+_\omega\right]\,+\,(\mathrm{left\,\,moving\,\,component})\,.
\ee
It follows that
\be
a_\omega^-|0\rangle_{\mathrm{K}}\,=\,0\,.
\ee
For an observer at rest located far away from the black hole, 
however, this state is not empty. In order to determine its condensation 
density, one can exploit the following similarities between the
formulae underpinning the Unruh and Hawking effects:
\begin{longtable}{lp{0.35\columnwidth}}
\caption*{}
\label{tab:longtable} \\
\toprule
\quad\qquad\quad\quad\hspace{-2mm} Unruh effect & \quad\quad\hspace{0mm} Hawking effect \\
\midrule
\endfirsthead
\multicolumn{2}{l}{\footnotesize\itshape
\qquad\qquad\qquad\qquad\qquad\qquad\quad\quad\qquad\quad Continua dalla pagina precedente}  \\
\toprule
\endhead
\bottomrule
\multicolumn{2}{l}{\footnotesize\itshape Continued on next page} \\
\endfoot
\bottomrule
\multicolumn{2}{r}{\footnotesize\itshape} \\
\endlastfoot
\hspace{13mm}Minkowski vacuum $|0\rangle_{\mathrm{M}}$ & \quad\hspace{0mm} Kruskal vacuum $|0\rangle_{\mathrm{K}}$\\ 
\hspace{15mm}Rindler vacuum $|0\rangle_{\mathrm{R}}$ & \quad\hspace{-2mm} Boulware vacuum $|0\rangle_{\mathrm{B}}$\\ 
\hspace{20mm}Acceleration $\alpha$ & \quad\hspace{-6mm} Surface gravity $\kappa={(4M)}^{-1}$\\ 
\hspace{22mm} $u(\alpha)$, $v(\alpha)$ &\quad \hspace{8mm} $u(\kappa)$, $v(\kappa)$
\end{longtable}
Using the above similarities and Eq.~\eqref{Usp}, we find the following 
expression for the thermal condensate
perceived by the remote observer, 
\be
\label{Hawkcond}
_{\mathrm{K}}\langle 0|\hspace{0.3mm}b_\Omega^+\,b_\Omega^{-}|0\rangle_{\mathrm{K}}
\,=\,
\frac{1}{e^{\frac{2\pi\Omega}{\kappa}}-1}\,\delta(0)\,,
\ee
corresponding to the black-hole temperature
\be
T_{\mathrm{H}}\,=\,\frac{\kappa}{2\pi}\,\equiv\,\frac{1}{8\pi M}\,.
\ee

Finally, we notice that an alternative way to derive the Hawking effect
is by use of the renormalized stress-energy tensor $T_{\mu\nu}$
of a quantum field in classical black hole. This has been
done for conformal fields in $1+1$-dimensional spacetime~\cite{Frolov:1981mz}, 
where it has been shown that, along with a local vacuum polarization contribution,
the stress tensor also contains a non-local contribution from the
Hawking thermal flux.

\newpage\null\thispagestyle{empty}\newpage

\chapter{Non-thermal signature of the Unruh effect in field mixing}
\label{Non-thermal signature}

\begin{flushright}
\emph{``The faster you run,\\the hotter you feel.''}\\[1mm]
- Anonymous Aphorism -\\[6mm]
\end{flushright}

Since Pontecorvo's revolutionary idea~\cite{Pontecorvo}, the theoretical basis of flavor mixing has been widely investigated. Although years of effort have been devoted to providing evidence for flavor oscillations, intriguing questions still remain open. Among these, for instance, the origin of this phenomenon within the Standard Model and the non-trivial condensate structure exhibited by the vacuum for mixed fields are the most puzzling problems. The latter aspect, in particular, has burst into the spotlight after the unitary inequivalence between mass and flavor vacua within the QFT framework was highlighted~\cite{Blasone:1995zc,bosonmix}.

Flavor mixing in QFT is notoriously a non-trivial issue \cite{GiuntiKimLam}, since it is related with the problem of inequivalent representations of the canonical (anti-)commutation relations \cite{BJV}. The origin of this result lies in the fact that mixing transformations, which act as pure rotations on massive particle states in Quantum Mechanics (QM), have a more complicated structure at level of field operators. Indeed, they include both rotations and  Bogoliubov transformations \cite{Gargiulo}, thus inducing a condensate of particle/antiparticle pairs into the vacuum for flavor fields. This has been pointed out in Minkowski spacetime first for Dirac fermions~\cite{Blasone:1995zc} and later for other fields \cite{bosonmix,Blasone:2002jv,neutral}, showing in both cases the limitations of the quantum mechanical approach in the treatment of flavor mixing (for a more detailed discussion, see Appendix~\ref{QFT of fm}). In particular, whilst the QFT formulation of boson mixing has proved to be successful 
for studying the phenomenology of mixed mesons, like the $K^0-\bar K^{0}$, $B^0-\bar B^{0}$ or $\eta-\eta'$ systems \cite{Ji:2001yd}, the corresponding analysis for fermionic fields may shed some new light on the longstanding issue of the dynamical generation of quark and neutrino mixing in the Standard Model~\cite{Smaldone}. Furthermore, once extended to non-trivial metrics, it might provide a starting point for the investigation of neutrino flavor oscillations in the QFT on curved background. The existing literature, indeed, deals with such a topic using several other approaches, \emph{e.g.} the WKB approximation \cite{Stodolsky:1978ks, WudkaJ2001}, the plane wave method \cite{Kim, Konno} or geometric treatments \cite{Cardall,Zhang}, and settling in various metrics. 
The question thus arises as to how the  genuinely Minkowski  formalism of Refs. \cite{Blasone:1995zc, bosonmix} gets modified in the presence of gravity.

In the present Chapter, a first step along this direction is taken by analyzing the QFT of both boson (Section~\ref{Hyperbrepr} and following) and fermion (Section~\ref{Non-thermal Unruh radiation for flavor neutrinos}) mixed fields for a uniformly accelerated observer. To make our analysis as transparent as possible, we focus on a simplified description with only two flavors (analogous considerations, however, can be promptly generalized to a more rigorous three-flavor analysis). Despite such a minimal setting, a rich mathematical framework arises due to the combination of the Bogoliubov transformation associated with mixing and the one related to the Rindler spacetime structure \cite{Hawking:1974sw,Unruh:1976db,Birrell,Fulling,Takagi:1986kn,Iorio:2002rr}. As a consequence, we find that the Unruh spectrum loses its characteristic thermal identity, thus opening new stimulating investigation scenarios along this line~\cite{Scardaus,Blasone:2018czm,BlasonePOS}.

Besides formal aspects, we stress that mixing transformations in non-inertial frames may serve as a tool for analyzing a number of other current intriguing questions in theoretical physics: the spin-down of a rotating star by neutrino emission~\cite{Dvornikov:2009rk} and the disagreement between the inverse $\beta$-decay rates of accelerated protons in comoving and inertial frames \cite{Ahluwalia:2016wmf} are just some of the most relevant problems appearing in this framework. The latter aspect, in particular, shall be thoroughly discussed in Chapter~\ref{The necessity of the Unruh effect in QFT: the proton}, providing a possible solution for the aforementioned incompatibility.
In this connection, a further interesting question to be potentially investigated is the Lorentz invariance violation in the context of mixed neutrinos~\cite{Di Mauro}. So far, indeed, flavor mixing in QFT has been analyzed only within the usual plane wave representation. Exploiting the hyperbolic scheme discussed in Section \ref{Hyperbrepr}, one could theoretically explore whether oscillation formulae are indeed sensitive to the effects of a boost on the source (detector).

The Chapter is organized as follows: in the next Section, 
as a basis for extending mixing transformations to the Rindler 
frame, we introduce the hyperbolic field quantization, i.e., the scheme in which the generator of Lorentz boost is diagonal. 
The obtained results are thus compared with the usual ones in the plane wave basis, showing their equivalence from the point of view of inertial observers. In Section III we review the Rindler-Fulling quantization for an accelerated observer and the related Unruh effect. Mixing transformations within Rindler background are derived in Section IV: the modified spectrum of Unruh radiation is explicitly calculated in the limit of small mass difference, highlighting its non-thermal character when mixed fields are involved. Conclusions are discussed in Section V. 

Throughout all the Chapter, we shall use the following notation for $4$-, $3$- and $2$-vectors:
\begin{equation}
\label{eqn:notation}
x\,=\,\{t,\bx\},\qquad \bx\,=\,\{x^1,\vec{x}\},\qquad \vec{x}\,=\,\{x^2,x^3\}.
\end{equation}

\section{Field quantization in hyperbolic (boost) modes}
\label{Hyperbrepr}

Let us consider a free complex scalar field $\phi$ with mass $m$ in a $1+3$-dimensional Minkowski spacetime. In the standard plane wave representation, the field expansion reads (see Eq.~(\ref{eqn:modes}) of Appendix~\ref{plane wave} with $n=4$)
\begin{equation}
\phi(x)\,=\,\int d^{3}{k}\, \Big\{a_{\bk}\, \uuu_{\bk}(x)\,+\, {\bar a_\bk}^\dagger\, \uuu_\bk^{\hspace{0.3mm}*}(x) \Big\},
\label{eqn:expans0}
\end{equation}
where
\begin{equation}
U_\bk(x)\,=\,{\big[2\omega_{\bk}{(2\pi)}^{3}\big]}^{-\frac{1}{2}}\, e^{i\left(\bk\cdot\bx-\omega_{\bk} t\right)}.
\label{eqn:modes0}
\end{equation}
As well known, the field quanta created by applying the operators $a_{\bk}^\dagger$ ($\bar a_{\bk}^\dagger$) on the Minkowski vacuum $|0\rangle_{\mathrm{M}}$ carry well defined momentum $\bk$ and frequency $\omega_{\bk}={\sqrt{m^2+|\bk|^2}}$ with respect to the Minkowski time $t$ (see Appendix \ref{plane wave}). We will refer to these quanta as Minkowski particles (antiparticles), in contrast to the Rindler quanta to be later defined.

In order to extend the field quantization to the Rindler framework, we now introduce the so-called hyperbolic representation, that is, the representation which diagonalizes the Lorentz boost operator. To check this, let us look at the expression of the Lorentz-group generators:
\begin{equation}
M^{(\alpha,\beta)}\,=\,\int d^{3}x\, \left(x^{\alpha}\, T^{(0,\beta)}\,-\,x^{\beta}\, T^{(0,\alpha)} \right).
\label{eqn:gener}
\end{equation}
The boost operator (for example along the $x^1$ axis) is the $(1,0)$ component of $M^{(\alpha,\beta)}$. Using the standard expression of the stress tensor $T_{\mu\nu}$ and replacing the field expansion Eq.~\eqref{eqn:expans0}, we obtain \cite{Itzykson}
\begin{equation}
M^{(1, 0)}\,=\,i\int\frac{d^{3}k}{2\omega_\bk}\,\Big(c_\bk^{\, \dagger}\, \omega_\bk\, \frac{\partial}{\partial k_1}c_\bk\,+\,\bar c_\bk^{\, \dagger}\, \omega_\bk\, \frac{\partial}{\partial k_1}\bar c_\bk\Big),
\label{eqn:boostgenerator}
\end{equation}
where $c_\bk\equiv\sqrt{2\omega_\bk} \, a_\bk$. The result in Eq.~(\ref{eqn:boostgenerator}) shows that $M^{(1, 0)}$ has a non-diagonal structure in the plane wave representation. With a straightforward calculation, however, it can be verified that such a task is carried out by the following operators \cite{Takagi:1986kn}:
\begin{equation}
\label{eqn:operat-d}
d_{\kappa}^{\,(\sigma)}\,=\,\int^{+\infty}_{-\infty}\!\!\!\! dk_1\, p_\Omega^{\,(\sigma)}(k_1)\, a_{\bk},\qquad \bar d_{\kappa}^{\,(\sigma)}\,=\,\int^{+\infty}_{-\infty}\!\!\!\! dk_1\, p_\Omega^{\,(\sigma)}(k_1)\, \bar a_{\bk},
\end{equation}
where  $\kappa$ stands for $(\Omega, \vec{k})$, $\sigma=\pm\,1$, $\Omega$ is a positive parameter and\footnote{The physical meaning of the parameters $\sigma$ and $\Omega$ will be explained in the next Section, where the quantization procedure will be analyzed from the point of view of a uniformly accelerated observer.}
\begin{equation}
p_\Omega^{\,(\sigma)}(k_1)\,=\,\frac{1}{\sqrt{2\pi\omega_\bk}}\, {\bigg(\frac{\omega_{\bk}+k_1}{\omega_{\bk}-k_1}\bigg)}^{i\sigma\Omega/2}\, .
\label{eqn:p}
\end{equation}
In terms of these operators, indeed, the boost generator $M^{(1, 0)}$ takes the form
\begin{equation}
M^{(1, 0)}\,=\,\int d^3 \kappa \sum_{\sigma}\sigma\,\Omega\left(d_{\kappa}^{\,(\sigma)\dagger}\,d_{\kappa}^{\,(\sigma)}\,+\,\bar d_{\kappa}^{\,(\sigma)\dagger}\,\bar d_{\kappa}^{\,(\sigma)}\right),
\end{equation}
which is clearly diagonal. 
For later use, it is worth noting that the functions $p_\Omega^{\,(\sigma)}$ in Eq.~(\ref{eqn:p}) form a complete and orthonormal set, i.e.
\begin{eqnarray}
\label{eqn:completeorthonorm-p}
&&\hspace{7mm}\sum_{\sigma,\hspace{0.4mm}\Omega}\, p_\Omega^{\,(\sigma)}(k_1)\, p_\Omega^{\,(\sigma)*}(k_1')\,=\,\delta(k_1-k_1'),\\[2mm]
\label{eqn:compl-p}
&&\int_{-\infty}^{+\infty}\!\!\!\! dk_1\,\, p_{\Omega}^{\,(\sigma)*}(k_1)\, p_{\Omega'}^{ (\sigma')}(k_1)\,=\,\delta_{\sigma\sigma'}\delta(\Omega-\Omega'),
\end{eqnarray}
where the following shorthand notation has been introduced:
\begin{equation}
\sum_{\sigma,\hspace{0.4mm}\Omega}\,\equiv\,\sum_{\sigma}\int_{0}^{+\infty}\hspace{-2.0mm}d\Omega\hspace{0.2mm}.
\label{eqn:simpnot}
\end{equation}
Since the operators $d_{\kappa}^{\,(\sigma)}$ ($\bar d_{\kappa}^{\,(\sigma)}$) in Eq.~(\ref{eqn:operat-d}) are linear combinations of Minkowski annihilators  $a_{\bk}$ ($\bar a_{\bk}$) alone, they also annihilate the Minkowski vacuum $|0\rangle_{\mathrm{M}}$ in Eq.~(\ref{eqn:Minkvac}):
\begin{equation}
d_{\kappa}^{\,(\sigma)}\,|0\rangle_{\mathrm{M}}\,=\,\bar d_{\kappa}^{\,(\sigma)}\,|0\rangle_{\mathrm{M}}\,=\,0\hspace{0.2mm},\qquad \forall \sigma,\, \kappa.
\label{eqn:annihil-d}
\end{equation}
In addition, by exploiting Eqs.~(\ref{eqn:completeorthonorm-p}), (\ref{eqn:compl-p}) and the commutation relations of $a_{\bk}$ and $\bar a_{\bk}$ in Eq.~(\ref{eqn:commutrelationster}), it is immediate to verify that the transformations Eq.~(\ref{eqn:operat-d}) are canonical, i.e.
\bea
\label{eqn:commut-d}
\left[d_{\kappa}^{\,(\sigma)}\,,\, d_{\kappa'}^{\,(\sigma')\dagger}\Big]\,=\,\Big[\bar d_{\kappa}^{\,(\sigma)}\,,\, \bar d_{\kappa'}^{\,(\sigma')\dagger}\right]\,=\,\delta_{\sigma\sigma'}\, \delta^3(\kappa-\kappa'),
\eea
with all other commutators vanishing. Therefore, Eqs.~(\ref{eqn:annihil-d}) and (\ref{eqn:commut-d}) allow  to state that, for inertial observers, the hyperbolic and plane wave quantizations are indeed
equivalent at level of ladder operators.

The hyperbolic wave functions associated with the operators $d_{\kappa}^{\,(\sigma)}$  can be now derived by inverting Eq.~(\ref{eqn:operat-d}) with respect to $a_{\bk}$ and $\bar a_{\bk}$ and substituting the resulting expressions into the field expansion Eq.~\eqref{eqn:expans0}. It follows that
\begin{equation}
\phi(x)\,=\,\sum_{\sigma,\hspace{0.4mm}\Omega}\,\int d^{2}k\,\Big\{d_{\kappa}^{\,(\sigma)}\;\widetilde{U}_{\kappa}^{\,(\sigma)}(x)\, +\, \bar d_{\kappa}^{\,(\sigma)\dagger}\;\widetilde{U}_{\kappa}^{\,(\sigma)*}(x)\Big\},
\label{eqn:expansionfieldutilde}
\end{equation}
where
\begin{equation}
\widetilde{U}_{\kappa}^{\,(\sigma)}(x)\,=\,\int_{-\infty}^{+\infty}\!\!\!\! dk_1\, p_\Omega^{\,(\sigma)*}(k_1)\,  U_\bk(x)\hspace{0.2mm}.
\label{eqn:Uwidetilde}
\end{equation}
The integral Eq.~(\ref{eqn:Uwidetilde}) can be more directly solved by introducing the Rindler coordinates $(\eta,\xi)$, related to the Minkowski ones by the following expressions
\begin{equation}
\label{eqn:rindlercoordinatesNONTHERMAL}
t\,=\,\xi\sinh\eta\,,\qquad x^1\,=\,\xi\cosh\eta\,,
\end{equation}
with $-\infty<\eta,\xi<\infty$ (note that $x^2$ and $x^3$ are common to both sets of coordinates). We have\footnote{The set of coordinates $(\eta,\xi,\vec{x})$ in Eq.~(\ref{eqn:wideutilde}) is denoted by $x$, as well as the corresponding set of Minkowski coordinates $(t,x^1,\vec{x})$ in Eq.~(\ref{eqn:modes0}). Therefore, according to our convention, the symbol $x$ refers to a spacetime point, rather than to its representation in a particular coordinate system.}~\cite{Proceeding}
\begin{equation}
\label{eqn:wideutilde}
\widetilde{U}_{\kappa}^{\,(\sigma)}(x)\,=\,\frac{\,e^{\sigma\pi\Omega/2}}{2 \sqrt{2}\,\pi^{2}}
\,K_{i\sigma\Omega}(\mu_k\xi)\,e^{i\left(\vec{k}\cdot\vec{x}-\sigma\Omega\eta\right)},
\end{equation}
where $K_{i\sigma\Omega}(\mu_k\xi)$ is the modified Bessel function of second kind and $\mu_{k}$ is the reduced frequency\footnote{Note the difference 
with the case of a circularly moving observer, Eq.~\eqref{pfminert}, 
for which field
modes are expressed in terms of the Bessel functions 
of the first kind rather than the second.}
\be
\mu_{k}\,=\,\sqrt{m^2+|\vec{k}|^2},
\label{eqn:muk}
\ee
with $\vec{k}\equiv \{k^2,k^3\}$. 

It is not difficult to show that the hyperbolic modes in Eq.~(\ref{eqn:Uwidetilde}) form a complete and orthonormal set with respect to the KG inner product defined in Appendix~\ref{plane wave},  i.e.
\bea
\hspace{-1mm}\Big(\widetilde{U}_{\kappa'}^{\,(\sigma')}\, ,\widetilde{U}_{\kappa}^{\,(\sigma)}\Big)\,=\,-\Big(\widetilde{U}_{\kappa'}^{\,(\sigma')*}\, ,\widetilde{U}_{\kappa}^{\,(\sigma)*}\Big)\,=\,\delta_{\sigma\sigma'}\delta^3(\kappa-\kappa'), \quad \Big(\widetilde{U}_{\kappa'}^{\,(\sigma')}\, ,\widetilde{U}_{\kappa}^{\,(\sigma)*}\Big)\hspace{0.1mm}=\hspace{0.1mm}0\hspace{0.2mm}.
\label{eqn:ortonormbis}
\eea

Before turning to discuss the Rindler quantization, it should be emphasized that, although the plane wave expansion Eq.~(\ref{eqn:expans0}) applies to all the points of the spacetime, the hyperbolic representation Eq.~(\ref{eqn:expansionfieldutilde}) is valid only on the Rindler manifolds $x^1>|t|\hspace{0.4mm}\cup\hspace{0.4mm} x^1<-|t|$. By analytically continuing the solutions (\ref{eqn:wideutilde}) across $x_1=\pm\, t$, one obtains the correct global functions, i.e. the Gerlach's Minkowski Bessel modes (see Ref.~\cite{Gerlach}). For our purpose, nevertheless, it is enough to consider the modes as above defined.

\section{The Unruh effect via the hyperbolic field quantization}
\label{Unruh}
The hyperbolic representation discussed above provides a springboard for analyzing the Rindler-Fulling quantization in a uniformly accelerated frame \cite{Fulling}.  As a first step for such an extension, exploiting the Rindler coordinates $(\eta, \xi, x^2, x^3)$ in Eq.~(\ref{eqn:rindlercoordinatesNONTHERMAL}), let us rewrite the line element $ds^2\,=\,\eta_{\mu\nu}dx{^\mu} dx{^\nu}$ in the form\footnote{To be consistent with the notation of Ref.~\cite{Blasone:2017nbf}, in the following we shall use the Rindler coordinates defined as in Eq.~(\ref{eqn:rindlercoordinatesNONTHERMAL}) rather than Eq.~(\ref{rc}). Clearly, our final result will be not affected by such a choice.}
\begin{equation}
ds^2\,=\,{(dt)}^2-{(dx^1)}^2-\sum_{j=2}^{3}{(dx^j)}^2\underset{\tiny{Rindler\; coord.}}{\longrightarrow}ds^2\,=\,\xi^2d\eta^2-d\xi^2-\sum_{j=2}^{3}{(dx^j)}^2\,.
\label{eqn:lineelementsectionunruh}
\end{equation}
Since the metric does not depend on $\eta$, the vector $B=\frac{\partial }{\partial\eta}$ is a timelike Killing vector. Using Eq.~(\ref{eqn:rindlercoordinatesNONTHERMAL}), it is a trivial matter to verify that $B$ coincides with the boost Killing vector along the $x^1$ axis.

The physical relevance of Rindler coordinates can be readily explained by considering the following world line
\begin{equation}
\xi(\tau)\,=\,\mathrm{const}\equiv a^{-1}\,,\quad x^2(\tau)\,=\,\mathrm{const}\,,\quad x^3(\tau)\,=\,\mathrm{const},
\label{eqn:lineadiuni}
\end{equation}
where $\tau$ is the proper time measured along the line. By inserting Eq.~(\ref{eqn:lineadiuni}) into the metric  Eq.~(\ref{eqn:lineelementsectionunruh}), we find that
\begin{equation}
\eta(\tau)\,=\,a\tau.
\label{eqn:rindlertime}
\end{equation}
Therefore, the proper time $\tau$ of an observer along the line (\ref{eqn:lineadiuni}) is the same as the Rindler time $\eta$, up to the scale factor $a$. We will refer to such an observer as Rindler observer.

Equation \eqref{eqn:rindlertime} plays a striking r\^ole; according to the above discussion on the Killing vector $\frac{\partial }{\partial\eta}$, indeed, it shows that the time-evolution for the Rindler observer is properly an infinite succession of infinitesimal Minkowski boost transformations. This is the reason why in the first Section we deeply insisted on the hyperbolic field representation as opposed to the more familiar plane wave expansion.

In Minkowski coordinates $(t, x^1, x^2, x^3)$, the world line Eq.~(\ref{eqn:lineadiuni}) takes the form
\begin{equation}
t(\tau)\,=\,a^{-1}\sinh a\tau,\quad x^1(\tau)\,=\,a^{-1}\cosh a\tau,\quad x^2(\tau)\,=\,\mathrm{const},\quad x^3(\tau)\,=\,\mathrm{const}.
\label{eqn:lineauniversomink}
\end{equation}
Equation \eqref{eqn:lineauniversomink} describes an hyperbola in the $(t,x^1)$ plane with asymptotes $t=\pm\, x^1$ (see Fig.~\ref{figure:RindlerNONTHERMAL}). It is not difficult to check that it represents the world line of a uniformly accelerated observer with proper acceleration $|a|$ \cite{Mukhanov}. In particular, for $a>0$ the observer is confined within the right wedge $R_+=\{x|x^1>|t|\}$, while for $a<0$  his motion occurs in the left wedge $R_-=\{x|x^1<-|t|\}$. 
The non-trivial causal structure of Rindler spacetime has been discussed in the previous Chapter;
here we only stress that the wedges $R_{+}$ and $R_-$ represent two causally disjoint universes.

On this basis, we are now ready to describe the field-quantization procedure for the Rindler observer. To this aim, let us observe that the solutions of the Klein-Gordon equation in Rindler coordinates can be written as (see Appendix~\ref{plane wave}) 
\begin{equation}
u_\kappa^{\,(\sigma)}(x)\,=\,\theta(\sigma\xi)\!\ {\left[2\Omega{(2\pi)^{2}}\right]}^{-\frac{1}{2}}\!\ h_\kappa^{\,(\sigma)}(\xi)\!\ e^{i\left(\vec{k}\cdot\vec{x}-\sigma\Omega\eta\right)},
\label{eqn:rindlermodesbis}
\end{equation}
where $\sigma=\pm\,1$ refers to the right/left wedges $R_\pm$, $\Omega$ is the frequency with respect to the Rindler time $\eta$ and $h_\kappa^{\,(\sigma)}$ is the modified Bessel function of second kind, up to a normalization factor (see Eq.~(\ref{eqn:besselmodified})). The Heaviside step function $\theta(\sigma\xi)$ has been inserted into Eq.~(\ref{eqn:rindlermodesbis}) in order to restrict the Rindler modes  $u_\kappa^{\,(\sigma)}$ to only one of the two causally separated wedges $R_\pm$.

Exploiting the completeness and orthonormality properties of the set $\{u_\kappa^{\,(\sigma)},\,u_{\kappa}^{\,(\sigma)\,*}\}$, we can expand the field in the Rindler framework as follows
\begin{equation}
\phi(x)\,=\,\sum_{\sigma,\hspace{0.4mm}\Omega}\,\int d^{2}k\,\Big\{{b^{\,(\sigma)}_\kappa}\, u_\kappa^{\,(\sigma)}(x)\,+\,{\bar b^{\,(\sigma)\dagger}_\kappa}\, u_\kappa^{\,(\sigma)*}(x) \Big\},
\label{eqn:espanrind}
\end{equation}
where $\kappa\equiv(\Omega, \vec{k})$ as already defined. The ladder operators ${b^{\,(\sigma)}_\kappa}$ and ${\bar b^{\,(\sigma)}_\kappa}$ are assumed to obey the canonical commutation relations:
\begin{equation}
\left[b_\kappa^{\,(\sigma)}, b_{\kappa'}^{ ( \sigma')\dagger}\Big]\,=\,\Big[{\bar b_\kappa}^{\,(\sigma)},\, {\bar b_{\kappa'}}^{\,(\sigma')\dagger}\right]\,=\,\delta_{\sigma\sigma'}\,\delta^3(\kappa-\kappa')\hspace{0.2mm},
\label{eqn:commutcanon2}
\end{equation}
with all other commutators vanishing. They can be interpreted as annihilation operators of Rindler-Fulling particles and antiparticles, respectively.  The Rindler-Fulling vacuum, denoted with $|0\rangle_{\mathrm{R}}$, is accordingly defined by
\begin{equation}
 b_\kappa^{\,(\sigma)}|0\rangle_{\mathrm{R}}\,=\,\bar b_\kappa^{\,(\sigma)}|0\rangle_{\mathrm{R}}\,=\,0,\qquad \forall\sigma, \kappa.
\label{eqn:vuotodirindler}
\end{equation}
In order to figure out the connection between the Minkowski and Rindler quantizations, let us now compare the two alternative field expansions on a spacelike hypersurface $\Sigma$ lying in the Rindler manifolds $R_\pm$ (for instance, we may consider an hyperplane of constant $\eta$). Due to the equivalence of plane wave and hyperbolic representations within the Minkowski framework, we could equally consider the relations Eqs.~(\ref{eqn:expans0}) and (\ref{eqn:expansionfieldutilde}) for the inertial observer. To simplify the calculations, we opt for the latter. Therefore, by equating  Eqs.~(\ref{eqn:expansionfieldutilde}) and (\ref{eqn:espanrind}) on the hypersurface $\Sigma$ and forming the KG inner product of both sides with the Rindler mode $u_\kappa^{\,(\sigma)}$, we have
\begin{equation}
b_{\kappa}^{\,(\sigma)}\,=\,\sqrt{1+N_{BE}\hspace{0.5mm}(\Omega)}\!\ d_{\kappa}^{\, (\sigma)}\,+\,\sqrt{N_{BE}\hspace{0.5mm}(\Omega)}\,\bar d_{\tilde\kappa}^{\!\ (-\sigma)\dagger}\,,
\label{eqn:newformbogotransform}
\end{equation}
where $\tilde\kappa\equiv(\Omega,-\vec{k})$ and
\begin{equation}
N_{BE}(\Omega)\,=\,\frac{1}{e^{2\pi\Omega}-1}\,,
\label{eqn:distrib}
\end{equation}
is the Bose-Einstein distribution. We will refer to Eq.~(\ref{eqn:newformbogotransform}) as \emph {thermal Bogoliubov transformation}.

The spectrum of Rindler particles in the Minkowski vacuum can be readily calculated by exploiting Eqs.~(\ref{eqn:annihil-d}) and (\ref{eqn:newformbogotransform}). It follows that
\begin{equation}
{}_{\mathrm{M}}\langle0|b_{\kappa}^{\,(\sigma)\dagger}\!\ b_{\kappa'}^{ (\sigma)}|0\rangle_{\mathrm{M}}\,=\, N_{BE}(\Omega)\!\ \delta^3{(\kappa-\kappa')},
\label{eqn:aspectval}
\end{equation}
which is the Planckian spectrum for radiation at the temperature $T_0=1/2\pi$. The temperature
$T_{\rm U}$ as measured by the accelerated observer is given by the Tolman relation~\eqref{Tolman}
\be
\label{Tolman}
T_{\rm U}\,=\,(g_{00})^{-\frac{1}{2}}\,T_0\,=\,\frac{a}{2\pi}\,,
\ee
according to what we found in Eq.~\eqref{billTemperature}.

\section{Flavor mixing transformations for an accelerated observer}
\label{flmixrind}
In the last two decades, the QFT of flavor mixing has been widely investigated first for fermions~\cite{Blasone:1995zc} and then for bosons~\cite{bosonmix}, showing in both cases the presence of a non-trivial condensate structure in the vacuum for fields with definite flavor (see Appendix~\ref{QFT of fm} for details). In Refs.~\cite{Blasone:1995zc,bosonmix} these studies have been carried out in the usual Minkowski spacetime;  the question thus arises as to how such a formalism would appear within more challenging frameworks and, in particular, from the viewpoint of Rindler observer.

To address this issue, let us consider the mixing transformations for scalar fields in a simplified two flavor model:
\begin{equation}
\begin{array}{l}
\label{Pontecorvoa}
\phi_{A}(x)\,=\,\phi_{1}(x)\, \cos\theta  \,+\, \phi_{2}(x)\, \sin\theta ,\\ [4mm]
\phi_{B}(x) \,=\, -  \phi_{1}(x)\, \sin\theta   \,+\,  \phi_{2}(x)\, \cos\theta,
\end{array}
\end{equation}
where $\phi_i$ $(i=1,2)$ are two free complex scalar fields with mass $m_i$, $\phi_\chi$ $(\chi=A,B)$ are the mixed fields and $\theta$ is the mixing angle. These are nothing but the usual Pontecorvo transformations~\cite{Pontecorvo} extended to the context of QFT.

Now, following the approach of Section~\ref{Unruh}, we can quantize both fields $\phi_i$ $(i=1,2)$  by use of Rindler-Fulling expansion, Eq.~\eqref{eqn:espanrind}. Similarly, exploiting the completeness and orthonormality properties of Rindler modes, the following free-field like expansions can be adopted  for flavor fields:
\be
\label{eqn:expancampmixrind}
\phi_{\ell}(x)\,=\, \sum_{\sigma,\hspace{0.4mm}\Omega}\,\int d^{2}k\,\Big\{b_{\kappa,\ell}^{\,(\sigma)}(\eta)\,{u}_{\kappa,j}^{\,(\sigma)}(x)\,+\,\bar b_{\kappa,\ell}^{\,(\sigma)\dagger}(\eta)\,{u}_{\kappa,j}^{\,(\sigma)*}(x)\Big\}\,,
\ee
where $(\ell,j)=(A,1), (B,2)$. Here $b_{\kappa,\ell}^{\,(\sigma)}$, $\bar b_{\kappa,\ell}^{\,(\sigma)}$ and their respective adjoints are the flavor operators for the Rindler observer\footnote{To simplify the notation, from now on we shall omit the time-dependence of flavor operators when no misunderstanding arises.} (by comparison, see the expansions of flavor fields in Minkowski spacetime, Eq.~\eqref{eqn:dynamicalmap2}).

In Appendix~\ref{QFT of fm} it has been remarked that a non-trivial Bogoliubov transformation connecting  ladder operators for definite flavor and mass fields lurks inside the mixing relations in QFT (see Eq.~(\ref{eqn:esbos0})). By virtue of Eq.~(\ref{eqn:newformbogotransform}), we thus expect that 
flavor operators for the Rindler observer are related to the mass operators for the inertial (Minkowski) observer  through  the combination of two Bogoliubov transformations, the one arising from the Rindler spacetime structure, the other associated to flavor mixing, precisely.
To investigate more directly such an interplay, we then follow the same shortcut of Section~\ref{Unruh}, that is, we firstly quantize mixed fields in terms of the boost modes Eq.~\eqref{eqn:wideutilde}, obtaining (see Appendix~\ref{QMBFHM})
\be
\label{eqn:expancampmixbulk}
\phi_{\ell}(x)\,=\, \sum_{\sigma,\hspace{0.4mm}\Omega}\,\int d^{2} {k}\,\Big\{d_{\kappa,\ell}^{\,(\sigma)}\,\widetilde{U}_{\kappa,j}^{\,(\sigma)}(x)\,+\,\bar d_{\kappa,\ell}^{\,(\sigma)\dagger}\,\widetilde{U}_{\kappa,j}^{\,(\sigma)*}(x)\Big\}\,,
\end{equation}
where $(\ell,j)=(A,1),(B,2)$. The operators $d_{\kappa, \ell}^{\,(\sigma)}$, $\bar d_{\kappa, \ell}^{\,(\sigma)}$ are the flavor annihilators in the hyperbolic field representation; their explicit expression is given in Eq.~\eqref{eqn:daoperator}.

Consistently with the case of a free field, the transformation between the $b_{\kappa,\chi}^{\,(\sigma)}$- and $d_{\kappa,\chi}^{\,(\sigma)}$-operators can be  found by equating the expansions Eqs.~\eqref{eqn:expancampmixrind} and~\eqref{eqn:expancampmixbulk} on a spacelike hypersurface at constant time, and multiplying both sides by the Rindler mode $u_{\kappa,1}^{\,(\sigma)}$. A rather awkward calculation leads to~\cite{Blasone:2017nbf}
\be
b_{\kappa,\chi}^{\,(\sigma)}\,=\,\sqrt{1+N_{BE}\hspace{0.5mm}(\Omega)}\; d_{\kappa,\chi}^{\,(\sigma)}\,+\,\sqrt{N_{BE}\hspace{0.5mm}\Omega}\; \bar d_{\tilde\kappa,\chi}^{\, (-\sigma)\dagger}\,,\,,
\label{eqn:newformbogotransform2}
\ee
where $\chi=A,B$ and $N_{BE}(\Omega)$ is the Bose-Einstein distribution in Eq.~\eqref{eqn:distrib}.

As argued in Appendix~\ref{QMBFHM}, the Bogoliubov coefficients appearing in the operators
$d_{\kappa,\chi}^{\,(\sigma)}$ have a non-trivial integral expression we are able to
manage only for $t=\eta=0$ and in the limit of small mass difference $\frac{\Delta m^2}{m_i^2}\equiv \frac{m_2^2-m_1^2}{m_i^2}\ll 1$ (see Eqs.~(\ref{eqn:coefficientunoapprox}), (\ref{eqn:coefficientdueapprox})). In such a regime, combining Eqs.~(\ref{eqn:newformbogotransform2}), (\ref{eqn:coefficientunoapprox}) and (\ref{eqn:coefficientdueapprox}), the Unruh condensate of mixed particles is found to be (to the leading order)
\bea
\label{eqn:divergenceexpectvalue}
{\cal N}(\theta)\Big|_{t=0}&\equiv& {}_{\mathrm{M}}\langle0|\,b_{\kappa,\chi}^{\,(\sigma)\dagger}\,
b_{\kappa',\chi}^{\,(\sigma)}\,|0\rangle_{\mathrm{M}}\Big|_{t=0}\,=\,\cos^2\theta\,N_{BE}(\Omega)\,\delta^3{(\kappa-\kappa')}\\[2mm]
\non
&&\hspace{-20mm}+\,\,\sin^2\theta\Bigg[\sqrt{N_{BE}(\Omega)}\,\sqrt{N_{BE}(\Omega')}\,\densityAA\,+\,\sqrt{1+N_{BE}{(\Omega)}}\,\sqrt{1+N_{BE}(\Omega')}\,\densityBB\\[2mm]\nonumber
&&\hspace{-20mm}+\,\,\sqrt{1+N_{BE}(\Omega)}\,\sqrt{N_{BE}(\Omega')}\,\densityBA\,+\,\sqrt{N_{BE}{(\Omega)}}\,\sqrt{1+N_{BE}(\Omega')}\,\densityAB\Bigg]\hspace{0.1mm}\delta^2(\vec{k}-\vec{k}')\,,
\eea
to be compared with the standard Unruh distribution Eq.~(\ref{eqn:distrib}). The following notation has been adopted in Eq.~(\ref{eqn:divergenceexpectvalue}):
\begin{eqnarray}
&&\hspace{-8mm}\densityAA\,\equiv\,\sum_{\hspace{0.6mm}\sigma',\hspace{0.1mm}\Omega''}{{\cal A}_{(\Om,\Om''),\,\vec{k}}^{(\si,\si')\,*}}\;\,{{\cal A}_{(\Om',\Om''),\,\vec{k}'}^{(\si,\si')}}\,,\qquad N_{\mathcal{BB}}\,\equiv\,\sum_{\hspace{0.6mm}\sigma',\hspace{0.1mm}\Omega''}\Betasiomst\;\,\Betasiom\,,\\[4mm]
&&\hspace{-8mm}\densityAB\,\equiv\,\sum_{\hspace{0.6mm}\sigma',\hspace{0.1mm}\Omega''}{{\cal A}_{(\Om,\Om''),\,\vec{k}}^{(\si,\si')\,*}}\;\,{{\cal B}_{(\Om',\Om''),\,\vec{k}'}^{(\si,-\si')}}\,,\qquad \densityBA\,\equiv\,\sum_{\hspace{0.6mm}\sigma',\hspace{0.1mm}\Omega''}{{\cal B}_{(\Om,\Om''),\,\vec{k}}^{(\si,\si')\,*}}\;\,{{\cal A}_{(\Om',\Om''),\,\vec{k}'}^{(\si,-\si')}}\hspace{0.2mm}\,,
\end{eqnarray}
with ${{\cal A}_{(\Om,\Om'),\,\vec{k}}^{(\si,\si')}}$ and ${{\cal B}_{(\Om,\Om'),\,\vec{k}}^{(\si,\si')}}$ given in Eqs.~(\ref{eqn:coefficientunoapprox}) and (\ref{eqn:coefficientdueapprox}), respectively.

From Eq.~(\ref{eqn:divergenceexpectvalue}),  exploiting the properties Eqs.~(\ref{eqn:primprop}), (\ref{eqn:secprop}) and defining:
\begin{eqnarray}
\label{eqn:F}
\hspace{-4mm}F(\Omega,\Omega')&\equiv&\sqrt{N_{BE}(\Omega)\,N_{BE}(\Omega')}\,+\,\sqrt{\left(1+N_{BE}(\Omega)\right)\left(1+N_{BE}(\Omega')\right)}\,,\\[5mm]
\hspace{-4mm}G(\Omega,\Omega')&\equiv&\sqrt{1+N_{BE}(\Omega)}\,\sqrt{N_{BE}(\Omega')}\,+\,\sqrt{N_{BE}(\Omega)}\,\sqrt{1+N_{BE}(\Omega')}\,,
\label{eqn:G}
\end{eqnarray}
we finally obtain
\begin{equation}
\label{eqn:finalresult}
{\cal N}(\theta)\Big|_{t=0}\,=\,N_{BE}(\Omega)\,\delta^3(\kappa-\kappa')\,+\,\sin^2\theta\Big[F(\Omega,\Omega')\,\densityBB+G(\Omega,\Omega')\,\densityAB\Big]\hspace{0.1mm}\delta^2(\vec{k}-\vec{k}')
\end{equation}
that is consistent with the outcome obtained in Ref.~\cite{Blasone:2017hal} via an alternative approach.

Therefore, due to the combination of thermal and mixing Bogoliubov transformations,  the radiation detected by the Rindler observer  gets significantly modified in the presence of mixed fields, resulting in the sum of the standard Unruh spectrum plus non-diagonal corrections. Here we sketch 
the calculation of these contributions; to this aim, using Eqs.~(\ref{eqn:completeorthonorm-p}), (\ref{eqn:compl-p}), let us rewrite $N_{\mathcal{BB}}$ and $N_{\mathcal{AB}}$ in Eq.~(\ref{eqn:finalresult}) as
\begin{eqnarray}
\label{eqn:nbb}
N_{\mathcal{BB}}&=&-\frac{1}{2}\,\delta\,(\Omega-\Omega')\,+\,J^{(-\Omega,\Omega')}\bigg(\frac{1}{\omega_{\bk,2}}\bigg)\,+\, K^{(-\Omega,\Omega')} \bigg(\frac{\omega_{\bk,2}}{\omega_{\bk,1}^{\hspace{0.3mm}2}}\bigg)\,,\\[3mm]
\label{eqn:nab}
\hspace{12mm}N_{\mathcal{AB}}&=&J^{(-\Omega,-\Omega')}\bigg(\frac{1}{\omega_{\bk,2}}\bigg)\,-\, K^{(-\Omega,-\Omega')} \bigg(\frac{\omega_{\bk,2}}{\omega_{\bk,1}^{\hspace{0.3mm}2}}\bigg)\,,
\end{eqnarray}
where
\begin{eqnarray}
\label{eqn:J}
J^{(-\Omega,\Omega')}\Big(\frac{1}{\omega_{\bk,2}}\Big)&\equiv&\int^{+\infty}_{-\infty}\frac{dk_1}{8\pi\omega_{\bk,2}}\,{\lf(\frac{\omega_{\bk,1}+k_1}{\omega_{\bk,1}-k_1}\ri)}^{-i\sigma\Omega/2} {\lf(\frac{\omega_{\bk,1}+k_1}{\omega_{\bk,1}-k_1}\ri)}^{i\sigma\Omega'/2},\\[5mm]
\label{eqn:K}
K^{(-\Omega,\Omega')} \Big(\frac{\omega_{\bk,2}}{\omega_{\bk,1}^{\hspace{0.3mm}2}}\Big)&\equiv&\int^{+\infty}_{-\infty}\frac{dk_1}{8\pi}\,\frac{\omega_{\bk,2}}{\omega_{\bk,1}^{\hspace{0.3mm}2}}\,{\lf(\frac{\omega_{\bk,1}+k_1}{\omega_{\bk,1}-k_1}\ri)}^{-i\sigma\Omega/2} {\lf(\frac{\omega_{\bk,1}+k_1}{\omega_{\bk,1}-k_1}\ri)}^{i\sigma\Omega'/2}.
\end{eqnarray}
For $\frac{\Delta m^2}{m_i^2}\equiv \frac{m_2^2-m_1^2}{m_i^2}\ll 1$, by expanding Eqs.~(\ref{eqn:j}), (\ref{eqn:K}) in Taylor series, it is a trivial matter to show that (to the leading order)~\cite{Blasone:2017nbf}
\begin{eqnarray}
\label{eqn:Nbbaprox}\non
N_{\mathcal{BB}}&\approx&\frac{1}{4}\,\int^{+\infty}_{-\infty}\frac{dk_1}{8\pi}\;\frac{{(\Delta m^2)}^2}{\omega_{\bk,1}^{\hspace{0.3mm}5}}\,{\lf(\frac{\omega_{\bk,1}+k_1}{\omega_{\bk,1}-k_1}\ri)}^{-i\sigma\Omega/2} {\lf(\frac{\omega_{\bk,1}+k_1}{\omega_{\bk,1}-k_1}\ri)}^{i\sigma\Omega'/2}\\[4mm]
&=&\frac{1}{192}\;\frac{{(\Delta m^2)}^2}{\mu_{k,1}^{\hspace{0.3mm}4}}\;\frac{\sigma\,(\Omega'-\Omega)}{\sinh\left[\frac{\pi}{2}\,\sigma\,(\Omega'-\Omega)\right]}\left[4\,+\,{(\Omega'-\Omega)}^2\right].
\end{eqnarray}
Similarly, for $N_{\mathcal{AB}}$ we have
\begin{eqnarray}
\label{eqn:Nabaprox}
\non
N_{\mathcal{AB}}&\hspace{-2mm}\approx\hspace{-2mm}&\int^{+\infty}_{-\infty}\frac{dk_1}{8\pi}\left(-\,\frac{\Delta m^2}{\omega_{\bk,1}^{\hspace{0.3mm}3}}\,+\,\frac{1}{2}\frac{{(\Delta m^2)}^2}{\omega_{\bk,1}^{\hspace{0.3mm}5}}\right)\,{\lf(\frac{\omega_{\bk,1}+k_1}{\omega_{\bk,1}-k_1}\ri)}^{-i\sigma\Omega/2} {\lf(\frac{\omega_{\bk,1}+k_1}{\omega_{\bk,1}-k_1}\ri)}^{-i\sigma\Omega'/2}
\\[5mm]
\nonumber
&&\hspace{-15mm}=\;-\,\frac{1}{8}\,\frac{\Delta m^2}{\mu_{k,1}^{\hspace{0.3mm}2}}\,\frac{\sigma\,(\Omega'+\Omega)}{\sinh\left[\frac{\pi}{2}\;\sigma\,(\Omega'+\Omega)\right]}\,+\,\frac{1}{96}\;\frac{{(\Delta m^2)}^2}{\mu_{k,1}^{\hspace{0.3mm}4}}\;\frac{\sigma\,(\Omega'+\Omega)}{\sinh\left[\frac{\pi}{2}\,\sigma\,(\Omega'+\Omega)\right]}\left[4+{(\Omega'+\Omega)}^2\right],
\\
\end{eqnarray}
where $\mu_{k,1}$ has been defined in Eq.~\eqref{eqn:muk}. Inserting Eqs.~(\ref{eqn:Nbbaprox}), (\ref{eqn:Nabaprox}) into~(\ref{eqn:finalresult}), we definitively obtain
\begin{eqnarray}
\label{eqn:spectrumapprox}
\non
{\cal N}(\theta)\Big|_{t=0}&\approx&N_{BE}(\Omega)\,\delta^3(\kappa-\kappa')\,-\,\sin^2\theta\,\Bigg\{\frac{\Delta m^2}{8\,\mu_{k,1}^{\hspace{0.3mm}2}}\;G(\Omega,\Omega')\;\frac{\sigma\,(\Omega'+\Omega)}{\sinh\left[\frac{\pi}{2}\,\sigma\,(\Omega'+\Omega)\right]}\\[4mm]
\nonumber
&&+\;\,\frac{{(\Delta m^2)}^2}{96\,\mu_{k,1}^{\hspace{0.3mm}4}}\,\bigg[\frac{F(\Omega,\Omega')}{2}\;\frac{\sigma\,(\Omega'-\Omega)}{\sinh\left[\frac{\pi}{2}\,\sigma\,(\Omega'-\Omega)\right]}\left[4\,+\,{(\Omega'-\Omega)}^2\right]\\[5mm]
&&+\;\,G(\Omega,\Omega')\;\frac{\sigma\,(\Omega'+\Omega)}{\sinh\left[\frac{\pi}{2}\,\sigma\,(\Omega'+\Omega)\right]}\left[4\,+\,{(\Omega'+\Omega)}^2\right]\bigg]\,.
\end{eqnarray}
Some comments are in order here. First, for $\theta\rightarrow 0$, Eq.~(\ref{eqn:spectrumapprox}) correctly reproduces the thermal distribution Eq.~(\ref{eqn:aspectval}), as it must be in the absence of mixing. Similar considerations hold for $m_1\rightarrow m_2$ and in the relativistic limit of large momenta, $|\vec{k}|\gg \sqrt{m_{1}^2+m_{2}^2}$, being $\frac{\Delta m^2}{\mu^2_{k,i}}\rightarrow 0$ in this regime.
Next, the total number of mixed particles with frequency $\Omega$ and $2$-momentum $\vec{k}$ can be obtained by integrating Eq.~(\ref{eqn:spectrumapprox}) over $\kappa'$. It is not difficult to verify that the higher the frequency, the more relevant the contribution of mixing corrections becomes.
Moreover, we emphasize that, although the characteristic Unruh distribution Eq.~(\ref{eqn:distrib}) does not carry the information about the mass of the field, the modified spectrum Eq.~(\ref{eqn:spectrumapprox}) explicitly depends on the squared mass difference of mixed fields. As a consequence, flavor mixing breaks the mass scale invariance of Unruh effect, giving rise to a non-thermal vacuum radiation. Of course, the extension of such a result to neutrinos can be potentially exploited to fix new constraints on the squared mass difference of these fields (see the next Section).
Last, as it can be seen from Eq.~\eqref{eqn:divergenceexpectvalue}, the condensation densities associated with flavor mixing and acceleration combine symmetrically in the evaluation of the (mixed) particle spectrum for the Rindler observer; since both of them mathematically arise from Bogoliubov transformations, the one associated with field mixing, the other with the particular structure of Rindler background, 
one finds it natural to describe mixing as an effect induced by the particular geometry of the spacetime in which we ``live''. Clearly, this is just our preliminary guess; more work is undoubtedly needed along this direction.

In closing, relying on the Hawking theory about black hole evaporation, we provide an heuristic interpretation for the obtained result:  in the absence of mixing, according to what we shown, inertial vacuum appears to the Rindler observer as a condensate of virtual particle/antiparticle pairs all of the same type. Typically, these pairs constantly pop into existence for an extremely short time and then annihilate each other. Nevertheless, close to the event horizon, it may happen that a particle loses its partner into the ``invisible'' region: unable to be annihilated, the virtual particle becomes ``real'', and then escapes as Unruh radiation. In other terms, for free unmixed fields, we can state that the thermal bath detected by the Rindler observer arises from the corresponding flux of \emph{one-type} antiparticles falling back into the horizon.

The above picture turns out to be deeply modified when flavor mixing is taken into account. In this context, indeed, vacuum is crowded with particle/antiparticle pairs of both the same and different flavors (see Appendix~\ref{QFT of fm} for details). Thus, Unruh radiation may be generated by \emph{both types} of antiparticles crossing the horizon. In other terms, if a $B$-flavor particle escapes, it could correspond to a $B$-flavor antiparticle fallen back into the horizon, as well as to an $A$-flavor antiparticle (see Fig.~\ref{figure:unruhmix}). As a consequence, the entropy of the system will be significantly increased, thus spoiling the original thermal spectrum.

\begin{figure}[t]
\centering
\resizebox{10cm}{!}{\includegraphics{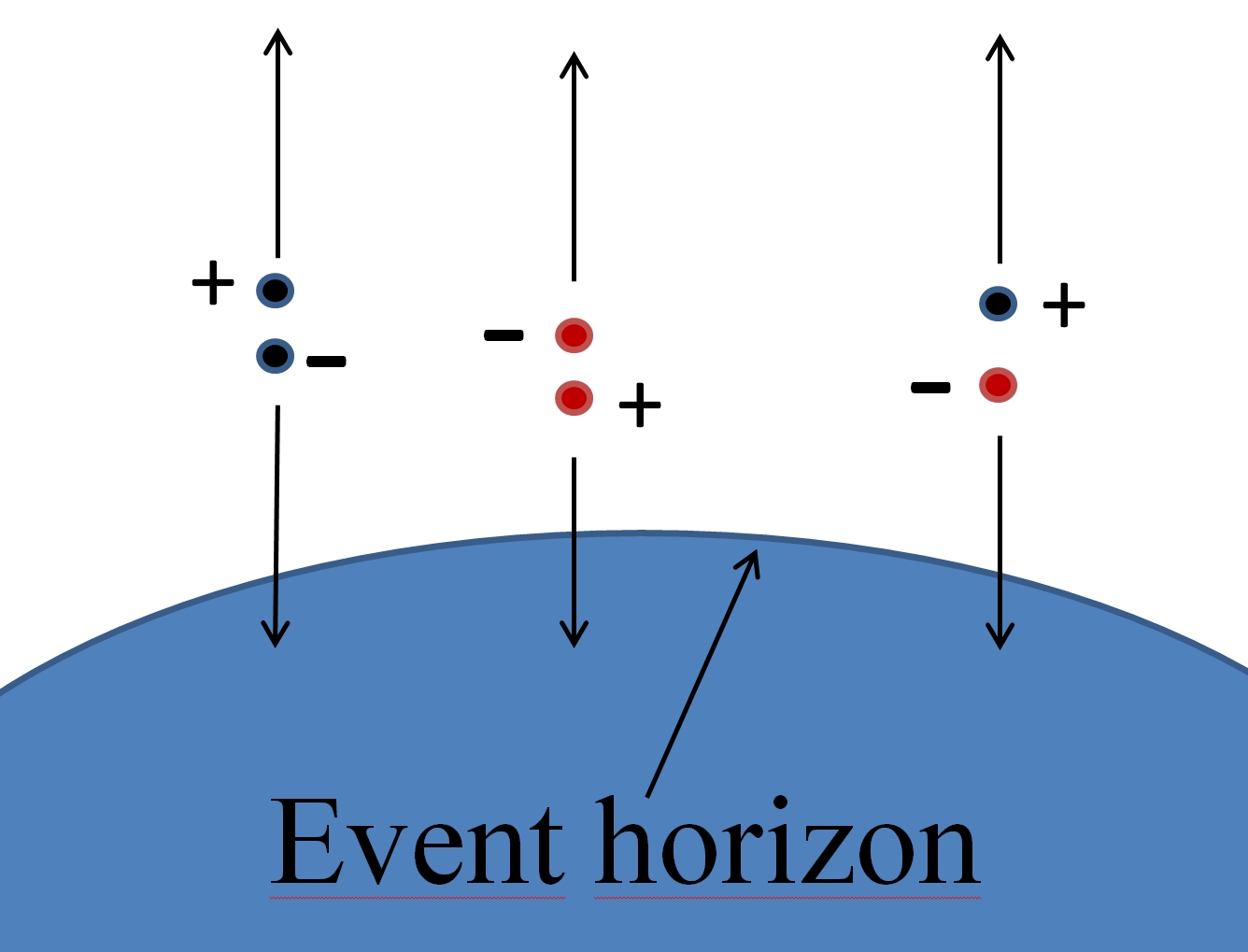}}
\caption{\small{Pictorial interpretation of the modified spectrum of mixed particles in  $|0\rangle_{\mathrm{M}}$ for a Rindler observer. Different dot colors correspond to different particle/antiparticle flavors. By contrast with the case free (unmixed) fields, vacuum is populated by particle/antiparticle pairs, both of the same (blue-blue, red-red) and different (blue-red) type.} }
\label{figure:unruhmix}
\end{figure}

\section{Unruh effect for mixed neutrinos}
\label{Non-thermal Unruh radiation for flavor neutrinos}

So far, the field theoretical treatment of flavor mixing for an accelerated observer has been analyzed for boson (scalar) fields; as discussed at the beginning of this Chapter, however, due to the largest number of possible applications, the situation becomes more and more appealing when dealing with Dirac fields, and in particular with neutrinos.
This is indeed the goal of the present Section: starting from the previous outcome, in what follows we investigate neutrino mixing transformations for a uniformly accelerated observer. In line with  Section~\ref{flmixrind}, we find that flavor mixing also spoils the thermality of Unruh radiation in this context, proving that the result achieved for bosons is actually independent of the spin structure. Besides its formal meaning, we show how the naive analysis developed here may be exploited in the future to fix more stringent constraints on the neutrino squared mass difference.

The remainder of this Section is organized as follows: in Section~\ref{Minkspace} we briefly review the second quantization of the free (Dirac) spinor field in Minkowski spacetime. 
 Section~\ref{Field quantization in Rindler spacetime: Unruh effect} is devoted to the discussion of the field quantization in Rindler spacetime (Rindler spacetime) and the related Unruh effect. The QFT of two mixed neutrinos is analyzed in Section~\ref{QFT of mixed fields: inertial and accelerated frames}, for both inertial and uniformly accelerated observers. Conclusions are shortly summarized in Section~\ref{Discandcon}. Note that, although throughout all the Section we shall refer to neutrinos, our results remain valid for any Dirac field.

\subsection{Dirac field in Minkowski spacetime}
\label{Minkspace}
It is well known that, in Minkowski spacetime, the standard plane wave expansion of the free Dirac field reads
\begin{equation}
\label{eqn:planewaveexpansion}
\psi(x) \,=\,\sum_{r=1,2} \int d^{\hspace{0.2mm}3}\hspace{0.1mm}k\left[\hspace{0.2mm}a^r_{\hspace{0.2mm}\bk}\hspace{0.8mm}\psi^{r\hspace{0.4mm}+}_{\hspace{0.2mm\bk}}(x)\,+\,b^{r\hspace{0.2mm}\dagger}_{\hspace{0.2mm}\bk}\hspace{0.8mm}\psi^{r\hspace{0.4mm}-}_{\hspace{0.2mm\bk}}(x)\hspace{0.2mm}\right],
\end{equation}
where $\psi^{r\hspace{0.4mm}+}_{\hspace{0.2mm\bk}}=N_k\hspace{0.5mm} u^r_{\hspace{0.2mm}\bk}\hspace{1mm}e^{-ik_\cdot x}$ ($\psi^{r\hspace{0.4mm}-}_{\hspace{0.2mm\bk}}=N_k\, v^r_{\hspace{0.2mm}\bk}\hspace{1.5mm}e^{+ik_\cdot x}$) are the positive (negative)  frequency plane waves, with $N_k={(2\pi)}^{-3/2}\sqrt{\frac{m}{\omega_{\bk}}}$ (see Ref.~\cite{Greiner:1996zu} for the convention used for the spinors $u^r_{\hspace{0.2mm}\bk}$ and $v^r_{\hspace{0.2mm}\bk}$) and $a^r_{\hspace{0.2mm}\bk}$, $b^r_{\hspace{0.2mm}\bk}$ are the canonical annihilatiors for the  Minkowski vacuum, 
\be
a^{\hspace{0.2mm}r}_{\hspace{0.2mm}\bk}|0\rangle_{\mathrm{M}}\,=\,b^{\hspace{0.2mm}r}_{\hspace{0.2mm}\bk}|0\rangle_{\mathrm{M}}\,=\,0\,.
\ee 
As an alternative, in analogy with Section~\ref{Hyperbrepr}, we can expand the field in terms of boost (fermion) modes as follows~\cite{Oriti:1999xq}:
\begin{equation}
\label{eqn:boostexpansion}
\psi(x)\,=\,\sum_{j=1,2}\int d^3\kappa\left[\hspace{0.2mm} c^{\hspace{0.3mm}j}_{\hspace{0.3mm}\kappa}\hspace{0.7mm}\Psi^{j\hspace{0.2mm}-}_{\kappa}(x)\,+\,d^{\hspace{0.2mm}j\hspace{0.3mm}\dagger}_{\hspace{0.2mm}\kappa}\hspace{0.7mm}\Psi^{j\hspace{0.2mm}+}_{\kappa}(x)\hspace{0.2mm}\right]\,,
\end{equation}
where
\begin{eqnarray}
\label{eqn:boostmodes}
\nonumber
\hspace{-3mm}\Psi^{j\hspace{0.2mm}\mp}_{\kappa}(x)&=&\frac{1}{2}\,N^\mp_\kappa\,\Bigg[X^{j}_{\hspace{0.2mm}k}\hspace{0.8mm} e^{\pm\hspace{0.2mm}\frac{i\pi}{2}\left(i\Omega-\frac{1}{2}\right)}\int_{-\infty}^{+\infty}\hspace{-0.9mm}d\theta\,e^{i\,\mu_{k}[x^1\hspace{0.1mm}\sinh\theta\,\mp\,t\,\cosh\theta]}\,e^{\mp\left(i\Omega-\frac{1}{2}\right)\theta}\\[2mm]
&&+\;Y^j_{\hspace{0.2mm}k}\hspace{0.8mm} e^{\pm\hspace{0.2mm}\frac{i\pi}{2}\left(i\Omega+\frac{1}{2}\right)}\int_{-\infty}^{+\infty}\hspace{-0.9mm}d\theta\,e^{i\,\mu_{k}[x^1\hspace{0.1mm}\sinh\theta\,\mp\,t\,\cosh\theta]}\,e^{\mp\left(i\Omega+\frac{1}{2}\right)\theta}\Bigg]\,e^{i\hspace{0.3mm}\vec{k}\cdot\vec{x}}
\end{eqnarray}
(see, for example, Ref.~\cite{Oriti:1999xq} for the definition of the spinors $X^{j}_{\hspace{0.2mm}k}$ and $Y^{j}_{\hspace{0.2mm}k}$).
These modes are indeed eigenfunctions of the Lorentz momentum operator $M_{01}$, with eigenvalue $\Omega$ and normalization  $N^\mp_\kappa=e^{\hspace{0.2mm}\pm\frac{1}{2}\pi\Omega}\hspace{0.2mm}/{(2\pi}\hspace{0.2mm}{\sqrt{\mu_k}})$ (as in the previous Section, $\kappa$ stands for $\{\Omega,\vec{k}\}$).

It is now a trivial matter to check that the unfamiliar quantization above introduced  is equivalent to the plane wave construction Eq.~(\ref{eqn:planewaveexpansion}), in the sense already illustrated for bosons. Indeed, we have~\cite{Oriti:1999xq}
\begin{equation}
\label{eqn:c}
c^{\hspace{0.3mm}j}_{\hspace{0.3mm}\kappa}\,=\,(\Psi^{j\hspace{0.2mm}-}_{\kappa}, \psi)
\,=\, \sum_{r=1,2}\int_0^{+\infty}\hspace{-0.9mm}{dk_1}\,F_{\hspace{0.2mm}j,r}(k_1,\Omega)\,a^r_{\hspace{0.2mm}\bk}
\end{equation}
where
\begin{equation}
F_{\hspace{0.2mm}j,r}\,=\,\frac{2\,\pi^2}{\omega_{\bk}}\,N^-_{\kappa}\,N_{k}\,\bigg[{\left(\frac{\omega_{\bk}+k_1}{\omega_{\bk}-k_1}\right)}^{i\hspace{0.1mm}\frac{\Omega}{2}+\frac{1}{4}}e^{i\frac{\pi}{2}\left(i\Omega+\frac{1}{2}\right)}\hspace{0.9mm} X^{j\hspace{0.2mm\dagger}}_{\hspace{0.2mm}k}\hspace{0.9mm}\hspace{0.7mm}+\hspace{0.7mm}{\left(\frac{\omega_{\bk}+k_1}{\omega_{\bk}-k_1}\right)}^{i\hspace{0.1mm}\frac{\Omega}{2}-\frac{1}{4}}e^{i\frac{\pi}{2}\left(i\Omega-\frac{1}{2}\right)}\hspace{0.9mm} Y^{j\hspace{0.2mm}\dagger}_{\hspace{0.2mm}k}\hspace{0.9mm}\bigg]u^r_{\hspace{0.2mm}\bk}
\end{equation}
(a similar connection between the $b^{r}_{\hspace{0.2mm}\bk}$- and $d^{\hspace{0.3mm}j}_{\hspace{0.3mm}\kappa}$-operators can be derived by projecting the field $\psi$ on the mode $\Psi^{j\hspace{0.2mm}+}_{\kappa}$). From Eq.~\eqref{eqn:c}, it thus arises that the $c^{\hspace{0.3mm}j}_{\hspace{0.3mm}\kappa}$- and $a^r_{\hspace{0.2mm}\bk}$-annihilators share the same vacuum, $|0\rangle_{\mathrm{M}}$. Additionally, the transformation Eq.~\eqref{eqn:c} is canonical.

\subsection{Dirac field in Rindler spacetime}
\label{Field quantization in Rindler spacetime: Unruh effect}
Let us now turn to the quantization of the Dirac field in the Rindler frame. 
To this aim, using the tetrad formalism, the Dirac equation in the wedge $R_+$ of Rindler spacetime (see Fig.~\ref{figure:RindlerNONTHERMAL}) can be cast into the form~\cite{Oriti:1999xq, Soffel:1980kx}
\begin{equation}
i\hspace{0.3mm}\partial_{\eta}\psi\,=\,\left(-i\hspace{0.4mm}\xi\hspace{0.4mm}\alpha^j\partial_j\,-\,\frac{1}{2}\hspace{0.4mm}i\hspace{0.4mm}\alpha^1\,+\,m\hspace{0.4mm}\xi\hspace{0.4mm}\beta\right)\psi\,,
\end{equation}
where we have used the Rindler coordinates introduced in Eq.~(\ref{eqn:rindlercoordinatesNONTHERMAL}), and  $\alpha^j\equiv\gamma^{\bar0}\,\gamma^{\bar j}$, $\beta\equiv\gamma^{\bar 0}$. Here $\gamma^{\bar\mu}$ are the analogous in Rindler spacetime of the Dirac matrices  $\gamma^{\mu}$; they satisfy the generalized condition $\gamma^{\bar\mu}\,\gamma^{\bar\nu}\,+\,\gamma^{\bar\nu}\,\gamma^{\bar\mu}=2g^{\bar\mu\bar\nu}$, with $g^{\bar\mu\bar\nu}$ given in Eq.~(\ref{eqn:lineelementsectionunruh}). We look for solutions of positive frequency with respect to the Rindler time $\eta$ in the form
\begin{equation}
\label{eqn:psirindcoord}
\Psi^{j\hspace{0.6mm}{R}}_{\kappa}\hspace{0.2mm}(\eta,\xi,\vec{x})\,=\,N_\Omega\left(X^{j}_{\hspace{0.2mm}k}\,K_{i\Omega\,-\,\frac{1}{2}}\left(\mu_k\hspace{0.2mm}\xi\right)\,+\,Y^{j}_{\hspace{0.2mm}k}\,K_{i\Omega\,+\,\frac{1}{2}}\left(\mu_k\hspace{0.2mm}\xi\right)\right)e^{-i\hspace{0.2mm}\Omega\hspace{0.2mm}\eta}\hspace{0.3mm}\hspace{0.5mm}e^{i\hspace{0.2mm}\vec{k}\cdot\hspace{0.2mm}\vec{x}},\quad\,\, j=1,2\,,
\end{equation}
where $N_\Omega=\frac{1}{4\hspace{0.2mm}\pi^2\hspace{0.2mm}{\sqrt{\mu_k}}}\,\sqrt{{\cosh{\left(\pi\hspace{0.2mm}\Omega\right)}}}$ and $K_{i\Omega\,\pm\,\frac{1}{2}}$ is the modified Bessel function.

Next, the relationship between the Minkowski and Rindler quantum constructions can be straightforwardly investigated by expressing the solutions of Eq.~(\ref{eqn:psirindcoord}) in Minkowski coordinates. In this connection, let us observe that, upon the coordinate transformation $(t,x^1)\rightarrow(\eta,\xi)$, the tetrads - and therefore the spinors - undergo an orthogonal rotation of the following form
\be 
\Psi (t, x^1)\,=\,\mathrm{exp}\left(\frac{1}{2}\hspace{0.2mm}\gamma^{\bar 0}\hspace{0.2mm}\gamma^{\bar1}\hspace{0.2mm}\eta\right)\Psi (\eta, \xi)\,.
\ee 
Equation~(\ref{eqn:psirindcoord})  then becomes
\begin{equation}
\label{eqn:rindlermodesmin}
\Psi^{j\hspace{0.6mm}{R}}_{\kappa}\hspace{0.2mm}(t, x^1, \vec{x})\,=\,N_\Omega\left(X^{j}_{\hspace{0.2mm}k}\,K_{i\Omega\,-\,\frac{1}{2}}\left(\mu_k\hspace{0.2mm}\xi\right)\,e^{-\left(i\hspace{0.2mm}\Omega-\frac{1}{2}\hspace{0.2mm}\right)\eta}\,+\,Y^{j}_{\hspace{0.2mm}k}\,K_{i\Omega\,+\,\frac{1}{2}}\left(\mu_k\hspace{0.2mm}\xi\right)e^{-\left(i\hspace{0.2mm}\Omega+\frac{1}{2}\hspace{0.2mm}\right)\eta}\right)\hspace{0.3mm}e^{i\hspace{0.2mm}\vec{k}\cdot\hspace{0.2mm}\vec{x}}.
\end{equation}
It must be clear that this equation is still defined only in the right wedge $R_+$. As shown in Ref.~\cite{Oriti:1999xq}, however, it could be linked to the solutions of the Dirac equation in the other sectors of Minkowski spacetime by analytically extending these functions across the horizons. The two possible paths of analytic continuation, one rotating clockwise  from $R_+$ to $R_-$, the other counterclockwise, correspond to the two different global representations of the boost eigenfunctions $\Psi^{j\hspace{0.2mm}\mp}_{\kappa}$ in Eq.~(\ref{eqn:boostmodes}).

Holding all the necessary tools, the Unruh effect for (free) spinor fields can now be promptly derived. Following the same procedure of Section~\ref{Unruh}, we then quantize the field in terms of solutions of the Dirac equation which are eigenfunctions of Rindler Hamiltonian (that is to say, of the Lorentz boost generator in Rindler coordinates), are orthonormal and analytic in the whole of Minkowski spacetime. The functions we are looking for are
\begin{eqnarray}
\label{eqn:R}
\mathcal{R}^{j\,(+)}_{\hspace{0.3mm}\kappa}&=&\frac{1}{\sqrt{2\cosh\left(\pi\hspace{0.2mm}\Omega\right)}}\left(e^{\frac{\pi\hspace{0.2mm}\Omega}{2}}\,\Psi^{j\hspace{0.2mm}-}_{\kappa}\,+\,e^{-\frac{\pi\hspace{0.2mm}\Omega}{2}}\,\Psi^{j\hspace{0.2mm}+}_{\kappa}\right)\,,\\[2mm]
\label{eqn:Rmeno}
 \mathcal{R}^{j\,(-)}_{\hspace{0.3mm}\kappa}&=&\frac{1}{\sqrt{2\cosh\left(\pi\hspace{0.2mm}\Omega\right)}}\left(e^{-\frac{\pi\hspace{0.2mm}\Omega}{2}}\,\Psi^{j\hspace{0.2mm}-}_{\kappa}\,-\,e^{\frac{\pi\hspace{0.2mm}\Omega}{2}}\,\Psi^{j\hspace{0.2mm}+}_{\kappa}\right)\,,
\end{eqnarray}
where $\Psi^{j\hspace{0.3mm}\pm}_{\kappa}$ are the globally defined modes in Eq.~(\ref{eqn:boostmodes}) and the superscripts $\pm$ on the l.h.s. are referred to the right/left wedges $R_{\pm}$, respectively. One can easily verify that the function $\mathcal{R}^{j\,(+)}_{\hspace{0.3mm}\kappa}$ vanishes in the left wedge, whereas it reduces to the Rindler solution of Eq.~(\ref{eqn:rindlermodesmin}) in the right sector, thereby providing the correct  global representation of the modes for a uniformly accelerated observer in Rindler spacetime (the inverse behavior, of course, is exhibited by $\mathcal{R}^{j\,(-)}_{\hspace{0.3mm}\kappa}$).

Now, inverting Eqs.~(\ref{eqn:R}), (\ref{eqn:Rmeno}) with respect to $\Psi^{j\hspace{0.3mm}\pm}_{\kappa}$ and inserting  into Eq.~(\ref{eqn:boostexpansion}), we have
\begin{equation}
\label{eqn:RLexpan}
\psi(x)\,=\,\sum_{j=1,2}\int d^3\kappa\left[\hspace{0.2mm} r^{\hspace{0.2mm}j}_{\hspace{0.2mm}\kappa}\,\mathcal{R}^{j\,(+)}_{\hspace{0.3mm}\kappa}\,+\,r^{\hspace{0.2mm}j\hspace{0.2mm}\dagger}_{\hspace{0.2mm}\kappa}\,\mathcal{R}^{j\,(+)}_{\hspace{0.3mm}\tilde\kappa}\,+\,l^{\hspace{0.2mm}j}_{\kappa}\,\mathcal{R}^{j\,(-)}_{\tilde{\kappa}}\,+\,l^{\hspace{0.2mm}j\hspace{0.2mm}\dagger}_{\kappa}\,\mathcal{R}^{j\,(-)}_{\hspace{0.2mm}\kappa}\right], 
\end{equation}
where $\tilde\kappa\equiv\{-\Omega,\vec{k}\}$ and 
\begin{equation}
\label{eqn:Bogo}
r^{\hspace{0.2mm}j}_{\hspace{0.2mm}\kappa}\,=\,\frac{c^{\hspace{0.3mm}j}_{\hspace{0.3mm}\kappa}\,e^{\frac{\pi\hspace{0.2mm}\Omega}{2}}\,+\,d^{\hspace{0.2mm}j\hspace{0.2mm}\dagger}_{\hspace{0.2mm}\kappa}\,e^{-\frac{\pi\hspace{0.2mm}\Omega}{2}}}{\sqrt{2\cosh\hspace{0.2mm} (\pi\hspace{0.2mm}\Omega)}}\,,
\end{equation}
(a similar expression can be derived for $l^{j}_{\hspace{0.2mm}\kappa}$).  The \emph{Bogoliubov transformation} so obtained allows to determine the spectrum of Rindler quanta in the inertial vacuum. A straightforward calculation leads to
\begin{equation}
\label{eqn:thermal}
_{\mathrm{M}}\langle0|\hspace{0.2mm}r^{i\hspace{0.3mm}\dagger}_{\kappa}\hspace{0.4mm}r^{j\hspace{0.2mm}}_{\kappa'}|0\rangle_{\mathrm{M}}\,=\,_{\mathrm{M}}\langle0|\hspace{0.2mm}l^{i\hspace{0.3mm}\dagger}_{\kappa}\hspace{0.4mm}l^{j\hspace{0.2mm}}_{\kappa'}|0\rangle_{\mathrm{M}}\,=\,\frac{1}{e^{2\pi\Omega}+1}\,\delta_{ij}\,\delta^{\hspace{0.2mm}3}(\kappa-\kappa')\equiv\frac{1}{e^{\frac{a\,\Omega}{T_{\rm U}}}+1}\hspace{0.3mm}\delta_{ij}\,\delta^{\hspace{0.2mm}3}(\kappa-\kappa'),
\end{equation}
It thus arises that the vacuum radiation perceived by the Rindler observer has a thermal spectrum, according to Fermi-Dirac statistics
\be
\label{FDstat}
N_{FD}(\Omega)=\frac{1}{e^{\frac{a\,\Omega}{T_{\rm U}}}+1}\,,
\ee 
with  $T_{\rm U}=\frac{a}{2\pi}$~\cite{Unruh:1976db}. In this way, we 
have recovered the result Eq.~\eqref{billTemperature} also for fermions.

\subsection{QFT of mixed neutrinos}
\label{QFT of mixed fields: inertial and accelerated frames}
In a simplified two flavor model, the mixing transformations for Dirac fields read (see Appendix~\ref{QFT of fm} for more details)
\begin{equation}
\label{eqn:Pontec1}
\begin{array}{l}
\psi_{e}(x)\,=\, \psi_{1}(x)\, \cos\theta  + \psi_{2}(x)\, \sin\theta\,,\\[4mm]
\psi_{\mu}(x) \,=\,-  \psi_{1}(x)\, \sin\theta   +  \psi_{2}(x)\, \cos\theta\,.
\end{array}
\ee
Here $\psi_\chi$ ($\chi=e,\mu$) are mixed fields with definite flavor $\chi$, $\psi_i$ ($i=1,2$) are free fields with definite mass $m_i$, and $\theta$ is the mixing angle. 

As argued in Appendix~\ref{QFT of fm}, Eqs.~\eqref{eqn:Pontec1} allow  to adopt  free field-like expansions for flavor fields (see Eq.~\eqref{flfieldexpans}), provided that a Bogoliubov transformation at level of ladder operators is introduced (see Eq.~\eqref{eqn:annihilator}). In particular, we have~\cite{Oriti:1999xq}
\begin{equation}
\label{eqn:planewavemixbulk}
\psi_\ell(x)\,=\,\sum_{r=1,2} \int d^{\hspace{0.2mm}3}\hspace{0.1mm}k\hspace{1mm} e^{i\bk\cdot\bx}\left[\hspace{0.2mm}a^{r}_{\bk,\ell}(t)\hspace{0.7mm}u^{r}_{	\bk,\sigma}(t)\,+\,b^{r\hspace{0.2mm}\dagger}_{-\bk,\ell}(t)\hspace{0.7mm}v^r_{-\bk,\sigma}(t)\hspace{0.2mm}\right],
\end{equation}
where $(\ell,\sigma)=(e,1)\,, (\mu,2)$,  $u^{r}_{\bk,\sigma}(t)\,\equiv\, u^{r}_{\bk,\sigma}\,e^{-i\omega_{\bk}\hspace{0.2mm}t}$, $v^{r}_{-\bk,\sigma}(t)\,\equiv\, v^{r}_{-\bk,\sigma}\,e^{+i\omega_{\bk}\hspace{0.2mm}t}$ and
\begin{equation}
\label{eqn:annihilatorelectron} 
a^r_{\bk,e}(t)=\cos\theta\,a^r_{\bk,1}\,+\,\sin\theta\;\left(\rho^{{\bk}\,*}_{12}(t)\; a^r_{\bk,2}\,-\,  \lambda^{{\bk}}_{12}(t)\; b^{-r\hspace{0.2mm}\dagger}_{-{\bk},2}\right),
\end{equation}
(similarly for the other ladder operators). The mixing Bogoliubov coefficients $\rho^{{\bk}}_{12}$ and  $\lambda^{{\bk}}_{12}$ are defined in Eqs.~\eqref{eqn:Uk}, \eqref{eqn:Vk}. 

The foregoing discussion is completely within the standard Minkowski framework. To extend the formalism to the Rindler metric, let us then quantize mixed fields in the hyperbolic scheme as in Eq.~(\ref{eqn:boostexpansion}):
\begin{equation}
\label{eqn:boostexpansionmix}
\psi_\ell(x)\,=\,\sum_{j=1,2}\int d^3\kappa\left[\hspace{0.2mm} c^{\hspace{0.3mm}j}_{\hspace{0.3mm}\kappa,\ell}\hspace{0.7mm}\Psi^{j\hspace{0.2mm}-}_{\kappa,\sigma}(x)\,+\,d^{\hspace{0.2mm}j\hspace{0.3mm}\dagger}_{\hspace{0.2mm}\kappa,\ell}\hspace{0.7mm}\Psi^{j\hspace{0.2mm}+}_{\kappa,\sigma}(x)\hspace{0.2mm}\right],
\end{equation} 
where $(\ell,\sigma)=(e,1), (\mu,2)$. Here $\Psi^{j\hspace{0.2mm}\mp}_{\kappa,\sigma}$ are the modes defined in Eq.~(\ref{eqn:boostmodes}) and
\begin{equation}
\label{eqn:cmix}
c^{\hspace{0.3mm}j}_{\hspace{0.3mm}\kappa,\ell}\,=\, \sum_{r=1,2}\int_0^{+\infty}\hspace{-0.9mm}{dk_1}\,F_{\hspace{0.2mm}j,r,\sigma}(k_1,\Omega)\,a^r_{\hspace{0.2mm}\bk,\sigma}\,,
\end{equation} 
(see the corresponding transformation for free fields, Eq.~\eqref{eqn:c}).

Equation~\eqref{eqn:boostexpansionmix} provides a relationship between the QFT of flavor mixing for an inertial observer and the corresponding treatment in accelerated frames. By extending the Unruh quantization Eq.~\eqref{eqn:RLexpan} to mixed fields, indeed, we can adopt the following expansions:
\begin{equation}
\label{eqn:RLexpanmix}
\psi_\ell(x)\,=\,\sum_{j=1,2}\int d^3\kappa\left[\hspace{0.2mm} r^{\hspace{0.2mm}j}_{\hspace{0.2mm}\kappa,\ell}\,\mathcal{R}^{j\,(+)}_{\hspace{0.3mm}\kappa,\sigma}\,+\,r^{\hspace{0.2mm}j\hspace{0.2mm}\dagger}_{\hspace{0.2mm}\kappa,\ell}\,\mathcal{R}^{j\,(+)}_{\hspace{0.3mm}\tilde\kappa,\sigma}\,+\,l^{\hspace{0.2mm}j}_{\kappa,\ell}\,\mathcal{R}^{j\,(-)}_{\tilde{\kappa},\sigma}\,+\,l^{\hspace{0.2mm}j\hspace{0.2mm}\dagger}_{\kappa,\ell}\,\mathcal{R}^{j\,(-)}_{\hspace{0.2mm}\kappa,\sigma}\right], 
\end{equation}
where $(\ell,\sigma)\,=\,(e,1), (\mu,2)$. Rindler flavor operators are given by~\cite{Oriti:1999xq}
\begin{equation}
r^{\hspace{0.2mm}j}_{\hspace{0.2mm}\kappa,\ell}=\frac{1}{\sqrt{2\cosh\hspace{0.2mm} (\pi\hspace{0.2mm}\Omega)}}\hspace{0.2mm}\,{\sum_{r=1,2}\int_0^{+\infty}\hspace{-0.9mm}{dk_1}\Bigg(e^{\frac{\pi\hspace{0.2mm}\Omega}{2}}\,F_{\hspace{0.2mm}j,r,\sigma}(\Omega, k_1)\,a^r_{\hspace{0.2mm}\bk,\sigma}\,+\,}\,e^{-\frac{\pi\hspace{0.2mm}\Omega}{2}}\,G_{j,r,\sigma}(\Omega,k_1)\,\,b^{\hspace{0.2mm}r\hspace{0.2mm}\dagger}_{\bk,\sigma}\Bigg)\,,
\end{equation}
where
\begin{equation}
G_{j,r}\,=\,\frac{2\,\pi^2}{\omega_{\bk}}\,N^+_{\kappa}\,N_{k}\bigg[{\left(\frac{\omega_{\bk}+k_1}{\omega_{\bk}-k_1}\right)}^{i\hspace{0.1mm}\frac{\Omega}{2}+\frac{1}{4}}e^{i\frac{\pi}{2}\left(i\Omega+\frac{1}{2}\right)}\hspace{0.9mm}\hspace{0.9mm} X^{j\hspace{0.2mm\dagger}}_{\hspace{0.2mm}k}\,+\,{\left(\frac{\omega'+k_1'}{\omega'-k_1'}\right)}^{i\hspace{0.1mm}\frac{\Omega}{2}-\frac{1}{4}}e^{i\frac{\pi}{2}\left(i\Omega-\frac{1}{2}\right)}\hspace{0.9mm} Y^{j\hspace{0.2mm\dagger}}_{\hspace{0.2mm}k}\hspace{0.9mm}\bigg]v^r_{\hspace{0.2mm}\bk}\,,
\end{equation}
and similar for $F_{j,r}$. 

As for the case of boson mixing, the  evaluation of the Bogoliubov coefficients is absolutely non-trivial; however, for $t=\eta=0$, calculations get significantly simplified, leading to the following expression for the Unruh condensate of mixed particles:
\begin{eqnarray}
\nonumber
\label{eqn:spectrum}
{\cal N}(\theta)\Big|_{t=0}&\equiv&_{\mathrm{M}}\langle0|\hspace{0.2mm}r^{i\hspace{0.3mm}\dagger}_{\kappa,\chi}\hspace{0.4mm}r^{j\hspace{0.2mm}}_{\kappa',\chi}|0\rangle_{\mathrm{M}}\Big|_{t=0}\\[2mm]\non
&=&N_{FD}(\Omega)\,\delta_{ij}\,\delta(\kappa-\kappa')\,+\,\sin^2\theta\bigg[\,\frac{e^{\frac{\pi\,(\Omega+\Omega')}{2}}}{2\hspace{0.1mm}\sqrt{\cosh\left(\pi\Omega\right)\,\cosh\left(\pi\Omega'\right)}}\,N_{F,F}^{\hspace{0.2mm}ij}(\Omega,\Omega')\\[2mm]
&&-\;\frac{e^{-\frac{\pi\,(\Omega+\Omega')}{2}}}{2\hspace{0.1mm}\sqrt{\cosh\left(\pi\Omega\right)\,\cosh\left(\pi\Omega'\right)}}\,N^{\hspace{0.2mm}ij}_{G,G}(\Omega,\Omega')\bigg]\hspace{0.2mm}\delta_{ij}\,,\qquad \chi=e,\mu\,,
\end{eqnarray}
where
\begin{eqnarray}
N_{F,F}^{\hspace{0.2mm}ij}&=&\sum_{r=1,2}\int\,F^{\hspace{0.2mm}*}_{\hspace{0.2mm}i,r}(k_1,\Omega)\,F_{\hspace{0.2mm}j,r}(k_1,\Omega')\,{|\lambda^{{\bk}}_{12}|}^2\,,\\[2mm]
N_{G,G}^{\hspace{0.2mm}ij}&=&\sum_{r=1,2}\int\,G^{\hspace{0.2mm}*}_{\hspace{0.2mm}i,r}(k_1,\Omega)\,G_{\hspace{0.2mm}j,r}(k_1,\Omega')\,{|\lambda^{{\bk}}_{12}|}^2\,,
\end{eqnarray} 
(a similar result is obtained for the vacuum expectation value of the $l^{i}_{\kappa}$-operators).

In analogy with the analysis carried out for bosons, we thus realize that the Unruh spectrum for Dirac fields loses its characteristic thermal behavior in the context of flavor mixing, acquiring off-diagonal (non-thermal) corrections. However, it is a trivial matter to verify that such additional contributions consistently vanish for $\theta\rightarrow 0$ and/or $m_1=m_2$, as it must be for vanishing mixing. Finally, let us observe that further manipulations of Eq.~(\ref{eqn:spectrum}) can be performed in the reasonable limit of small mass difference $\frac{\Delta m^2}{m^2}\equiv \frac{|m_2^2-m_1^2|}{m^2}\ll 1$, yielding
\begin{equation}
\label{eqn:approxunruhdensity}
{\cal N}(\theta)\Big|_{t=0}\,\approx\,N_{FD}(\Omega)\,\delta_{ij}\,\delta(\kappa-\kappa')\,+\,\sin^2\theta\;\mathcal{O}\hspace{-0.8mm}\left({\frac{{\Delta m^2}}{{m}^2}}\right)\,,
\end{equation}
to be compared with the corresponding spectrum for bosons, Eq.~\eqref{eqn:spectrumapprox}.\footnote{We remark that the heuristic interpretation of the result Eq.~\eqref{eqn:approxunruhdensity} is the same as the one illustrated in Fig.~\ref{figure:unruhmix} for bosons.} 
Beyond the theoretical relevance of the obtained result, we stress that, if one day technology will be developed enough to make the Unruh radiation phenomenologically accessible,  Eq.~(\ref{eqn:approxunruhdensity}) may be applied to the case of neutrinos, in order to fix new potential constraints on their squared mass differences. An outlook at future developments of the present analysis can be found in Section~\ref{Discandcon}.

\section{Inertial effects on neutrino oscillations}
We conclude this Chapter with a brief digression on the effects of a linear acceleration on neutrino oscillations in vacuum~\cite{noepl}. Due to technical difficulties, however,
we shall focus on a purely quantum mechanical treatment of the problem, reserving 
the extension to the  QFT framework for a future publication.

As stated above, neutrino oscillations in  flat spacetime have been largely analyzed
since Pontecorvo's pioneering idea of non-degenerate neutrino mass-matrix~\cite{Pontecorvo}.
Over the years, however, alternative mechanisms have
been proposed, among which the most renowned are the ones by Gasperini~\cite{Gasperini:1989rt}
and Liu~\cite{Liu:1997km}.
Although they have  both been rejected by experiments, these solutions
represent a first attempt to accommodate gravity effects
 into the standard picture of neutrino oscillations.

A systematic treatment of flavor oscillations in curved spacetime has
been discussed by a number of authors in
Refs.~\cite{Cardall,Ahluwalia:1996ev}.
In Ref.~\cite{Cardall}, in particular,
the authors provide a simple framework to demonstrate
that gravitational effects are closely related to the redshift of
neutrino energy. The framework becomes even richer in astrophysical regimes, where the presence of strong gravitational and magnetic fields (provided that neutrinos possess a non-vanishing magnetic moment) significantly affects the oscillation probability~\cite{Lambiase:2004qk}.
Due to the equivalence principle,  similar results are expected to be valid
 also in accelerated frames.
%
Along this line, a pilot analysis
of phenomenological aspects of neutrino oscillations for an accelerating and rotating observer
has been performed in Ref.~\cite{Capozziello:1999ww}.
Recently, mixing transformations in Rindler (uniformly accelerated) background
have also been studied in Quantum Field Theory (QFT)~\cite{Blasone:2017nbf,Blasone:2018byx}, showing that non-thermal corrections to the Unruh radiation may arise due to the interplay between the Bogoliubov transformation related to the structure of Rindler spacetime and the one hiding in field mixing~\cite{Blasone:1995zc}.

Apart from phenomenological implications, we stress that a deeper
understanding of inertial effects on flavor mixing and oscillations
may shed some light on a number of intriguing issues
at a theoretical level. Recently, for instance, the r\^ole of neutrino mixing
in the decay of accelerated protons (inverse $\beta$-decay) has been investigated
with controversial results~\cite{Ahluwalia:2016wmf,Blasone:2018czm,Cozzella:2018qew}.
Specifically, in Refs.~\cite{Matsas:1999jx} it was pointed out that the Unruh effect is necessary  to maintain the general covariance of QFT when considering the
inverse $\beta$-decay rate in the laboratory and comoving
frames, respectively. Subsequently, it was noted that
neutrino mixing can spoil this agreement~\cite{Ahluwalia:2016wmf}, and
further discussion~\cite{Blasone:2018czm,Cozzella:2018qew}
has narrowed down possible causes to the true nature of asymptotic neutrinos (we remand the
reader to the next Chapter for a more detailed discussion on this). Cleary, such an
ambiguity affects flavor oscillations too.
In particular, since the oscillation probability calculated in the ordinary QFT
by means of the exact flavor states~\cite{Blasone:1998hf} contains extra-terms
with respect to the usual quantum mechanical formula,
one expects corrections to arise also for the non-inertial case.

In the present work, a preliminary step
along this direction is taken by analyzing the effects of a linear acceleration on neutrino oscillations
in the context of Quantum Mechanics.
The obtained result should be regarded as a benchmark
for the field theoretical treatment of the problem, for which work is in progress.

The remainder of this Section is organized as follows: in Section~\ref{Neutrino oscillations in flat spacetime}, we briefly review the standard treatment of neutrino oscillations in  flat spacetime using the plane wave formalism.
Section~\ref{Inertial effects on neutrino oscillations: a heuristic treatment} is devoted to a heuristic derivation of the oscillation probability  for a uniformly accelerated observer. The same result is recovered in Section~\ref{Inertial effects on neutrino oscillations: a geometric viewpoint}
by solving the Dirac equation in accelerated  frames. The obtained expression is critically compared with the one in Ref.~\cite{Capozziello:1999ww}, where corrections are calculated in a more geometric framework. 
As possible applications, in Section~\ref{Applications} we discuss how Earth's gravity affects the oscillation probability of atmospheric neutrinos. Furthermore, we propose a gedanken experiment in which an ideal detector is used for testing the obtained result in proximity of high-density astrophysical objects. Subection~\ref{Discandcon} contains a discussion and final remarks.

\subsection{Neutrino oscillations in  flat spacetime}
\label{Neutrino oscillations in flat spacetime}
We start by reviewing the standard theory
of neutrino oscillations in Minkowski spacetime. 
For the sake of simplicity, we focus on a  
 model with only two flavor generations (for a more rigorous three-flavor description, we remand the reader to Ref.~\cite{Maki:1962mu}).
In the conventional matrix notation, indicating by $|\nu_\alpha\rangle$ ($\alpha=e,\mu$) and
 $|\nu_k\rangle$ ($k=1,2$) neutrino flavor and mass 
 eigenstates, respectively, the following relation holds~\cite{KimPev}
\begin{equation}
\label{eqn:U}
\begin{pmatrix}
|\nu_e\rangle  \\
|\nu_\mu\rangle
\end{pmatrix}\,=\,U(\theta)\begin{pmatrix}
|\nu_1\rangle  \\
|\nu_2\rangle
\end{pmatrix},
\end{equation}
where $\theta$ is the mixing angle and $U(\theta)$ the Pontecorvo  unitary
matrix,
\begin{equation}
\label{PMMmatrix}
U(\theta)\,=\,
\begin{pmatrix}
\cos\theta&\sin\theta\\
-\sin\theta&\cos\theta
\end{pmatrix}\,.
\end{equation} 
In what follows, we shall describe the propagation of the
mass eigenstates by plane waves, i.e.
\begin{equation}
|\nu_k(t,\bx)\rangle\,=\,\exp[-i\Phi_k(t, \bx)]\,|\nu_k\rangle,\quad (k\,=\,1,2)\,,
\end{equation}
where 
\begin{equation}
\Phi_k\,=\,E_k\,t-\bp_k\cdot\bx
\end{equation} 
is the quantum-mechanical phase 
of the $k^{th}$ neutrino state, with $E_k$ and $\bp_k$ being its energy and momentum, respectively. Mass, energy and momentum are related by the mass-shell condition as
\begin{equation}
E^2_k\,=\,m^2_k\,+\,{|\bp_k|}^2\,.
\end{equation}
In the relativistic approximation, labelling with $A\hspace{0.1mm}(t_A, \bx_A)$ and $B\hspace{0.1mm}(t_B, \bx_B)$  the spacetime points in which neutrinos are produced and
detected, respectively, the phase acquired by the $k^{th}$ eigenstate after propagating 
over the distance  $L_p\,\equiv\,\left|\bx_B-\bx_A\right|$ reads\footnote{In order for the interference pattern not 
to be destroyed, we remark that neutrinos
must be produced coherently and  measured at 
the same spacetime point.}
\begin{equation}
\label{eqn:phase}
\Phi_{k}\,=\,E_k\hspace{0.1mm}(t_B-t_A)\,-\,|\bp_k| |\bx_B-\bx_A|\,\approx\,\frac{m^2_k}{2E_0}L_p\,.
\end{equation} 
Note that, in the second step of Eq.~(\ref{eqn:phase}), we have exploited the relativistic condition $m_k\ll  E_k$, so that 
\begin{equation}
\label{eqn:approx}
t_B\,-\,t_A\,\simeq\,\left|\bx_B\,-\,\bx_A\right|,
\end{equation}
and, to the first order,
\begin{equation}
\label{ap}
E_k\,\simeq\, E_0 \,+\, \mathcal {O}\left(\frac{m^2_k}{2E_0}\right)\,,
\end{equation} 
with $E_0$ being the energy for a massless 
neutrino. The last equation amounts to require that mass eigenstates are also energy eigenstates with a common energy $E_0$.

Let us now consider an electron neutrino $|\nu_e\rangle$ emitted via weak interaction at the  point $A\hspace{0.1mm}(t_A, \bx_A)$.  Using Eq.~(\ref{eqn:U}), the probability that it is revealed as muon neutrino $|\nu_\mu\rangle$ at the point $B\hspace{0.1mm}(t_B, \bx_B)$ is given by
\begin{equation}
\label{pontecdetector}
P_{\nu_e\rightarrow\nu_\mu}\,\equiv\,{\big|\langle\nu_\mu(t_B, \bx_B)\hspace{0.1mm}|\hspace{0.1mm}\nu_e(t_A, \bx_A)\rangle\big|}^2\,=\,\sin^2(2\theta)\sin^2\left(\frac{\Phi_{12}}{2}\right),
\end{equation}
where, according to Eq.~(\ref{eqn:phase}), the phase-shift $\Phi_{12}\equiv \Phi_1-\Phi_2$ takes the form
\begin{equation}
\label{eqn:relativephase}
\Phi_{12}\,\simeq\, \frac{\Delta m^2_{12}}{2E_0}\hspace{0.2mm}L_p\hspace{0.1mm}\,\equiv\,  \frac{m_1^2\,-\,m_2^2}{2E_0}\hspace{0.2mm}L_p\hspace{0.1mm}\,. 
\end{equation}

It should be noted that, in the case where at least one
of the states $|\nu_k\rangle$ is non-relativistic, a wave packet approach is required instead of the above plane wave formalism~\cite{Giunti:1991ca}. 
For our purposes, however, such an analysis would show that the 
approximations Eqs.~(\ref{eqn:approx}) and (\ref{ap}) are adequate, leading to the 
formula Eq.~\eqref{pontecdetector} for the oscillation probability. We also stress that this equation represents the quantum-mechanical limit of a more general formula derived within the framework of QFT. For a thorough analysis of this treatment, see Refs.~\cite{Blasone:1995zc,bosonmix}.

The foregoing considerations hold for an observer at rest or moving inertially
with respect to the oscillation experiment. Nevertheless, due to gravity, 
any stationary laboratory on Earth experiences a
linear acceleration (in the present analysis, we do not take care of rotational effects. A thorough discussion of this subject can be found in 
Ref.~\cite{Capozziello:1999ww}). To show how acceleration 
affects flavor oscillations, 
let us cast then recast the quantum mechanical phase Eq.~\eqref{eqn:phase} 
into a covariant form, according to~\cite{Stodolsky:1978ks}
\begin{equation}
\Phi_k\,=\,\int_A^{B}p_\mu^{(k)}\hspace{0.1mm}dx^{\mu}, 
\label{eqn:covphase}
\end{equation}
where
\begin{equation}
\label{eqn:moment}
p_\mu^{(k)}\,=\,m_k\hspace{0.3mm}g_{\mu\nu}\hspace{0.3mm}\frac{dx^\nu}{ds}
\end{equation}
is the canonical four-momentum conjugated to the coordinates $x^\mu$
and $ds$, $g_{\mu\nu}$ are the line element and the metric
tensor, respectively. The integration in Eq.~\eqref{eqn:covphase} 
has to be performed  along the light-ray trajectory linking the spacetime points 
$A$ and $B$. For $g_{\mu\nu}$ corresponding to the Minkowski flat
metric, it is easy to show that Eqs.~(\ref{eqn:covphase}), (\ref{eqn:moment}) 
 reproduce the standard result Eq.~\eqref{eqn:phase}, as it should be.

\subsection{Inertial effects on oscillations: a heuristic treatment}
\label{Inertial effects on neutrino oscillations: a heuristic treatment}
We now turn to the discussion of neutrino oscillations for a uniformly accelerated
observer. In order to apply the covariant formalism above described, let us recall that the line element in an accelerated frame can be written as (we neglect the effects of the spacetime curvature)~\cite{Misner}
\begin{equation}
\label{eqn:linelementacframe}
ds^2\,=\,f(\textbf{a},\textbf{x})\hspace{0.1mm}\hspace{0.1mm}{(dt)}^2
\,-\,d\textbf{x}\cdot d\textbf{x}\,,
\end{equation}
where
\begin{equation}
\label{eqn:funct}
f(\textbf{a},\textbf{x})\,\equiv\,\left(1\,+\,\textbf{a}\cdot\textbf{x}\right)^2\,.
\end{equation}
Here, $\textbf{a}$ is the proper three-acceleration and  $x^\mu=(t,\textbf{x})$ are
the Fermi coordinates for an accelerated observer~\cite{Misner}, whose range of validity is limited by the requirement $|\textbf{x}|\ll {|\textbf{a}|}^{-1}$. This occurs because the above reference frame is conceived to describe a neighborhood of the observer's world line as long as the previous condition holds. However, the confinement on the spatial region does not affect the relevance of our considerations, since typical oscillation lengths of neutrino experiments allow us to deal with even considerable values of $\textbf{a}$. For instance, for acceleration of the order of Earth's gravity, the metric is valid within a range of one light-year.

Without loss of generality, we can restrict our analysis to $1+1$-dimensions,
assuming the acceleration to be antiparallel to the direction of neutrino propagation (see Fig~\ref{fig1}). 
According to Eq.~\eqref{eqn:moment}, the components of the neutrino
canonical momentum $p_\mu^{(k)}$ are
\begin{eqnarray}
p_t^{(k)}&=&m_k\hspace{0.3mm}f(a, x)\hspace{0.2mm}\frac{dt}{ds}\,, 
\\[1mm]
p_x^{(k)}&=&-m_k\frac{dx}{ds}\,.
\end{eqnarray}
They are related to each other and to the mass $m_k$ by the 
generalized mass-shell condition
\begin{equation}
m^2_k\,=\,g^{\mu\nu}\hspace{0.1mm}p^{(k)}_\mu\hspace{0.1mm}p^{(k)}_\nu,
\label{eqn:genmasscond}
\end{equation}
with $g_{\mu\nu}$ given in Eq.~(\ref{eqn:linelementacframe}). Since the metric does not depend on the coordinate $t$, the timelike momentum component 
$p_t^{(k)}$ is conserved along the geodesic trajectory
of the $k^{th}$ neutrino eigenstate.
We define such a constant to be $p_t^{(k)}\equiv E_k$.
It represents the energy measured by an observer at rest 
at the origin. Due to the metric Eq.~(\ref{eqn:linelementacframe}), however,
it differs from the energy at any other 
spacetime point. The local energy, defined as the energy measured by an
observer at rest at the generic position $x$, is related to $E_k$ by~\cite{Misner}
\begin{equation}
\label{eqn:localenergy}
E_k^{(loc)}(x)\,=\,\left|g_{tt}\right|^{-1/2}\hspace{0.1mm}E_k\,=\,{f(a, x)}^{-1/2}E_k\,.
\end{equation}
\begin{figure}[t]
\resizebox{15.6cm}{!}{\includegraphics{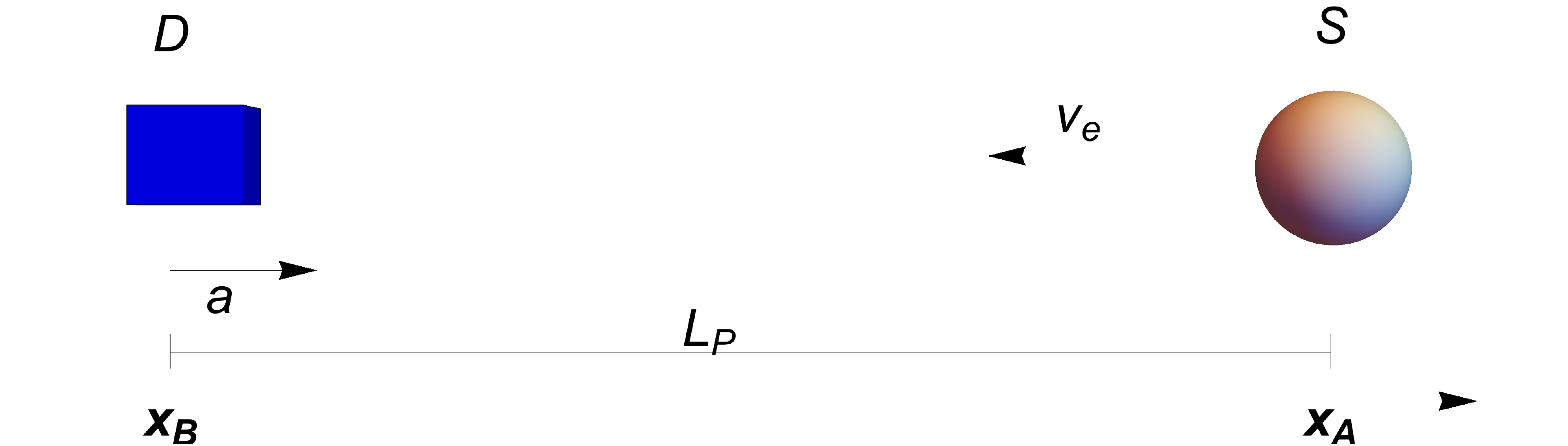}}
\caption{Neutrino emission from a source $S$ at the point $x_A$. After propagating 
over the distance $L_p$, neutrinos reach the detector $D$ at the point $x_B\,<\,x_A$. Note that the direction of neutrino motion is assumed to be antiparallel to the acceleration of the detector.}
\label{fig1}
\end{figure}
Next, by use of Eq.~(\ref{eqn:covphase}), the phase of the $k^{th}$ neutrino eigenstate reads
\begin{equation}
\label{eqn:covariantphase}
\Phi_k\,=\,\int_{A}^{B}\left[E_k\left(\frac{dt}{dx}\right)_{\!0}\hspace{-0.5mm}\,-\,p_{k}(x)\right]dx\,,
\end{equation}
where the momentum  $p_k(x)\equiv- p_x{(k)}$ is obtained from the generalized mass-shell condition Eq.~(\ref{eqn:genmasscond}) as
\begin{equation}
\label{eqn:momentum}
p_k(x)\,=\,-{\left(\frac{E^2_k}{f(a, x)}\,-\,m^2_k\right)}^{1/2},
\end{equation}
and the light-ray differential $\left(dt/dx\right)_0$ is given by
\begin{equation}
\label{eqn:lightray}
\left(\frac{dt}{dx}\right)_{\!0}\,=\, -f(a, x)^{-1/2}\,.
\end{equation}
where the minus sign in Eqs.~(\ref{eqn:momentum}) and (\ref{eqn:lightray})
is due to the fact that neutrinos propagate antiparallel to the $x$-axis.
By inserting Eqs.~(\ref{eqn:momentum}) and (\ref{eqn:lightray})
into Eq.~(\ref{eqn:covariantphase}), we get
\begin{equation}
\widetilde\Phi_k\,=\,-\int_{A}^{B}E_k\left\{1\,-\,{\left[1-f(a, x)\frac{m^2_k}{E_k^2}\right]}^{1/2}\right\}{f(a, x)}^{-1/2}\hspace{0.3mm}dx\,,
\label{eqn:simplcovphase}
\end{equation}
where the tilde has been introduced to distinguish the above expression of the phase from the standard one in Eq.~(\ref{eqn:phase}).

Now, since detecting non-relativistic neutrinos is an extremely hard task, 
it is reasonable to require that
\begin{equation}
\label{eqn:relativcond}
\frac{m^2_k}{{\left[E_k^{(loc)}(x_B)\right]}^2}\,=\,f(a, x_B)\frac{m^2_k}{E_k^2}\,\ll\, 1\,.
\end{equation}
This amounts to restrict our analysis to neutrinos that are relativistic 
at the detector position $B\hspace{0.1mm}(t_B, x_B)$, and thus along all their path
$(x_B\le x\le x_A)$, for we have
\begin{equation}
\label{eqn:localenergyrelation}
f(a, x)\frac{m^2_k}{E_k^2}\,=\,\frac{f(a, x)}{f(a, x_B)}
\frac{m^2_k}{{\big[E_k^{(loc)}(x_B)\big]}^2}
\,\equiv\,\frac{1-ax}{1-ax_B}\,\frac{m_k^2}{{\big[E_k^{(loc)}(x_B)\big]}^2}\,\le\,\frac{m^2_k}{{\big[E_k^{(loc)}(x_B)\big]}^2}\,\ll\, 1\,.
\end{equation}
Equations~(\ref{eqn:relativcond}) and (\ref{eqn:localenergyrelation}) allow
us to approximate the covariant phase Eq.~(\ref{eqn:simplcovphase})
as follows
\begin{equation}
\label{eqn:covphaseapprox}
\widetilde\Phi_k\,\simeq\,-\int_{A}^{B}\frac{m^2_k}{2\hspace{0.3mm}E_0}\hspace{0.2mm}{f(a, x)}^{1/2}dx\,,
\end{equation}
where, as in the absence of acceleration, we have used the first-order
approximation $E_k\,\simeq\, E_0\, +\, 
\mathcal {O}\left(\frac{m^2_k}{2E_0}\right)$, $E_0$ being the energy 
at the origin for a massless particle. 
Since this energy is constant along the light-ray trajectory between $A$ and $B$, the integration in Eq.~(\ref{eqn:covphaseapprox}) 
can be readily performed, obtaining
\begin{equation}
\label{eqn:phaseapprox}
\widetilde\Phi_k\,\simeq\,\frac{m^2_k}{2E_0}|x_B\,-\,x_A|\left(1\,-\,\phi_a\right),
\end{equation}
where we have introduced the shorthand notation
\begin{equation}
\label{eqn:phia}
\phi_a\,\equiv\,\frac{a}{2}\left(x_B\,+\,x_A\right). 
\end{equation}
The phase-shift responsible for the oscillation thus takes the form
\begin{equation}
\label{eqn:phaseshiftapprox}
\widetilde\Phi_{12}\,\simeq\,\frac{\Delta m^2_{12}}{2E_0}|x_B\,-\,x_A|\left(1\,-\,\phi_a\right)\,=\,\frac{\Delta m^2_{12}}{2E_0}\hspace{0.3mm}L_p\left(1\,-\,\phi_a\right), 
\end{equation}
where we have used the definition of proper distance at constant time $dl\,\equiv\,\sqrt{-g_{xx}}\hspace{0.5mm}dx\,=\,dx$.

We remark that Eq.~(\ref{eqn:covphaseapprox})
does not match with the corresponding result Eq.~(25) of Ref.~\cite{Capozziello:1999ww}.
In that case, indeed, the correction on the neutrino phase-shift depends logarithmically on the acceleration.
We suspect that such a discrepancy arises because of an incorrect derivation of the final expression of the phase-shift in Ref.~\cite{Capozziello:1999ww} from the corresponding
formula in Ref.~\cite{Cardall}.

Now, in order to compare Eq.~(\ref{eqn:phaseshiftapprox})
with the standard result Eq.~(\ref{eqn:relativephase}),
let us rewrite $E_0$ in 
terms of the neutrino local energy at the detector position $B\hspace{0.1mm}(t_B, \bx_B)$. Using Eq.~(\ref{eqn:localenergy}), it  follows that 
\begin{equation}
\label{eqn:phaseshiftapprox2}
\widetilde\Phi_{12}\,\simeq\,\frac{\Delta m^2_{12}\,L_p}{2E^{(loc)}_0(x_B)\hspace{0.2mm}{f(a, x_B)}^{1/2}}\left(1\,-\,\phi_a\right).
\end{equation}
By virtue of the condition on the range of validity of the adopted metric (namely $ax\ll1$), Eq.~\eqref{eqn:phaseshiftapprox2} can be further manipulated obtaining 
\begin{equation}
\label{eqn:finalform}
\widetilde\Phi_{12}\,\simeq\,\frac{\Delta m^2_{12}\hspace{0.3mm}L_p}{2E^{(loc)}_0(x_B)}\left[1\,-\,\frac{a}{2}\hspace{0.2mm}L_p\right],
\end{equation}
where we have neglected higher order terms in the acceleration.
The first term on the r.h.s. is the 
only surviving contribution for vanishing acceleration.
As expected, it corresponds
to the standard oscillation phase in Eq.~(\ref{eqn:relativephase}).
The remaining term provides the correction induced by a uniform linear
acceleration on the neutrino oscillation phase. 

\subsection{Inertial effects on oscillations: a geometric viewpoint}
\label{Inertial effects on neutrino oscillations: a geometric viewpoint}
In the previous Section, we have derived inertial effects
on neutrino oscillations in a simple heuristic way. Using a more geometric treatment, now we want to prove that the same result  can be obtained by solving the Dirac equation 
in an accelerated frame.  As a first step, let us write down the
covariant Dirac equation in curved spacetime~\cite{Weinberg},
\begin{equation}
\label{eqn:covdireq}
\Big[i\gamma^{a}\hspace{0.2mm}e^{\mu}_{a}\left(\partial_\mu\,-\,\Gamma_\mu\right)\,-\,M\Big]\psi\,=\,0\,,
\end{equation}
where $M$ is the neutrino mass matrix and $\psi$ is a column vector of spinors
of different neutrino mass\footnote{In this Section, greek (latin) indices refer 
to general curvilinear (locally inertial) coordinates.}. The vierbein fields $e^{\mu}_a$ 
connect the general curvilinear
and locally inertial sets of coordinates. The spinorial connection $\Gamma_{\mu}$ is
accordingly defined by
\begin{equation}
\label{eqn:spinconn}
\Gamma_{\mu}\,=\,\frac{1}{8}\left[\gamma^b,\gamma^c\right]\hspace{-0.5mm}e^{\nu}_{b}\nabla_\mu e_{c\hspace{0.2mm}\nu}\,.
\end{equation}
Using the relation
\begin{equation}
\label{tot}
\gamma^a\hspace{-0.9mm}\left[\gamma^b, \gamma^c\right]\,=\,2\eta^{ab}\gamma^c\,-\,2\eta^{ac}\gamma^b\,-\,
2\hspace{0.1mm}i\hspace{0.1mm}\epsilon^{abcd}\gamma^5\gamma_d\,,
\end{equation}
we find that the only non-vanishing contribution from the spin connection is 
\begin{equation}
\gamma^a\hspace{0.1mm}e^{\mu}_a\Gamma_\mu\,=\,\gamma^a\hspace{0.1mm}e^{\mu}_a\left\{i\hspace{0.1mm}A_{G\mu}\left[-\,{\left(-g\right)}^{-1/2}\frac{\gamma^5}{2}\right]\right\}\,,
\end{equation}
where  we have denoted by $\epsilon^{abcd}$ the totally antisymmetric tensor with $\epsilon^{0123}\,=\,+1$ and
\begin{equation}
\label{eqn:A}
A_{G}^{\mu}\,=\,\frac{1}{4}{\left(-g\right)}^{1/2}e^{\mu}_a\epsilon^{abcd}\left(e_{b\sigma,\nu}\,-\,e_{b\nu,\sigma}\right)e^{\nu}_c e^{\sigma}_d\,,
\end{equation}
with $g\,\equiv\,\mathrm{det}\,g_{\mu\nu}$. 
%
Following Ref.~\cite{Cardall}, the generalized flavor neutrino state can 
be now written as
\begin{equation}
\label{eqn:genstate}
|\psi_\alpha\rangle\,=\,\sum_{k=1,2} U_{{\alpha k}}\hspace{0.1mm}(\theta)\hspace{0.1mm}e^{i\Phi}|\nu_k\rangle\,,\quad \alpha\,=\,e, \mu\,,
\end{equation}
where $U_{{\alpha k}}\hspace{0.1mm}(\theta)$ is the generic element of the Pontecorvo matrix in Eq.~(\ref{PMMmatrix}). The
neutrino oscillation phase is given by
\begin{equation}
\label{eqn:phaseneu}
\Phi\,=\,\int_{A}^{B} P_\mu\hspace{0.3mm} \frac{dx^\mu_{null}}{d\lambda}\hspace{0.3mm}d\lambda\,,
\end{equation}
where $P_\mu$ is the four-momentum operator that generates the spacetime
translation of the mass eigenstates and $\frac{dx^\mu_{null}}{d\lambda}$ is the null tangent vector to the
neutrino worldine $x^{\mu}$, parameterized by $\lambda$. For diagonal
metrics, denoting with $dl$ the differential proper distance at constant time, 
we have
\begin{equation}
\label{eqn:lambd}
d\lambda\,=\,dl{\left[g_{00}\hspace{0.2mm}{\left(\frac{dx^0}{d\lambda}\right)}^2\right]}^{-1/2}.
\end{equation}
The momentum operator $P_\mu$ can be derived from the
generalized mass-shell condition,
\begin{equation}
\label{eqn:genmascond}
\left(P^{\mu}\,-\,A^\mu_{G}\gamma^5\right)\left(P_{\mu}\,-\,A_{G\mu}\gamma^5\right)\,=\,M^2.
\end{equation}
Next, by requiring neutrino mass eigenstates to be energy eigenstates with a common energy $E_0$ and assuming the spatial components of $P^\mu$ and $\frac{dx^\mu_{null}}{d\lambda}$ to be antiparallel (as in Section~\ref{Inertial effects on neutrino oscillations: a heuristic treatment}), one has in the relativistic approximation~\cite{Cardall}
\begin{equation}
\label{eqn:pmudxmu}
P_\mu\hspace{0.3mm}\frac{dx^\mu_{null}}{d\lambda}\,=\,-\left(\frac{M^2}{2}\,+\,\frac{dx^\mu_{null}}{d\lambda}A_{G\mu}\gamma^5\,\right),
\end{equation}
where we have neglected terms of  $\mathcal{O}(A_G^2)$ and $\mathcal{O}{(A_GM^2)}$.

Let us now apply Eqs.~(\ref{eqn:A}) and (\ref{eqn:phaseneu})
to the particular case of a uniformly accelerated frame. In the same fashion as the previous heuristic analysis, we restrict to 
$1+1$-dimensions; with reference to the metric tensor Eq.~(\ref{eqn:linelementacframe}), the only non-trivial component
of the vierbein fields is thus given by
\begin{equation}
\label{eqn:vierb}
e^{\mu}_a\,=\,{f(a,x)}^{-1/2},\quad \mathrm{for}\hspace{1.2mm} \mu=a=0\,,
\end{equation}
where ${f(a,x)}$ is defined as in Eq.~(\ref{eqn:funct}).
Inserting Eq.~(\ref{eqn:vierb}) into Eq.~(\ref{eqn:A}), one directly
obtains $A^{\mu}_G\,=\,0$, yielding
\begin{equation}
\label{P}
P_\mu\hspace{0.3mm}\frac{dx^\mu_{null}}{d\lambda}\,=\,-\frac{M^2}{2}\,.
\end{equation}
By use of  Eqs.~(\ref{eqn:lambd}) and (\ref{P}), 
the phase in Eq.~(\ref{eqn:phaseneu}) then becomes
\begin{equation}
\label{tildphi}
\widetilde\Phi\,=\,\int_{A}^{B} P_\mu\hspace{0.3mm} \frac{dx^\mu_{null}}{d\lambda}\hspace{0.2mm}d\lambda\,=\,-\int_{A}^{B} \frac{M^2}{2\hspace{0.3mm}E_0}\hspace{0.2mm}{f(a, x)}^{1/2} dx\,,
\end{equation}
where we have exploited Eq.~(\ref{eqn:localenergy}) 
and the definition of proper distance introduced above.
After the mass operator in Eq.~(\ref{eqn:genstate}) has acted on $|\nu_k\rangle$,
we obtain
\begin{equation}
\widetilde\Phi_k\,=\,-\int_{A}^{B}\frac{m^2_k}{2\hspace{0.3mm}E_0}\hspace{0.2mm}{f(a, x)}^{1/2}dx\,, \quad k\,=\,1,2\,,
\end{equation}
that is the same expression derived in Eq.~(\ref{eqn:covphaseapprox}).

\subsection{Applications}
\label{Applications}
In this Section, we analyze some illustrative physical applications of our result. We begin by discussing the phenomenological implications of Eq.~(\ref{eqn:finalform}) in the framework of atmospheric neutrinos. In that case, mimicking the metric of a stationary observer on Earth with the one in Eq.~(\ref{eqn:linelementacframe}) and exploiting the equivalence principle, we can estimate the corrections induced by gravity on the probability of neutrino oscillations (we stress again that we are not concerned with effects of Earth's rotation). Then, we present a gedanken experiment in which these corrections are evaluated in more exotic regimes.

\subsubsection{Earth's gravity effects on atmospheric neutrinos}
In the context of the atmospheric neutrino problem, it is known that flavor oscillations
can be faithfully analyzed using a simplified two-generations model, since they largely occur between muonic 
and tauonic flavors ($\nu_\mu\leftrightarrow\nu_\tau$).\footnote{This happens because the mixing angle $\theta_{13}$ 
is much smaller than the others, and 
two of the neutrino
mass states are very close in mass compared to the third ($\Delta m^2_{21}\ll\Delta m^2_{32}\approx\Delta m^2_{31}$ in the normal mass hierarchy)~\cite{Tanabashi}.} 

Atmospheric neutrinos are produced in hadronic showers resulting from the interaction of cosmic rays with nuclei in the atmosphere.
Typical flight distances in experiments involving these neutrinos range from $10^2$ km (for neutrinos \emph{downward-going} from an interaction 
above the detector) to more than $10^4$ km 
(for neutrinos \emph{upward-going} from collisions on the other side of the Earth).  
We restrict to the first case, where no background matter effect occurs. 

Consider a detector comoving with the Earth: by restoring proper units in Eqs.~(\ref{pontecdetector}) and (\ref{eqn:finalform}), a straightforward calculation then leads to
\begin{equation}
\label{Pvalue}
\big|P_{\nu_\mu\rightarrow\nu_\tau}\,-\,\widetilde P_{\nu_\mu\rightarrow\nu_\tau}\big|\,=\,\mathcal{O} \hspace{-0.3mm}\left(10^{-15}\right),
\end{equation}
where we have indicated with $P$ ($\widetilde P$) the oscillation probability as measured by the inertial (accelerated) observer. To numerically evaluate Eq.~(\ref{Pvalue}), we have set a neutrino mean flight path $L_p\sim10^2\,{\mathrm{km}}$, an acceleration of the order of Earth's gravity, $a\sim10\,{\mathrm{m/s}}^2$,  $\Delta m_{\mathrm{atm}}^2\sim 10^{-3}\,{\mathrm{eV}}^2$, $E_0\sim 1\,\mathrm{GeV}$ and maximal mixing $\theta_{atm}\approx\ang{45}$~\cite{Tanabashi}.

 The obtained correction is far below the uncertainty on the current best-fit value of the oscillation probability $P_{\nu_\mu\rightarrow\nu_\tau}$, thus preventing any possibility to detect gravitational effects on atmospheric neutrino oscillations at present. Future experiments, however, may give new insights in this direction.

\subsubsection{Neutrino oscillations in extreme acceleration regime: a gedanken experiment}
We now propose a gedanken experiment in order to test our formalism in astrophysical regimes. In this framework, it is  reasonable to expect a larger contribution of gravitational effects on the oscillation probability, due to the extremely high accelerations that might be reached in this case. 

As proof of this, let us deal with an ideal accelerated detector in proximity of a high-density object; for instance, consider the case of Sirius B, 
the nearest (known) white dwarf to the Earth.
It is known that the gravity on the surface of this star is of the order of $10^6\, \mathrm{m/s^2}$~\cite{wd}. For such an acceleration, using Eq.~(\ref{pontecdetector}) and (\ref{eqn:finalform}), we obtain 
\begin{equation}
\label{resu}
\big|P_{\nu_\alpha\rightarrow\nu_\beta}\,-\,\widetilde P_{\nu_\alpha\rightarrow\nu_\beta}\big|\,=\,\mathcal{O}\hspace{-0.3mm} \left(10^{-4}\right), 
\end{equation}
where we have set $L_p\sim10^4\,{\mathrm{km}}$ and, as for the previous case, $\Delta m^2\sim10^{-3}\,{\mathrm{eV}}^2$, $E_0\sim 1\,\mathrm{GeV}$ and maximal mixing $\theta\approx \ang{45}$.

As predicted, inertial effects may not be completely negligible in this case. However, it is worth saying that experiments like the one  considered above are far from being viable nowadays. Indeed, it would be technically cumbersome to build a detector able to resist those extreme accelerations without breaking: on the other hand, even if it were possible, then the problem would arise of how to send and retrieve a probe from the surface of such remote sources (Sirius B, for example, lies at a distance of $8.60$ light years away from the Sun).

Notwithstanding these technical difficulties, some of the implications of the result Eq.~\eqref{resu} in the physics of neutrino oscillations will be discussed in the next Section.

\section{Discussion and Conclusions}
\label{Discandcon}

In this Chapter we have addressed the topic of flavor mixing for an accelerated observer within the framework of QFT, discussing both cases of scalar and  fermionic (Dirac) fields. Due to the combination of two Bogoliubov transformations -- the one hiding in flavor mixing, the other associated with the Rindler spacetime structure -- we have found that the spectrum of Unruh radiation get significantly modified, resulting in the sum of the standard Bose-Einstein (Fermi-Dirac) distribution plus non-trivial corrections arising from mixing. The explicit calculation of these additional terms has been performed in the limit of small mass difference, showing how the Unruh radiation does indeed lose its characteristic thermality when mixed fields are involved.

As an aside, we have investigated the effects of a uniform linear
acceleration on the  oscillation probability formula. Because of some 
both technical and conceptual difficulties
in the QFT framework, however,
such an analysis has been preliminarily performed in the context of Quantum Mechanics.
Corrections on the standard result 
have been derived by use of Stodolsky covariant definition 
of neutrino quantum phase.
Relying on phenomenological considerations,
we have restricted our discussion to relativistic neutrinos,
so that a plane wave treatment could be applied.
In order to realize how acceleration affects the well-known Pontecorvo achievement, 
the formalism of neutrino 
oscillations in curved spacetime has been used.
Within such a framework, we have found that inertial effects 
are intimately related to the redshift of neutrino energy, 
according to Ref.~\cite{Cardall}. Furthermore, it has 
been pointed out that a separate ``acceleration phase" can be 
extracted from the standard result only for small accelerations.

As a possible application of our analysis, we have
calculated the correction induced by Earth's gravity on the oscillation
probability of atmospheric neutrinos. In that case, simulating the metric of an observer comoving with the Earth with the one in Eq.~(\ref{eqn:linelementacframe}),  we have found that the contribution to the neutrino phase-shift is negligible, 
thus leading to unmeasurable effects
at present.
It is clear that the origin of this outcome can be traced back to the difficulty of detecting gravitational effects on oscillations in the weak-field regime, as it is near the Earth.
On the other hand, in astrophysical regimes, \emph{e.g.} outside a black hole or in proximity of pulsars,
we expect these corrections to be far more relevant (as also suggested by the analysis carried out for a white dwarf), resulting in a modification of the oscillation probability induced by gravity (see also Ref.~\cite{Miller:2013wta} for a quasi-classical treatment of neutrino oscillations in
the gravitational field of an heavy astrophysical object). If confirmed, such a resonant  effect could be exploited for investigating the gravity-induced interactions that neutrinos may have experienced during their travel throughout the Universe, and thus the mass distribution of the Universe itself. Moreover, oscillations of neutrinos from supernovae and active galactic nuclei may be valuable to search footprints of space-time quantum foam~\cite{KlapdorKleingrothaus:2000fr}.
Non-trivial implications also arise in the context of supernova nucleosynthesis, and, in particular, in the enhancement of the production of heavy elements due to neutrino emission from proto-neutron stars and in the nucleosynthesis in outer shells of supernovae (see, for example, Ref.~\cite{Wu:2014kaa} and therein).
In light of the central role played by supernovae in the production of various heavy nuclei, such an analysis may provide insights towards a better understanding of the evolution of the chemical composition of the Universe. A further interesting scenario to be addressed is the relevance of neutrino oscillations in the generation of the rotational pulsar velocity in the presence of intense magnetic fields~\cite{Kusenko:1996sr}.
These aspects, however, will be matter of future investigation.

In addition to these phenomenological aspects, we emphasize
that the analysis of gravity/inertial effects on neutrino mixing and oscillations may be 
enlightening in a variety of controversial issues at a theoretical level. 
In a series of works, for instance, it has been shown neutrino oscillations can be equivalently described in terms of \emph{entanglement} of neutrino flavor modes, within both the quantum mechanical and field theoretical  frameworks (for a more detailed discussion on this, see the end of Appendix~\ref{QFT of fm}). So far, however, such an analysis has been only performed in the  standard Minkowski framework for inertial observers: the question thus arises as to how this formalism gets modified in the Rindler background, where, according to Refs.~\cite{FuentesSchuller:2004xp,Hwang:2010ib} and therein, the loss of information due to the horizon structure would lead to a degradation in the entanglement properties.  Along this line, relevant processes in the presence of gravity or in non-inertial frames have also been addressed in Refs.~\cite{LambiasePapini,Dvornikov:2006ji,Dvornikov:2015tza}. In Ref.~\cite{Dvornikov:2015tza}, in particular, it has been shown that the \emph{oscillation resonance condition} of neutrinos propagating in  a rotating background matter (\emph{e.g.} a millisecond pulsar) turns out to be non-trivially modified due to interaction terms. Work is clearly in progress to settle these
(and many other) issues concerning neutrino mixing and oscillations in accelerated frames.

\newpage\null\thispagestyle{empty}\newpage

\chapter{The necessity of the Unruh effect in QFT: \emph{the floor to} the proton}
\label{The necessity of the Unruh effect in QFT: the proton}
\begin{flushright}
\emph{``Think like a proton,\\ always positive.''}\\[1mm]
- Anonymous Aphorism -\\[6mm]
\end{flushright} 
In Chapter~\ref{The fascinating puzzle of quantum vacuum: the Hawking-Unruh effect} it has been argued that macroscopic observations of the Unruh effect are extremely challenging, since accelerations of the order of $10^{20}\,\mathrm{m/s^2}$ corresponding to temperature of $1\,\mathrm{K}$ are hard to be reached for ordinary objects. Therefore, from the very beginning, all hopes of detecting 
the Unruh radiation have been pinned on elementary
particle experiments. Even within such a framework, however, matters appear 
somewhat convoluted, although an increasing number of creative tests has been proposed during the last decades (see, for example, Ref.~\cite{Rosu:1994ug} for a full list).

In spite of these experimental failings, it is not questionable that the Unruh effect is   
is a \emph{key ingredient} of QFT, at least at a theoretical level. 
This has been shown in different contexts in Refs.~\cite{Matsas:1999jx,Matsas:1999jx2,Matsas:1999jx3,Higuchi:1992we}. In Refs.~\cite{Matsas:1999jx,Matsas:1999jx2,Matsas:1999jx3}, in particular, the attention was first called on the fact that inertial and uniformly accelerated observers would measure incompatible proper lifetimes for non-inertial protons if the Unruh thermal condensate did not exist, thus violating the general covariance  of QFT.

In the present Chapter, after a brief \emph{historical excursus} on the concerning literature, we investigate in detail the inverse $\beta$-decay of accelerated protons,
\be
\label{protondecaylab}
p\,\rightarrow\, n\, +\, e^{+}\,+\,\nu_e\,,
\ee
clarifying some ambiguities recently raised in literature about the agreement of the decay rates in the laboratory and comoving frames when flavor mixing for neutrinos is taken into account~\cite{Ahluwalia:2016wmf,Blasone:2018czm, BlasonePOS, Cozzella:2018qew}.
 
\section{Historical excursus}
It was pointed out by M\"{u}ller~\cite{Muller:1997rt} that the decay properties of particles can be changed by acceleration. In particular, it was shown that usually forbidden processes such as the decay of the proton
become kinematically possible under the
influence of acceleration, thus leading to a finite lifetime for even supposedly
stable particles.
Drawing inspiration from this idea, Matsas and Vanzella~\cite{Matsas:1999jx,Matsas:1999jx2,Matsas:1999jx3} analyzed the decay of uniformly accelerated protons in both the laboratory and comoving frames, showing that the two rates perfectly agree only when one considers Minkowski vacuum to appear as a thermal bath of neutrinos and electrons for the accelerated observer (comoving frame). This has been exhibited as a ``theoretical check" of the Fulling-Davies-Unruh (FDU) effect\footnote{Similar arguments apply, for example, to the analysis of the QED bremsstrahlung radiation. In this case, it has been shown that the emission of a photon by accelerated charges in the inertial frame can be seen  as either
the emission or the absorption of a zero-energy photon in the FDU thermal bath of the comoving observer~\cite{Higuchi:1992we, Crispino:2007eb}. A closely related discussion about
whether or not uniformly accelerated charges emit radiation  according to inertial observers can be found in Refs.~\cite{Fult}, where the problem is addressed in a classical context.}, the  implications of which in QFT are still matter of study~\cite{Becattini:2017ljh, Goto:2018ijz}. For technical reasons, the analysis of Refs.~\cite{Matsas:1999jx,Matsas:1999jx2,Matsas:1999jx3} was performed in $1+1$-dimensions and taking the neutrino to be massless. Subsequently, Suzuki and Yamada~\cite{Suzuki:2002xg} confirmed these results by extending the discussion to the $1+3$-dimensional case with massive neutrino. 

Recently, Ahluwalia, Labun and Torrieri~\cite{Ahluwalia:2016wmf} made the very intriguing observation that neutrino mixing can have non-trivial consequences in this context. They indeed found that the decay rates in the two frames could possibly not agree due to mixing terms: in particular, this  happens when  neutrino mass eigenstates are taken as asymptotic states in the comoving frame, a choice which is compatible with the Kubo-Martin-Schwinger (KMS) condition for thermal states~\cite{Crispino:2007eb,Haag:1967sg}. On the other hand, the authors of Ref.~\cite{Ahluwalia:2016wmf} also affirm that the choice of flavor states in the above calculation would lead to an equality of the two decay rates, but in that case the accelerated neutrino vacuum would not be thermal, contradicting the essential characteristic of the FDU effect.  They finally conclude  that such a contradiction has to be solved experimentally.

Motivated by the idea that the above question must instead be settled at a theoretical level in order to guarantee the consistency of the theory in conformity with the principle of general covariance, in this Chapter we carefully analyze the proton decay process in the presence of neutrino mixing. We show that, in Ref.~\cite{Ahluwalia:2016wmf}, calculations performed in the laboratory frame neglected an important contribution which here is explicitly evaluated. Then, we prove that the choice of neutrinos with definite mass as asymptotic states (in the comoving frame) inevitably leads to a disagreement of the two decay rates. Finally, we consider the case  when flavor states are taken into account: here technical difficulties arise which do not allow for an exact evaluation of the decay rates. However, adopting an appropriate approximation, we show that they perfectly match again. 

These conclusions are consistent with results on the quantization for mixed neutrino fields, for which flavor states are rigorously defined as eigenstates of the leptonic charge operators~\cite{Blasone:2001qa}. Although the usual Pontecorvo states turn out to be a good approximation of the exact flavor eigenstates in the ultrarelativistic limit, the Hilbert space associated to flavor neutrinos is actually orthogonal to the one for massive neutrinos~\cite{Blasone:1995zc}. Consistency with the Standard Model (SM) requires conservation (at tree level) of leptonic number in the charged current weak interaction vertices, thus ruling out the choice of neutrino mass eigenstates as asymptotic states. 

Results of the present Chapter corroborate this view, although further investigation is needed to go beyond the aforementioned approximation. In particular, when employing the exact neutrino flavor states, one should take into account the non-thermal character of Unruh radiation, as discussed in previous Chapters.

The Chapter is organized as follows. Section~\ref{Decay of accelerated protons: a brief review} is devoted to briefly review the standard calculation of the proton decay rate in both  the inertial and comoving frames. For this purpose, we closely follow Ref.~\cite{Suzuki:2002xg}. In Section~\ref{prdec} we analyze the same process in the context  of neutrino flavor mixing. Working within a suitable framework, we show the decay rates to agree with each other. Our results shall be critically compared with the ones of Ref.~\cite{Ahluwalia:2016wmf}, where a dichotomy is instead highlighted. Furthermore, as an aside, in Section~\ref{Cozzella} we comment on the recent work by Cozzella \emph{et al.}~\cite{Cozzella:2018qew}, in which a similar subject, even though with a different approach, is addressed. Section~\ref{Conclusionsproton} contains conclusions and an outlook at future developments of the present analysis.

\section{Decay of accelerated protons: a brief review}
\label{Decay of accelerated protons: a brief review}
In this Section we discuss the decay of accelerated protons in both  the laboratory and comoving frames. Throughout the whole analysis, neutron $|n\rangle$ and proton $|p\rangle$ are considered as excited and unexcited states of the nucleon, respectively. Moreover, we assume that they are energetic enough to have a well defined trajectory. As a consequence,  the current-current interaction of Fermi theory can be treated with a classical hadronic current ${J}^{\mu}_{\ell}{J}_{h,\mu}\,\rightarrow\,{J}^{\mu}_{\ell}{J}^{(cl)}_{h,\mu}$, where 
\be
\label{Jcl}
{J}^{(cl)}_{h,\mu}\ =\ {q}(\tau)\hspace{0.2mm}u_{\mu}\hspace{0.2mm}\delta(x)\hspace{0.2mm}\delta(y)\hspace{0.2mm}\delta(u-a^{-1})\,.
\ee
Here $u\,=\,a^{-1}\,=\,\mathrm{const}$ is the spatial Rindler coordinate describing the world line of the uniformly accelerated nucleon with proper acceleration $a$, and $\tau\,=\,v/a$ is its proper time, with $v$ being the Rindler time coordinate. The nucleon's four-velocity $u^{\mu}$ is given by
\be
u^{\mu}\ =\ (a, 0, 0, 0), \qquad u^{\mu}\ =\ (\sqrt{a^2t^2+1}, 0, 0, a\hspace{0.2mm}t)\,,
\ee  
in Rindler and Minkowski coordinates, respectively\footnote{To keep the notation
of Ref.~\cite{Blasone:2018czm} untouched, throughout this Chapter we shall denote
the Rindler time by $v$ rather than $\eta$; additionally, 
we will assume the proton to be accelerated along the 
$z$-direction. Hence, in what follows,
the Rindler coordinates $(v, x, y, u)$ will be related to
the Minkowski ones $(t, x, y, z)$ by: $t=u\sinh{v}$, $z=u\cosh{v}$, 
with $x$ and $y$ left unchanged.}. 
According to Refs.~\cite{Birrell,Matsas:1999jx,Matsas:1999jx2,Matsas:1999jx3}, the Hermitian monopole ${q}(\tau)$  is defined as
\be
{q}(\tau)\ \equiv\ e^{i{H}\tau}\hspace{0.2mm} {q}_0\hspace{0.2mm}e^{-i{H}\tau},
\ee 
where ${H}$ is the nucleon Hamiltonian and $ {q}_0$ is related to the Fermi constant $G_F$ by
\be
G_F\ \equiv\ \langle p\hspace{0.2mm}|\hspace{0.2mm} q_0\hspace{0.2mm}|\hspace{0.2mm}n\rangle.
\ee
Next, the minimal coupling of the electron ${\Psi}_{e}$ and neutrino $\Psi_{\nu_e}$ fields to the nucleon current ${J}^{(cl)}_{h,\mu}$ can be expressed through the Fermi action
\be
\label{eqn:Fermiaction}
{S}_{I}\ =\ \hspace{-0.5mm}\int d^{4}x\hspace{0.2mm}\sqrt{-g}\hspace{0.3mm}{J}^{(cl)}_{h,\mu}\left({\overline{\Psi}}_{\nu_e}\gamma^{\mu}{\Psi}_{e}\, +\, {\overline{\Psi}}_{e}\gamma^{\mu}{\Psi}_{\nu_e}\right)\hspace{-0.2mm},
\ee
where $g\equiv \mathrm{det}(g_{\mu\nu})$ and $\gamma^\mu$ are the gamma matrices in Dirac representation (see, \emph{e.g.}, Ref.~\cite{Itzykson}).

\subsection{Inertial frame calculation}
\label{subsect1}
Let us firstly analyze the decay process in the inertial frame. In this case, the proton is accelerated by an external field and is converted into a neutron by emitting a positron and a neutrino, according to 
\be
p\,\rightarrow\, n\, +\, e^{+}\,+\,\nu_e\,.
\ee
In order to calculate the transition rate, we quantize fermionic fields in the usual way \cite{Matsas:1999jx,Matsas:1999jx2,Matsas:1999jx3}:
\be
\label{inertexp}
{\Psi}(t,\bx)\ =\ \sum_{\sigma=\pm}\int d^3k\left[{b}_{\bk\sigma}\psi_{\bk\sigma}^{(+\omega)}(t,\bx)\,+\,{d}_{\bk\sigma}^{\dagger}\psi_{-\bk-\sigma}^{(-\omega)}(t,\bx)\right],
\ee
where $\bx\,\equiv\,(x,y,z)$. Here we have denoted by ${b}_{\bk\sigma}$ (${d}_{\bk\sigma}$) the canonical annihilation operators of fermions (antifermions) with momentum $\bk\,\equiv\,(k^x, k^y, k^z)$, polarization $\sigma\,=\,\pm$ and frequency $\omega\,=\,\sqrt{\bk^{2}+m^{2}}\,>\,0$, $m$ being the mass of the field.  The modes $\psi_{\bk\sigma}^{(\pm\omega)}$ are positive and negative energy solutions of the Dirac equation in Minkowski spacetime:
\be
\big(i\gamma^{\mu}\partial_{\mu}\,-\,m\big)\psi_{\bk\sigma}^{(\pm\omega)}(t, \bx)\ =\ 0\,.
\ee
In the adopted representation for the $\gamma$ matrices, they take the form~\cite{Matsas:1999jx,Matsas:1999jx2,Matsas:1999jx3}
\be
\label{modes}
\psi_{\bk\sigma}^{(\pm\omega)}(t, \bx) \ =\ \frac{e^{i(\bk\cdot\bx\,\mp\,\omega t)}}{2^2\hspace{0.2mm}\pi^{\frac{3}{2}}}\hspace{0.2mm}\hspace{0.2mm}u_{\sigma}^{(\pm\omega)}(\bk),
\ee 
where
\be
u_{+}^{(\pm\omega)}(\bk)\ =\ \frac{1}{\sqrt{\omega(\omega\pm m)}}
\begin{pmatrix}
m\,\pm\,\omega \\
0 \\
k^z \\
k^x\,+\,ik^y
\end{pmatrix},\qquad
u_{-}^{(\pm\omega)}(\bk)\ =\ \frac{1}{\sqrt{\omega(\omega\pm m)}}
\begin{pmatrix}
0 \\
m\,\pm\,\omega \\
k^x\,-\,ik^y \\
-\hspace{0.2mm}k^z
\end{pmatrix}.
\ee
It is easy to show that the modes Eq.~(\ref{modes}) are orthonormal with respect to the inner product 
\be
\label{innprod}
\left\langle\psi_{\bk\sigma}^{(\pm\omega)},\psi_{\bk'\sigma'}^{(\pm\omega')}\right\rangle=\int_{\Sigma}d\Sigma_{\mu}\,\overline{\psi}_{\bk\sigma}^{(\pm\omega)}\gamma^{\mu}\psi_{\bk'\sigma'}^{(\pm\omega')}\ =\ \delta_{\sigma\sigma'}\,\delta^3(k-k')\,\delta_{\pm\omega\pm\omega'},
\ee
where $\overline{\psi}\,=\,\psi^{\dagger}\gamma^{0}$, $d\Sigma_{\mu}\,=\,n_{\mu}d\Sigma$, with $n_{\mu}$ being a unit vector orthogonal to the arbitrary spacelike hypersurface $\Sigma$ and pointing to the future. 

Next, by using the definition Eq.~(\ref{eqn:Fermiaction}) of the Fermi action and expanding leptonic fields according to Eq.~(\ref{inertexp}), we obtain the following expression for the tree-level transition amplitude:
\begin{equation}
\mathcal{A}^{p\rightarrow n}_{in}\ \equiv\ \langle n|\otimes\langle e_{k_{e}\sigma_{e}}^{+},\nu_{k_{\nu}\sigma_{\nu}}|{S}_{I}|0\rangle\otimes|p\rangle\ =\ \frac{G_F}{2^4\pi^3}\,\mathcal{I}_{\sigma_\nu\sigma_e}(\omega_\nu, \omega_e),
\label{tramp}
\end{equation}
where 
\be
\label{processI}
\mathcal{I}_{\sigma_\nu\sigma_e}(\omega_\nu, \omega_e)\ =\ \hspace{-0.5mm}\int_{-\infty}^{+\infty}\hspace{-2.3mm}d\tau\, e^{i\big[\Delta m\hspace{0.2mm}\tau\,+\,a^{-1}\left(\omega_\nu\,+\,\omega_{e}\right)\sinh a\tau\,-\,a^{-1}\left(k_\nu^z\,+\,k^z_{e}\right)\cosh a\tau\big]}u_{\mu}\left[\bar{u}_{\sigma_\nu}^{(+\omega_\nu)}\gamma^{\mu}{u}_{-\sigma_e}^{(-\omega_e)}\right].
\ee
Here $\Delta m$ is the difference between the nucleon masses. By defining the differential transition rate as
\begin{eqnarray}
\label{dtr}
\frac{d^{6}\mathcal{P}_{in}^{p\rightarrow n}}{d^3k_{\nu}\,d^3k_{e}}&\equiv&\sum_{\sigma_{\nu},\sigma_e}\left|\mathcal{A}_{in}^{p\rightarrow n}\right|^{2} \\[3mm]
\non
&=&\frac{G_F^2}{2^8\pi^6}\int_{-\infty}^{+\infty}\hspace{-2mm}d\tau_1\int_{-\infty}^{+\infty}\hspace{-2mm}d\tau_2\hspace{0.3mm}u_{\mu}u_{\nu}\sum_{\sigma_\nu,\sigma_e}\left[\bar{u}_{\sigma_\nu}^{(+\omega_\nu)}\gamma^{\mu}{u}_{-\sigma_e}^{(-\omega_e)}\right]\left[\bar{u}_{\sigma_\nu}^{(+\omega_\nu)}\gamma^{\nu}{u}_{-\sigma_e}^{(-\omega_e)}\right]^*\\[3mm]
\non
&&\times \;e^{i\big[\Delta m(\tau_1-\tau_2)+a^{-1}\left(\omega_\nu+\omega_{e}\right)(\sinh a\tau_1-\sinh a\tau_2)-a^{-1}\left(k_\nu^z+k^z_{e}\right)(\cosh a\tau_1-\cosh a\tau_2)\big]}\,,
\end{eqnarray}
the total transition rate is simply given by
\be
\label{ttr}
\Gamma_{in}^{p\rightarrow n}\ =\ \mathcal{P}_{in}^{p\rightarrow n}/T\,,
\ee 
where $T\,=\,\int_{-\infty}^{+\infty}ds$ is the nucleon proper time. The above integrals can be  solved by introducing the new variables 
\be
\label{newvariab}
\tau_1\ =\ s\,+\,\xi/2,\qquad \tau_2\ =\ s\,-\,\xi/2\,,
\ee
and using the spin sum
\begin{eqnarray}
\label{usumspin}
\non
&&\hspace{-10mm}u_{\mu}u_{\nu}\sum_{\sigma_\nu,\sigma_e}\left[\bar{u}_{\sigma_\nu}^{(+\omega_\nu)}\gamma^{\mu}{u}_{-\sigma_e}^{(-\omega_e)}\right]\left[\bar{u}_{\sigma_\nu}^{(+\omega_\nu)}\gamma^{\nu}{u}_{-\sigma_e}^{(-\omega_e)}\right]^*\\[2mm]\non
&&\hspace{-5mm}\,=\,\frac{2^2}{\omega_{\nu}\omega_e}\,\bigg[\big(\omega_\nu\omega_e\,+\,k^z_\nu k^z_e\big)\cosh 2as\,-\,\big(\omega_\nu k^z_e\,+\,\omega_ek^z_\nu\big)\sinh 2as\\[2mm]
&&\hspace{0.1mm}\,+\,\big(k^x_\nu k^x_e\,+\,k^y_\nu k^y_e\,-\,m_\nu m_e\big)\cosh a\xi\bigg].
\end{eqnarray}
By explicit calculation, we obtain
\begin{eqnarray}\label{gammain}
\non
\Gamma_{in}^{p\rightarrow n}&=& \frac{G_F^2}{a\,\pi^6e^{\pi\Delta m/a}}\int d^3 k_\nu\int d^3 k_e \left[K_{2i\Delta m/a}\left(\frac{2(\omega_\nu\,+\,\omega_e)}{a}\right)\right.\\[3mm]
&&\left.+\;\frac{m_\nu m_e}{\omega_\nu\omega_e}\,\mathrm{Re}\left\{K_{2i\Delta m/a+2}\left(\frac{2(\omega_\nu\,+\,\omega_e)}{a}\right)\right\}\right].
\end{eqnarray}
The analytic evaluation of the integral Eq.~(\ref{gammain}) can be found in Ref.~\cite{Suzuki:2002xg}.

\subsection{Comoving frame calculation}
We now analyze the same decay process in the proton comoving frame. As well known, the natural manifold to describe phenomena for uniformly accelerated observers is the Rindler wedge, i.e., the Minkowski spacetime region defined by $z>\left|t\right|$. Within such a manifold, fermionic fields are expanded in terms of the positive and negative frequency solutions of the Dirac equation with respect to the boost Killing vector $\partial/\partial v$~\cite{Suzuki:2002xg}:
\be 
\label{Rindexp}
{\Psi}(v,\bx)\ =\ \sum_{\sigma=\pm}\int_{0}^{+\infty}\hspace{-2mm}d\omega\int d^2k\left[{b}_{\bw\sigma}\psi^{(+\omega)}_{\bw\sigma}(v,\bx)\,+\,{d}_{\bw\sigma}^{\dagger}\psi^{(-\omega)}_{\bw-\sigma}(v,\bx)\right],
\ee
where now $\bx\,\equiv\,(x,y,u)$ and $\bw\,\equiv\,(\omega, k^x, k^y)$. We recall that the Rindler frequency $\omega$ may assume arbitrary positive real values. In particular, unlike the inertial case, there are massive Rindler particles with zero frequency. 

The modes $\psi_{\bk\sigma}^{(\pm\omega)}$ in Eq.~(\ref{Rindexp}) are positive and negative energy solutions of the Dirac equation in Rindler spacetime:
\be
\label{DR}
(i\gamma^{\mu}_R\widetilde\nabla_\mu\,-\,m)\hspace{0.2mm}\psi_{\bw\sigma}^{(\omega)}(v, \bx)\ =\ 0,
\ee
where
\begin{eqnarray}
&\gamma^{\mu}_R\ \equiv\ (e_\nu)^{\mu}\gamma^{\nu},\qquad (e_0)^{\mu}\ =\ u^{-1}\delta_0^{\mu},\qquad (e_i)^{\mu}\ =\ \delta_i^{\mu},&\\[3mm]
&\tilde\nabla_{\mu}\ \equiv\ \partial_{\mu}\,+\,\frac{1}{8}\left[\gamma^{\alpha}, \gamma^\beta\right]\hspace{-0.3mm}{\left(e_\alpha\right)}^\lambda\nabla_\mu(e_\beta)_\lambda.&
\label{DiracRind}
\end{eqnarray}
By virtue of these relations and using the Rindler coordinates, Eq.~(\ref{DR}) becomes
\be
i\frac{\partial \psi^{(\omega)}_{\bw\sigma}(v, \bx)}{\partial v}\ =\ \left(\gamma^0mu\,-\,\frac{i\alpha^3}{2}\,-\,iu\alpha^i\partial_i\right)\hspace{-0.2mm}\psi^{(\omega)}_{\bw\sigma}(v, \bx),\qquad \alpha^i\,=\,\gamma^0\gamma^i, \,\,\,i=1,2,3\,,
\ee
the solutions of which can be written in the form~\cite{Suzuki:2002xg}
\be
\label{modesRind}
\psi_{\bw\sigma}^{(\omega)}(v, \bx)\ =\ \frac{e^{i(k_\alpha x^\alpha\,-\,\omega v/a)}}{{(2\pi)}^{\frac{3}{2}}}\hspace{0.4mm}u_{\sigma}^{(\omega)}\hspace{0.2mm}(u,\bw),\qquad \alpha=1,2,
\ee 
with
\begin{eqnarray}
u_{+}^{(\omega)}(u,\bw)&=&
N\begin{pmatrix}
i\hspace{0.2mm}l\hspace{0.1mm}K_{i\omega/a-1/2}(u\hspace{0.2mm}l)\,+\,m\hspace{0.2mm}K_{i\omega/a+1/2}(ul) \\[3mm]
-(k^x\,+\,ik^y)K_{i\omega/a+1/2}(ul) \\[3mm]
i\hspace{0.2mm}l\hspace{0.1mm}K_{i\omega/a-1/2}(u\hspace{0.2mm}l)\,-\,m\hspace{0.2mm}K_{i\omega/a\,+\,1/2}(ul) \\[3mm]
-(k^x+ik^y)K_{i\omega/a+1/2}(ul)
\end{pmatrix}\hspace{-0.6mm},\\[8mm]
u_{-}^{(\omega)}(u,\bw)&=&
N\begin{pmatrix}
(k^x\,-\,ik^y) K_{i\omega+1/2}(ul) \\[3mm]
i\hspace{0.2mm}l\hspace{0.1mm}K_{i\omega/a-1/2}(u\hspace{0.2mm}l)\,+\,m\hspace{0.2mm}K_{i\omega/a+1/2}(ul) \\[3mm]
-(k^x\,-\,ik^y) K_{i\omega+1/2}(ul) \\[3mm]
-i\hspace{0.2mm}l\hspace{0.1mm}K_{i\omega/a-1/2}(u\hspace{0.2mm}l)\,+\,m\hspace{0.2mm}K_{i\omega/a+1/2}(ul) 
\end{pmatrix}.
\end{eqnarray}
Here we have denoted by $K_{i\omega/a+1/2}(ul)$ the modified Bessel function of the second kind with complex order,  $N\,=\,\sqrt{\frac{a\cosh(\pi\omega/a)}{\pi l}}$ and $l\,=\,\sqrt{m^2+(k^x)^2+(k^y)^2}$. Again, one can verify that the modes in Eq.~(\ref{modesRind}) are normalized with respect to the inner product Eq.~(\ref{innprod}) expressed in Rindler coordinates.
\smallskip

As it will be shown, in the comoving frame the proton decay is represented in terms of Rindler particles  as the combination of the three following processes~\cite{Matsas:1999jx,Matsas:1999jx2,Matsas:1999jx3}:
\be
\label{threeprocesses}
(i)\quad p^{+}\,+\,e^{-}\,\rightarrow\, n\,+\,\nu_e, \qquad (ii)\quad p^{+}\,+\,\overline{\nu}_e\,\rightarrow\, n\,+\,e^{+}, \qquad (iii)\quad p^{+}\,+\,e^{-}\,+\,\overline{\nu}_e\,\rightarrow\, n.
\ee   
These processes are characterized by the conversion of
protons in neutrons due to the absorption of $e^-$ and $\bar{\nu}_e$, and
emission of $e^+$ and $\bar{\nu}_e$ from and to the FDU thermal bath~\cite{Unruh:1976db}. Since the strategy for calculating the transition amplitude is the same for each of these processes, by way of illustration we shall focus on the first. 

By exploiting the Rindler expansion Eq.~(\ref{Rindexp}) for the electron and neutrino fields, it can be shown that
\begin{equation}
\label{first}
\mathcal{A}^{p\rightarrow n}_{(i)}\ \equiv\ \left\langle n\right|\otimes\langle\nu_{\omega_\nu\hspace{0.2mm}\sigma_{\nu}}|\hspace{0.2mm}{S}_{I}\hspace{0.2mm}|e^{-}_{\omega_{e^-}\hspace{0.2mm}\sigma_{e^-}}\rangle\otimes\left|p\right\rangle\ =\ \frac{G_F}{(2\pi)^2}\,\mathcal{J}_{\sigma_\nu\sigma_e}(\omega_\nu, \omega_e),
\end{equation}
where ${S}_I$ is given by Eq.~(\ref{eqn:Fermiaction}) with $\gamma^\mu$ replaced by the Rindler gamma matrices $\gamma^{\mu}_R$ defined in Eq.~(\ref{DiracRind}) and
\be
\label{eqn:j}
\mathcal{J}_{\sigma_\nu\sigma_e}(\omega_\nu, \omega_e)\ =\ \delta\big(\omega_e-\omega_{\nu}-\Delta m\big)\,\bar{u}_{\sigma_{\nu}}^{(\omega_{\nu})}\gamma^0 u_{\sigma_{e}}^{(\omega_{e})}.
\ee
Now, bearing in mind that the probability for the proton to absorb (emit) a particle of frequency $\omega$ from (to) the thermal bath is $N_{FD}(\omega)$ $\big(1\,-\,N_{FD}(\omega)\big)$, with $N_{FD}$ given in Eq.~\eqref{FDstat}, the differential transition rate per unit time for the process $(i)$ can be readily evaluated, yielding
\begin{eqnarray}
\label{eqn:diftransraterind}
\frac{1}{T}\frac{d^{6}\mathcal{P}^{p\rightarrow n}_{(i)}}{d\omega_{\nu}\hspace{0.2mm}d\omega_{e}\hspace{0.2mm}d^2k_{\nu}\hspace{0.2mm}d^2k_e}&\equiv&\frac{1}{T}\sum_{\sigma_{\nu},\sigma_e}\big|\mathcal{A}^{p\rightarrow n}_{(i)}\big|^{2}\,N_{FD}(\omega_{e})\big[1\,-\,N_{FD}(\omega_\nu)\big]\\[3mm]
\nonumber
&=&\frac{\hspace{0.2mm}G_{F}^{2}}{2^7\pi^{5}}\hspace{0.2mm}\frac{\sum_{\sigma_{\nu},\sigma_e}\big|\bar{u}_{\sigma_{\nu}}^{(\omega_{\nu})}\gamma^0 u_{\sigma_{e}}^{(\omega_{e})}\big|^2\hspace{0.2mm}\delta\left(\omega_e-\omega_{\nu}-\Delta m\right)}{e^{\pi\Delta m/a}\cosh(\pi\omega_{\nu}/a)\cosh(\pi\omega_{e}/a)},
\end{eqnarray}
where $T=2\pi\delta(0)$ is the total proper time of the proton. To finalize the evaluation of the transition rate, we observe that
\bea
\label{spinsum2}
\non
\sum_{\sigma_{\nu},\sigma_e}\big|\bar{u}_{\sigma_{\nu}}^{(\omega_{\nu})}\gamma^0 u_{\sigma_{e}}^{(\omega_{e})}\big|^2&=&\frac{2^4}{(a\,\pi)^2}\cosh(\pi\omega_{\nu}/a)\cosh(\pi\omega_{e}/a)\\[2mm]
\non
&&\times\;\Bigg[l_{\nu}l_e\bigg|K_{i\omega_{\nu}/a+1/2}\left(\frac{l_{\nu}}{a}\right)K_{i\omega_{e}/a+1/2}\left(\frac{l_{e}}{a}\right)\bigg|^2
\\[3mm]
\non
&&+\;\left(k^x_\nu k^x_e\,+\,k^y_\nu k^y_e\,+\,m_{\nu}m_e\right)\\[3mm]
&&\times\;\mathrm{Re}\left\{K^2_{i\omega_{\nu}/a-1/2}\left(\frac{l_{\nu}}{a}\right)K^2_{i\omega_{e}/a+1/2}\left(\frac{l_{e}}{a}\right)\right\}\Bigg].
\eea
Using this equation, the differential transition rate for the process $(i)$ takes the form
\bea\nonumber
\label{difftransrate}
\hspace{-2mm}\frac{1}{T}\frac{d^{6}\mathcal{P}^{p\rightarrow n}_{(i)}}{d\omega_{\nu}d\omega_{e}d^2k_{\nu}d^2k_e}&\equiv&\frac{G_F^2}{2^3\,a^2\,\pi^7\,e^{\pi\Delta m/a}}\,\delta\hspace{-0.2mm}\left(\omega_e-\omega_{\nu}-\Delta m\right)\\[3mm]
\non
&&\times\,\left[l_{\nu}l_e\biggl|K_{i\omega_{\nu}/a+1/2}\left(\frac{l_{\nu}}{a}\right)K_{i\omega_{e}/a+1/2}\left(\frac{l_{e}}{a}\right)\bigg|^2\right.\\[3mm]
&&\left.+\;m_{\nu}m_e\mathrm{Re}\left\{K^2_{i\omega_{\nu}/a-1/2}\left(\frac{l_{\nu}}{a}\right)K^2_{i\omega_{e}/a+1/2}\left(\frac{l_{e}}{a}\right)\right\}\right].
\eea
Next, by performing similar calculation for the processes $(ii)$ and $(iii)$ and adding up the three contributions, we end up with the following integral expression for the total decay rate in the comoving frame:
\be\label{gammaacc}
\Gamma_{acc}^{p\rightarrow n}\ \equiv\ \Gamma_{(i)}^{p\rightarrow n}\,+\,\Gamma_{(ii)}^{p\rightarrow n}\,+\,\Gamma_{(iii)}^{p\rightarrow n}\ =\ \frac{2\hspace{0.2mm}G_{F}^{2}}{a^2\hspace{0.2mm}\pi^7\hspace{0.2mm}e^{\pi\Delta m/a}}\int_{-\infty}^{+\infty}\hspace{-2mm}d\omega\,\mathcal{R}(\omega),
\ee 
where
\bea\nonumber
\label{R}
\hspace{-12mm}\mathcal{R}(\omega)&\hspace{-1.5mm}=\hspace{-1.5mm}&\int d^2k_{\nu}\,l_{\nu}\biggl|K_{i(\omega-\Delta m)/a+1/2}\left(\frac{l_{\nu}}{a}\right)\bigg|^2\int d^2k_{e}\,l_{e}\bigg|K_{i\omega/a+1/2}\left(\frac{l_{e}}{a}\right)\bigg|^2\\[3mm]
&&+\;m_{\nu}m_e\mathrm{Re}\hspace{0.2mm}\left\{\int d^2k_{\nu}\,K^2_{i(\omega-\Delta m)/a-1/2}\left(\frac{l_{\nu}}{a}\right)\int d^2k_e\,K^2_{i\omega/a+1/2}\left(\frac{l_{e}}{a}\right)\right\}\hspace{-0.9mm}.
\eea
The analytic resolution of the integral Eq.~(\ref{gammaacc}) is performed in Ref.~\cite{Suzuki:2002xg}. Comparing this result to the one in the inertial frame (Eq.~(\ref{gammain})), it is possible to show that the resulting expressions for the decay rates perfectly agree with each other, thus corroborating the necessity of the FDU effect for the consistency of QFT.

\section{Proton decay involving mixed neutrinos}
\label{prdec}
So far, in the evaluation of the transition amplitude, we have treated the electron neutrino as a particle with definite mass $m_\nu$. However, it is well known that neutrinos exhibit flavor mixing: in a simplified two-flavor model, by denoting with $\theta$ the mixing angle, the transformations relating the flavor eigenstates $|\nu_{\ell}\rangle$ ($\ell\,=\,e, \mu$) and mass eigenstates $|\nu_i\rangle$ ($i\ =\ 1, 2$) are determined by the Pontecorvo unitary mixing matrix\footnote{Note that the number of neutrino generations does not affect the results of our analysis.}, Eq.~\eqref{PMMmatrix}:
\be
\begin{pmatrix}
\label{Pontec}
|\nu_e\rangle\\
|\nu_\mu\rangle
\end{pmatrix}\,=\, \begin{pmatrix}
\cos\theta&\sin\theta\\
-\sin\theta&\cos\theta
\end{pmatrix}
\begin{pmatrix}
|\nu_1\rangle\\
|\nu_2\rangle
\end{pmatrix}.
\ee

Along the line of Ref.~\cite{Ahluwalia:2016wmf}, the question thus arises whether such a transformation is consistent with the framework of Section~\ref{Decay of accelerated protons: a brief review}.

\subsection{Inertial frame calculation}
\label{inerframemix}
Let us then implement the Pontecorvo rotation Eq.~(\ref{Pontec}) on both neutrino fields and states appearing in Eq.~(\ref{tramp}).  Note that in Ref.~\cite{Ahluwalia:2016wmf} this step is missing in the inertial frame calculation since $\Psi_{\nu_e}$ is treated as a free-field even when taking into account flavor mixing, and indeed the same result as in the case of unmixed fields is obtained.  We  explicitly show that the decay rate exhibits a dependence on  $\theta$ in the inertial frame, a feature which is not present in the analysis of Ref.~\cite{Ahluwalia:2016wmf}. 

By assuming equal momenta and polarizations for the two neutrino mass eigenstates, the transition amplitude Eq.~(\ref{tramp}) now becomes
\begin{equation}
\mathcal{A}_{in}^{p\rightarrow n}\ =\ \frac{G_F}{2^4\pi^3}\Big[\cos^2\theta\, \mathcal{I}_{\sigma_\nu\sigma_e}(\omega_{\nu_1},\omega_e)\,+\, \sin^2\theta\, \mathcal{I}_{\sigma_\nu\sigma_e}(\omega_{\nu_2},\omega_e)\Big],
\end{equation}
where $\mathcal{I}_{\sigma_\nu\sigma_e}(\omega_{\nu_j},\omega_e)$, $j\,=\,1,2$, is defined as in Eq.~(\ref{processI}) for each of the two mass eigenstates, and we have rotated the electron neutrino field according to 
\be
\label{rotfield}
\Psi_{\nu_e}(t,\bx)\ =\ \cos\theta\hspace{0.4mm}\Psi_{\nu_1}(t,\bx)\,+\,\sin\theta\hspace{0.4mm}\Psi_{\nu_2}(t,\bx).
\ee 
Using Eq.~(\ref{dtr}), the differential transition rate takes the form
\begin{eqnarray}\nonumber
\frac{d^6\mathcal{P}_{in}^{p\rightarrow n}}{d^3k_\nu\hspace{0.2mm} d^3k_e}&=&\sum_{\sigma_\nu,\sigma_e}\frac{{G_F}^2}{2^8\pi^6}\hspace{-0.5mm}\left\{\cos^4\theta\,{\big|\mathcal{I}_{\sigma_\nu\sigma_e}(\omega_{\nu_1},\omega_e)\big|}^2\,+\,\sin^4\theta\,{\big|\mathcal{I}_{\sigma_\nu\sigma_e}(\omega_{\nu_2},\omega_e)\big|}^2\right.\\[3mm]
&&+\;\cos^2\theta\sin^2\theta\Big[\mathcal{I}_{\sigma_\nu\sigma_e}(\omega_{\nu_1},\omega_e)\,\mathcal{I}^{\hspace{0.2mm}*}_{\sigma_\nu\sigma_e}(\omega_{\nu_2},\omega_e)\,+\,\mathrm{c.c.}\Big]\Big\}\,.
\end{eqnarray}
The total decay rate $\Gamma_{in}^{p\rightarrow n}$  is obtained after inserting this equation  into the definition Eq.~(\ref{ttr}):
\be
\label{eqn:inertresultat}
\Gamma^{p\rightarrow n}_{in}\ =\ \cos^4\theta\, \Gamma^{p\rightarrow n}_{1}\,+\,\sin^4\theta\,\Gamma^{p\rightarrow n}_{2}\,+\,\cos^2\theta\sin^2\theta\,\Gamma^{p\rightarrow n}_{12},
\ee
where we have introduced the shorthand notation
\begin{equation}
\label{integral}
\Gamma^{p\rightarrow n}_{j}\ \equiv\ \frac{1}{T}\sum_{\sigma_\nu,\sigma_e}\frac{{G_F}^2}{2^8\pi^6}\int d^3k_\nu\int d^3k_e\,{\big|\mathcal{I}_{\sigma_\nu\sigma_e}(\omega_{\nu_j},\omega_e)\big|}^2,\qquad j=1,2,
\end{equation}
and 
\begin{equation}
\label{integral12}
\Gamma^{p\rightarrow n}_{12}\ \equiv\ \frac{1}{T}\sum_{\sigma_\nu,\sigma_e}\frac{{G_F}^2}{2^8\pi^6}\int d^3k_\nu\int d^3k_e\Big[\mathcal{I}_{\sigma_\nu\sigma_e}(\omega_{\nu_1},\omega_e)\,\mathcal{I}^{\hspace{0.2mm}*}_{\sigma_\nu\sigma_e}(\omega_{\nu_2},\omega_e)\,+\,\mathrm{c.c.}\Big].
\end{equation}
We observe that, for $\theta\rightarrow 0$, the obtained result correctly reduces to Eq.~(\ref{gammain}), as it should be in the absence of mixing. Unfortunately, due to technical difficulties in the evaluation of the integral Eq.~(\ref{integral12}), at this stage we are not able to give the exact expression  of the inertial decay rate Eq.~(\ref{eqn:inertresultat}). A preliminary result, however, can be obtained in the limit of small neutrino mass difference ${\delta m}\,\equiv\,{m_{\nu_2}\,-\,m_{\nu_1}}\,\ll\, 1$. In this case, indeed, we can expand  $\Gamma^{p\rightarrow n}_{12}$ according to
\be
\label{approxima}
\Gamma^{p\rightarrow n}_{12}\ =\ 2\Gamma^{p\rightarrow n}_1\,+\,\Gamma^{(1)}\,{\delta m}\,\,+\, \mathcal{O}\left({\delta m^2}\right),
\ee
where $\Gamma^{p\rightarrow n}_1$ is defined as in Eq.~(\ref{integral}) and we have denoted by $\Gamma^{(1)}$ the first-order term of the Taylor expansion. The explicit expression of $\Gamma^{(1)}$ is rather awkward to exhibit. Nevertheless, for $m_{\nu_1}\rightarrow 0$, it can be  substantially simplified, thus giving
\bea
\label{firstorder}
{\Gamma^{(1)}}&=&\frac{1}{T}\frac{{G_F}^2\,m_e}{2^7\pi^6}\int \frac{d^3k_\nu}{|k_\nu|}\int \frac{d^3 k_e}{\omega_e} \int_{-\infty}^{+\infty}\hspace{-2mm}ds\\[3mm]
\non
&&\times\,\int_{-\infty}^{+\infty}\hspace{-2mm}d\xi \cosh a\xi\left[e^{i\big\{\Delta m\,\xi+\frac{2\sinh a\xi/2}{a}\big[\left(|k_\nu|+\omega_{e}\right)\cosh as-\left(k_\nu^z+k^z_{e}\right)\sinh as\big]\big\}}\,+\,\mathrm{c.c.}\right],
\eea
where $s$ and $\xi$ are defined in Eq.~(\ref{newvariab}). By performing a boost along the $z$-direction: 
\be
{k'}^{\hspace{0.2mm}x}_{\hspace{0.2mm}\ell}\ =\ {k}^{\hspace{0.2mm}x}_{\hspace{0.2mm}\ell},\qquad {k'}^{\hspace{0.2mm}y}_{\hspace{0.2mm}\ell}\ =\ {k}^{\hspace{0.2mm}y}_{\hspace{0.2mm}\ell},\qquad {k'}^{\hspace{0.2mm}z}_{\hspace{0.2mm}\ell}\, =\,- \hspace{0.01mm}\omega_\ell\sinh as\,+\,{k}^{\hspace{0.2mm}z}_{\hspace{0.2mm}\ell}\cosh as,\qquad \ell\,=\,\nu_1,e\,,
\ee
Eq.~(\ref{firstorder}) can be cast in the form
\be\label{besselk}
{\Gamma^{(1)}}\, =\, \lim_{\varepsilon\rightarrow 0}\hspace{0.3mm}\frac{2\,G_F^2\,m_e}{a\,\pi^6\,e^{\pi\Delta m/a}}\int\frac{d^3k_\nu}{\omega_\varepsilon}\int\frac{d^3k_e}{\omega_e}\mathrm{Re}\left\{K_{2i\Delta m/a+2}\left(\frac{2(\omega_\varepsilon\,+\,\omega_e)}{a}\right)\right\},
\ee
where $\omega_\varepsilon\,=\,\sqrt{\bk^2_\nu\,+\,\varepsilon^2}$, with $\varepsilon$ acting as a regulator. In order to perform $k$-integration, we use the following representation of the modified Bessel function:
\be
K_\mu(z)\ =\ \frac{1}{2}\int_{C_1}\frac{ds}{2\pi i}\Gamma(-s)\Gamma(-s-\mu)\left(\frac{z}{2}\right)^{2s+\mu},
\ee
where $\Gamma$ is the Euler Gamma function. $C_1$ is the path in the complex plane including all the poles of $\Gamma(-s)$ and $\Gamma(-s-\mu)$, chosen in such a way that the integration with respect to the momentum variables does not diverge~\cite{Suzuki:2002xg}.

Using spherical coordinates, Eq.~(\ref{besselk}) becomes
\bea\nonumber
\frac{\Gamma^{(1)}}{m_{\nu_1}}&=&\lim_{\varepsilon\rightarrow 0}\hspace{0.3mm}\frac{2^3\,G_F^2\,m_e}{a\,\pi^4\,e^{\pi\Delta m/a}}\int_{0}^{+\infty}\hspace{-2mm}dk_\nu\hspace{0.3mm}\frac{k^2_\nu}{\omega_\varepsilon}\int_{0}^{+\infty}\hspace{-2mm}dk_e\hspace{0.3mm}\frac{k^2_e}{\omega_e}\int_{C_s}\frac{ds}{2\pi i}\hspace{-0.2mm}\left(\frac{\omega_\varepsilon\,+\,\omega_e}{a}\right)^{2s}\\[3mm]\non
&&\times\hspace{0.4mm}\left[\Gamma\left(-s\hspace{0.2mm}+\hspace{0.2mm}\frac{i\Delta m}{a}\hspace{0.2mm}+\hspace{0.2mm}1\right)\Gamma\left(-s\hspace{0.2mm}-\hspace{0.2mm}\frac{i\Delta m}{a}\hspace{0.2mm}-\hspace{0.2mm}1\right)\right.\\[3mm]
&&\left.+\;\Gamma\left(-s\hspace{0.2mm}+\hspace{0.2mm}\frac{i\Delta m}{a}\hspace{0.2mm}-\hspace{0.2mm}1\right)\Gamma\left(-s\hspace{0.2mm}-\hspace{0.2mm}\frac{i\Delta m}{a}\hspace{0.2mm}+\hspace{0.2mm}1\right)\right].
\eea
Let us observe at this point that~\cite{Suzuki:2002xg}
\be
\left(\frac{\omega_\varepsilon\,+\,\omega_e}{a}\right)^{2s}\, =\, \int_{C_2}\frac{dt}{2\pi i}\frac{\Gamma(-t)\Gamma(t\hspace{0.2mm}-\hspace{0.2mm}2s)}{\Gamma(-2s)}\left(\frac{\omega_\varepsilon}{a}\right)^{-t+2s}\left(\frac{\omega_e}{a}\right)^t,
\ee
where $C_2$ is the contour in the complex plane separating the poles of $\Gamma(-t)$ from the ones of $\Gamma(t-2s)$. Exploiting this relation and properly redefining the integration variables, we finally obtain
\bea
\label{lastin}
\non
{\Gamma^{(1)}}&=&\lim_{\varepsilon\rightarrow 0}\hspace{0.3mm}\frac{G_F^2\,m_e\,a^3}{\pi^3\,e^{\pi\Delta m/a}}\int_{C_s}\frac{ds}{2\pi i}\int_{C_t}\frac{dt}{2\pi i}\left(\frac{\varepsilon}{a}\right)^{2s+2}\left(\frac{m_e}{a}\right)^{2t+2}\\[3mm]
\non
&&\times\;\frac{\Gamma(-2s)\Gamma(-2t)\Gamma(-t\,-\,1)\Gamma(-s\,-\,1)}{\Gamma(-s\,+\,\frac{1}{2})\Gamma(-t\,+\,\frac{1}{2})\Gamma(-2s\,-\,2t)}\\[3mm]\nonumber
&&\times\hspace{0.3mm}\left[\Gamma\left(-s\,-\,t\,+\,1\,+\,i\hspace{0.2mm}\frac{\Delta m}{a}\right)\Gamma\left(-s\hspace{0.2mm}-\hspace{0.2mm}t\hspace{0.2mm}-\hspace{0.2mm}1\hspace{0.2mm}-\hspace{0.2mm}i\hspace{0.2mm}\frac{\Delta m}{a}\right)\right.\\[3mm]
&&\left.+\;\Gamma\left(-s\hspace{0.2mm}-\hspace{0.2mm}t\hspace{0.2mm}+\hspace{0.2mm}1\hspace{0.2mm}-\hspace{0.2mm}i\hspace{0.2mm}\frac{\Delta m}{a}\right)\Gamma\left(-s\hspace{0.2mm}-\hspace{0.2mm}t\hspace{0.2mm}-\hspace{0.2mm}1\hspace{0.2mm}+\hspace{0.2mm}i\hspace{0.2mm}\frac{\Delta m}{a}\right)\right].
\eea
where the contour $C_{s(t)}$ includes all poles of gamma functions in $s$ ($t$) complex plane\footnote{Strictly speaking, the expansion in Eq.~\eqref{approxima} should 
be performed around a dimensionless parameter; a good choice, for example, would be
to rescale the mass difference $\delta m$
with respect to the acceleration $a$. In so doing, however, the same result
as in Eq.~\eqref{lastin} would be obtained, up to a scale factor.}.

From Eqs.~(\ref{approxima}) and (\ref{lastin}), we thus infer that the off-diagonal term $\Gamma_{12}^{p\rightarrow n}$ is non-vanishing, thereby leading to a structure of the inertial decay rate Eq.~(\ref{eqn:inertresultat}) that is different from the corresponding one in Ref.~\cite{Ahluwalia:2016wmf}.

\subsection{Comoving frame calculation}
\label{sect2}
Let us now extend the above discussion to the proton comoving frame. As done in the inertial case, we require the asymptotic neutrino states to be flavor eigenstates (the choice of mass eigenstates would inevitably lead to a contradiction, as discussed in Appendix~\ref{Cozzella}). Note that the same assumption is also contemplated in Ref.~\cite{Ahluwalia:2016wmf}. In spite of this, those authors exclude such an alternative on the basis of the KMS condition, claiming that  the accelerated neutrino vacuum must be a thermal state of neutrinos with definite mass rather than with definite flavor. Actually, this argument does not apply, at least  within the first-order approximation we are dealing with (see Eq.~(\ref{approxima})). Indeed, as shown in Refs.~\cite{Blasone:2017nbf, Blasone:2018byx}, non-thermal corrections to the Unruh spectrum for flavor (mixed) neutrinos only appear at orders higher than $\mathcal{O}\left({\delta m}\right)$. 

Relying on these considerations, let us evaluate the decay rate in the comoving frame. A straightforward calculation leads to the following expression for the transition amplitude Eq.~(\ref{first}):
\be
\mathcal{A}_{(i)}^{p\rightarrow n} \ =\ \frac{G_{F}}{(2\pi)^2}\left[\cos^2\theta\mathcal{J}^{(1)}_{\sigma_\nu\sigma_e}(\omega_\nu, \omega_e)\,+\, \sin^2\theta\mathcal{J}^{(2)}_{\sigma_\nu\sigma_e}(\omega_\nu, \omega_e)\right],
\ee
where $\mathcal{J}^{(j)}_{\sigma_\nu\sigma_e}(\omega_\nu, \omega_e)$, $j\,=\,1,2$, is defined as in Eq.~(\ref{eqn:j}) for each of the two neutrino mass eigenstates. The differential transition rate per unit time thus reads
\begin{eqnarray}
\label{difmixrindright}
\nonumber
\hspace{-4mm}\frac{1}{T}\frac{d^{6}\mathcal{P}^{p\rightarrow n}_{(i)}}{d\omega_{\nu}\hspace{0.2mm}d\omega_{e}\hspace{0.2mm}d^2k_{\nu}\hspace{0.2mm}d^2k_e}&=&\frac{1}{T}\frac{\hspace{0.2mm}G_{F}^{2}}{2^6\pi^{4}}\hspace{0.2mm}\frac{1}{e^{\pi\Delta m/a}\cosh(\pi\omega_{\nu}/a)\cosh(\pi\omega_{e}/a)}\\[3mm]
\non
&&\times\;\sum_{\sigma_{\nu},\sigma_e}\biggl\{\cos^4\theta\,\big|\mathcal{J}^{(1)}_{\sigma_\nu\sigma_e}(\omega_\nu, \omega_e)\big|^2\,+\,\sin^4\theta\,\big|\mathcal{J}^{(2)}_{\sigma_\nu\sigma_e}(\omega_\nu, \omega_e)\big|^2\\[3mm]
&&+\;\cos^2\theta\sin^2\theta\left[\mathcal{J}^{(1)}_{\sigma_\nu\sigma_e}(\omega_\nu, \omega_e)\,\mathcal{J}^{(2)\hspace{0.2mm}*}_{\sigma_\nu\sigma_e}(\omega_\nu, \omega_e)\,+\,\mathrm{c.c.}\right]\biggr\}\hspace{-0.5mm}.
\end{eqnarray}
The spin sum for the process $(i)$ is given by
\bea\label{mixspinsum}
\non
&&\frac{1}{T}\sum_{\sigma_\nu,\sigma_e}\left[\mathcal{J}^{(1)}_{\sigma_\nu\sigma_e}(\omega_\nu, \omega_e)\,\mathcal{J}^{(2)\hspace{0.2mm}*}_{\sigma_\nu\sigma_e}(\omega_\nu, \omega_e)\,+\,\mathrm{c.c.}\right]\,=\,\\[3mm] 
\non
&&\hspace{1cm}\frac{2^3\,\delta(\omega_e-\omega_\nu-\Delta m)}{a^2\,\pi^3\,\sqrt{l_{\nu_1}l_{\nu_2}}}\,\cosh\left(\pi\omega_\nu/a\right)\cosh\left(\pi\omega_e/a\right)\Biggl[l_e\left(\kappa_\nu^2\,+\,m_{\nu_1}m_{\nu_2}\,+\,l_{\nu_1}l_{\nu_2}\right)\\[3mm]
\non
&&\hspace{1cm}\times\;\biggl|K_{i\omega_e/a+1/2}\left(\frac{l_e}{a}\right)\bigg|^2\,\mathrm{Re}\left\{K_{i\omega_\nu/a+1/2}\left(\frac{l_{\nu_1}}{a}\right)K_{i\omega_\nu/a-1/2}\left(\frac{l_{\nu_2}}{a}\right)\right\}\\[3mm]\nonumber
&&\hspace{1cm}+\;\Big[\hspace{-0.2mm}\left(k^x_\nu k^x_e\hspace{0.2mm}+\hspace{0.2mm}k^y_\nu k^y_e\right)\left(l_{\nu_1}\hspace{0.2mm}+\hspace{0.2mm}l_{\nu_2}\right)\hspace{0.2mm}+\hspace{0.2mm}m_e\left(l_{\nu_1}m_{\nu_2}\hspace{0.2mm}+\hspace{0.2mm}l_{\nu_2}m_{\nu_1}\right)\hspace{-0.2mm}\Big]\\[3mm]
&&\hspace{1cm}\times\;\mathrm{Re}\left\{K^2_{i\omega_e/a+1/2}\left(\frac{l_e}{a}\right)K_{i\omega_\nu/a+1/2}\left(\frac{l_{\nu_1}}{a}\right)K_{i\omega_\nu/a+1/2}\left(\frac{l_{\nu_2}}{a}\right)\right\}\Bigg]\,,
\eea
where $\kappa_\nu\equiv(k_\nu^x,k_\nu^y)$.

Next, by performing similar calculations for the other two processes and adding up the three contributions, we finally obtain the total transition rate in the comoving frame:
\begin{equation}
\label{gammacc}
\Gamma_{acc}^{p\rightarrow n}\ =\ 
\cos^4\theta\, \widetilde\Gamma^{p\rightarrow n}_{1}\,+\,\sin^4\theta\,\widetilde\Gamma^{p\rightarrow n}_{2}\,+\,\cos^2\theta\sin^2\theta\,\widetilde\Gamma^{p\rightarrow n}_{12},
\end{equation}
where $\widetilde\Gamma^{p\rightarrow n}_{j}$, $j\,=\,1,2$, is defined as
\begin{equation}
\label{integralbis}
\widetilde\Gamma^{p\rightarrow n}_{j}\ \equiv\ \frac{2\,G_{F}^{2}}{a^2\,\pi^7\,e^{\pi\Delta m/a}}\int_{-\infty}^{+\infty}\hspace{-1.5mm}d\omega\hspace{0.4mm}R_{j}(\omega),\qquad j\,=\,1,2,
\end{equation}
with $\mathcal{R}_j(\omega)$ being defined as in Eq.~(\ref{R}) for each of the two neutrino mass eigenstates, and
\begin{eqnarray}
\label{int1212}
\widetilde\Gamma^{p\rightarrow n}_{12}&=&\frac{2\,G_{F}^{2}}{a^2\,\pi^7\,\sqrt{l_{\nu_1}l_{\nu_2}}\,e^{\pi\Delta m/a}}\int_{-\infty}^{+\infty}\hspace{-2mm}d\omega\,\Biggl\{\,\int d^2k_e\,l_e\bigg|K_{i\omega/a+1/2}\left(\frac{l_e}{a}\right)\bigg|^2\\[3mm]\non
&&\times\;\int d^2k_\nu\,\big(\kappa_\nu^2\,+\,m_{\nu_1}m_{\nu_2}\,+\,l_{\nu_1}l_{\nu_2}\big)\\[2mm]\nonumber
&&\times\;\mathrm{Re}\left\{K_{i(\omega-\Delta m)/a+1/2}\left(\frac{l_{\nu_1}}{a}\right)K_{i(\omega-\Delta m)/a-1/2}\left(\frac{l_{\nu_2}}{a}\right)\right\}\\[3mm]\non
&&+\; m_e\int d^2k_e\int d^2k_\nu\big(l_{\nu_1}m_{\nu_2}\,+\,l_{\nu_2}m_{\nu_1}\big)\\[3mm]
\non
&&\times\;\mathrm{Re}\left\{K^2_{i\omega/a+1/2}\left(\frac{l_e}{a}\right)K_{i(\omega-\Delta m)/a-1/2}\left(\frac{l_{\nu_1}}{a}\right)K_{i(\omega-\Delta m)/a-1/2}\left(\frac{l_{\nu_2}}{a}\right)\right\}\Biggr\}.
\end{eqnarray}
It is now possible to verify that 
\be
\label{equality}
\Gamma^{p\rightarrow n}_{j}\ =	\ \widetilde\Gamma^{p\rightarrow n}_{j}\qquad j\,=\,1,2.
\ee 
By comparing Eqs.~(\ref{eqn:inertresultat}) and~(\ref{gammacc}) and using the above equality, we thus realize that inertial and comoving calculations would match, provided that the integrals Eqs.~(\ref{integral12}) and~(\ref{int1212}) coincide. As in the inertial case, however, the treatment of the $\widetilde\Gamma^{p\rightarrow n}_{12}$ is absolutely non-trivial. A clue to a preliminary solution can be found by expanding $\widetilde\Gamma^{p\rightarrow n}_{12}$ in the limit of small neutrino mass difference\footnote{See note 4.}, as in Section~\ref{inerframemix}:
\be\label{approxrind}
\widetilde\Gamma^{p\rightarrow n}_{12}\ =\ 2\widetilde\Gamma^{p\rightarrow n}_1\,+\,\widetilde\Gamma^{(1)}\,{\delta m}\,+\, \mathcal{O}\left({\delta m^2}\right),
\ee 
where $\widetilde\Gamma^{p\rightarrow n}_1$ is defined in Eq.~(\ref{integralbis}) and we have denoted by $\widetilde\Gamma^{(1)}$ the first-order term of the expansion. For $m_{\nu_1}\rightarrow 0$, it is possible to show that
\be\label{firstordrind}
{\widetilde\Gamma^{(1)}}\ =\ \lim_{\varepsilon\rightarrow 0}\hspace{0.3mm}\frac{2^2\,G_F^2\,m_e}{a^2\,\pi^7\,e^{\pi\Delta m/a}}\int_{-\infty}^{+\infty}\hspace{-2mm}d\omega\,\mathrm{Re}\left\{\int d^2k_\nu\hspace{0.3mm} K^2_{i(\omega-\Delta m)/a-1/2}\left(\frac{l_\varepsilon}{a}\right)\hspace{-2mm}\int d^2k_e\hspace{0.3mm} K^2_{i\omega/a+1/2}\left(\frac{l_e}{a}\right)\right\},
\ee
where $l_\varepsilon\ =\ \sqrt{(k^x_\nu)^2\,+\,(k^y_\nu)^2\,+\,\varepsilon^2}$, with $\varepsilon$ acting as a regulator. 

Equation~(\ref{firstordrind}) can be now further manipulated by introducing the following relation involving Meijer G-function (see, \emph{e.g.}, Ref.~\cite{Grad}):
\bea\label{meijer}
x^{\sigma}K_{\nu}(x)K_{\mu}(x)&\hspace{-1mm}=& \\[3mm]
\non
&&\hspace{-35mm}\frac{\sqrt{\pi}}{2}G_{24}^{40}\,\Biggl(x^{2}\Biggr|\begin{matrix}\frac{1}{2}\sigma,\,\frac{1}{2}\sigma\,+\,\frac{1}{2}\\[2mm]\frac{1}{2}(\nu\,+\,\mu\,+\,\sigma),\frac{1}{2}(\nu\,-\,\mu\,+\,\sigma),\frac{1}{2}(-\nu\,+\,\mu\,+\,\sigma),\frac{1}{2}(-\nu\,-\,\mu\,+\,\sigma)
\end{matrix}\Biggr)\,.
\eea 
A somewhat laborious calculation then leads to
\bea\label{bigrind}\nonumber
{\widetilde\Gamma^{(1)}}&=&\lim_{\varepsilon\rightarrow 0}\hspace{0.3mm}\frac{2\,G_F^2\,m_e}{a^2\,\pi^4\,e^{\pi\Delta m/a}}\int_{-\infty}^{+\infty}\hspace{-2mm}d\omega\int_{C_s}\frac{ds}{2\pi i}\int_{C_t}\frac{dt}{2\pi i}\int_{0}^{+\infty}\hspace{-2mm}dk_\nu\,k_\nu\,l_\varepsilon^{2s}\int_{0}^{+\infty}\hspace{-2mm}dk_e\,k_e\,l_e^{2t}\\[2mm]\nonumber
&&\times\;\Bigg[\frac{\Gamma\left(-s\right)\Gamma\left(-t\right)\Gamma\left(\frac{i\omega}{a}\hspace{0.2mm}+\hspace{0.2mm}\frac{1}{2}\hspace{0.2mm}-\hspace{0.2mm}t\right)\Gamma\left(-\frac{i\omega}{a}\hspace{0.2mm}-\hspace{0.2mm}\frac{1}{2}\hspace{0.2mm}-\hspace{0.2mm}t\right)}{\Gamma\left(-s\hspace{0.2mm}+\hspace{0.2mm}\frac{1}{2}\right)\Gamma\left(-t\hspace{0.2mm}+\hspace{0.2mm}\frac{1}{2}\right)}\\[3mm]\non
&&\times\;\frac{\Gamma\left(\frac{i(\omega\hspace{0.2mm}-\hspace{0.2mm}\Delta m)}{a}\hspace{0.2mm}-\hspace{0.2mm}\frac{1}{2}\hspace{0.2mm}-\hspace{0.2mm}s\right)\Gamma\left(-\frac{i(\omega\hspace{0.2mm}-\hspace{0.2mm}\Delta m)}{a}\hspace{0.2mm}+\hspace{0.2mm}\frac{1}{2}\hspace{0.2mm}-\hspace{0.2mm}s\right)}{a}\\[3mm]\non
&&+\;\frac{\Gamma\left(-s\right)\Gamma\left(-t\right)\Gamma\left(\frac{i\omega}{a}\hspace{0.2mm}-\hspace{0.2mm}\frac{1}{2}\hspace{0.2mm}-\hspace{0.2mm}t\right)\Gamma\left(-\frac{i\omega}{a}\hspace{0.2mm}+\hspace{0.2mm}\frac{1}{2}\hspace{0.2mm}-\hspace{0.2mm}t\right)\Gamma\left(\frac{i(\omega\hspace{0.2mm}-\hspace{0.2mm}\Delta m)}{a}\hspace{0.2mm}+\hspace{0.2mm}\frac{1}{2}\hspace{0.2mm}-\hspace{0.2mm}s\right)}{\Gamma\left(-s\hspace{0.2mm}+\hspace{0.2mm}\frac{1}{2}\right)\Gamma\left(-t\hspace{0.2mm}+\hspace{0.2mm}\frac{1}{2}\right)}\\[3mm]
&&\times\;{\Gamma\left(-\frac{i(\omega\hspace{0.2mm}-\hspace{0.2mm}\Delta m)}{a}\hspace{0.2mm}-\hspace{0.2mm}\frac{1}{2}\hspace{0.2mm}-\hspace{0.2mm}s\right)}{\Gamma\left(-s\hspace{0.2mm}+\hspace{0.2mm}\frac{1}{2}\right)\Gamma\left(-t\hspace{0.2mm}+\hspace{0.2mm}\frac{1}{2}\right)}\Bigg].
\eea
In order to perform the integration with respect to $\omega$, let us use the first Barnes lemma, according to which~\cite{Grad}
\be\label{barnes}
\int_{-i\infty}^{+i\infty}\hspace{-2mm}d{\omega}\,\Gamma(a\,+\,{\omega})\Gamma(b\,+\,{\omega})\Gamma(c\,-\,{\omega})\Gamma(d\,-\,{\omega})\ =\ 2\pi i\,\,\frac{\Gamma(a\,+\,c)\Gamma(a\,+\,d)\Gamma(b\,+\,c)\Gamma(b\,+\,d)}{\Gamma(a\,+\,b\,+\,c\,+\,d)}.
\ee
Inserting this relation into Eq.~(\ref{bigrind}), it follows that
\bea
\label{lastacc}
\frac{\widetilde\Gamma^{(1)}}{m_{\nu_1}}&=&\lim_{\varepsilon\rightarrow 0}\frac{G_F^2\,m_e\,a^3}{\pi^3\,e^{\pi\Delta m/a}}\int_{C_s}\frac{ds}{2\pi i}\int_{C_t}\frac{dt}{2\pi i}\left(\frac{\varepsilon}{a}\right)^{2s+2}\left(\frac{m_e}{a}\right)^{2t+2}\\[3mm]\non
&&\times\;\frac{\Gamma(-2s)\Gamma(-2t)\Gamma(-t-1)\Gamma(-s-1)}{\Gamma(-s+\frac{1}{2})\Gamma(-t+\frac{1}{2})\Gamma(-2s-2t)}\\[2mm]\nonumber
&&\times\hspace{0.4mm}\left[\Gamma\left(-s\hspace{0.2mm}-\hspace{0.2mm}t\hspace{0.2mm}+\hspace{0.2mm}1\hspace{0.2mm}+\hspace{0.2mm}i\hspace{0.2mm}\frac{\Delta m}{a}\right)\Gamma\left(-s\hspace{0.2mm}-\hspace{0.2mm}t\hspace{0.2mm}-\hspace{0.2mm}1\hspace{0.2mm}-\hspace{0.2mm}i\hspace{0.2mm}\frac{\Delta m}{a}\right)\right.\\[3mm]
&&\left.+\;\Gamma\left(-s\hspace{0.2mm}-\hspace{0.2mm}t\hspace{0.2mm}+\hspace{0.2mm}1\hspace{0.2mm}-\hspace{0.2mm}i\hspace{0.2mm}\frac{\Delta m}{a}\right)\Gamma\left(-s\hspace{0.2mm}-\hspace{0.2mm}t\hspace{0.2mm}-\hspace{0.2mm}1\hspace{0.2mm}+\hspace{0.2mm}i\hspace{0.2mm}\frac{\Delta m}{a}\right)\right]\hspace{-0.5mm},
\eea
which is the same expression obtained in the inertial frame (see Eq.~(\ref{lastin})).

\section{Conclusions}
\label{Conclusionsproton}
In this Chapter we have discussed the decay of uniformly accelerated protons. Following the line of reasoning of Refs.~\cite{Matsas:1999jx,Matsas:1999jx2,Matsas:1999jx3,Suzuki:2002xg}, we have reviewed the calculation of the total decay rate in both the laboratory and comoving frames, highlighting the incompatibility between the two results when taking into account neutrino flavor mixing~\cite{Ahluwalia:2016wmf}. Such an inconsistency would not be striking if the underlying theory were not generally covariant, but this is not the case, since the fundamental ingredients for analyzing the process, namely the SM and QFT in curved space-time, are by construction generally covariant. On the other hand, the authors of Ref.~\cite{Ahluwalia:2016wmf} argue their result claiming that mixed neutrinos are not
representations of the Lorentz group with a well-defined
invariant $P^2$, and that the mathematical
origin of the disagreement arises from the non-commutativity of weak and energy-momentum currents. 
Furthermore, they propose the experimental investigation as the only way to resolve
such a controversial issue. Even assuming there are no flaws in this reasoning, we believe the last statement to be basically incorrect: an experiment, indeed, should not be used as a tool for checking the internal consistency of theory against a theoretical paradox.

Led by these considerations, we have thus revised calculations of Ref.~\cite{Ahluwalia:2016wmf} modifying some of the key assumptions of that work. In particular, we have required the asymptotic neutrino states to be flavor rather than mass eigenstates. Within this framework, by comparing the obtained expressions for the two decay rates, it has been shown that they would coincide, provided that the off-diagonal terms Eqs.~(\ref{integral12}) and~(\ref{int1212}) are equal to each other. In order to check whether this is the case, we have performed the reasonable approximation of small neutrino mass difference, pushing our analysis up to the first order in ${\delta m}$. However, due to computational difficulties, the further assumption of vanishing neutrino mass $m_{\nu_1}\rightarrow 0$ has proved to be necessary for getting information about these terms. In such a regime, we have found that Eqs.~(\ref{lastin}) and~(\ref{lastacc}) are perfectly in agreement,  thus removing the aforementioned ambiguity at a purely theoretical level.

Relying on these considerations, one can state that the theoretical framework underlying our outcome is the correct one in the context of neutrino mixing, provided that flavor neutrinos are considered to be the fundamental ``objects'' in the thermal FDU state. For the sake of completeness, in Appendix~\ref{Cozzella} we expose our criticism on some recent findings of Ref.~\cite{Cozzella:2018qew}, where the choice of fundamental mass neutrinos is instead adopted. We show how the treatment of flavor mixing there proposed is missing of some basic conceptual points, discussing their formalism in connection with the results of this Chapter.

Finally, we stress that further aspects of the fascinating problem firstly addressed in Ref.~\cite{Ahluwalia:2016wmf} can be investigated only when an exact evaluation of the two decay rates will be available: work is in progress along this direction~\cite{In prepaproton}, also in view of the non-trivial nature 
of neutrino flavor states in QFT~\cite{Blasone:1995zc}.

\newpage\null\thispagestyle{empty}\newpage

\chapter{Modified Unruh effect from Generalized Uncertainty Principle}
\label{GUPScard}
\begin{flushright}
\emph{``Uncertainty" is NOT "I don't know."\\ It is "I can't know." \\"I am uncertain" does not mean "I could be certain.''}\\[1mm]
- Werner Heisenberg -\\[6mm]
\end{flushright}
In the last thirty years, many studies have converged on the idea that the Heisenberg uncertainty
principle (HUP) \cite{Heisenberg} should be modified when gravitation is taken into account.
In microphysics, gravity is usually neglected on the ground of its weakness,
when compared with the other fundamental interactions.
However, this argument should not apply when one wants to address fundamental questions in Nature.
In this perspective, gravity should be included, especially when we discuss the formulation of fundamental principles
like Heisenberg's principle.
And in fact, gravitation has always played a pivotal r\^ole in the generalization of the HUP,
from the early attempts~\cite{GUPearly}, to the more recent proposals, like those in string theory, loop quantum gravity,
deformed special relativity, non-commutative geometry, and studies of black hole 
physics~\cite{MM,FS,VenezGrossMende,Adler2,CGS,SC2013}.
\par
A possible way for this generalization is to reconsider the well-known classical argument of the
Heisenberg microscope~\cite{Heisenberg}. The size $\delta x$ of the smallest detail of an object,
theoretically detectable with a beam of photons of energy $E$, is roughly given by (if we assume
the dispersion relation $E=p$)\footnote{We shall always
work with $c=1$, but explicitly show the Newton constant $\gn$ and the Planck constant 
$\hbar$. The Planck length is defined as $\lp=\sqrt{\gn\,\hbar/c^3}\simeq 10^{-35}\,$m,
the Planck energy as $\Ep\,\lp = \hbar\, c/2$, and the Planck mass
as $\mpl=\Ep/c^2\simeq 10^{-8}\,$kg, so that $\lp=2\gn\,\mpl$ and $2\,\lp\,\mpl= \hbar$.
The Boltzmann constant $k_{\rm B}$ will be shown explicitly, unless otherwise stated.}
\be
\delta x
\,\simeq\,
\frac{\hbar}{2\, E}
\,,
\label{HS}
\ee
so that increasingly large energies are required to explore decreasingly small details.
In its original formulation, Hei\-senberg's gedanken experiment ignores gravity.
Nevertheless, gedanken experiments involving formation of gravitational instabilities
in high energy scatterings of strings~\cite{VenezGrossMende}, or gedanken experiments taking
into account the possible formation, in high energy scatterings, of micro black holes with a
gravitational radius $R_S=R_S(E)$ proportional to the (centre-of-mass) scattering energy $E$ 
(see Ref.~\cite{FS}), suggest that the usual uncertainty relation should be modified as
\be
\delta x
\,\simeq\,
\frac{\hbar}{2\, E}
\,+\,
\beta\, R_S(E)
\, ,
\ee
where $\beta$ is a dimensionless parameter.
Recalling that $R_S\simeq 2\,\gn\, E = 2\, \lp^2\, E/\hbar$, we can write
\be
\delta x
\,\simeq\,
\frac{\hbar}{2\, E}
\,+\,
2\beta\, \lp^2\frac{E}{\hbar}
\,=\,
\lp
\left(
\frac{\mpl}{E}
\,+\,
\beta\, \frac{E}{\mpl}
\right).
\label{He}
\ee
This kind of modification of the uncertainty principle was also proposed in Ref.~\cite{Adler2}.
\par
The dimensionless deforming parameter $\beta$ is not (in principle) fixed by the theory,
although it is generally assumed to be of order one.
This happens, in particular, in some models of string theory (see again for instance
Ref.~\cite{VenezGrossMende}), and has been confirmed by an explicit calculation in Ref.~\cite{SLV}.
However, many studies have appeared in literature, with the aim to set bounds on $\beta$
(see, for instance, Refs.~\cite{brau}).
\par
Equation~(\ref{He}) can be recast into the form of an uncertainty relation, namely a deformation
of the standard HUP, usually referred to as Generalized Uncertainty Principle (GUP),
\be
\Delta x\, \Delta p
\,\geq\,
\frac{\hbar}{2}
\left[1
\,+\,\beta
\left(\frac{\Delta p}{\mpl}\right)^2
\right].
\label{gup}
\ee
For mirror-symmetric states (with $\langle {p} \rangle = 0$), the inequality Eq.~(\ref{gup}) is equivalent to the commutator
\be
\left[{x},{p}\right]
\,=\,
i \hbar \left[
1
\,+\,\beta
\left(\frac{{p}}{\mpl} \right)^2 \right]
,
\label{gupcomm}
\ee
since $\Delta x\, \Delta p \geq (1/2)\left|\langle [{x},{p}] \rangle\right|$.
Vice-versa, the commutator Eq.~(\ref{gupcomm}) implies the inequality Eq.~(\ref{gup}) for any state.
The GUP is widely studied in the context of Quantum Mechanics~\cite{Pedram},
QFT~\cite{Husain:2012im}, quantum gravity~\cite{Hossain:2010wy},
and for various deformations of the quantization rules~\cite{Hossain:2010wy,Jizba:2009qf}.

The above $\beta$-deformed commutator Eq.~(\ref{gupcomm}) will be the starting point of the present
investigation. 
In what follows, using Eq.~(\ref{gupcomm}), we shall describe the Unruh effect,
thereby calculating corrections to the Unruh temperature to first order in $\beta$
(for a direct derivation of the standard Unruh effect from the HUP, see Ref.~\cite{FS9506}).
Furthermore, non-trivial modifications to the Unruh spectrum have also been pointed out,
in the GUP context, in Refs.~\cite{Nicolini,Majhi:2013koa,Husain,Gim}, and in different contexts
in Refs.~\cite{Blasone:2017nbf,Blasone:2018byx,Achim} (in Chapter~\ref{Non-thermal signature}, for instance, we have
discussed the non-thermality of
the Unruh spectrum within the framework of flavor mixing of fields).

\section{Heuristic derivation of the Unruh effect from uncertainty relations}
\label{Heuristic derivation of Unruh Effect from uncertainty relations}
In this Section we derive the canonical Unruh temperature, Eq.~\eqref{Tolman}, starting directly
from the HUP. Simple classical physics relations will be used together with the quantum principle,
following closely Ref.~\cite{FS9506} (see also the recent Ref.~\cite{gine}).
This procedure will then allow  to estimate what kind of corrections are induced by a GUP.

\subsection{Unruh temperature from Heisenberg's principle}
Let us consider some elementary particles, for example electrons, kept at rest in an uniformly
accelerated frame.
The kinetic energy acquired by each of these particles while the accelerated frame moves a
distance $\delta x$ will be given by
\be
E_k
\,=\,
m\,a\,\delta x
\, ,
\ee
where $m$ is the mass of the particle and $a$ the acceleration of the frame, and therefore
of the particle.
Now, suppose this energy is sufficient to create $N$ pairs of the same kind of particles from the quantum
vacuum.
Namely, we set
\be
E_k
\,\simeq\,
2\,N\, m
\, ,
\ee
and find that the distance along which each particle must be accelerated in order to create $N$ pairs is
\be
\delta x
\,\simeq\,
2\,\frac{N}{a}
\, .
\label{dx}
\ee
The original particles and the pairs created in this way are localized inside a spatial region of width $\delta x$,
therefore the fluctuation in energy of each single particle is
\be
\delta E
\,\simeq\,
\frac{\hbar}{2\, \delta x}
\,\simeq\,
\frac{\hbar\, a}{4\, N}
\,.
\ee
If we interpret this fluctuation as a classical thermal agitation of the particles, we can write
\be
\frac{3}{2}\,k_{\rm B}\,T
\,\simeq\,
\delta E
\,\simeq\,
\frac{\hbar\, a}{4\,N}
\ ,
\label{dE}
\ee
or
\be
T
\,=\,
\frac{\hbar\, a}{6\,N\,k_{\rm B}}
\, .
\ee
On comparing with the well-known Unruh temperature, Eq.~\eqref{billTemperature},
\be
T_{\rm U}
\,=\,
\frac{\hbar\, a}{2\,\pi\, k_{\rm B}}
\, ,
\label{Tu}
\ee
we can set the arbitrary parameter $N$ and obtain an effective number of pairs $N=\pi/3\simeq 1$.

\subsection{Unruh temperature from GUP}
We now repeat the same argument as above in the context of the GUP.
Upon replacing Eq.~\eqref{dx} into Eq.~\eqref{He} and interpreting the energy fluctuation $\delta E$
in terms of a classical thermal bath, we find 
\be
2\,\frac{N}{a}
\,\simeq\,
\frac{\hbar}{3\, k_{\rm B}\, T}
\,+\,
\beta\, \lp^2\, \frac{3\, k_{\rm B}\, T}{\hbar}
\, .
\label{approx}
\ee
Requiring that the $T$ equals the Unruh temperature~\eqref{Tu} for $\beta \to 0$ again fixes $N=\pi/3\simeq 1$,
and we finally obtain
\be
\frac{2\,\pi}{a}
\,\simeq\,
\frac{\hbar}{k_{\rm B}\, T}
\,+\,
9\beta\, \lp^2\, \frac{k_{\rm B}\, T}{\hbar}
\,=\,
\lp\left(
\frac{2\mpl}{k_{\rm B}\,T}
\,+\,
9\,\beta\,\frac{k_{\rm B}\, T}{2\mpl}
\right) .
\label{acctemp}
\ee
This relation can be easily inverted for $T=T(a)$.
However, it is reasonable to assume that $\beta\,k_{\rm B} T/\mpl\sim \beta \,m/\mpl$ is very small for any
fundamental particle with $m\ll \mpl$.
We can therefore expand in $\beta\,m/\mpl$ and find
\be
T
\,\simeq\,
T_{\rm U}
\left(1
\,+\,
\frac{9\,\beta}{4}\,\frac{\lp^2\,a^2}{\pi^2}
\right)
\,=\,
T_{\rm U}
\left[1
\,+\,
\frac{9\,\beta}{4}
\left(\frac{k_{\rm B}\,T_{\rm U}}{\mpl}
\right)^2
\right].
\label{newTHeuristic}
\ee
We also note an interesting physical property suggested by Eq.~(\ref{acctemp}), that is, by the GUP.
In order to maintain the inverted relation $T=T(a)$ physically meaningful (i.e., the temperature must be
a real number), there will be a maximal value for the acceleration, namely
\be
a
\,\lesssim\,
\frac{\pi}{3 \, \sqrt{\beta}\, \lp}
\ ,
\ee
and a corresponding maximal value for the Unruh temperature,
\be
k_{\rm B}\,T_{\rm U}
\,\lesssim\,
\frac{\mpl}{3\,\sqrt{\beta}}
\ .
\ee
These ideas and estimates naturally make contact with those reported, for example,
in Refs.~\cite{caianiello}.

\section{Field theoretical derivation of the Unruh effect from GUP}
\label{GUP}
In Chapter~\ref{Non-thermal signature}, the Unruh temperature Eq.~(\ref{Tu}) has  been derived within the framework of canonical QFT.  At this stage, one may wonder that result gets modified when starting from
the GUP commutator in Eq.~(\ref{gupcomm}).
To answer this question, it comes in handy an intermediate step concerning the effects of GUP on a quantum $1$-dimensional harmonic oscillator.

\subsection{GUP for the $1$-dimensional harmonic oscillator}
It is well-known that the ladder operators $A$ and $A^\dagger$ for
the $1$-dimensional harmonic oscillator can be expressed 
in terms of  the position and momentum operators
${x}={x}^\dagger$ and ${p}={p}^\dagger$, respectively, by
means of the usual relations
\be
\begin{array}{l}
	A \,=\, \strut\displaystyle\frac{1}{\sqrt{2\,m\,\hbar\,\omega}}
	\left(m\omega{x} + i \,{p}\right),
	\\[6mm]
	A^\dagger\, =\, \strut\displaystyle\frac{1}{\sqrt{2\,m\,\hbar\,\omega}}
	\left(m\omega{x} - i\, {p}\right),
\end{array}
\ee
and their inverse relations
\be
\begin{array}{l}
	{x}\, =\, \strut\displaystyle\sqrt{\frac{\hbar}{2\,m\,\omega}}
	\left(A^\dagger + A\right),
	\\[6mm]
	{p}\, =\, \strut\displaystyle i\,\sqrt{\frac{m\,\hbar\,\omega}{2}}
	\left(A^\dagger - A\right),
\end{array}
\ee
It is then easy to see that 
\be
[A,A^\dagger]\,=\,\frac{1}{i\hbar}\,[ x,  p]\,.
\ee
Therefore, due to the modified commutator Eq.~(\ref{gupcomm}) between 
$ x$ and $ p$, the deformed algebra for the $1$-dimensional harmonic oscillator should be written as
\begin{equation}
\label{eqn:oscillharm}
\left[A, A^\dagger\right]
\,=\,
\frac{1}{1\,-\,\alpha}
\left[1
\,-\,\alpha
\left(A^\dagger\, A^\dagger
\,+\,A\,A
\,-\,2\,A^\dagger\, A
\right)
\right],
\end{equation}
where 
\be
\alpha\,=\,\beta\,\frac{m\,\hbar\, \omega}{2\,\mpl^2}
\, ,
\ee
with $m$ and $\omega$ being the mass and frequency of the harmonic oscillator, respectively.

\subsection{GUP effects on the Unruh temperature}
The modified quantization rules Eq.~(\ref{eqn:oscillharm}) can be now extended in a natural way
to a scalar field in the plane wave representation, if we consider that, for a given momentum $k$,
the energy $\hbar \,\omega_{k}$ of the scalar field plays the r\^ole of the mass $m$ of the harmonic oscillator.
The deformation parameter $\alpha$ can then be suitably redefined as
\be 
\tilde{\alpha}\,=\,
\beta\,\frac{\hbar^2 \omega_{k}^2}{2\,\mpl^2}\,=\,
2\,\beta\,\lp^2\,\omega_{k}^2\,,
\ee
and the commutator between ladder operators becomes
\be
[A_k, A_{k'}^\dagger] \,=\, 
\frac{1}{1-\tilde{\alpha}}
\left[1
\,-\,
\tilde{\alpha}
\left(A_{k}^\dagger\, A_{k'}^\dagger
\,+\,A_{k}\, A_{k'}
\,-\,2\,A_{k}^\dagger\, A_{k'}
\right)
\right]
\delta(k-k')
\, .
\label{modplanwav}
\ee
\par
In Section~\ref{Hyperbrepr} of Chapter~\ref{Non-thermal signature}, we have seen that the scalar field in Minkowski spacetime can be equivalently quantized using plane waves and boost-modes (see Eqs.~(\ref{eqn:expans0})
and~(\ref{eqn:expansionfieldutilde}), respectively).
In that context, the choice between these two representations is just a matter of convenience,
since the corresponding sets of ladder operators $a_k$ and $d_{\Omega}^{(\sigma)}$
are related by a canonical transformation, Eq.~(\ref{eqn:operat-d}).
With deformed quantization rules, however, Lorentz invariance is violated and such an equivalence is not guaranteed.
Nevertheless, in the limit of very small deformation (that is, $\beta\, p^2\ll \mpl^2$), it appears reasonable
to assume the same structure of the modified algebra for the two sets of operators.
According to this argument, we thus conjecture the following deformation for the commutator in the
boost-mode representation
\bea
\label{eqn:d-D}
\left[D_{\Omega}^{(\sigma)},\,D_{\Omega'}^{(\sigma')\dagger}\right]
&=&
\frac{1}{1-\gamma}\left[
1\,-\,\gamma\,\left(
D_{\Omega}^{(\sigma)\dagger}\,D_{\Omega'}^{(-\sigma')\dagger}
\,+\,
D_{\Omega}^{(\sigma)}\,D_{\Omega'}^{(-\sigma')} \right. \right.
\nonumber
\\[3mm]
&&
\left. \left.
-\;D_{\Omega}^{(\sigma)\dagger}\,D_{\Omega'}^{(\sigma')}
\,-\,D_{\Omega}^{(-\sigma)\dagger}\,D_{\Omega'}^{(-\sigma')}
\right)
\right]
\delta_{\sigma\sigma'}\,
\delta(\Omega-\Omega')\,,
\qquad
\eea
where $D_{\Omega}^{\,(\sigma)}$ and $D_{\Omega}^{\,(\sigma)\dagger}$ are the ladder operators
in the deformed algebra and the deforming parameter $\gamma$ is defined by
\be
\label{ips}
\gamma
\,=\,
\beta\,
\frac{\hbar^2 \omega^2}{2\,\mpl^2}
\,=\,
\beta\,
\frac{\hbar^2 a^2\,\Omega^2}{2\,\mpl^2}
\,=\, 2\,\beta\,\lp^2\,a^2\,\Omega^2
\, ,
\ee
being $\omega=a\Omega$ the Rindler frequency.
\par
Some comments about Eq.~(\ref{eqn:d-D}) are needed.
First, in order to adapt the deformed commutator Eq.~(\ref{modplanwav}) to the boost operators $D$,
we have modified \emph{ad hoc} the definition of the deforming parameter $\tilde{\alpha}$
by replacing the plane-frequency $\omega_{k}$ with the  boost-mode frequency
$\omega=a\,\Omega$ [see Eq.~(\ref{ips})].
Furthermore, the commutator Eq.~(\ref{eqn:d-D}) has been multiplied by $\delta_{\sigma\sigma'}$
to ensure that the ladder operators in the right wedge $R_+$ are still commuting with the
corresponding operators in the left wedge $R_-$.
In addition, we have symmetrized it with respect to $\sigma$ and $-\sigma$, so that
\begin{equation}
\label{eqn:equalcomm}
\left[D_{\Omega}^{(\sigma)},D_{\Omega'}^{(\sigma')\dagger}\right]
\,=\,
\left[D_{\Omega}^{(-\sigma)},D_{\Omega'}^{(-\sigma')\dagger}\right].
\end{equation}
By exploiting this property and recasting the Bogoliubov transformation Eq.~(\ref{eqn:newformbogotransform})
into the form
\begin{equation}
B_{\Omega}^{(\sigma)}
\,=\,
{\sqrt{1+{N_{BE}}(\Omega)}}
\,D_{\Omega}^{(\sigma)}
\,+\,
\sqrt{{N_{BE}}(\Omega)}\,
D_{\Omega}^{(-\sigma)\dagger}
\,,
\label{eqn:newformbogotr}
\end{equation}
one can verify that the deformation Eq.~(\ref{eqn:d-D}) induces an identical modification to the
algebra of Rindler operators $B$.
\par
GUP effects on the Unruh temperature can now be investigated by calculating the distribution
of $B$-quanta in the Minkowski vacuum $|0\rangle_{\mathrm{M}}$.
By use of the transformation Eq.~(\ref{eqn:newformbogotr}), it can be shown that
\begin{equation}
\label{eqn:modexpecnum}
{}_{\mathrm{M}}\langle0|\,
B_{\Omega}^{(\sigma)\dagger}\,B_{\Omega'}^{ (\sigma')}\,
|0\rangle_{M}
\,=\,
\frac{1}{\left(e^{2\pi\Omega}-1\right)\left(1-\gamma\right)}\,
\delta_{\sigma\sigma'}\,
\delta(\Omega-\Omega')
\, ,
\end{equation}
to be compared with the standard Bose-Einstein distribution in Eq.~(\ref{eqn:aspectval}).
As expected, the Unruh spectrum gets non-trivially modified when using the deformed 
algebra Eq.~(\ref{eqn:d-D}), losing its characteristic thermal behavior.
However, for Rindler frequencies $\Omega$ such that $\gamma\ll1$,
namely (since $\beta\sim 1$) for $\hbar\,\omega\ll\mpl$, we have $e^{-\gamma}\simeq 1-\gamma$,
and Eq.~\eqref{eqn:modexpecnum} can be approximated as 
\begin{equation}
\label{eqn:approxexpecnum}
{}_{\mathrm{M}}\langle0|\,
B_{\Omega}^{(\sigma)\dagger}\,B_{\Omega'}^{ (\sigma')}\,
|0\rangle_{M}
\,\simeq\,
\frac{1}{e^{2\pi\Omega-\gamma}-1}\,
\delta_{\sigma\sigma'}\,
\delta(\Omega-\Omega')\,,
\end{equation}
where we neglected the term linear in $\gamma$ in the denominator of the r.h.s.
We can interpret Eq.~(\ref{eqn:approxexpecnum}) as a shifted Bose-Einstein thermal distribution
by introducing a shifted Unruh temperature $T$ such that the term $(2\pi\Omega-\gamma)$
can be rewritten as
\be
\label{newT}
2\pi\Omega-\gamma 
\,= \,
\frac{\hbar \,a\, \Omega}{k_{\rm B}\,T_{\rm U}} \,-\, \gamma
\, \equiv \,
\frac{\hbar \,a\, \Omega}{k_{\rm B}\,T}
\ .
\ee
We thus find for the shifted Unruh temperature
\be
T
\,=\,
\frac{T_{\rm U}}{1-\beta\,\pi\,\Omega\,k_{\rm B}^2\,T_{\rm U}^2/\mpl^2}
\,\simeq\,
T_{\rm U}
\left[1\, +\, \beta\, \pi\, \Omega \left(\frac{k_{\rm B}T_{\rm U}}{\mpl}\right)^2 \right]
\,=\,
T_{\rm U}
\left(1\,+\,\beta\,\pi\,\Omega\,\frac{\lp^2\,a^2}{\pi^2}\right)
\,.
\label{eqn:newT}
\ee
We note that such a modified temperature $T$ contains an explicit dependence on the Rindler
frequency $\Omega$.
This is due to the deformed structure of the
commutator Eq.~(\ref{gupcomm}), which explicitly depends on ${p}^2$, that is, essentially,
on the energy of the considered quantum mode. So, it is not surprising to recover such an
explicit dependence in the final formulae. Nevertheless, a simple thermodynamic argument
allows to get rid of this $\Omega$-dependance.
In fact, for small deformations, we are still close to the thermal black body spectrum.
Therefore the vast majority of the Unruh quanta will be emitted around a Rindler frequency
$\omega$ such that $\hbar\,\omega\simeq k_{\rm B}\,T_{\rm U}$, which means $\Omega\approx 1/2\pi$.
For this typical frequency, Eq.~\eqref{eqn:newT} reproduces quite closely the heuristic 
estimate Eq.~\eqref{newTHeuristic}. In fact
\begin{equation}
T
\simeq
T_{\rm U}
\left[1 \,+\, \frac{\beta}{2}  \left(\frac{k_{\rm B}T_{\rm U}}{\mpl}\right)^2 \right]
\,=\,
T_{\rm U}
\left(1 \,+\, \frac{\beta}{2}\frac{\lp^2\,a^2}{\pi^2}\right).
\label{eqn:newTQFT}
\end{equation}
\par
\par 
As a final remark, it should be noted that the deformation of the algebra Eq.~(\ref{eqn:d-D}) should affect, in principle, also the Hamiltonian.
Therefore, the Rindler frequency $\Omega$ in Eq.~(\ref{newT}) should be accordingly modified.
In the present analysis, however, since we are considering only small deformations of the quantization rules,
we have reasonably neglected those corrections, thus approximating the modified Rindler Hamiltonian 
with the original one.
\par
Concluding, for small deviations from the canonical quantization, we have found that the Unruh distribution
maintains its original thermal spectrum, provided that a new temperature $T$ is defined as in Eq.~(\ref{eqn:newT}).
\section{Conclusions}
In the context of the Generalized Uncertainty Principle, we have computed the correction induced on the
Unruh temperature by a deformed fundamental commutator.
This result has been obtained by following two independent paths.
First, we have proceeded in a heuristic way, using very general and reasonable physical considerations.
Already at this stage, however, we have been able to point out a dependence of the deformed Unruh
temperature on the cubic power of the acceleration. These considerations have been substantiated
and confirmed by means of a full-fledged QFT calculation.
This has been achieved by taking into account modified commutation relations for the ladder operators
compatible with the GUP in Eq.~(\ref{gupcomm}). In the limit of a small deformation of the commutator,
we have obtained again a dependence of the first correction term on the third power of the acceleration.
Besides, the more refined formalism of QFT has helped us to point out an explicit dependence of the
deformed Unruh temperature on the Rindler frequency $\Omega$, which, on the other hand, was
reasonably expected.
A simple and effective thermodynamic argument has then been used to identify the values of
the Rindler frequency $\Omega$ corresponding to the most probable emission.
As a consequence, the QFT calculation has been seen to match the heuristic estimate, indeed with almost the
same numerical coefficients. 
\par
Of course, many avenues for further investigations appear now in front of us. 
On a technical side, for example, one would like to check whether the QFT corrections to the Unruh
temperature are left unchanged when deforming the algebra at the level of field rather than ladder operators, 
or when adopting different modified uncertainty relations. Some possibilites, for instance, would be to 
work with a Generalized Uncertainty Principle containing a linear term in 
momentum~\cite{Ali:2009zq}, or with classes of non-commutative geometries which 
imply ultraviolet and infrared modifications in the form of 
nonzero minimal uncertainties in both positions and momenta~\cite{Kempf:1996ss}. 
A direct correspondence between  deformations of the uncertainty principle at Planck scale 
and corrections to the Unruh temperature might, in principle, clarify 
to what extent the existence of a minimal observable length does really affect
the standard QFT scenario.

Moreover, further light should be thrown on the relation between the deviation from thermality of the Unruh radiation
discussed in this Chapter and those found in different contexts
(\emph{e.g.}, the non-thermal like behavior of the Unruh effect derived in Chapter~\ref{Non-thermal signature}).
It could also be interesting to extend our formalism to the Hawking effect, for which an heuristic derivation
of the modified temperature has already been performed in Ref.~\cite{FS9506}.
\par
Finally, two examples (among the many possible) of applications that could be affected by the results
of this Chapter are, broadly speaking, the field of relativistic quantum
information theory (\emph{e.g.}~entanglement degradation, entanglement satellite experiments~\cite{brukner}) and
the corrections induced on analogue gravity experiments (\emph{e.g.} analogue Unruh radiation in fluids,
in BEC, etc.~\cite{silke}). Much work is still in progress along these directions.

\newpage\null\thispagestyle{empty}\newpage

\chapter{Thermal Quantum Field Theories and potential connections with the Hawking and Unruh effects} 
\label{Unified formalism for Thermal Quantum Field Theories: a geometric viewpoint}
\begin{flushright}
\emph{``The black holes of nature \\are the most perfect macroscopic objects\\ there are in the universe:\\ the only elements in their construction \\are our concepts of space and time.''}\\[1mm]
- Subrahmanyan Chandrasekhar -\\[6mm]
\end{flushright}
We have learned from previous Chapters that, net of
higher-order corrections coming out from superstructures (\emph{e.g.}
the mixing of fields with different masses or a modified algebra), 
the thermal character  of quantum vacuum 
is at the heart of Hawking and Unruh effects. 
In this sense, it is self-evident that the
such phenomena provide a natural bridge between the standard QFT in curved
background and field theories at finite 
temperature and density~\cite{UM0,BJV,DAS,LEB,LW87,KAP}. This undoubtedly 
offers new insights  into the analysis of the yet nebulous  relationship
among general covariance, temperature and
QFT.

To understand this connection
more deeply, in what follows, after briefly discussing the
various existent approaches to Thermal Quantum Field Theory (TQFT), 
we focus specifically on Thermo Field Dynamics (TFD)~\cite{UM0,UM01,UM1}, 
also in light of the close parallel between the \emph{doubling}
of degrees of freedom this formalism is built on
and the augmenting naturally arising in spacetimes
endowed with event horizon(s) (among which the Rindler and Schwarzschild
backgrounds are the most eloquent examples). 
As it will be shown in Section~\ref{TFDform},
the basic idea of TFD is the transposition of thermal averages
(which are traces in statistical mechanics) to expectation values on 
a temperature-dependent ``vacuum'' living in a wider domain than
the original Hilbert space. This ad-hoc framework is obtained  by
augmenting the degrees of freedom with the introduction
of a dynamical replica of the physical system, 
the so-called \emph{tilde} Fock space:  the effective thermal representation
is then generated from the ensuing doubled space via
a Bogoliubov transformation 
analogous to those encountered so far. 

Enlarging this view to the QFT in backgrounds with non-trivial topology,
one finds it natural to describe
particle states on the hidden side of horizon(s) in terms of  
``additional'' degrees of freedom to trace over~\cite{Israel:1976}: therefore, in this naive picture, 
the loss of information associated with event horizon(s) ends up 
playing the same r\^ole in the appearance of temperature in QFT
as the doubled Hilbert space in TFD.

The relationship between geometric properties of the spacetime and
thermal quantum effects makes 
the analysis of QFTs at finite temperature
(and, in particular, of TFD)
all the more essential at this stage. This is indeed what
we intend to analyze in Sections~\ref{TQFTabo} and~\ref{TFDform},
also in view of extending our focus to the investigation of the 
thermodynamic of black holes
and quantum entanglement in thermal systems 
(see, for example, Ref.~\cite{Hashizume} and therein
for a preliminary treatment of the latter issue in both
equilibrium and non-equilibrium situations).
Furthermore, in order to develop a unified theoretical perspective
on the various existent TQFT formalisms, in Section~\ref{UFTQFT}
we illustrate how these seemingly different approaches can
be merged via the introduction of a flat complex manifold,
 the so-called $\eta$-$\xi$ spacetime. By performing field quantization,
it will be shown that the zero temperature vacuum
corresponds to a usual
thermal state for Minkowski inertial observers in this spacetime.
Besides, the intimate connection  
between backgrounds with a non-trivial horizon structure and 
laws of QFT and thermodynamics will be found to naturally
arise, thereby providing precious insights 
towards a complete understanding of the
common thread running through them.

\section{Thermal Quantum Field Theories: a brief oveview}
\label{TQFTabo}
Under the generic name Thermal Quantum Field Theory~\cite{UM0,BJV,DAS,LEB,LW87,KAP} one collects all formalisms of QFT at finite temperature and density, i.e., the Matsubara or Imaginary Time (IT) approach~\cite{DAS,LEB,LW87}, the
Thermo Field Dynamics (TFD)~\cite{UM0,DAS,UM01,UM1} and the Path Ordered Method (POM)~\cite{BJV,Mills}. The latter includes, for instance, the familiar Closed Time Path (CTP) formalism of Keldysh and Schwinger~\cite{DAS,LEB,LW87} as a special case.
The existence of these distinct approaches results from conceptually different efforts to introduce a temperature within the framework of QFT. For instance, in the IT formalism one exploits the analogy between the (inverse) temperature and imaginary time in calculating the partition function. Within this approach, the two-point Green's function is given by the Matsubara propagator and the sum over Matsubara frequencies (alongside with various summation techniques) must be invoked when dealing with multi-loop thermal diagrams. Because time is traded for (inverse) temperature, the IT formalism can not directly address field dynamics within a heat bath and it is thus suitable basically only for QFT at thermal equilibrium.
In principle, real-time quantities can be obtained by analytic continuation to the real axis, but in practice this procedure is plagued with ambiguities and further delimitations are typically needed~\cite{LEB,LW87,Dolan}. Aforementioned ambiguities typically appear in higher point Green's functions, 
\emph{e.g.}, in three-point thermal Green's functions~\cite{Baier,Evans,Frenkel}. Ambiguities were also reported in the $\beta$-function calculations at the one-loop level~\cite{Braten,LandsmanIII}.

In contrast, both the POM formalism and TFD accommodate time and temperature on equal footing and no extra analytic continuation is required. On the one hand, this provides a powerful theoretical platform allowing to address such issues as a temporal dynamics of quantum fields in a thermal heat bath~\cite{PVL,GUI92ferm,JT}, dynamics of phase-transition processes~\cite{BJV} or linear responses~\cite{LEB}. On the other hand, the price to be paid for working with the real time is the doubling of the field degrees of freedom which is reflected in a $2 \times 2$ matrix structure of thermal propagators and self-energies. Consequently, in higher-loop orders a much larger number of diagrams has to be taken into account as compared to the vacuum (i.e., zero temperature) theory. Surprisingly, the POM and TFD approaches lead to the same matrix form for thermal propagator in equilibrium~\cite{H95}. This is a quite intriguing fact since  POM and TFD  have very different conceptual underpinnings.
In the POM formalism one introduces the temperature by adding a pure imaginary number to the real time and chooses a special time path in the complex-time plane which involves the use of both time- and anti-time-ordered Green's functions. In comparison with this, the TFD
is a method for describing mixed states as pure states in an enlarged Hilbert space (akin to a purification procedure used in quantum optics~\cite{Cirac}).
It is characterized by a doubling of the field algebra and its mathematical underpinning is provided by algebraic QFT ($C^*$-algebras,
the Haag-Hugenholtz-Winnink (HHW) formalism and Tomita-Takesaki modular theory). In the TFD  the temperature is contained explicitly in the resulting ``vacuum''
pure state, which is referred to as the ``thermal vacuum'' state. Ensuing field propagators are then expressed as expectation values of time-ordered products of quantum
fields with respect to such a thermal state. From the above considerations, it appears that a doubling of the degrees of freedom is necessary in order to be able to calculate real-time
Green's functions. This doubling is, however, absent in the Matsubara formalism, and thus it could a mere mathematical artifact. Yet, such a doubling of the field degrees of freedom
also appears in the axiomatic formulation of quantum statistical mechanics~\cite{Haag:1967sg,HaagII,Sewel}.
This indicates that a two-component extension is essential for a consistent Minkowski-space field theory at finite temperature and density: in light of this, let us investigate in detail such a viewpoint. 

\section{Thermo Field Dynamics}
\label{TFDform}
In this Section we present essential notions of TFD,  taking special care 
of the characteristic doubling of degrees of freedom it is built on. To this end, we closely
follows Ref.~\cite{UM0}, where TFD was originally developed.

Thermo Field Dynamics is an operatorial, real time formalism for field theory at finite
temperature and density.  Consider, for simplicity, a quantum
system of Hamiltonian ${H}$ and discrete eigenstates 
$|n\rangle$, with  $H|n\rangle=E_n|n\rangle$ and $\langle n|m\rangle=\delta_{nm}$.\footnote{Similar
considerations can be generalized to systems with continuous energy spectrum (details can be found
in Refs.~\cite{H95,Landsman:1988ta}).}
The key idea of TFD is the possibility 
to express the statistical average of a generic
observable $A$ in a state with inverse temperature $\beta=1/T$ 
as vacuum expectation value in a suitable Fock space via the
assumption
\be
\label{medthermTFDform}
\langle {A}\rangle_{\beta}\,=\,Z^{-1}(\beta)\sum_n e^{-\beta E_n}\,\langle n| {A}|n\rangle
\,=\, Z^{-1}(\beta)\, \mathrm{Tr}(e^{-\beta H}{A})\,,
\ee
where $Z(\beta)=\sum_n e^{-\beta E_n}=\mathrm{Tr}(e^{-\beta{H}})$ 
is the grand-canonical partition function and the trace is taken over the full
Hilbert space.

The point is thus to construct a temperature-dependent state $|0(\beta)\rangle$ (i.e., the
``thermal vacuum'') satisfying
\be
\label{TFDcruceq}
\langle 0(\beta)|A|0(\beta)\rangle\,=\,Z^{-1}(\beta)\,\sum_n \langle n|A|n\rangle\,e^{-\beta E_n}\,.
\ee 
As noted by Takahashi and Umezawa, 
such a state cannot be constructed as long as one remains in the 
Fock space spanned by $|n\rangle$.  To see this, let us
consider  the following expansion:
\be
\label{thermvacTFD}
|0(\beta)\rangle\,=\,\sum_nf_n(\beta)|n\rangle\,.
\ee
By inserting into Eq.~(\ref{medthermTFDform}), one obtains
\be
\label{obtTFD}
f^*_n(\beta)\,f_m(\beta)\,=\,Z^{-1}(\beta)\,e^{-\beta E_n}\,\delta_{nm}\,,
\ee
which indeed cannot be satisfied by any $c$-number function $f_n(\beta)$.
The obtained relation, however,  can be reinterpreted
as the orthogonality condition
for vectors in a Hilbert space. We may thus think of  $|0(\beta)\rangle$
as a state living in a larger space than the original one $\{|n\rangle\}$:
as anticipated in the introductory Section, 
such an enlarged representation  is obtained by introducing 
a fictitious dynamical system identical to the one under consideration.
The quantities associated
with this dual system are denoted by the \emph{tilde} in the usual 
TFD notation~\cite{UM0}, namely $\tilde{H}|\tilde{n}\rangle=E_n|\tilde{n}\rangle$ 
and $\langle \tilde{n}|\tilde{m}\rangle=\delta_{nm}$, with $E_n$ being defined
as for the physical states. Furthermore, it is assumed that 
non-tilde and tilde operators are (anti-)commuting among themselves for (fermions) bosons.

In this naive picture, the thermal ground state 
satisfying Eq.~\eqref{medthermTFDform} 
takes the form
\begin{equation}
|0(\beta)\rangle\,=\,Z^{-1/2}(\beta)\sum_n\,e^{-\beta E_n/2}|n,\tilde{n}\rangle\,,
\label{eqn:0beta}
\end{equation}
where $|n,\tilde{n}\rangle=|n\rangle\otimes|\tilde{n}\rangle$ belongs to the doubled 
space. 
The states $|n\rangle$ and $|\tilde{n}\rangle$ thus appear
as a pair in the thermal vacuum $|0(\beta)\rangle$.

Physical interpretations for the  ``fictitious'' states
$|\tilde{n}\rangle$ can be provided by observing
that their introduction does allow to 
pick up the diagonal matrix elements of $A$ in the calculation
of expectation values. For example, thinking to the 
r\^ole of the environment in the QM decoherence processes~\cite{Zurek:1981xq},
which is to reduce the density matrix of the system to its diagonal
form, the $|\tilde{n}\rangle$ vectors are indeed susceptible to being interpreted as
a sort of environment degrees of freedom (see below for more comments on 
this). On the other hand, in the context of QFT in backgrounds endowed
with event horizon(s), these vectors may  be  regarded as
particle states living in the hidden side of  the spacetime (in
the Rindler framework (Fig.~\ref{figure:RindlerNONTHERMAL}),
for instance, an accelerated observer
confined inside the right wedge will find
it natural to associate the tilde degrees of freedom
to ``unobservable'' particles in the left sector).

In order to show the most relevant features of the thermal space
$\{|0(\beta)\rangle\}$, let us now contextualize the 
the above considerations within two emblematic examples.

\subsection{TFD for bosons}
Consider two  sets of commuting ladder operators for bosonic systems, 
\begin{equation}
\label{eqn:commutrelations}
[a_{\bk}, a^\dagger_{\bp}]\,=\,[\tilde a_{\bk},\tilde a^\dagger_{\bp}]\,=\,\delta_{\bk,\bp}\,,
\end{equation}
with all other commutators vanishing and $a_{\bk}$, $\tilde a_{\bk}$ commuting 
among themselves. Let us denote by $H=\sum_{\bk}\omega_k\, a^\dagger_{\bk}a_{\bk}$,
$\tilde{H}=\sum_{\bk}\omega_k\, \tilde{a}^\dagger_{\bk}\tilde{a}_{\bk}$ 
the corresponding (free) Hamiltonians and be $|0\rangle$ 
their common vacuum state, i.e. $a_{\bk}|0\rangle=\tilde a_{\bk}|0\rangle=0$.

We want to show that the thermal
Fock space $\{|0(\beta)\rangle\}$ can be generated 
from the doubled Fock space $\{|0,0\rangle\}$
via a non-trivial Bogoliubov transformation; to this end, let us define
the thermal operators by means of the following transformations\footnote{In 
Sec.~\ref{sec.4} it will be shown that the form of the Bogoliubov
transformations is not unique; although this arbitrariness 
is irrelevant in {\em thermal equilibrium}, since it does not affect 
physical quantities, the situation
becomes much more complicated 
in non-equilibrium conditions, where
the choice of the parameterization is 
indeed related to a particular
form of a transport equation~\cite{H95}.}:
\begin{eqnarray}
a_{\bk}(\theta)&=&e^{-iG}a_{\bk}e^{iG}\,=\,a_{\bk}\cosh\theta_{\bk}\,-\,\tilde a^\dagger_{\bk}\sinh\theta_{\bk}\,,\label{eqn:discretbogoa}\\[2mm]
\tilde a_{\bk}(\theta)&=&e^{-iG}\tilde a_{\bk}e^{iG}\,=\,\tilde a_{\bk}\cosh\theta_{\bk}\,-\,a^\dagger_{\bk}\sinh\theta_{\bk}\,,\label{eqn:discretbogob}
\end{eqnarray}
where $\theta_{\bk}=\theta_{\bk}(\beta)$ is the temperature-dependent
Bogoliubov parameter to be determined and $G$ is the hermitian generator
given by
\begin{equation}
G\,=\,i\sum_{\bk}\theta_{\bk}\left[a^\dagger_{\bk}\tilde a^\dagger_{\bk}-\tilde{a}_{\bk}a_{\bk}\right].
\end{equation}
The ``hyperbolic'' relation
\begin{equation}
\cosh^2\theta_{\bk}\,-\,\sinh^2\theta_{\bk}\,=\,1\,,
\label{eqn:guaranteerelation}
\end{equation}
guarantees that the thermal operators
are still canonical.

In terms of the original ground state $|0\rangle$, the  vacuum 
$|0(\theta)\rangle$ annihilated by $a_{\bk}(\theta)$ and $\tilde a_{\bk}(\theta)$ 
takes the form
\begin{equation}
|0(\theta)\rangle=e^{-i{G}}|0\rangle=\prod_{\bk}\,\frac{1}{\cosh{\theta_{\bk}}}\,\exp\Big[\tanh\theta_{\bk}\,a^\dagger_{\bk}\tilde{a}^\dagger_{\bk}\Big]|0\rangle\,,
\label{eqn:|0(theta)rangle}
\end{equation}
with $a_{\bk}(\theta)|0(\theta)\rangle=\tilde{a}_{\bk}|0(\theta)\rangle=0$, 
as expected. Note that this state is the vacuum (zero energy state) of
neither $H$ nor $\tilde H$; it is however the zero energy eigenstate of the
total Hamiltonian 
\be
 H=H-\tilde H\,,
\ee
i.e. $ H|0(\theta)\rangle=0$.

In order to identify the state $|0(\theta)\rangle$ with the
thermal vacuum $|0(\beta)\rangle$ defined in
Eq.~(\ref{eqn:0beta}), the relation between $\theta$
and $\beta$ must be made explicit. To this aim, let 
us recast Eq.~(\ref{eqn:0beta}) into the form
\begin{equation}
|0(\theta)\rangle\,=\,e^{-S/2}\exp\bigg[\sum_{\bk}a^\dagger_{\bk}\tilde{a}^\dagger_{\bk}\bigg]|0\rangle\,=\,e^{-\tilde{S}/2}\exp\bigg[\sum_{\bk}a^\dagger_{\bk}\tilde{a}^\dagger_{\bk}\bigg]|0\rangle\,,
\end{equation}
where
\begin{equation}
S\,=\,-\sum_{\bk}\bigg[a^{\dagger}_{\bk}a_{\bk}\log\sinh^2\theta_{\bk}\,-\,a_{\bk}a^\dagger_{\bk}\log\cosh^2\theta_{\bk}\bigg],\quad \tilde S\,=\, S\,(a_\bk\rightarrow \tilde a_\bk)\,,
\label{eqn:entropy}
\end{equation}
(see below for a possible thermodynamical interpretation of this operator).

As widely discussed in the previous Chapters, due to the 
thermal Bogoliubov transformation in Eq.~\eqref{eqn:|0(theta)rangle},
the vacuum state $|0(\theta)\rangle$ acquires a non-trivial condensate structure; 
specifically, it turns out to be a 
$SU(1,1)$ generalized coherent state~\cite{Perepere}. Notice also that,
for each $\theta$, one has a copy of the original algebra 
$\{a_\bk(\theta), a^\dagger_\bk(\theta),
\tilde a_\bk(\theta), \tilde a^\dagger_\bk(\theta);\, |0(\theta)\rangle |\hspace{1mm} \forall \bk\}$
induced by the Bogoliubov generator $G$, which thus can be 
regarded as a generator of the group of automorphisms of $\otimes_\bk su(1,1)_\bk$
parametrized by $\theta_\bk$. In this connection, let us
observe that the operator $N_\bk-\tilde{N}_\bk\equiv a^\dagger_\bk a_\bk-\tilde a^\dagger_\bk \tilde a_\bk$,
where $N_\bk$ ($\tilde{N}_\bk$) is the number operator for physical (tilde) particles, 
is invariant under the Bogliubov transformations Eqs.~\eqref{eqn:discretbogoa}-\eqref{eqn:discretbogob},
i.e.:
\be
a^\dagger_\bk a_\bk-\tilde a^\dagger_\bk \tilde a_\bk\,=\,a^\dagger_\bk(\theta) a_\bk(\theta)\,-\,\tilde a^\dagger_\bk(\theta) \tilde a_\bk(\theta),\qquad \forall \theta\,.
\ee
Therefore, $\frac{\delta}{\delta\theta}(N_\bk(\theta)-\tilde{N}_\bk(\theta))=0$,
with $N_\bk(\theta)-\tilde N_\bk(\theta)\equiv a^\dagger_\bk(\theta) a_\bk(\theta)\,-\,\tilde a^\dagger_\bk(\theta) \tilde a_\bk(\theta)$, in accordance with the fact
that $\frac{1}{4}{(N_\bk-\tilde N_\bk)}^2$ is the Casimir
operator of the $su(1,1)_\bk$ algebra.

By use of Eqs.~\eqref{eqn:discretbogoa}-\eqref{eqn:discretbogob}, the condensation
density of physical particles in the thermal vacuum can be  easily calculated, yielding
\begin{equation}
n_{\bk}\,\equiv\,\langle0(\theta)|a^\dagger_{\bk}a_{\bk}|0(\theta)\rangle\,=\,\sinh^2\theta_{\bk}\,,
\end{equation}
with an analogous result for the tilde particles. 

At the same time, by minimizing with respect to $\theta_\bk$
the quantity
\begin{equation}
\label{ome0}
\langle\Omega\rangle_0\,\equiv\,\langle0(\theta)|\bigg[-\frac{1}{\beta} S\,+\,H\,-\,\mu N\bigg]|0(\theta)\rangle\,,
\end{equation}
we obtain (redefining the energies $\omega_\bk$ by including $\mu$)
\begin{equation}
n_{\bk}\,=\,\frac{1}{e^{\beta\omega_k}\,-\,1}\,,
\label{eqn:mediaattesa}
\end{equation}
which is the correct (Bose-Einstein) thermal average for a bosonic system 
at inverse temperature $\beta$. By virtue of this, the vacuum $|0(\theta)\rangle$ can be properly identified
with the (fundamental) thermal ground state $|0(\beta)\rangle$: thus, 
we conclude that the thermal Fock space $\{|0(\beta)\rangle\}$ and the 
free (enlarged) Fock space $\{|0,0\rangle\}$ are indeed related
via the Bogoliubov transformations Eqs.~\eqref{eqn:discretbogoa}-(\ref{eqn:discretbogob}).

The obtained result also allows for a possible thermodynamical 
interpretation of the quantities $S$ ($\tilde S$) and
$\Omega$ introduced above: exploiting the thermality of the Bose-Einstein
distribution Eq.~\eqref{eqn:mediaattesa}, indeed, one finds it natural
to describe $S$ ($\tilde S$) in Eq.~\eqref{eqn:entropy} as the entropy operator for the physical (tilde)
system, while $\Omega$ in Eq.~\eqref{ome0} is the Landau free energy, with $\mu$ being the
chemical potential. 

In this picture, the physical meaning of the tilde degrees
of freedom is made possible, for example, by looking at the relation
$a_\bk|0(\theta)\rangle=0=\tilde a_\bk|0(\theta)\rangle$;
this shows that the creation (annihilation) of a physical $a$-particle
in the vacuum $|0(\theta)\rangle$ is equivalent (up to an irrelevant
$c$-number factor) to the annihilation (creation) of a tilde $\tilde 
a$-particle in the vacuum itself (cf., e.g., the transformations 
in Eqs.~\eqref{eqn:discretbogoa}-(\ref{eqn:discretbogob}). The $\tilde 
a$-particles thus can be interpreted as \emph{holes}, or anti-particles,
for the physical quanta. Additionally, when dealing with either
non-equilibrium or dissipative systems, it can be seen that the 
energy flowing out of the physical system is exchanged with the dual tilde system~\cite{UM01,UM1,Celeghini:1991yv}, which may therefore represent the
environment into which the $a$-particles live.

In closing, we stress that the Bogoliubov transformation Eq.~(\ref{eqn:|0(theta)rangle})
has merely a formal meaning, holding at finite volume; in the infinite volume limit, 
using the continuous limit relation $\sum_\bk\rightarrow \frac{V}{{(2\pi)}^3}\int d^3k$, 
one can show that~\cite{BJV} 
\be
\langle 0(\theta)|0\rangle\rightarrow 0\,,\quad \forall \theta\,\equiv\,\{\theta_\bk\}\neq 0\,,
\ee 
and, more general, that
\be
\langle 0(\theta)|0(\theta')\rangle\rightarrow 0\,,
\ee 
as $V\rightarrow\infty$, 
$\forall \theta'\neq\theta$. Thus, for each $\theta\equiv\{\theta_\bk\}$,
one has a representation $\{|0(\theta)\rangle\}$ of the canonical 
commutation relations that is unitarily inequivalent
to any other representation $\{|0(\theta')\rangle,\, \forall \theta'\neq\theta\}$ 
in the infinite volume limit (this is a well-known feature of
QFT~\cite{UM01} reflecting into the non-unitary nature of
the generator of Bogoliubov transformations in the infinite volume limit. We remand 
the reader to Appendix~\ref{QFT of fm} for a discussion on this issue in the context of flavor mixing in QFT).
As usual, one works at
finite volume and only at the end of the computations the limit $V\rightarrow\infty$ is
performed.

The analysis of these (and other) features of TFD will be taken up in Section~\ref{Sec.4.3}, 
where we are also going to comment on the \emph{tilde-conjugation symmetry}
(i.e. the symmetry between physical and tilde worlds)  in connection
with the geometric properties of the \EX spacetime.

\subsection*{Boson thermal propagators}
\label{bthpr}
In order to illustrate the propagator structure 
of a thermal theory, consider the simplest situation of a (real) scalar
field in thermal equilibrium. We have\footnote{As shown in Ref.~\cite{BJV},
a perturbative approach based on (on-shell) free fields does not hold at
finite temperature; for our purpose, however, we shall assume
that the temperature is low enough so that a description
of excitations in terms of free fields is approximatively valid.}
\begin{eqnarray}
\phi(x)&\hspace{-1mm}=\hspace{-1mm}&\int 
\frac{d^{3}{k}}{(2\pi)^{\frac{3}{2}}
(2\omega_{\bk})^{\frac{1}{2}}}\,
\left[a_{{\bk}}\hspace{0.2mm}e^{i\left({\bk\cdot\bx}\,-\,\omega_{\bk} t\right)} \, + \,
a^{\dag}_{{\bk}}\hspace{0.2mm}e^{i\left(-{\bk\cdot\bx}\,+\,\omega_{\bk} t\right)}\right],
\mlab{4.21a} \\[2mm]
{\tilde \phi}(x) &\hspace{-1mm}=\hspace{-1mm}& \int
\frac{d^{3}{k}}{(2\pi)^{\frac{3}{2}}
(2\omega_{\bk})^{\frac{1}{2}}}\,
\left[{\tilde a}_{{\bk}}\hspace{0.2mm}e^{i\left(-{\bk\cdot\bx}\,+\,\omega_{\bk} t\right)}\,+
\, {\tilde a}^{\dag}_{{\bk}}\hspace{0.2mm}
e^{i\left({\bk\cdot\bx\,-\,\omega_{\bk} t}\right)}\right].
\label{4.21b}
\end{eqnarray}
where $\phi(x)$ and $\tilde \phi(x)$ commute with each other and
\begin{eqnarray}
\Big[\phi(t,\bx),\partial_t\phi(t,\bx')\Big]&\hspace{-1mm}=\hspace{-1mm}&i\delta^3(x-x')\,,\\[2mm]
\left[\tilde\phi(t,\bx),\partial_t\tilde\phi(t,\bx')\right]&\hspace{-1mm}=\hspace{-1mm}&-i\delta^3(x-x')\,.
\end{eqnarray}
In TFD it is well-known that  the propagators exhibit a matrix
structure, arising from the
various possible combinations of physical and tilde fields in the (thermal)
vacuum expectation value (although the physical and tilde
degrees of freedom are not coupled in the total hamiltonian $ H$, indeed,
they do couple in the thermal vacuum state $|0(\theta)\rangle$).
The thermal causal propagator for 
a free scalar field $\phi(x)$ is calculated as
\begin{equation}
\mlab{4.36bubu}
\Delta_{ab}(x,y) \, = \, 
\langle 0(\theta)|\hspace{0.2mm} T\big[ \phi^a(x) \phi^{b\dagger}(y)\big]\hspace{0.2mm}|0(\theta)\rangle\, ,
\end{equation}
where $T$ is the time ordering symbol and 
the $a,b$ indices refer to the thermal doublet 
$\phi^1 \,= \,\phi$ and $\phi^2 \,=\, \tilde{\phi^{\dagger}}$. 
Since we are considering a 
real scalar field, we should use in the above 
definition $\phi^2 \,=\, \tilde{\phi}$.

Remarkably, in the momentum representation, the
above propagator takes the form~\cite{UM01}
\be
\Delta_{ab}(k_0,\bk)\,=\,{\left(B_\bk^{-1}\left[\frac{1}{k_0^2\,-\,{(\omega_\bk-i\varepsilon\tau_3)}^2}\right]B_\bk\tau_3
\right)}_{\hspace{-0.5mm}ab}\,,
\ee
where is the usual Pauli matrix $\mathrm{diag}(1,-1)$ and $\varepsilon>0$.
Note that the internal (or ``core'') matrix is diagonal
and coincides with the standard (zero temperature)
Feynman propagator. It thus arises that the thermal 
propagator is obtained from the standard one via the action
of the Bogoliubov matrix Eqs.~\eqref{eqn:discretbogoa}-\eqref{eqn:discretbogob}, 
here rewritten as
\be
B_\bk
\,=\,\begin{pmatrix}
\cosh\theta_{\bk}&-\sinh\theta_{\bk}\\
-\sinh\theta_{\bk}&\cosh\theta_{\bk}
\end{pmatrix}\,.
\ee
Therefore, the thermal Bogoliubov transformation 
only acts on the imaginary part of the propagator; for instance,
the $(1,1)$ component of $\Delta_{ab}(k_0,\bk)$ turns out to be
\be
\Delta_{11}(k_0,\bk)\,=\,{\left[k_0^2-{\left(\omega_\bk-i\varepsilon\right)}^2\right]}^{-1}\,-\,
2\pi i n_\bk\delta(k_0^2\,-\,\omega_\bk^2)\,,
\ee
where $n_\bk$ is the number of $a_\bk$-quanta.

A further relevant property of the finite temperature
propagator Eq.~\eqref{4.36bubu} is that only three out of the four elements
are independent, since one has
\be
\Delta_{11}\,+\,\Delta_{22}\,-\,\Delta_{12}\,-\,\Delta_{21}\,=\,0\,,
\ee
which is valid for any different (gauge) parameterizations of the
thermal Bogoliubov matrix, as discussed in the next Section.

\subsection{TFD for fermions}
A construction similar to the above can be developed in the case
of two sets of anti-commuting fermionic ladder operators~\cite{UM0}, 
here denoted as $\alpha$, $\tilde\alpha$:
\begin{equation}
\label{eqn:commutrelationsbis}
\left\{\alpha_{\bk},\alpha^\dagger_{\bp}\right\}\,=\,\left\{\tilde \alpha_{\bk},\tilde \alpha^\dagger_{\bp}\right\}=\delta_{\bk,\bp}\,,
\end{equation}
with all other anti-commutators vanishing and $\alpha_{\bk}$, $\tilde \alpha_{\bk}$ anti-commuting 
among themselves. 

In analogy with Eqs.~\eqref{eqn:discretbogoa}-\eqref{eqn:discretbogob}, 
the thermal operators are now given by
\begin{eqnarray}
\alpha_{\bk}(\theta)&=&e^{-i{G_f}}\alpha_{\bk}e^{i{G_f}}=\alpha_{\bk}\cos\theta_{\bk}\,-\,\tilde \alpha^\dagger_{\bk}\sin\theta_{\bk}\, ,\label{eqn:discretbogoa2}\\[2mm]
\tilde \alpha_{\bk}(\theta)&=&e^{-i{G_f}}\tilde a_{\bk}e^{i{G_f}}=\tilde \alpha_{\bk}\cos\theta_{\bk}\,+\,\alpha^\dagger_{\bk}\sin\theta_{\bk}\, ,\label{eqn:discretbogob2}
\end{eqnarray}
where the generator $G_f$ has the form
\begin{equation}
{G_f}\,=\,i\sum_{\bk}\theta_{\bk}\left[\alpha^\dagger_{\bk}\tilde \alpha^\dagger_{\bk}\,-\,\tilde{\alpha}_{\bk}\alpha_{\bk}\right]\,,
\end{equation}
and it is non-unitary in the infinite volume limit, as discussed above. 
The ``circular'' relation
\begin{equation}
\cos^2\theta_{\bk}\,+\,\sin^2\theta_{\bk}\,=\,1\,,
\end{equation} 
guarantees that the thermal operators are still canonical.

Unlike bosons, the Bogoliubov transformation Eqs.~\eqref{eqn:discretbogoa2}-\eqref{eqn:discretbogob2}
should now be thought of as inner automorphism of the algebra $\otimes su(2)_\bk$ parameterized by $\theta_\bk$; thus, the fermionic thermal vacuum acquires the structure of a $SU(2)$ 
generalized coherent state~\cite{Perepere}, 
\begin{equation}
|0(\theta)\rangle\,=\,e^{-iG_f}|0\rangle\,=\,\prod_{\bk}\left[\cos\theta_{\bk}+\sin\theta_{\bk}\,\alpha^\dagger_{\bk}\tilde\alpha^\dagger_{\bk}\right]|0\rangle\,,
\end{equation}
with
\begin{equation}
\alpha_{\bk}(\theta)|0(\theta)\rangle=\tilde\alpha_{\bk}(\theta)|0(\theta)\rangle=0\,,
\end{equation}
and entropy operator 
\begin{equation}
S\,=\,-\sum_{\bk}\left[\alpha^\dagger_{\bk}\alpha_{\bk}\,\log\sin^2\theta_{\bk}\,+\,\alpha_{\bk}\alpha^\dagger_{\bk}\log\cos^2\theta_{\bk}\right],\quad \tilde S\,=\, S\,(a_\bk\rightarrow \tilde a_\bk)\,.
\end{equation}
The condensation density of physical particles  is given by
\begin{equation}
n_{\bk}\equiv\langle 0(\theta)|\alpha^\dagger_{\bk}\alpha_{\bk}|0(\theta)\rangle
\,=\,\sin^2\theta_{\bk}\,,
\end{equation}
with an analogous result for the tilde particles. By a procedure similar 
to that of Eq.~\eqref{ome0}, we also obtain the Fermi-Dirac distribution
\be
n_{\bk}\,=\,\frac{1}{e^{\beta\omega_\bk}+1}\,,
\ee 
to be compared with the Bose-Einstein distribution, Eq.~\eqref{eqn:mediaattesa}.

We further remark that the same propagator structure 
as the one described in the previous Section still holds for
fermions (the reader is remanded to the literature quoted above
for a more thorough description of this).

\section{Unified formalism for TQFTs: a geometric viewpoint}
\label{UFTQFT}
So far, we have discussed analogies and differences 
of various existent TQFT formalisms, focusing specifically on Thermo Field
Dynamics. In view of the obtained results, an interesting question
 is whether these seemingly different approaches have some
roots in common or, in other words, if their features can be understood in a deeper way
so that they appear to be unified. A clue to an answer may be found in the well-known
Hawking's discovery~\cite{Hawking:1974sw} that temperature may arise in a quantum theory as a
result of a non-trivial background endowed with event horizon(s).
In the case of the Hawking effect, the background in question is any
asymptotically flat black hole spacetime, such as  Schwarzschild~\cite{Hawking:1974sw,Page},
Reissner-Nordstr\"{o}m~\cite{PageII,Sorkin} or Kerr~\cite{PageI} spacetime.
Rindler spacetime, as widely discussed in the previous Chapters, 
 is also known to exhibit thermal features~\cite{Birrell,Rindler:1966zz}, being the ``theater'' 
of the Unruh effect.
The same logical scheme can be further extended to cosmological
horizons, like the event horizon in de Sitter spacetime, where the ensuing (de Sitter) temperature is
$T_{\rm dS}=H/2\pi$ ($H^{-1}$ is the radius of the horizon -- de Sitter radius)~\cite{Guo,Bousso}.

In all these cases the (zero temperature) vacuum state of an inertial observer is perceived as a thermal state
by a certain kind of ``non-inertial'' observers, \emph{e.g.},
a black hole spacetime (Hartle-Hawking vacuum) represents a thermal state for a static (i.e., non-inertial) observer in Schwarzschild
spacetime~\cite{Hawking:1974sw,Israel:1976}, and similarly, Minkowski vacuum agrees with a thermal state for an accelerated (i.e.,
non-inertial) observer~\cite{Unruh:1976db}. It has been known for some time that the aforesaid
concept of an observer-dependent vacuum, or more precisely, an observer-dependent notion of particle being
emitted from the horizon (alongside with the ensuing concepts of a heat bath and temperature) offers an interesting
route towards unification of some of TQFT formalisms~\cite{GUI90}.
The merger can be achieved when one constructs a new spacetime in which the (zero temperature) vacuum corresponds to a usual
thermal state for Minkowski inertial observer, i.e, where the Minkowski observer
is an appropriately chosen non-inertial observer from the point of view of the new spacetime.
In addition, such a spacetime should have more than $1+3$-dimensions to allow for analytic continuation between
Minkowski and Euclidean spacetimes~\cite{GH}.
Along these lines, a flat background with a non-trivial horizon structure providing desired thermal features -- the so-called \EX spacetime -- has been constructed~\cite{GUI90,ZG95,ZG98}.

In its essence, the \EX spacetime is a flat complex manifold with complexified $S^1\times R^3$ topology.
Its Lorentzian section consists of four copies of Minkowski spacetime glued together
along their past or future null hyperplanes. Since in Kruskal-like coordinates the metric is singular
on these hyperplanes, we shall call them formally event horizons. Their existence leads to the doubling of the
degrees of freedom of the fields. The vacuum propagator on this section is found to be equal to
the real-time thermal matrix propagator. On the other hand, in the Euclidean section of
\EX spacetime, the time coordinate is periodic and the field
propagator can be identified with the conventional Matsubara propagator.

Our aim here is to show that the \EX spacetime is structurally richer than previously thought and,
in doing so, we point out the existence and relevance of other complex sections of the \EX
spacetime aside from the already known Lorentzian and Euclidean ones. In this context, it is worth recalling that TQFT formalisms
have a number of physically equivalent (though technically distinct) parameterizations~\cite{BJV,LEB,LW87,KAP}.
For instance, the real-time TQFTs are characterized by a freedom in the parameterization of
the thermal matrix propagator. In the POM formalism, this freedom in parameterization is related to the
choice of a specific path in the complex-time plane going from $t=0$ to $t=-i\beta$, which is not
unique~\cite{NS}. This, so-called Niemi-Semenoff time path, is depicted in Fig.~\ref{fig1NS}.

\begin{figure}[t]
\begin{center}
\resizebox{11.5cm}{!}{\includegraphics{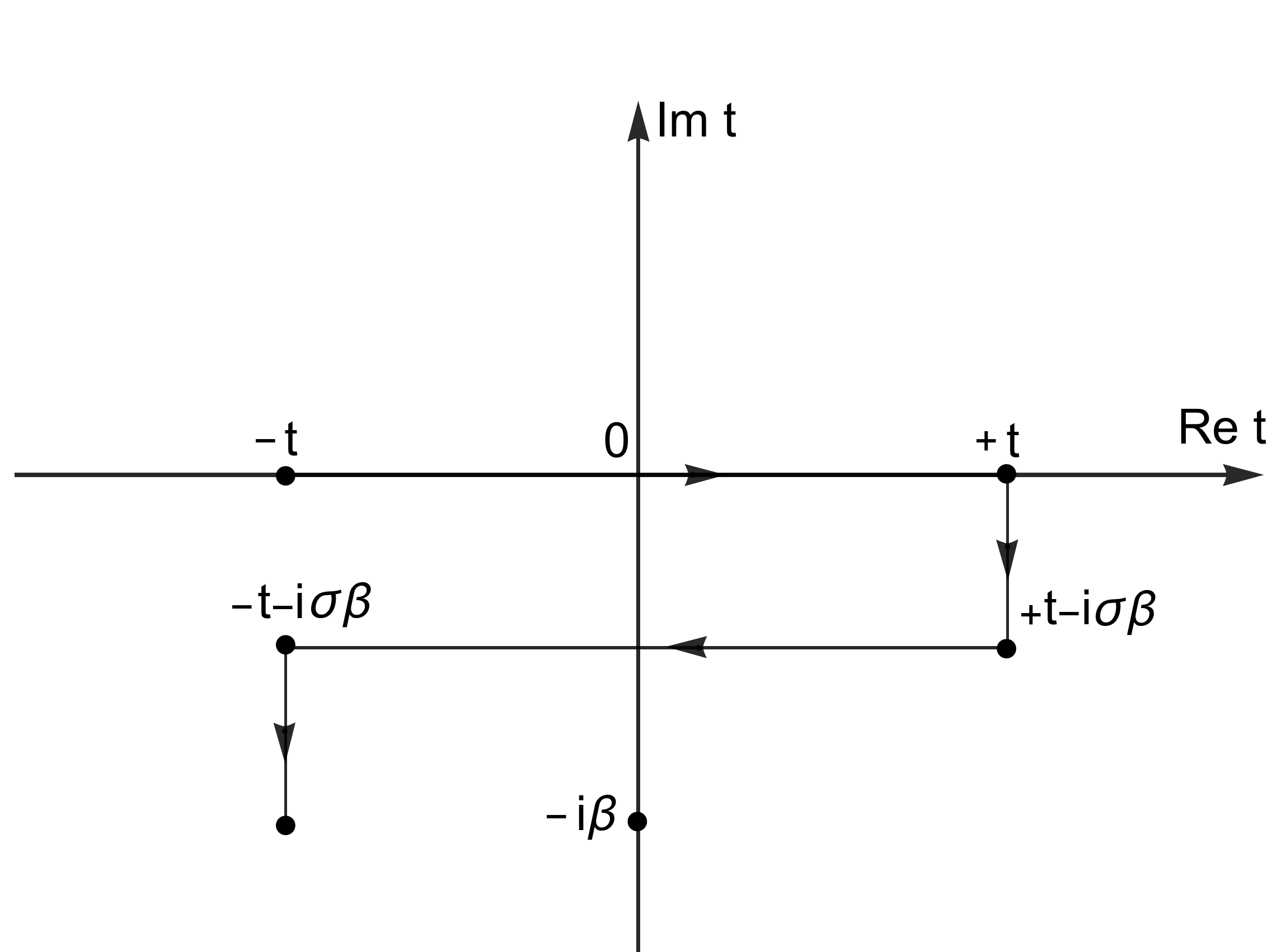}}
\end{center}
\caption{The Niemi-Semenoff time path used in POM. The parameter $\sigma$ ranges from
the value $\sigma=0$ (Closed Time Path) to $\sigma=1$.}
\label{fig1NS}
\end{figure}
In TFD, different parameterizations of the Bogoliubov thermal matrix are permitted~\cite{H95}.
We stress that although the choice of the parameterization (in both  POM and TFD) is irrelevant in {\em thermal equilibrium}
since it does not affect physical quantities~\cite{BJV,H95},
it plays a crucial r\^{o}le in non-equilibrium situations, where
the choice of a closed-time path in POM or the {\em left} and {\em right} statistical states in TFD relate to a particular
form of a transport equation~\cite{H95}.
As yet, such distinct parameterizations of TQFTs have not been considered in the context of the \EX spacetime.
Here we will see that all the aforementioned parameterizations can be realized within
the \EX spacetime framework and can be interpreted in purely geometric terms.
The geometric picture for TQFTs is consequently enlarged.  We will also show how the
unification of the different formalisms of TQFT arises naturally
within this framework. The generalization introduced here could be useful
in order to extend the geometric picture of \EX spacetime to systems
out of thermal equilibrium, for which the typical choice of equilibrium parametrization is not convenient~\cite{H95}.

The remainder of following Sections is organized as follows. Section~\ref{sec.2} provides theoretical
essentials of the \EX spacetime. In doing so, we focus our attention on Euclidean and
Lorentzian sections alongside with their respective complex extensions. As a tool for our analysis, in Subection~\ref{sec.3} we show how to perform field quantization in \EX
spacetime. To keep our discussion as transparent as possible, we consider only a scalar quantum field.  Section~\ref{sec.4} is devoted to examining the relationships
between \EX spacetime and TQFTs, and to the unification of various existent TQFTs
in the framework of \EX spacetime. Furthermore we discuss some of the salient features of the extended Lorentzian
section. Various remarks and generalizations are addressed in the concluding Section~\ref{sec.5}.

\subsection{\EX spacetime -- essentials and beyond }\label{sec.2}

The \EX spacetime was originally introduced in Refs.~\cite{GUI90}. It represents a $1+3$-dimensional complex 
manifold defined by the line element
\begin{equation}
 \mlab{guimetric}
ds^2 \, = \, \frac{-d\eta^2 \, + \,  d\xi^2}{\alpha^2\left(\xi^2-
\eta^2\right)} \, + \, dy^2 \, + \, dz^2\, ,
\end{equation}
where $\alpha =  2\pi/\beta$ is a real constant and
$(\eta,\xi,y,z)\in{\mathbb {C}^4}$.  In the following, we will use the symbol
$\xi^\mu = (\eta,\xi,y,z)$ to denote the entire set of \EX
coordinates, but for simplicity we will often drop the index
$\mu$ when no confusion with the space-like coordinate occurs.

\subsection*{Euclidean section}

One of the key sections of the \EX spacetime is the Euclidean section. The associated metric is obtained from Eq.~(\ref{guimetric})  by assuming that
$(\sigma, \xi, y,z)\in{\Bbb R^4}$, where $\sigma\equiv-i\eta$. A straightforward substitution leads to
\begin{equation} \mlab{euclidmetric}
ds^2 \, = \, \frac{d\sigma^2 \, + \, d\xi^2}{\alpha^2\left(\sigma^2 \, + \, \xi^2\right)}
 \, + \, dy^2 \, + \, dz^2\, .
\end{equation}
By use of the transformations
\begin{eqnarray} \mlab{euclidTrans}
\sigma &\hspace{-1mm} =\hspace{-1mm}&  (1/\alpha)\,
\exp\left(\alpha x\right) \sin\left(\alpha\tau\right),  \\[2mm]
\xi &\hspace{-1mm} =\hspace{-1mm}& (1/\alpha)\,
\exp\left(\alpha x\right) \cos\left(\alpha\tau\right),
\label{euclidTrans2}
\end{eqnarray}
the metric becomes that of a cylindrical (Euclidean) spacetime, i.e.
\begin{equation}
ds^2 \, = \, d\tau^2 \, + \, dx^2 \, + \, dy^2 \, + \, dz^2\, ,
\end{equation}
where the temporal direction $\alpha\tau$  is $2\pi$-periodic. In our following considerations,
we will restrict the basic temporal interval to $0 \leq \tau \leq \beta$.
On the QFT level, this setting together with the single-valuedness of quantum fields
will yield the typical periodicity property of the Euclidean propagator.

\subsection*{Lorentzian section} \label{subsecAC}
\begin{figure}[t]
\begin{center}
\resizebox{8.2cm}{!}{\includegraphics{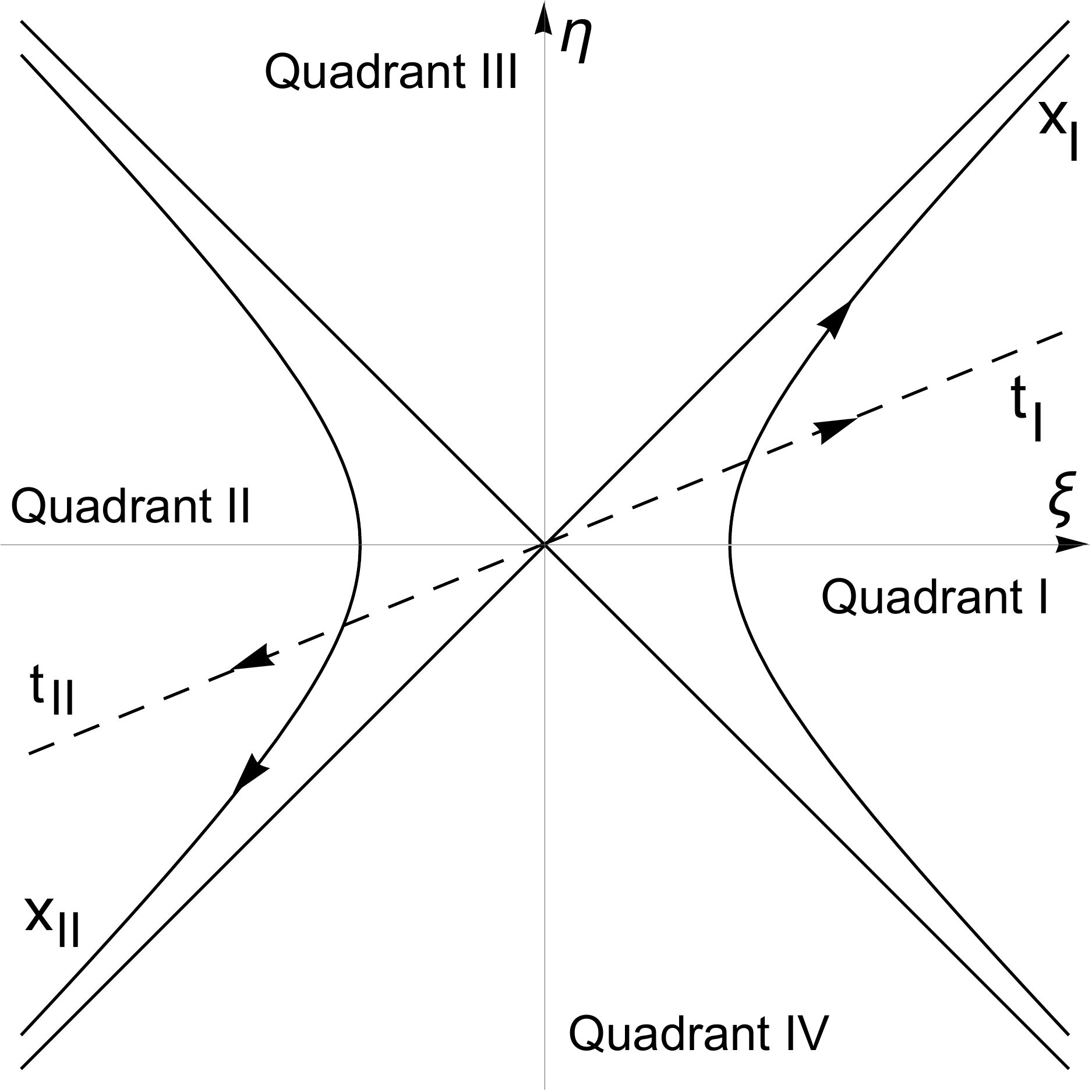}}
\end{center}
\caption{Lorentzian section of \EX spacetime: the solid lines
represent the singularities at $\xi^2-\eta^2=0$. On the (dashed) straight lines
time is constant, while on the hyperbolas the Minkowski coordinate $x$
is constant.
Note that times $t_\I$ and $t_\II$ flow in opposite directions as a consequence of opposite orientations
of ensuing time-like Killing vectors in regions $R_\I$ and $R_\II$.}
\label{fig2}
\end{figure}

The second important section is the Lorentzian section. In this case, the line element is given by Eq.~(\ref{guimetric})
with $(\eta, \xi, y,z)\in{\Bbb R^4}$.  The ensuing metric is singular on the
two hyperplanes $\eta=\pm\,\xi$, which we will call
``event horizons'' (see comment after Eq.~\eqref{ACt}). These divide \EX spacetime into four regions
denoted by $R_\I, R_\II,R_\III$ and $R_\IV$ (see Fig.~\ref{fig2}).
In the first two regions, one can define two sets of tortoise-like
coordinates $x_\IQ^{\mu}\in\R^4$ by $x_\IQ^{\mu}=(t_\IQ,x_\IQ,y,z)$ where
\begin{eqnarray}
&\mbox{in $R_\I$:} & \quad \left\{
\begin{array}{rcl}
\eta & \hspace{-1mm}=&\hspace{-1mm} +(1/\alpha)\,
\exp\left(\alpha x_\I\right)\sinh\left(\alpha t_\I\right), \\[2mm]
\xi &\hspace{-1mm}=&\hspace{-1mm} +(1/\alpha)\,
\exp\left(\alpha x_\I\right)\cosh\left(\alpha t_\I\right),
\end{array}\right.
\label{TransI}\\[3mm]
&\mbox{in $R_\II$:} & \quad \left\{
\begin{array}{rcl}
\eta &\hspace{-1mm}=&\hspace{-1mm} - (1/\alpha)\,
\exp\left(\alpha x_\II\right)\sinh\left(\alpha t_\II\right), \\[2mm]
\xi &\hspace{-1mm}=&\hspace{-1mm} - (1/\alpha)\,
\exp\left(\alpha x_\II\right) \cosh\left(\alpha t_\II\right).
\end{array}
\right.
\label{TransII}
\end{eqnarray}

Similar transformations also hold in regions $R_\III$ and $R_\IV$
(see Refs.~\cite{GUI90,FV,FV1}) but for our purposes it will be sufficient to consider only the first two regions.  In terms of the new coordinates $x_\IQ^{\mu}$, the metric in Eq.~(\ref{guimetric}) becomes the usual Minkowski one with the metric signature in the spacelike convention
\begin{equation}
ds^2 \, = \, -dt^2_\IQ \, + \, dx_\IQ^2 \, + \, dy^2 \, + \, dz^2\, ,
\end{equation}
(and similarly for $R_\III$ and $R_\IV$). It thus arises that regions $R_\I$ to $R_\IV$ represent four copies of Minkowski spacetime glued
together along the ``event horizons''~\footnote{See comment at the end of the Section}, making up together the Lorentzian section of \EX spacetime.

Although Eqs.~(\ref{TransI})-(\ref{TransII}) formally correspond to Rindler transformations~\cite{Rindler:1966zz}, the $(t_{\I,\II},x_{\I,\II},y,z)$
coordinates should not be confused with the Rindler ones, since in those coordinates the metric takes the standard Minkowski form.
Indeed, an observer whose world-line is the branch of hyperbola described, for example, by $(x_\I(t_\I),y(t_\I),z(t_\I))=(x_0,y_0,z_0)$ moves in the Lorentzian wedge $R_\I$ inertially and thus cannot be identified with an accelerated observer as in the Rindler case. The r\^{o}le of the inertial and non-inertial coordinates in Eqs.~(\ref{TransI})-(\ref{TransII}) are actually reversed with respect to the Rindler case~\cite{GUI90,Rindler:1966zz}.

We now study the analytic properties of the transformations in
Eqs.~(\ref{TransI})-(\ref{TransII}). In the null coordinates $\xi^\pm=\eta\pm\xi$, we can
rewrite these equations as
\begin{eqnarray}
&\mbox{in $R_\I$:}& \;
\; \left\{\begin{array}{rcl}
\xi^+_>(t_\I,x_\I) &\hspace{-1.5mm}=\hspace{-1.5mm}&
+(1/\alpha)\exp\left[+\alpha\left(t_\I\,+\,x_\I\right)\right],
\\[2mm]
\xi^-_<(t_\I,x_\I) &\hspace{-1.5mm}=\hspace{-1.5mm}&
-(1/\alpha)\exp\left[-\alpha\left(t_\I\,-\,x_\I\right)\right],
\end{array}\right.
\\[2mm]
&\hspace{0.5mm}\mbox{in $R_\II$:}& \; \; \left\{\begin{array}{rcl}
\xi^+_<(t_\II,x_\II) &\hspace{-1.5mm}=\hspace{-1.5mm}&
-(1/\alpha)\exp\left[+\alpha\left(t_\II\,+\,x_\II\right)\right],
\\[2mm]
\xi^-_>(t_\II,x_\II) &\hspace{-1.5mm}=\hspace{-1.5mm}&
+(1/\alpha)\exp\left[-\alpha\left(t_\II\,-\,x_\II\right)\right],
\end{array}\right.
\end{eqnarray}
where  the subscripts $<$ and $>$ have been added to the variables
$\xi^\pm$ to indicate their ranges, i.e.~one has $\xi^\pm_>>0$ and
$\xi^\pm_<<0$. The reciprocals of these transformations are
\begin{eqnarray}
&\mbox{in $R_\I$:}& \;
\; \left\{\begin{array}{rcl}
t_\I(\xi^+_>,\xi^-_<) &\hspace{-1.5mm}=\hspace{-1.5mm}& \displaystyle
\frac{1}{2\alpha}\ln\left(\displaystyle-\frac{\xi^+_>}{\xi^-_<}\right),
\\[4mm]
x_\I(\xi^+_>,\xi^-_<) &\hspace{-1.5mm}=\hspace{-1.5mm}&\displaystyle
\frac{1}{2\alpha}\ln\left(-\alpha^2\xi^+_>\xi^-_<\right),
\end{array}\right.
\label{RecI} \\[2mm]
&\mbox{in $R_\II$:}& \; \;\left\{\begin{array}{rcl}
t_\II(\xi^+_<,\xi^-_>) &\hspace{-1.5mm}=\hspace{-1.5mm}& \displaystyle
\frac{1}{2\alpha}\ln\left(\displaystyle-\frac{\xi^+_<}{\xi^-_>}\right),
\\[4mm]
x_\II(\xi^+_<,\xi^-_>) &\hspace{-1.5mm}=\hspace{-1.5mm}&\displaystyle
\frac{1}{2\alpha}\ln\left(-\alpha^2\xi^+_<\xi^-_>\right).
\end{array}\right.
\label{RecII}
\end{eqnarray}
Equations (\ref{RecI})-(\ref{RecII}) are defined in regions $R_\I$
and $R_\II$, respectively. We now would like to analytically continue
these expressions to obtain the functions $t_\I(\xi)$, $t_\II(\xi)$
and $x_\I(\xi)$, $x_\II(\xi)$ defined in $R_\I\cup R_\II$. This
amounts to extend these expressions from positive or negatives values
of $\xi^\pm$ to their negative or positive values respectively.  In
order to do this, we choose to perform the analytic extension
in the {\it lower} half-planes of both the $\xi^+$ and $\xi^-$
complex planes for reasons which will become clear below. In other
words, we assume that $-\pi<\arg\xi^\pm\leq\pi$, or equivalently that
the cuts in the $\xi^\pm$ complex planes are given by $\R_-+i0_+$.
\begin{figure}[t]
\begin{center}
\resizebox{10.4cm}{!}{\includegraphics{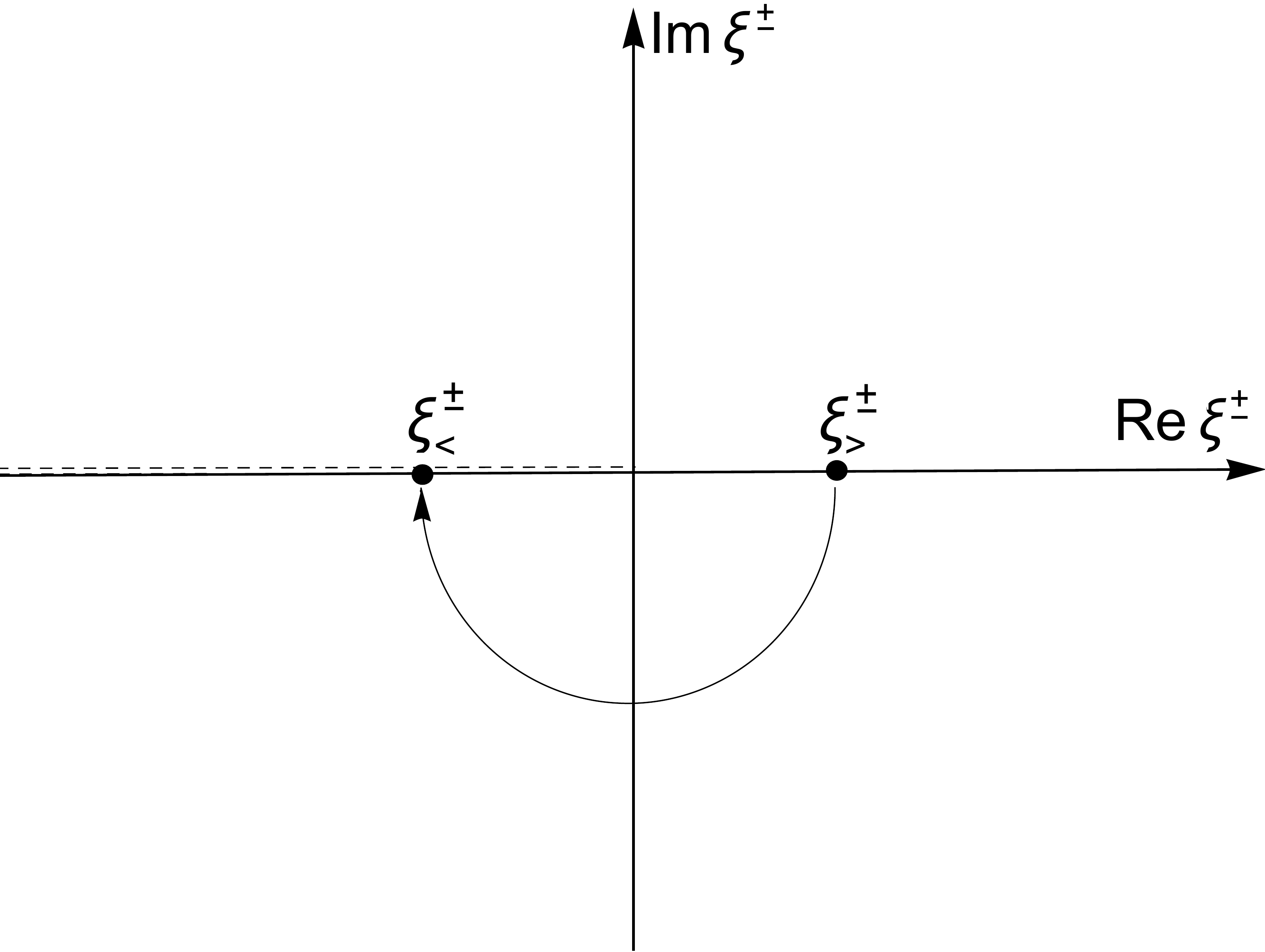}}
\end{center}
\caption{Analytic extension in the \emph{lower} half-planes of the $\xi^+$
and $\xi^-$ complex planes. The dashed line $\R_-+i0_+$ is the branch cut of the logarithm
in Eq.~(\ref{analcont1}).
 }
\label{fig3}
\end{figure}
It is not possible to perform the analytic continuation with respect to
the two variables $\xi^\pm$ {\it at once}, otherwise an erroneous
result would be obtained. To fix the ideas, we choose to perform the
extension first in the $\xi^+$ variable and then in $\xi^-$ (the
choice of the opposite order gives the same result for our particular
purposes).  If $\xi^\pm_<$ is the analytic continuation of $\xi^\pm_>$
from positive to negative values, one has (see Fig.~\ref{fig3})
\begin{eqnarray}
\label{analcont1}
\ln\left(-\xi^\pm_<\right) &\hspace{-1.5mm}=\hspace{-1.5mm}& \ln\left(+\xi^\pm_>\right)\,+\,i\pi\, ,
\\[2mm]
\ln\left(+\xi^\pm_>\right) &\hspace{-1.5mm}=\hspace{-1.5mm}& \ln\left(-\xi^\pm_<\right)\,-\,i\pi\, .
\label{analcont2}
\end{eqnarray}
This implies
\begin{eqnarray}
\hspace{5.5mm}\ln\left(-\frac{\xi^+_<}{\xi^-_>}\right) &\hspace{-1.5mm}=\hspace{-1.5mm}&
\ln\left(-\frac{\xi^+_>}{\xi^-_<}\right) \,+\, i2\pi\, ,
\\[2mm]
\ln\left(-\alpha^2\xi^+_<\xi^-_>\right) &\hspace{-1.5mm}=\hspace{-1.5mm}&
\ln\left(-\alpha^2\xi^+_>\xi^-_<\right),
\end{eqnarray}
which means that these expressions are the analytic continuations of
each other. Consequently, by using Eqs.~(\ref{RecI})-(\ref{RecII}), one obtains
in $R_\I\cup R_\II$
\begin{eqnarray}
\label{ACt}
t_\II(\xi)  &\hspace{-1.5mm}=\hspace{-1.5mm}&  t_\I(\xi) \, + \, i\,\beta/2, \\[2mm]
x_\II(\xi) &\hspace{-1mm} =\hspace{-1mm}& x_\I(\xi)\, .
\label{ACtx}
\end{eqnarray}

We note in passing that we can call the hypersurfaces $\xi^2 \,-\, \eta^2 =0$
``horizons'' only in the sense that an inertial observer in
region $R_{\I}$ cannot receive any signal sent from $\xi^- \,=\,  0$, and
cannot send any signal to $\xi^+ \,=\,  0$. So the hypersurface
$\xi^-  \,=\, 0$ or $\xi^+ \,=\, 0$ can formally be called a ``future horizon''
$\mathcal{H}^+$ or ``past horizon'' $\mathcal{H}^-$, respectively for an inertial
observer in region $R_{\I}$. Analogous conclusion holds, of course also for an inertial observer 
in region $R_{\II}$.

\subsection*{Extended Lorentzian section} \label{sub:ELS}

Let us now consider a class of complex sections of \EX spacetime generated
from the Lorentzian section by shifting the Minkowski time coordinate
in the imaginary direction but {\it only} in the region $R_\II$, namely
\begin{eqnarray}\mlab{shift}
\begin{array}{lll}
&\mbox{in $R_\I\cup R_\III\cup R_\IV$:}
\qquad & t_q \rightarrow {t_q}_{_\delta}\,=\,t_q, \;
\;\;\; q\, =\, I,II \mbox{~and~} III\, ,
\\[2mm] &\mbox{in $R_\II$:} \qquad & t_\II
\rightarrow t_{\IId}\,=\, t_\II\, +\, i\beta\delta\, ,
\end{array}
\label{delta}
\end{eqnarray}
where $\delta\in[-1/2,1/2]$. The reason for this interval will become clear shortly.
We shall call this one-parametric class of sections as ``extended
Lorentzian section'' and denote by $R_\IId$ the image of the region
$R_\II$ resulting from this shift. In $R_\IId$ the \EX coordinates become
complex variables and are transformed according to $(\eta,\xi)
\rightarrow (\etad,\xid)$ where, from Eq.~(\ref{shift}), we have
\begin{eqnarray}\mlab{mex2a}
{\mbox{in}}~R_\IId: \;\;\; \left\{
\begin{array}{rcl}
\etad &\hspace{-1.5mm}=\hspace{-1.5mm}& -(1/\alpha)\,\exp\left(\alpha x_\II\right)
\sinh\left[\alpha\left(t_\II\,+\,i\beta\delta\right)\right],
\\[2mm]
\xid &\hspace{-1.5mm}=\hspace{-1.5mm}& -(1/\alpha)\,\exp\left(\alpha x_\II\right)
\cosh\left[\alpha\left(t_\II\,+\, i \beta \delta\right)\right].
\end{array}
\right.
\end{eqnarray}
In terms of the real \EX variables, one can write
\begin{eqnarray}
\label{mex31}
\etad &\hspace{-1.5mm}=\hspace{-1.5mm}& +\eta \, \cos\left(2\pi\delta\right)
\,+\, i\xi\,\sin\left(2\pi\delta\right), \\[2mm]
i\xid &\hspace{-1.5mm}=\hspace{-1.5mm}& -\eta \, \sin\left(2\pi\delta\right)
\,+\, i\xi\,\cos\left(2\pi\delta\right),
\mlab{mex3}
\end{eqnarray}
or, equivalently, in terms of null coordinates
\begin{eqnarray}\mlab{mex3b}
\xid^+ &\hspace{-1.5mm}=\hspace{-1.5mm}& \exp\left(+i2\pi\delta\right)\,\xi^+, \\[2mm]
\xid^- &\hspace{-1.5mm}=\hspace{-1.5mm}& \exp\left(-i2\pi\delta\right)\,\xi^-.
\label{defxipm}
\end{eqnarray}
The time shift in Eq.~(\ref{shift}) thus induces a rotation in the $(\eta,i\xi)$ plane of
$R_\II$.  By using the rotated coordinates, the metric can then be recast into the form
\begin{equation}\mlab{mex1}
ds^2 \, = \, \frac{-d\eta_\delta^2 \, + \, d\xi_\delta^2}
{\alpha^2\left(\xi_\delta^2 \, - \, \eta^2_\delta\right)} \,+ \, dy^2 \, + \, dz^2\, ,
\end{equation}
and is thus unchanged by the time shift Eq.~(\ref{shift}), which is therefore an isometry
of the \EX spacetime. After the time shift, Eqs.~(\ref{ACt})-(\ref{ACtx}) become 
\begin{eqnarray}
\label{ACtxd1}
t_\IId(\xid) &\hspace{-1.5mm}=\hspace{-1.5mm}&\ \!
 t_\I(\xid) \,+\, i\,\beta\left(1/2 \,+\, \delta\right), 
\\[2mm]
x_\IId(\xid) & \hspace{-1mm}=\hspace{-1mm}&\ \! x_\I(\xid)\, .
\label{ACtxd}
\end{eqnarray}

Let us finally comment on Killing fields in
(extended) Lorentzian section.
Since a timelike Killing vector defines a preferred time coordinate in a time-independent spacetime~\cite{Straumann}, we can expect
(with some foresight) that its structure is pertinent  for the understanding of the connection between the (extended) Lorentzian section and the POM.
It will also prove useful when we will discus TFD in Section~\ref{Sec.4.3}.
%
%

In order to find the Killing vector field $\kappa$ in various sections of \EX spacetime, we need to solve the Killing equation
$({\mathcal{L}}_{\kappa} \ \! g)_{\mu\nu} = \kappa^{\lambda}g_{\mu\nu, \lambda} + g_{\lambda\nu} \kappa^{\lambda}_{\ \! , \mu}
+ g_{\mu \lambda} \kappa^{\lambda}_{\ \! , \nu}
= 0$. Here ${\mathcal{L}}_{\kappa}$ is the Lie derivative along the vector field $\kappa$ and $g_{\mu\nu}$ is the pullback metric on a given section. For instance,
in the Lorentzian section we have
\begin{equation}
\kappa \, = \, \alpha\left(\xi \ \! \frac{\partial}{\partial \eta} \, + \, \eta \ \! \frac{\partial}{\partial \xi} \right),
\label{Lorentz.case.a}
\end{equation}
which is clearly timelike in $R_\I \cup R_\IId$ as there $\kappa^2 = g_{\mu \nu} \kappa^{\mu} \kappa^{\nu} = -1$. The parameter $\alpha$ was introduced in Eq.~(\ref{Lorentz.case.a}) so that  the components of $\kappa$ become dimensionless and normalized to $-1$. Along the same lines, we see that, in the extended Lorentzian section, we obtain from Eq.~(\ref{mex1}) 
\begin{equation}
\kappa \, = \, \alpha\left(\xi_{\delta} \ \! \frac{\partial}{\partial \eta_{\delta}} \, + \, \eta_{\delta} \ \! \frac{\partial}{\partial \xi_{\delta}} \right).
\label{Lorentz.case.b}
\end{equation}
Again, $\kappa$ is timelike in $R_\I \cup R_\IId$. The integral curves of the above Killing vector fields satisfy  equations ${d \xi_{\delta}}/{d s} \,=\, \alpha \eta_{\delta}$ and ${d \eta_{\delta}}/{d s} \,=\, \alpha \xi_{\delta}$, which yield a parametric representation of orbits in the form
\begin{eqnarray}
\label{orbits1}
\xi_{\delta}(s) &\hspace{-1.5mm}=\hspace{-1.5mm}& \ \! \xi_{\delta}^0 \ \! \cosh[\alpha(s \,+\, i\beta \delta)]\, ,  \\[2mm]
\eta_{\delta}(s) &\hspace{-1.5mm}=\hspace{-1.5mm}& \ \! \eta_{\delta}^0 \ \! \sinh[\alpha(s \,+\, i\beta \delta)]\, ,
\label{orbits}
\end{eqnarray}
where $\delta$ is the shift parameter introduced in Eq.~(\ref{delta}), in particular, $\delta \,=\, 0$ in $R_\I$. 
The integration constants $\xi_{\delta}^0$ and $\eta_{\delta}^0$ depend on an actual position $x_{\I}$ or $x_{\II}$ of the observer.  
Equations~(\ref{orbits1})-(\ref{orbits}) precisely coincide with the world-lines of a static observer in respective Minkowski wedges. Note that the Killing 
vectors in regions $R_\I$ and $R_\IId$  have mutually opposite orientations of their real parts.

Analogous reasonings yield in the Euclidean section the Killing vector
\begin{equation}
\kappa \, = \, \alpha\left(\xi \ \! \frac{\partial}{\partial \sigma} \, - \, \sigma \ \! \frac{\partial}{\partial \xi} \right),
\label{Eucl.cas.a}
\end{equation}
which is the Euclidean analogue of the timelike Killing vector.
The integral curves of the above $\kappa$ are circles around the origin with radius $R = e^{\alpha x}/\alpha$ (here $R^2 = \sigma^2 + \xi^2$). 
Hence these orbits can be identified with preferred inverse-temperature coordinate in the Euclidean section.

Note, finally, that the coordinates $x_\IQ^{\mu}=(t_\IQ,x_\IQ,y,z)$, $x_{\I,\IId}^{\mu}=(t_{\I,\IId},x_{\I,\IId},y,z)$ and $x^{\mu} = (\tau,x,y,z)$ 
are adapted to the Killing fields in Eqs.~(\ref{Lorentz.case.a}), (\ref{Lorentz.case.b})-(\ref{Eucl.cas.a}), respectively.

\subsection{Field quantization in \EX spacetime}\label{sec.3}
%
For sake of simplicity, let us now consider a free scalar field with a support in \EX spacetime.
The corresponding generalization to the fermionic sector is quite straightforward and one may, for instance,
follow the line of reasonings presented in Ref.~\cite{GUI92ferm}.

We denote the ``global'' scalar field in \EX coordinates as $\Phi(\xi)$. It satisfies
the Klein-Gordon equation
\begin{equation} \mlab{kge}
(\Box \,-\, m^2) \Phi(\xi) \, = \,  0\, ,
\end{equation}
where $\Box \,=\, {g}^{-1/2}\partial_\mu\hspace{0.2mm} g^{\mu\nu}{g}^{-1/2}\partial_\nu$ is the Laplace-Beltrami operator, with $g(\xi) \,=\, \left|\mathrm{det}\hspace{0.4mm} g_{\mu\nu}\right|$. We define the inner product of two Klein-Gordon fields by
\begin{equation}
(\Phi_1,\Phi_2) \, = \, -i\int_\Sigma\hspace{-1mm} d\Sigma\, {\left(g(\xi)\right)}^{1/2}\hspace{0.1mm}
\Phi_1(\xi)\,n^\nu\stackrel{\leftrightarrow}{\partial}_\nu \Phi_2^*(\xi)\, ,
\label{scalarproduct}
\end{equation}
where the integral is taken over a Cauchy hypersurface $\Sigma$ and $n^\nu$ is an orthonormal
vector to this hypersurface.

\subsection*{Euclidean section}

In the Euclidean section of \EX spacetime, one has $\Phi=\Phi(\sigma,\xi,y,z)$. The
field in the $\tau$-$x$ coordinates in Eqs.~(\ref{euclidTrans})-(\ref{euclidTrans2})
shall be denoted by $\phi=\phi(\tau,x,y,z)$. These two fields are
related by
\begin{equation}
\phi(\tau,x,y,z) \, = \, \Phi(\sigma(\tau,x),\xi(\tau,x),y,z)\, .
\end{equation}
Because of the periodic nature of the time $\tau$ and presumed single-valuedness of the field, the following condition must be satisfied
\begin{equation}
\phi(\tau,x,y,z) \, = \, \phi(\tau+\beta,x,y,z)\, .
\label{periodicfield}
\end{equation}
This periodic boundary condition will prove to be important in Section~4.
Note that Eq.~(\ref{periodicfield}) is nothing but the familiar Kubo-Martin-Schwinger (KMS) boundary condition for Euclidean fields~\cite{BJV,LW87}.

\subsection*{Lorentzian section} \label{Lorentzian section}

Lorentzian section, as we have seen, is made up of four different
regions, each of them being a complete Minkowski spacetime.  Since our primary
interest is only in regions $R_\I$ and $R_\II$, we shall limit ourselves to
consider the quantum field over these two regions.  Our aim is to find an
expansion for the global field $\Phi$ in the joining $R_\I \cup
R_\II$.

Let us start by defining the ``local'' fields $\phi^\I(x_\I)$ and
$\phi^\II(x_\II)$ by
\begin{eqnarray} \label{defphi}
\Phi(\xi) \ = \ \left\{
\begin{array}{lll}
\phi^\I(x_\I(\xi)), &\quad& \mbox{when $\xi\in R_\I$}\, , \\[2mm]
\phi^\II(x_\II(\xi)), && \mbox{when $\xi\in R_\II$}\, .
\end{array}\right.
\end{eqnarray}
They have support in $R_\I$ and $R_\II$, respectively.
By choosing $\Sigma$ to be any of the one-parametric class of hypersurfaces $\eta = a\xi$ (with $-1<a<1$), 
we obtain from Eq.~(\ref{scalarproduct}) that the global inner product assumes the form
\begin{equation}
(\Phi_1,\Phi_2) \, = \, \langle \phi^\I_1,\phi^\I_2 \rangle \, + \, \langle \phi^\II_1,\phi^\II_2 \rangle \, ,
\label{SP}
\end{equation}
where $\langle \,,\, \rangle$ is the local inner product in Minkowski spacetime
\begin{equation}
\langle \phi_1,\phi_2 \rangle \, = \, -i\int_{\R^3}\hspace{-1.9mm} d^3x\,
\phi_1(x)\stackrel{\leftrightarrow}{\partial}_t\phi_2^*(x)\, .
\label{SPM}
\end{equation}
This is nothing but the usual Klein-Gordon inner product known from relativistic quantum theory~\cite{Itzykson}.

In \EX spacetime covered by the $x_\IQ^{\mu}$ coordinates defined by
Eqs.~(\ref{TransI})-(\ref{TransII}), the solutions of the
Klein-Gordon equation are just plane waves restricted to a given
region. So, we can write explicitly
\begin{eqnarray}
u_\Vk(x_\I) &\hspace{-1.5mm}=\hspace{-1.5mm}& \left(4\pi\omega_\Vk\right)^{-\frac{1}{2}}\,
e^{i\left(\Vk\cdot\Vx_\I\,-\,\omega_\Vk\,t_\I\right)},
\mlab{MMu} \\[2mm]
v_\Vk(x_\II) &\hspace{-1.5mm}=\hspace{-1.5mm}& \left(4\pi\omega_\Vk\right)^{-\frac{1}{2}}\,
e^{i\left(\Vk\cdot\Vx_\II\,+\,\omega_\Vk\,t_\II\right)},
\mlab{MMv}
\end{eqnarray}
where $\omega_\Vk= \sqrt{{|\Vk|}^2+m^2}$. Starting from these Minkowski modes,
one defines the two wave functions $U_\Vk(\xi)$ and $V_\Vk(\xi)$ with
support in $R_\I$ and $R_\II$, respectively, by
\begin{eqnarray}
U_\Vk(\xi) &=&  \left\{
\begin{array}{lcl}
u_\Vk(x_\I(\xi)), &\quad&
\mbox{when $\xi\in R_\I$}, \\[2mm]
0, && \mbox{when $\xi\in R_\II$},
\end{array}\right.
\label{defU}\\[2mm]
V_\Vk(\xi) &=&
\left\{\begin{array}{lcl}
0, && \mbox{when $\xi\in R_\I$}, \\[2mm]
v_\Vk(x_\II(\xi)), &\quad& \mbox{when $\xi\in R_\II$}.
\end{array}\right.
\label{defV}
\end{eqnarray}
Their power spectra with respect to the momenta conjugated to $\xi^+$
and $\xi^-$ contain negative contributions, which are furthermore not
bounded from below. The sets of functions $\left\{
U_\Vk(\xi), U^*_{-\Vk}(\xi)\right\}_{\Vk\in\R^3}$ and $\left\{
V_\Vk(\xi), V^*_{-\Vk}(\xi)\right\}_{\Vk\in\R^3}$ defined on $R_\I$
and $R_\II$, respectively, are thus over-complete, since the same energy
contribution (i.e.~momentum contribution conjugate to $\eta$) can
appear twice in these sets. In other words, the energy spectra of
$U_\Vk$ and $U^*_{-\Vk}$ overlap, and so do the spectra of $V_\Vk$ and
$V^*_{-\Vk}$. Therefore, these sets can not be used as a basis in their
respective supports and their joining is clearly not a
basis in $R_\I\cup R_\II$.
In order to construct a basis in $R_\I\cup R_\II$, we could solve, for instance, the
Klein-Gordon equation in \EX coordinates to obtain the field modes in
these coordinates. The Bogoliubov transformations resulting
from this choice of basis would be, however, quite complicated. So instead of following this route,
we shall construct basis elements with positive energy
spectra from the wave functions $u_\Vk(x_\I(\xi))$
and $v^*_{-\Vk}(x_\II(\xi))$. We will further demand these basis elements should be analytic functions in
the lower complex planes of $\xi^+$ and $\xi^-$, so that their
spectra with respect to the momenta conjugate to $\xi^+$ and $\xi^-$ contain only positive contributions.
Consequently, they will have positive energy spectra.

To this end, we analytically extend the two wave functions $u_\Vk(x_\I(\xi))$
and $v^*_{-\Vk}(x_\II(\xi))$ in the lower complex planes of $\xi^+$
and $\xi^-$ (as in Section \ref{subsecAC}, the cut in the complex
planes is represented by $\R_-+i0_+$). By applying Eqs.~(\ref{ACt})-(\ref{ACtx}),
we obtain directly
\begin{eqnarray}
u_\Vk(x_\I(\xi)) &\hspace{-1.5mm}=\hspace{-1.5mm}& e^{-\frac{\beta}{2}\omega_\Vk}\,v^*_{-\Vk}(x_\II(\xi))\, ,
\label{ACuv} \\[2mm]
v_\Vk(x_\II(\xi)) &\hspace{-1.5mm}=\hspace{-1.5mm}& e^{-\frac{\beta}{2}\omega_\Vk}\,u^*_{-\Vk}(x_\I(\xi))\, .
\label{ACvu}
\end{eqnarray}
The expressions on the right and left hand sides of these
equations are analytic continuations of each other. We are therefore led
to introduce the following two normalized linear combinations
\begin{eqnarray}
\label{globalmodes1}
F_{\Vk}(\xi) &\hspace{-1.5mm}=\hspace{-1.5mm}&\left(1-f_\Vk\right)^{-\frac{1}{2}}
\left[ U_\Vk(\xi) \,+\, f_\Vk^{\frac{1}{2}}\,V^*_{-\Vk}(\xi)\right],
\\[3mm]
{\widetilde F}_{\Vk}(\xi) &\hspace{-1.5mm}=\hspace{-1.5mm}& \left(1-f_\Vk\right)^{-\frac{1}{2}}
\left[ V_{\Vk}(\xi) \,+\, f_\Vk^{\frac{1}{2}}\,U^*_{-\Vk}(\xi)\right],
\label{globalmodes}
\end{eqnarray}
where $f_\Vk=e^{-\beta\omega_\Vk}$ and $U_\Vk(\xi)$ and $V_\Vk(\xi)$ are defined in Eqs.~(\ref{defU})-(\ref{defV}), respectively.
These global wave functions are still solutions of the Klein-Gordon equation. Moreover, they
are analytic in $R_\I\cup R_\II$ and in particular at the origin
$\xi^+=\xi^-=0$.  Since they are analytic complex functions in the
lower complex planes of $\xi^+$ and $\xi^-$, their energy spectra are positive. The set $\{F_{\Vk}, F^*_{-\Vk},
{\widetilde F}_\Vk, {\widetilde F}^*_{-\Vk} \}_{\Vk\in\R^3}$ is thus
complete (but not over-complete) over the joining $R_\I\cup
R_\II$. Furthermore it is an orthogonal set since
\begin{eqnarray}
(F_\Vk,F_\Vp)& \hspace{-1.3mm}=\hspace{-1.3mm} & (\widetilde{F}^*_\Vk,\widetilde{F}^*_\Vp)
\,=\,  +\,\delta^3(k-p)\, ,
\\[2mm]
(F^*_\Vk,F^*_\Vp)& \hspace{-1.3mm}=\hspace{-1.3mm} &(\widetilde{F}_\Vk,\widetilde{F}_\Vp)
\,=\,  -\delta^3(k-p)\, ,
\end{eqnarray}
with all the other scalar products vanishing.

On the one hand, we can expand the local scalar fields in terms of Minkowski modes given
in Eqs.~(\ref{MMu})-(\ref{MMv}) as follows
\begin{eqnarray}
\phi^\I(x_\I) &\hspace{-1.5mm}=\hspace{-1.5mm}&\int\hspace{-1mm} d^3k \left[ a^\I_\Vk\,u_\Vk(x_\I)\,+\,
a^{\I\dagger}_\Vk\,u^*_\Vk(x_\I)\right],
\label{ExpI}\\[1mm]
\phi^\II(x_\II) &\hspace{-1.5mm}=\hspace{-1.5mm}& \int\hspace{-1mm} d^3k \left[ a^\II_\Vk\,v_\Vk(x_\II)\,+\,
a^{_\II\dagger}_\Vk\,v^*_\Vk(x_\II)\right].
\label{ExpII}
\end{eqnarray}
On the other hand, the global scalar field can be expanded in terms of the $F$-modes given in Eqs.~(\ref{globalmodes1})-(\ref{globalmodes}) as
\begin{equation}
{ \Phi(\xi)\,=\, \int\hspace{-1mm} d^3k \Big[
b_\Vk\,F_\Vk(\xi) \,+\, b^\dagger_\Vk\,F^*_{\Vk}(\xi)
 }\, +\,\tilde{b}_\Vk\,\widetilde{F}_{-\Vk}(\xi)
\,+\,\tilde{b}_\Vk^\dagger\,\widetilde{F}^*_{-\Vk}(\xi) \Big]\hspace{-0.3mm}.
\mlab{Exp}
\end{equation}
These three expansions define the local and global creation and
annihilation operators, which are connected to each other by Bogoliubov
transformations. In order to obtain these transformations, we use the definition Eq.~
(\ref{defphi}) relating the local and global fields and the
field expansions Eqs.~(\ref{ExpI}), (\ref{ExpII}) and (\ref{Exp}). A straightforward calculation
leads tof
\begin{eqnarray}\label{bogol}
b_\Vk &\hspace{-1mm} =\hspace{-1mm} &  a^\I_\Vk\,\cosh\theta_\Vk\, -\, a^{\II\dagger}_{-\Vk}\,\sinh\theta_\Vk\, ,
\\[2mm]
{\tilde b}_\Vk &\hspace{-1mm} =\hspace{-1mm} &
a^\II_{-\Vk}\,\cosh\theta_\Vk\,-\,a^{\I\dagger}_\Vk\,\sinh\theta_\Vk\, ,
\label{bogol2}
\end{eqnarray}
where $\sinh^2\theta_\Vk = n(\omega_\Vk)=(e^{\beta\omega_\Vk } -1)^{-1}$ is the Bose-Einstein distribution.

\subsection*{Extended Lorentzian section}

By following the above outlined procedure, we can now construct a set of
positive energy modes in the extended Lorentzian section
introduced in Section~\ref{sub:ELS}. We start by considering the region
$R_\II$, where the plane wave set
$\{v_\Vk(x_\II),v^*_{-\Vk}(x_\II)\}_{\Vk\in\R^3}$ has the form
\begin{eqnarray}
v_\Vk(x_\II) &\hspace{-1.5mm}=\hspace{-1.5mm}& \left(4\pi\omega_\Vk\right)^{-\frac{1}{2}}\,
e^{i\left(\Vk\cdot\Vx_\II\,+\,\omega_\Vk\,t_\II\right)},
\\[2mm]
v_{-\Vk}^*(x_\II) &\hspace{-1.5mm}=\hspace{-1.5mm}& \left(4\pi\omega_\Vk\right)^{-\frac{1}{2}}\,
e^{i\left(\Vk\cdot\Vx_\II\,-\,\omega_\Vk\,t_\II\right)}.
\end{eqnarray}
Under the time shift Eq.~(\ref{shift}), this set is transformed into
$\{v_\Vk(x_{\IId}),v^\sharp_{-\Vk}(x_{\IId})\}_{\Vk\in\R^3}$, where the symbol
 $\ast$ has been replaced by $\sharp$ because
$v^\sharp_{-\Vk}(x_{\IId})$ is no longer the complex conjugate of
$v_\Vk(x_{\IId})$. In fact, one has
\begin{eqnarray}
v_\Vk(x_{\IId}(x_\II)) &\hspace{-1.5mm}=\hspace{-1.5mm}& e^{-\beta\omega_\Vk\delta}\,v_\Vk(x_\II),
\label{vII}\\[2mm]
v^\sharp_{-\Vk}(x_{\IId}(x_\II)) &\hspace{-1.5mm}=\hspace{-1.5mm}&
e^{+\beta\omega_\Vk\delta}\,v_{-\Vk}^*(x_\II).
\label{vIIs}
\end{eqnarray}
We emphasize that the complex conjugation and the time shift do not commute.
Indeed, the mode $v^\sharp_{-\Vk}(x_{\IId}(x_\II))$ can be obtained from
$v_\Vk(x_{\IId}(x_\II))$ by complex conjugation {\it and} by the
replacement $\delta\rightarrow-\delta$. This rule actually defines the
$\sharp$-conjugation.

Similarly as in Eq.~(\ref{defphi}), one defines
\begin{eqnarray} \label{defphid}
\Phi(\xi)\ =\ \left\{
\begin{array}{lll}
\phi^\I(x_\I(\xi)), &\quad& \mbox{when $\xi\in R_\I$}, \\[2mm]
\phi^\IId(x_\IId(\xi)), && \mbox{when $\xi\in R_\IId$}.
\end{array}\right.
\end{eqnarray}
In $R_\I\cup R_\IId$ the Klein-Gordon-like inner product Eq.~(\ref{SP}) takes the form
\begin{equation}
(\Phi_1,\Phi_2)_{_\delta} \, = \,
\langle \phi^\I_1,\phi^\I_2\rangle \, + \, \langle \phi^\IId_1,\phi^\IId_2\rangle_{_\delta}\, ,
\end{equation}
where the local Minkowski inner product $\langle \ ,\ \rangle_{_\delta}$ in region
$R_\IId$ is given by
\begin{equation}
\langle \phi_1,\phi_2\rangle_{_\delta}\, = \, -i\int_{\R^3}\hspace{-2.2mm} d^3x_\IId\,
\phi_1(x_\IId)\stackrel{\leftrightarrow}{\partial}_t \phi_2^\sharp(x_\IId)\, .
\end{equation}
Equations (\ref{ACuv}), (\ref{ACvu}), (\ref{vII}) and (\ref{vIIs})
imply
\begin{eqnarray}
&&u_\Vk(x_\I(\xi_\delta))  \ = \
e^{-\beta\omega_\Vk(1/2+\delta)}\,v^\sharp_{-\Vk}(x_{\IId}(\xi_\delta))\, ,
\label{ACuvd} \\[2mm]
&&v_\Vk(x_\IId(\xi_\delta)) \ = \
e^{-\beta\omega_\Vk(1/2+\delta)}\,u^\sharp_{-\Vk}(x_\I(\xi_\delta))\, .
\label{ACvud}
\end{eqnarray}
With this, we can also write
\begin{eqnarray}
&&u^\sharp_\Vk(x_\I(\xi_\delta)) \ = \ e^{-\beta\omega_\Vk(1/2-\delta)}\,
v_{-\Vk}(x_{\IId}(\xi_\delta))\, ,
\label{ACuvds}\\[2mm]
&&v^\sharp_\Vk(x_\IId(\xi_\delta))  \ = \
e^{-\beta\omega_\Vk(1/2-\delta)}\,u_{-\Vk}(x_\I(\xi_\delta))\, .
\label{ACvuds}
\end{eqnarray}
The expressions on the left and right hand sides of these equations
are thus analytic continuations of each other. If we now define
\begin{eqnarray}
U_\Vk(\xid) &=&
\left\{\begin{array}{lcl}
u_\Vk(x_\I(\xid)), &\quad&
\mbox{when $\xid\in R_I$}, \\[2mm]
0, && \mbox{when $\xid\in R_\IId$},
\end{array}\right.
\\[3mm]
V_\Vk(\xid) &=&
\left\{\begin{array}{lcl}
0, && \mbox{when $\xid\in R_I$}, \\[2mm]
v_\Vk(x_{\IId}(\xid)), &\quad&
\mbox{when $\xid\in R_\IId$},
\end{array}\right.
\end{eqnarray}
the global modes in the extended Lorentzian section $R_\I\cup R_\IId$ are
be written as
\begin{eqnarray} \mlab{ex11}
\label{GM1}
&&G_{\Vk}(\xi_\delta) \ = \ (1\,-\,f_\Vk)^{-\frac{1}{2}} \left[ U_\Vk(\xi_\delta)
\,+\,f_\Vk^{\frac{1}{2}+\delta}\,V^\sharp_{-\Vk}(\xi_\delta) \right], 
\\[2mm]
&&{\widetilde G}_{\Vk}(\xi_\delta) \ = \
(1\,-\,f_\Vk)^{-\frac{1}{2}}\left[ V_\Vk(\xi_\delta) \,+\,
f_\Vk^{\frac{1}{2}+\delta}\,U^\sharp_{-\Vk}(\xi_\delta) \right],
\label{GM}
\end{eqnarray}
and
\begin{eqnarray} \mlab{ex11bis}
\label{GMs1}
&&G^\sharp_{\Vk}(\xi_\delta) \ = \
(1\,-\,f_\Vk)^{-\frac{1}{2}} \left[ U^\sharp_\Vk(\xi_\delta) \,+\,
f_\Vk^{\frac{1}{2}-\delta}\,V_{-\Vk}(\xi_\delta) \right],
\\[2mm]
&&{\widetilde G}^\sharp_{\Vk}(\xi_\delta) \ = \
(1\,-\,f_\Vk)^{-\frac{1}{2}} \left[ V^\sharp_\Vk(\xi_\delta)\,+\,f_\Vk^{\frac{1}{2}-\delta}\,U_{-\Vk}(\xi_\delta) \right],
\label{GMs}
\end{eqnarray}
where $f_\Vk\,=\,e^{-\beta\omega_\Vk}$ is the conventional Boltzmann factor. 

Equations~(\ref{ACuvd})-(\ref{ACvuds}) allow  to state that the global modes Eqs.~(\ref{GM1})-(\ref{GMs}) are analytic in $R_\I\cup
R_\IId$, in particular at the origin $\xi^+_\delta=\xi^-_\delta=0$.
Since they are analytic complex functions in the lower complex
planes of $\xi^+_\delta$ and $\xi^-_\delta$, their energy spectra have only
positive contributions. We might note that, for $\delta=0$, they reduce to the expressions
Eqs.~(\ref{globalmodes1})-(\ref{globalmodes}), as it should be expected.

Let us stress that the global modes $G^*_\Vk$ and
$\widetilde{G}^*_\Vk$ are {\em not} analytic in the extended
Lorentzian section, contrary to the non-Hermitian combinations
$G^\sharp_\Vk$ and $\widetilde{G}^\sharp_\Vk$.  Non-Hermitian
conjugation operations such as our ``sharp'' conjugation $\sharp$ are
actually common in TQFT, see for example Ref.~\cite{BESV98} for a
formally similar situation. In Ref.~\cite{OST}, the
necessity of the so-called Osterwalder-Schrader (reflection) positivity as
opposed to the Hermiticity property is shown in Euclidean field theories
even when the temperature vanishes.

The set $\{G_{\Vk},{\widetilde G}_{\Vk},G^\sharp_{\Vk}, {\widetilde
G}^\sharp_{\Vk}\}_{\Vk\in\R^3}$ is thus complete over $R_\I\cup
R_\IId$. It is furthermore an orthogonal set since
\begin{eqnarray}
&&(G_\Vk,G_\Vp)_\delta \ = \
(\widetilde{G}^\sharp_\Vk,\widetilde{G}^\sharp_\Vp)_\delta
\ = \ \hspace{-1.3mm} \delta^3(k-p)\,,  \\[2mm]
&&(G^\sharp_\Vk,G^\sharp_\Vp)_\delta \ = \
(\widetilde{G}_\Vk,\widetilde{G}_\Vp)_\delta
\ = \ -\delta^3(k-p)\, ,
\end{eqnarray}
with all the other inner products vanishing.

Following the procedure of Section~\ref{Lorentzian section}, we now expand the local fields in the Minkowski
modes over regions $R_\I$ and $R_\IId$ as
\begin{eqnarray} \mlab{expansionydelta1}
\phi^\I(x_\I) &\hspace{-1.5mm}=\hspace{-1.5mm}&
\int\hspace{-1mm} d^3k \left[ a^\I_\Vk\,u_\Vk(x_\I)\,+\,
a^{\I\dagger}_\Vk\,u^*_\Vk(x_\I)\right],
\\[1mm]\mlab{expansionydelta2}
\phi^\IId(x_\IId) &\hspace{-1.5mm}=\hspace{-1.5mm}&
\int\hspace{-1mm} d^3k \left[ a^\IId_\Vk\,v_\Vk(x_\IId)\,+\,
a^{_\IId\dagger}_\Vk\,v^*_\Vk(x_\IId)\right].
\end{eqnarray}
On the other hand, the expansion of the global field in the $G$ modes
over the region $R_\I\cup R_\IId$ reads as
\begin{equation}
\Phi(\xi_\delta) \,=\, \int\hspace{-1mm} d^3k\left[ c_\Vk\, G_\Vk(\xi_\delta)
\,+\, c^\sharp_\Vk\,G^\sharp_{\Vk}(\xi_\delta)
\,+\, {\tilde c}_\Vk\,{\tilde G}_{-\Vk}(\xi_\delta)\,
\,+\, {\tilde c}^\sharp_\Vk\,{\tilde G}^\sharp_{-\Vk}(\xi_\delta) \right].
\mlab{ex12}
\end{equation}
From these last expansions,  by using Eqs.~(\ref{GM1})-(\ref{GMs}), one finds the Bogoliubov transformations
\begin{eqnarray}
\label{BEc1}
c_\Vk & \hspace{-1mm}=\hspace{-1mm} & (1-f_\Vk)^{-\frac{1}{2}}
\left(a^\I_\Vk\,-\,f_\Vk^{\frac{1}{2}-\delta}\,a^{\IId\dag}_{-\Vk}\right),
\\[1.5mm]
{\tilde c}_\Vk & \hspace{-1mm}=\hspace{-1mm} & (1-f_\Vk)^{-\frac{1}{2}}
\left(a^\IId_{-\Vk}\,-\,f_\Vk^{\frac{1}{2}-\delta}\,a^{\I\dag}_\Vk\right),
\label{BEc}
\end{eqnarray}
and their $\sharp$-conjugate duals
\begin{eqnarray}
\label{BEcs1}
c^\sharp_\Vk &\hspace{-1mm} =\hspace{-1mm} & (1-f_\Vk)^{-\frac{1}{2}}
\left( a^{\I\dag}_\Vk \,-\, f_\Vk^{\frac{1}{2}+\delta}\,a^\IId_{-\Vk}\right),
\\[1.5mm]
{\tilde c}^\sharp_\Vk & \hspace{-1mm}=\hspace{-1mm} & (1-f_\Vk)^{-\frac{1}{2}}
\left( a^{\IId\dag}_{-\Vk}\,-\,f_\Vk^{\frac{1}{2}+\delta}\,a^\I_\Vk\right).
\label{BEcs}
\end{eqnarray}
Note that the Bogoliubov transformations Eqs.~(\ref{bogol})-(\ref{bogol2}) in the Lorentzian section are recovered
for $\delta\,=\,0$.

\subsection{Relationship between \EX spacetime and TQFTs}\label{sec.4}

We are now ready to formulate and to prove the connection
between QFTs in \EX spacetime and TQFTs. In particular, we will show
that in the aforesaid sections of \EX spacetime, QFT
naturally reproduces all the known formalisms of TQFTs inasmuch as the
correct thermal Green's functions are recovered in respective sections.
Without loss of generality, we will carry out our discussion
in terms of a self-interacting real scalar field.

We start first by recalling the known result~\cite{GUI90} which states that in the Euclidean section of \EX
spacetime, QFT reproduces the imaginary time formalism. This can be seen on both the level
of generating functional and ensuing two-point Green's functions. The latter turn out to be
nothing but thermal Green's functions.
In the next step, we shall see that in the extended
Lorentzian section, QFT corresponds to the two known real-time TQFT formalisms,
namely the POM and TFD.
Moreover, we shall identify the parameter $\delta$ of the extended
Lorentzian section with the parameter $\sigma$ that naturally parametrizes both POM and TFD
formalisms. 

In order to show this, let us start from the general form of the Lagrange density in the full \EX spacetime for the real scalar theory with a Schwinger-type source term ${\rm J}$. This is given by
\begin{equation}
{\cal L}[\Phi,{\rm J}] \, = \, - \sqrt{g} \left(
\frac{1}{2}\,\partial_\mu\Phi\partial{\hspace{0.19mm}^\mu}\Phi \,+\,
\frac{m^2}{2}\,\Phi^2 \,+\, V(\Phi) \,-\, {\rm J} \Phi\right),
\end{equation}
where $V(\Phi)$ is an arbitrary local self-interaction which might be further restricted in its form, \emph{e.g.} by requiring the renormalizability of the theory.

\subsection*{Matsubara formalism}

In the Euclidean section of \EX spacetime, the generating functional of Green's functions is
given by~\cite{GUI90}
\begin{equation}
Z_E[J] \, = \, N \int \hspace{-0.5mm}\mathcal{D}\Phi \exp\left\{-\hspace{-0.5mm}\int\hspace{-0.7mm} d\sigma d\xi dydz\
{\cal L}_{\sigma,\xi}[\Phi,{\rm J}]\right\},
\end{equation}
where
\begin{eqnarray}
\lefteqn{
{\cal L}_{\sigma,\xi}[\Phi,{\rm J}] \,=\,
\frac{1}{2}\,\left[\left(\partial_\sigma \Phi\right)^2
\,+\, \left(\partial_\xi \Phi\right)^2\right]
} \mlab{euclexz} \nonumber\\ [1mm]
 &&\hspace{3mm}
\hspace{10mm}+\, \frac{1}{\alpha^2\left(\sigma^2 \,+\,\xi^2\right)} \left\{
\frac{1}{2}\left(\nabla_{\!\perp}\Phi\right)^2
\,+\, \frac{m^2}{2}\,\Phi^2 \,+\, V(\Phi) \,-\, {\rm J}\Phi \right\},
\end{eqnarray}
is the corresponding pullback of the Lagrange density of the full \EX spacetime to the Euclidean section.

By performing the change of coordinates in Eqs.~(\ref{euclidTrans})-(\ref{euclidTrans2}), the generating functional takes the form
\begin{equation}\mlab{euclz}
Z_E[J] \, = \, N \int\hspace{-0.5mm} \mathcal{D}\Phi
\exp\left\{ \!-\!\int_0^\beta\!\!\!\!d\tau\hspace{-1.0mm}\int_{\R^3}\!\!\!\!
dx\hspace{0.1mm}dy\hspace{0.1mm}dz\,{\cal  L}_{\tau,x}[\phi,J] \right\},
\end{equation}
where the functional integration is taken over fields satisfying the Euclidean KMS boundary condition Eq.~(\ref{periodicfield}) and
\begin{equation}\mlab{euclzL}
{\cal L}_{\tau,x}[\phi,J]
\, = \, \frac{1}{2} \left[ \left( \partial_\tau\phi \right)^2
\, + \, \left( \nabla\phi \right)^2
\, + \, m^2\,\phi^2 \right] \,+\, V(\phi) \,-\, J\phi\, ,
\end{equation}
with $J(\tau,x,y,z)\,=\,{\rm J}(\sigma,\xi,y,z)$.
By differentiation of Eq.~(\ref{euclz}) with respect to the source $J$,
we obtain the Matsubara propagator, the Fourier transform of which is
\begin{equation}
G_\beta({\bk}, \omega_n) \, = \, \frac{1}{\omega_n^2 \,+\, {\bk}^2 \,+\, m^2}\, .
\end{equation}
Here the (bosonic) Matsubara frequencies $\omega_n$ are given by
$\omega_n \,=\, 2\pi n/\beta$ ($n\in{\Bbb N}$).

\subsection*{Real time formalism -- POM}\label{Sec.4.2}

Let us now consider the extended Lorentzian section.
The generating functional of Green's functions can be written as
\begin{equation}\mlab{lorentzexz}
Z[J] \, = \, {\cal N}\hspace{-0.5mm} \int \mathcal{D}\Phi \exp \left\{\!i\int\hspace{-0.7mm}
d\etad d\xid dydz\, {\cal L}_{\etad,\xid}[\Phi,{\rm J}] \right\},
\end{equation}
where
\begin{eqnarray}
\lefteqn{{\cal L}_{\etad,\xid}[\Phi,{\rm J}] \, = \,
\frac{1}{2} \left[ \left( \partial_{\etad}\!\Phi \right)^2
\,-\, \left( \partial_{\xid}\!\Phi \right)^2 \right]
} && \label{lorentzact} \\[1mm] \nonumber
&& \hspace{16mm}+ \, \frac{1}{\alpha^2\left|\xid^2\,-\,\etad^2\right|}
\left\{ - \frac{1}{2}\left(\nabla_{\!\perp}\Phi\right)^2
\,-\,\frac{m^2}{2}\Phi^2 \,-\, V(\Phi) \,+\, {\rm J}\Phi \right\}.
\end{eqnarray}
Since we are interested only in Green's functions with spacetime
arguments belonging to $R_\I\cup R_\IId$, we can set the
source to zero in regions $R_\III$ and $R_\IV$, i.e.
${\rm J}(x)  \,=\,  0$ when $x\in R_\III\cup R_\IV$.
This amounts to reducing Eq.~(\ref{lorentzexz}) to
\begin{equation}\mlab{lorentzexz2}
Z[J] \, = \, {\cal N} \int\hspace{-0,5mm} \mathcal{D}\Phi \exp \left\{ i\int_{R_\I\cup R_\IId}
\hspace{-6mm} d\etad d\xid dydz\
{\cal L}_{\etad,\xid}[\Phi,{\rm J}] \right\}.
\end{equation}
By using the transformations in Eq.~(\ref{mex2a}), the fields in regions $R_\I$ and $R_\IId$
can now be expressed in terms of the local Minkowskian coordinates as
\begin{eqnarray}
&&\mbox{\hspace{-1.3cm}}Z[J] \ = \ {\cal N} \int\hspace{-0.5mm} \mathcal{D}\phi \ \! \exp \left\{i\int\hspace{-0.7mm} dt_\I dx_\I dydz \ \! {\cal L}_{t,x}[\phi,J](t_\I, {\bx}_\I)\right\}
\nonumber \\[1mm]
&&\mbox{\hspace{0mm}} \times \ \! \exp \left\{i\int\hspace{-0.7mm}  dt_{\IId} dx_{\IId} dydz\ {\cal L}_{t,x}[\phi,J](t_{\IId}, {\bx}_{\IId}) \right\},
\label{lorentzexz3}
\end{eqnarray}
where $\phi$ is the local field, the integration is taken over the Minkowski spacetime and
\begin{equation}\mlab{lorentzL}
{\cal L}_{t,x}[\phi,J]
\, = \, \frac{1}{2} \left[ \left( \partial_t\phi \right)^2
\,-\,  \left( \nabla\phi \right)^2
 \,-\,  m^2\,\phi^2 \right] - V(\phi) \,+\, J\phi\, .
\end{equation}
We now use Eqs.~(\ref{ACtxd1})-(\ref{ACtxd}) to further manipulate the generating
functional. It then follows that
\begin{eqnarray}
\label{lorentzexz4}
&&\mbox{\hspace{0mm}}Z[J] \, = \, {\cal N} \int \hspace{-0.5mm}\mathcal{D}\phi \ \! \exp
\left\{ i\int\hspace{-0.7mm} dt\hspace{0.3mm}dx\hspace{0.3mm}dy\hspace{0.3mm}dz
 \ \! {\cal L}_{t,x}[\phi,J]\,(t,{\bx})\right\}
 \\[1mm]
\nonumber&&\mbox{\hspace{1.2cm}} \times \,  \exp \left\{- i\int\hspace{-0.7mm} dt\hspace{0.3mm}dx\hspace{0.3mm}dy\hspace{0.3mm}dz
 \ \! {\cal L}_{t,x}[\phi,J]\,\big(t \,+\, i\beta (1/2 \,+\, \delta), {\bx}\big) \, \big] \right\}\hspace{-1mm},
\end{eqnarray}
where in the last step we have dropped the subscript $I$ and employed
the fact that the time direction (epitomized by time-like Killing vector) is mutually opposite in regions $R_\I$ and $R_\IId$.

Let us now consider the expression for the generating functional as given
in the POM formalism~\cite{LEB, Rivers}
\begin{equation}\mlab{POMgf}
Z_{\mbox{\tiny POM}}[J] \, = \, {\cal N}' \int\hspace{-0.5mm} \mathcal{D}\phi
\exp \left\{ i\int_C\hspace{-0.7mm} dt\hspace{0.3mm}dx\hspace{0.3mm}dy\hspace{0.3mm}dz\ {\cal L}_{t,x}[\phi,J]\right\}\, ,
\end{equation}
where the time path $C$ is the Niemi-Semenoff time path depicted in Fig.~\ref{fig1NS}. The 
path integration is over all fields satisfying periodicity condition $\phi(t_i, \bx)\,=\,\phi(t_i\hspace{0.3mm}-\hspace{0.3mm}i\beta, \bx)$, $t_i$ being the initial time.
For most practical purposes (though not for all, see note in Discussion and Conclusions) one can disregard the contribution from 
the vertical parts of the path contour and assimilate it
in the normalization factor ${\cal N}'$~\cite{mabilat}.
In so doing, the generating functionals Eqs.~(\ref{lorentzexz4})-(\ref{POMgf}) can be identified provided that
\begin{equation}
\mlab{sigma}
\delta  \, = \, \sigma \,-\, 1/2\, .
\end{equation}
Therefore, we see that the time path used in the POM formalism is directly related
to the ``rotation angle'' between the two regions $R_\I$ and
$R_\IId$.
From the quadratic sector (i.e., free-field part) of the Lagrangian in Eq.~(\ref{POMgf}), we can read-off
the thermal-matrix propagator, which in the momentum space acquires the familiar Mills form~\cite{DAS,LEB,LW87,Mills}
\begin{eqnarray}
\label{matprop1}
&&\Delta_{11}(k) \, = \, \displaystyle
\frac{i}{k^2-m^2+i0_+}\, +\, 2\pi\,n(k_0)\,\delta(k^2-m^2)\, ,
\\[3mm]
&&\Delta_{22}(k) \, = \, \Delta_{11}^*(k)\, ,
\\[4mm]
\label{matprop2}
&&\Delta_{12}(k) \, = \, e^{\sigma\beta k_0}
\left[\,n(k_0)+\theta(-k_0)\,\right] 2\pi\,\delta(k^2-m^2)\, ,
\\[4mm]
&&\Delta_{21}(k)\, = \, e^{-\sigma\beta k_0}
\left[\,n(k_0)+\theta(k_0)\,\right] 2\pi\,\delta(k^2-m^2)\, ,
\mlab{matprop}
\end{eqnarray}
where $n(k_0)\,=\,(e^{\beta |k_0|} \,-\,1)^{-1}$. It is worth noting that the parameter $\sigma$
explicitly appears only in the off-diagonal components of the matrix
propagator.

\subsection*{Real time formalism -- TFD \label{Sec.4.3}}

As argued in Section~\ref{TFDform}, 
there is yet another formalism for real-time TQFT, namely the
Thermo Field Dynamics~\cite{UM0,BJV,UM01,UM1,Nair}.
In this approach, it has been said  that a crucial r\^{o}le is played by the
Bogoliubov transformations relating the zero temperature
annihilation and creation operators with the thermal ones. In the
Takahashi-Umezawa representation~\cite{UM0,UM01,UM1}, these transformations 
are given by Eqs.~\eqref{eqn:discretbogoa}-\eqref{eqn:discretbogob}, here rewritten as
\begin{eqnarray}
a_{\bk}(\theta)&=&{\left[1+n(\omega_{\bk})\right]}^{1/2}\,a_{\bk}\,-\,{n(\omega_{\bk})}^{1/2}\,\tilde a^\dagger_{\bk}\,,\label{eqn:discretbogoanewformh1}\\[2mm]
\tilde a_{\bk}(\theta)&=&{\left[1+n(\omega_{\bk})\right]}^{1/2}\,\tilde a_{\bk}\,-\,{n(\omega_{\bk})}^{1/2}\,a^\dagger_{\bk}\,,\label{eqn:discretbogobnewformh2}
\end{eqnarray}
where $n(\omega_{\bk}) = (e^{\beta \omega_{\bk}}-1)^{-1}$. 
%
%
%
%

The form of the thermal Bogoliubov matrix, however, is not unique. Indeed, Eqs.~\eqref{eqn:discretbogoanewformh1}-\eqref{eqn:discretbogobnewformh2}
can be generalized to a non-Hermitian
superposition of the form~\cite{UM1,H95}
\begin{eqnarray}
\label{ex31}
\zeta_\Vk & \hspace{-1mm}= \hspace{-1mm}& (1\,-\,f_\Vk)^{-\frac{1}{2}}
\left(a_\Vk\,-\,f_\Vk^{1-\sigma}\,{\tilde a}^\dagger_\Vk \right), 
\\[2mm]
{\tilde\zeta}_\Vk & \hspace{-1mm}= \hspace{-1mm}& (1\,-\,f_\Vk)^{-\frac{1}{2}}
\left({\tilde a}_\Vk\,-\,f_\Vk^{1-\sigma}\,a^\dagger_\Vk \right).
\mlab{ex3}
\end{eqnarray}
The non-Hermitian property of
the last transformation implies that the canonical conjugates of
$\zeta$ and ${\tilde\zeta}$ are not $\zeta^\dagger$ and
${\tilde\zeta}^\dagger$, but are rather the combinations
\begin{eqnarray}
\label{ex41}
\zeta^\sharp_\Vk &\hspace{-1mm} =\hspace{-1mm} &(1\,-\,f_\Vk)^{-\frac{1}{2}}
\left( a^\dagger_\Vk \,-\,f_\Vk^\sigma\,{\tilde a}_\Vk \right), 
\\[2mm]
{\tilde\zeta}^\sharp_\Vk & \hspace{-1mm}=\hspace{-1mm} & (1\,-\,f_\Vk)^{-\frac{1}{2}}
\left( {\tilde a}^\dagger_\Vk \,-\,f_\Vk^\sigma\, a_\Vk \right),
\mlab{ex4}
\end{eqnarray}
which give the correct canonical commutators,
$[\zeta_\Vk,\zeta^\sharp_{\bp}]\,=\,\delta^3(k-p)$, $[\zeta_\Vk,\zeta_{\bp}]\,=\, 0$, $[\zeta^\sharp_\Vk,\zeta^\sharp_{\bp}]
\,=\, 0$ and similarly for ${\tilde\zeta}_\Vk$ and ${\tilde\zeta}^\sharp_\Vk$.
Here the $\sharp$-conjugation is defined as the usual Hermitian
conjugation {\em together} with the replacement
$\sigma\rightarrow 1-\sigma$. The Hermitian representation Eqs.~(\ref{eqn:discretbogoanewformh1})-(\ref{eqn:discretbogobnewformh2}) is recovered when $\sigma \,=\, 1/2$.
Thermal averages are now expressed as~\cite{BJV,UM1,H95,Nair}
%
%
\begin{equation}\mlab{4.31}
\langle A \rangle  \, = \,
\frac{(\hspace{-0.9mm}( \rho^L|\hspace{-0.5mm}| A |\hspace{-0.5mm}| \rho^R)\hspace{-0.9mm})}{(\hspace{-0.9mm}(\rho^L|\hspace{-0.5mm}|\rho^R)\hspace{-0.9mm})},
\end{equation}
where $A$ is a generic operator acting on the Liouville space and
\begin{equation}
\label{eqn:vacua}
|\hspace{-0.5mm}|\rho^R)\hspace{-1mm}) \, = \, 
\exp\left(\prod_{\bk}f_{\bk}^{\hspace{0.4mm}\sigma}\hspace{0.2mm}a^\dagger_{\bk}\tilde{a}^\dagger_{\bk}\right)\hspace{-1mm}|0,\tilde{0}\rangle, 
\quad\,\,\,
(\hspace{-1mm}(\rho^L|\hspace{-0.5mm}| \, = \, 
\langle 0,\tilde{0}|\exp\hspace{-1mm}\left(\prod_{\bk}f_{\bk}^{\hspace{0.4mm}(1-\sigma)}\hspace{0.2mm}a_{\bk}\tilde{a}_{\bk}\right)\hspace{-1mm}.
\end{equation}
In the special case when $A \,\equiv\, A\otimes 1  \mbox{\hspace{-1mm}} {\rm{I}}$, then $\langle A \rangle$ reduces to the standard thermal average of an observable $A$.
Again, for $\sigma\,=\,1/2$, the states $|\hspace{-0.5mm}|\rho^R)\hspace{-1mm})$ and $(\hspace{-1mm}(\rho^L|\hspace{-0.5mm}|$ become Hermitian conjugates. 
Furthermore, by employing Eqs.~(\ref{ex31})-(\ref{ex4}) and (\ref{eqn:vacua}), it can be seen that
\begin{equation}\mlab{4.35}
\left.\begin{array}{c}
\zeta\\{\tilde \zeta}
\end{array}\right\}\hspace{-0.5mm}
|\hspace{-0.5mm}|\rho^R)\hspace{-1mm})\,=\,0 \,=\,
(\hspace{-1mm}(\rho^L|\hspace{-0.5mm}|
\left\{\begin{array}{c}
\zeta^{\sharp} \\{\tilde \zeta}^{\sharp}
\end{array}\right..
\end{equation}
The thermal propagator for a scalar field in TFD is calculated as
\begin{equation}\mlab{4.36}
\Delta_{ab}(x,y) \, = \, 
\langle T\left[ \phi^a(x) \phi^{b\dagger}(y) \right] \rangle\, ,
\end{equation}
where $T$ is the time ordering symbol and the $a,b$ indices refer to the thermal doublet $\phi^1 \,= \,\phi$ and $\phi^2 \,=\, \tilde{\phi^{\dagger}}$. In the present case of a 
real scalar field we should use in the above definition $\phi^2 \,=\, \tilde{\phi}$.
Quite remarkably, the propagator Eq.~(\ref{4.36}) is equal to the one given by Eqs.~(\ref{matprop1})-(\ref{matprop}), as it can
be easily verified by employing the definitions given above.  

The connection of TFD with the geometric picture of \EX spacetime is
immediate by making the identification
\begin{equation}
\hspace{-1mm}\left(\begin{array}{c}
\phi \\ {\tilde \phi}
\end{array}\right)
\, \equiv \,
\left(\begin{array}{c}
\phi^\I \\ \phi^\IId
\end{array}\right).
\end{equation}

Let us now analyze some other salient features of \EX spacetime, which
are directly related the rotation Eqs.~(\ref{mex31})-(\ref{mex3}).  Along the lines
of Ref.~\cite{ZG98}, we consider the analytic extension of the
imaginary time thermal propagator to real times within the framework of \EX
spacetime. In Ref.~\cite{ZG98} it was shown that the geometric
structure of this spacetime plays a central r\^ole in obtaining the
matrix real-time propagator from the Matsubara one.
In order to see how this works, we consider the simple case of
a massless free scalar field in $1+1$-dimensions. In the Euclidean
section, the equation for the propagator has the form
\begin{equation}\mlab{of1}
\left(\frac{\partial^2}{\partial \sigma^2}
\,+\,\frac{\partial^2}{\partial \xi^2} \right)\hspace{-0.2mm}\Delta_E(\xi^{\mu}\,-\,\xi^{\mu'})
\, = \, - (g_E)^{-\frac{1}{2}}\delta(\xi^{\mu}\,-\,\xi^{\mu'})\, ,
\end{equation}
where $(\xi^{\mu}, \xi^{\mu'})$ denotes a couple of points, $g_E$ stands for the determinant of
the pullback metric in the Euclidean section and $\Delta_E$ is the
imaginary time thermal propagator. Let us now continue Eq.~(\ref{of1}) to the
extended Lorentzian section. This is achieved by first replacing
$\sigma$ by $-i\eta$ and then performing the rotation in Eqs.~(\ref{mex31})-(\ref{mex3}). If $\Delta$ is the real time propagator, we have
\begin{equation}\mlab{of2}
\left(-\frac{\partial^2}{\partial \etad^{2}} \,+\,
\frac{\partial^2}{\partial \xid^{2}} \right)\hspace{-0.2mm}\Delta(\xi^\mu\,-\,\xi^{\mu'})
\,=\, - (g_{L})^{-\frac{1}{2}}\delta(\xi^\mu\,-\,\xi^{\mu'})\, ,
\end{equation}
where $g_{L}$ is the absolute value of the determinant of the pullback metric in the
Lorentzian section. Due to the presence of different disconnected
regions in the Lorentzian section, the propagator exhibits a matrix
structure, since now the points $\xi^\mu$ and $\xi^{\mu'}$ can belong either to
region $R_{\I}$ or $R_{\IId}$ ($R_{\III}$ and $R_{\IV}$, as mentioned earlier, are excluded
from our consideration). 
In terms of Minkowski coordinates, Eq.~(\ref{of2}) reads
\begin{equation}\mlab{of3}
\left(-\frac{\partial^2}{\partial t^{2}}
\, + \, \frac{\partial^2}{\partial x^{2}} \right) \hspace{-0.2mm}\Delta(\xi^{\mu} \,-\, \xi^{\mu'})
\, = \, - \delta_C(\xi^{\mu} \,-\, \xi^{\mu'}),
\end{equation}
where the $\delta_C$ is defined as derivative of a contour step function  
\begin{equation}
\theta_C(t-t')\,=\,\theta(s-s'),
\end{equation}
so that 
\begin{equation}
\delta_C(t-t')\, =\, {\left(\frac{dz}{ds}\right)}^{-1}\delta(s-s').
\end{equation}
Here $t\,=\,z(s)$, with $s\in \mathbb{R}$ monotonically increasing parametrization of the time path $C$.
This path coincides with the Niemi-Semenoff time path of Fig.~\ref{fig1NS}
when the identification Eq.~(\ref{sigma}) is made.  By using Eqs.~(\ref{ACtxd1})-(\ref{ACtxd}),
we obtain, for example, for the component $\Delta_{12}$, the equation
\begin{equation}
{ \left(-\frac{\partial^2}{\partial t^{2}}
\, + \, \frac{\partial^2}{\partial x^{2}}\right)\hspace{-0.2mm}\Delta(t\hspace{0.5mm}-\hspace{-0.5mm}t'\hspace{0.5mm}+\hspace{0.5mm}i\sigma\beta,x\hspace{0.5mm} -\hspace{0.5mm}x') }
\, = \, - \delta_C(t\hspace{0.5mm}-\hspace{0.5mm}t'\hspace{0.5mm}+\hspace{0.5mm}i\sigma\beta) \,\delta(x \hspace{0.5mm}-\hspace{0.5mm}x'),
\mlab{of5}
\end{equation}
which gives us the solution propagator $\Delta_{12}$ given in Eq.~(\ref{matprop2}).

We finally consider the {\em tilde conjugation} within the framework of the
\EX spacetime. The tilde conjugation rules are postulated in TFD in
order to connect the physical and the tilde operators.  Due to the
geometric structure of \EX spacetime, these rules are there seen as
coordinate transformations. This result, which was first discussed in
Ref.~\cite{ZG95}, is here enlarged to the extended Lorentzian
section of \EX spacetime.

Let us recall the tilde conjugation rules as originally defined in TFD \cite{UM0,UM01,UM1} (we
restrict for simplicity to bosonic operators):
\begin{eqnarray}
\begin{array}{rclcrcl}
\left(A B\right)\tilde{} &\hspace{-1.5mm}=\hspace{-1.5mm}& \tilde{A} \tilde{B}, &&\quad
\left(c_1 A \,+\, c_2 B \right)\tilde{} &\hspace{-1.5mm}=\hspace{-1.5mm}& c_1^*\tilde{A} \,+\, c_2^*\tilde{B}\, ,
\\
\left(\tilde{A}\right)\tilde{} &\hspace{-1.5mm}=\hspace{-1.5mm}& A, &&\quad
\hspace{0.2mm}\left(\tilde{A}\right)^\dagger &\hspace{-1.5mm}=\hspace{-1.5mm}& \left(A^\dagger\right)\tilde{},
\end{array}\mlab{of6}
\end{eqnarray}
where $A, B$ are two generic operators and $c_1, c_2$ are $c$-numbers.  In order to
reproduce this operation in the extended Lorentzian section, let us first introduce the following $M$ operation as defined in Ref.~\cite{ZG95}:
\begin{equation}\mlab{of7}
M\,\Phi(\eta, \xi)\,M^{-1} \, \equiv \, \Phi(-\eta, -\xi)\, .
\end{equation}
By expressing the field in terms of Minkowskian coordinates, the $M$ operation can be cast into
\begin{equation}\mlab{of8}
M \,\phi(t, x)\, M^{-1} \, = \, \phi(t\,-\,i\beta/2, x)\, .
\end{equation}
Note that the $M$ operation is anti-linear, since it induces a time inversion
together with the shift $t\rightarrow t\,-\,i\beta/2$. This is clear when we consider its action on the
conjugate momentum $\pi(t,x)\,=\, \partial_t \phi^\dagger(t,x)$, then
\begin{equation}\mlab{of8b}
M\,\pi(t, x)\,M^{-1} \,= \, - \pi(t\,-\,i\beta/2, x)\, .
\end{equation}
Next we perform a rotation by an angle $\delta$ transforming the
$\eta,\xi$ coordinates according to Eqs.~(\ref{mex31})-(\ref{mex3}). The field then
becomes
\begin{equation}\mlab{of8c}
R_\delta\,\Phi(\eta,\xi)\,R_\delta^{-1} \, \equiv \, \Phi(\etad,\xid)\, .
\end{equation}
Finally, we introduce a $\delta$-conjugation operation, which is similar to a charge conjugation, by
\begin{equation}\mlab{of9}
C_\delta\,\phi(t, x)\,C_\delta^{-1} \, \equiv \, \phi^\sharp(t, x)\, .
\end{equation}
Here the change $\delta\rightarrow -\delta$ (or equivalently
$\sigma\rightarrow 1-\sigma $) has to be performed together with
usual charge conjugation.

The combination of these three operations results in the tilde
conjugation. By defining the combined transformation $G_\delta \,\equiv\,
C_\delta\,R_\delta\,M$, we have
\begin{equation}\mlab{o10}
G_\delta\,\phi(t, x)\,G_\delta^{-1} \, = \,  \phi^\sharp(t\,-\,i\sigma \beta, x)
\,= \, \phi^\dagger(t\,-\,i(1\,-\,\sigma)\beta,x)\, .
\end{equation}
In order to reproduce the tilde rules of Eq.~(\ref{of6}),  we can now omit for simplicity the space dependence of the field. Then we have
\begin{eqnarray}
&&\mbox{\hspace{-10mm}}G_\delta\, \phi_1(t)\,\phi_2(t')\,G_\delta^{-1}
\,= \, \phi_1^\sharp(t\,-\,i\sigma \beta)\,\phi_2^\sharp(t'\,-\,i\sigma\beta)\, ,
\\[2mm]
&&\mbox{\hspace{-10mm}}G_\delta \left[G_\delta\,\phi(t)\,G_\delta^{-1}\right] G_\delta^{-1}
\, = \, \phi(t)\, ,
\\[2mm] \mlab{of11}
&&\mbox{\hspace{-10mm}}G_\delta \left[ B_1\,\phi_1(t) \, +  \, B_2\,\phi_2(t') \right]
G_\delta^{-1} \, = \,
B_1^*\,\phi_1^\sharp(t\,-\,i\sigma \beta) \, + \,
B_2^*\,\phi_2^\sharp(t'\,-\,i\sigma\beta)\, ,
 \\[2mm]
&&\mbox{\hspace{-10mm}}G_\delta\left[ \phi^\dagger(t) \right] G_\delta^{-1}
 \, = \, \left[ G_\delta\,\phi(t)\,G_\delta^{-1}\right]^\dagger.
\end{eqnarray}
The $c$-numbers are conjugated since the $M$ operation is
anti-linear. The second of the above relations follows form the fact that
\begin{eqnarray}
\mbox{\hspace{-6mm}}G_\delta \left[G_\delta \,
\phi(t)\, G_\delta^{-1}\right] G_\delta^{-1}
&\hspace{-1.5mm}=\hspace{-1.5mm}& C_\delta\,R_\delta\,M\,\phi^\dagger(t\,-\,i(1\,-\,\sigma) \beta)\,
M^{-1}\,R_\delta^{-1}\, C_\delta^{-1} \nonumber
\\[2mm]
&\hspace{-1.5mm}=\hspace{-1.5mm}& C_\delta\,\phi^\dagger(t\,-\,i(1\,-\,\sigma)\beta \,-\, i\sigma\beta)
\, C_\delta^{-1} \nonumber
\\[2mm] \mlab{of12}
&\hspace{-1.5mm}=\hspace{-1.5mm}& \ \phi(t\,-\,i\beta) \, = \, \phi(t)\, .
\end{eqnarray}
On the last line we have used the periodicity boundary condition for fields in the Lorentzian section, cf. Section~2.
In this way the tilde rules of Eq.~(\ref{of6}) for generic $\sigma$ are directly reproduced.

\section{Discussion and Conclusions}\label{sec.5}

In this Chapter we have addressed all formalisms of QFT at finite temperature and density, 
i.e., the Matsubara or Imaginary Time (IT) approach~\cite{DAS,LEB,LW87}, the
Thermo Field Dynamics (TFD)~\cite{UM0,DAS,UM01,UM1} and the Path Ordered Method (POM)~\cite{BJV,Mills}, including the Closed Time Path (CTP) formalism of Keldysh and Schwinger~\cite{DAS,LEB,LW87} as a special case.
Searching for roots in common among all these seemingly unrelated techniques, we
have also introduced the so-called \EX spacetime, discussing its
thermal properties in connection with other non-trivial backgrounds
endowed with event horizon(s).
Our particular focus was on specific complex sections of this spacetime
which could be identified with the general geometric background for real-time TQFTs
at equilibrium. More specifically, we have shown that there is a one-to-one relationship
between the {\em vacuum} Green's functions in the respective sections of \EX spacetime
and generic mathematical representation (the so-called Mills representation) of {\em thermal} Green's functions in Minkowski spacetime.
Complex sections discussed here can be regarded as an extension of the Lorentzian section
of Gui~\cite{GUI90} by means of a rotation of region $R_\II$ with respect
to $R_\I$ in the complex \EX spacetime. In terms of Minkowski coordinates, this rotation is shown
to be an  isometry: it is equivalent to a constant time shift, leaving the metric invariant.
The angle between the two regions turns out to be related to the
$\sigma$ parameter of the time path as used in the POM formalism. It
also reproduces Umezawa's characteristic parameter appearing in the Bogoliubov thermal matrix
of TFD, when the relation between modes belonging to different regions
is considered.
All in all, we have shown that the full \EX spacetime is versatile enough to allow for
analytic extension from the imaginary-time (Matsubara) propagator to generic $2\times 2$ thermal matrix
propagator  of the real-time formalism -- feat impossible in fixed $1+3$-dimensional spacetime, and for a
consistent prescription of the tilde conjugation rule in TFD.

In the course of our analysis, we have seen that the geometric framework of \EX spacetime allows to understand the various existent 
formalisms of TQFT in a unified way. In particular, with regard to the real-time
methods, i.e. the POM formalism and TFD, one can draw the following  geometric picture.
In the Lorentzian section of \EX spacetime there are two different regions $R_\I$ and $R_\IId$ over
which the global field is defined.  For a global observer this field propagates in a zero temperature heat bath.
However, when one restricts itself to one of the two regions (say $R_\I$),
temperature arises as a consequence of the loss of information (increase in entropy) about the other region.
In order to calculate the corresponding thermal propagator, one needs then to compare the fields
defined in different regions. This can be done essentially in two conceptually distinct ways.

i) By the analytic continuation of the field $\phi^\IId(x_\IId)$ defined in
region $R_\IId$ to region $R_\I$. In this case the time argument gets shifted
by $i\beta(1/2\,+\,\delta)$, as described in Section III. One thus ends up
with {\em one} field and  {\em two} possible time arguments, which can be
either $t$ or $t\,-\,i\beta(1/2\,+\,\delta) $. The generating functional
defined in \EX spacetime by following this procedure turns out to be
the same of the one defined in the POM formalism. The matrix structure of the two-point
thermal Green's function is obtained by functionally differentiating the generating functional
with respect to Schwinger sources with four possible combinations of time arguments.

ii) One can attach the information about the region to the field
operator rather than putting it in the time argument. In this case the
identification $\phi^\I(x)\,\equiv\, \phi(x)$ and $\phi^\IId(x)\,\equiv\,
{\tilde \phi}(x)$ can be made and one obtains the formalism of TFD,
which consists of {\em two} commuting field operators and a {\em single}
time argument. The matrix structure of the two-point thermal Green's function
arises then from the four possible combinations of {\em physical} and {\em tilde}
fields in the (thermal) vacuum expectation value.

These two pictures lead to the same physics which is manifested by the same matrix form of
thermal Green's functions in generic Mills representation.
After all, this could be expected since both pictures represent just a different
``viewpoints'' of an inertial local observer in the context of full \EX spacetime.
In this connection it is interesting to mention the r\^{o}le of regions $R_{\III}$ and $R_{\IV}$  of the extended
Lorentzian section. These regions were intentionally omitted from our discussion in the main text.
This omission was motivated, partially by technical simplifications, but mainly by the fact that
for most applications  the vertical parts of the time path in POM can be neglected.
In particular, for computation of correlation functions in TQFT, the regions $R_{\I}$ and $R_{\IId}$ of
the extended Lorentzian section fully suffice. It is, however, known that when vacuum-bubble diagrams are
important (\emph{e.g.}, when vacuum pressure, effective action or Casimir effect are considered) then the full
Niemi-Semenoff POM with vertical time paths included is obligatory~\cite{mabilat}. In such cases
the partition function $Z_{\mbox{\tiny POM}}$  cannot be factorized (as assumed in Section~\ref{Sec.4.2}), i.e.
$Z_{\mbox{\tiny POM}} = Z_{R_{\I}\cup R_{\II}\cup R_{\III}\cup R_{\IV}} \neq Z_{R{\I}\cup R_{\II}} Z_{R_{\III}\cup R_{\IV}}$,
or in other words, the regions $R_{\III}$ and $R_{\IV}$  must be correlated with regions $R_{\I}$ and $R_{\II}$. Similar conclusion holds
also for the extended Lorentzian section. Note that the aforesaid cross-horizon correlation has purely quantum-mechanical origin
(presence of vacuum-bubble diagrams is required) and hence it cannot be explained by classical means.
One can estimate the correlation between $R_{\III} \cup R_{\IV}$ and $R_{\I} \cup R_{\II}$, for instance, by checking how much the vertical time paths
contribute  in observable quantities (such as pressure). 
In this connection the Casimir effect at finite temperature  with oscillating plates (or other geometries) is a particularly pertinent system. 
Conceptually a similar issue was recently considered in the context of Rindler spacetime in order to
explain the origin of Unruh radiation in terms of vacuum entanglement among
all four different regions of that spacetime~\cite{Higuchi:2017gcd}.

Another interesting question to ask is to what extent the connection between \EX spacetime and TQFTs
can be generalized to out-of-thermal-equilibrium situations. This would be highly desirable endeavor as in
the last two decades  there has been a demand for a new set of tools and concepts from QFT
to treat the non-equilibrium dynamics of relativistic many-body systems and for understanding of further ensuing issues
like dissipation, entropy, fluctuations, noise and decoherence in these systems.
The catalyst has been the infusion of (mostly) experimental data from nuclear particle physics in the relativistic
heavy-ion collision experiments (at LHC and RHIC), early-universe cosmology in the wake of high-precision
observations (such as WMAP, Planck probe or BICEP3), cold atom (such as Bose-Einstein) condensation physics in highly
controllable environments, quantum mesoscopic processes and collective phenomena in condensed matter systems
(topological insulators, spintronics or out-of equilibrium phase transitions), etc.

While the \EX spacetime connection can certainly be applied in the Linear-Response-Theory
(i.e., near-to-equilibrium situations), as there one still employs the real-time (equilibrium) thermal
Green's functions and concomitant Keldysh-Schwinger (or Niemi-Semenoff) POM~\cite{LEB}, the situation far-from-equilibrium is considerably less clear.
The major difficulty that hinders the applicability of the outlined geometric picture  to generic non-equilibrium QFT systems
is the lack of any (asymptotically time-like) Killing vector field in the geometry of a dynamical time-dependent spacetime.
This leave us without a preferred time coordinate with which we could study the problem. One possible way to proceed is to employ
the {Kodama vector} as a substitute for the Killing vector,
that it is parallel to the timelike Killing vector in the static case (as well as at spatial infinity if one
assumes the evolving spacetime is asymptotically flat)~\cite{helou,wiser}. This ``preferred'' time coordinate is also known as the 
Kodama time~\cite{wiser}. Whether this route can lead to a new conceptual paradigm remains yet to be seen. Work in the direction is presently under active investigation.

\newpage\null\thispagestyle{empty}\newpage

\chapter{Summary, conclusions and future
prospects}
\begin{flushright}
\emph{``Before I came here, \\I was confused abuot this subject.\\ Having listened to your lecture, I am still confused.\\ But on a higher level.''}\\[1mm]
- Enrico Fermi -\\[6mm]
\end{flushright}
The connection among gravity, quantum theory and
thermodynamics underlies a cluster of ongoing debated issues
and  surprising theoretical phenomena, 
of which the Hawking and Unruh effects are the most eloquent examples. 
The discovery that temperature may arise in quantum frameworks as a
result of non-trivial backgrounds endowed with event horizon(s)  has 
disclosed the existence of tantalizing and not yet entirely
investigated scenarios, in which general relativity, thermodynamics and QFT
are intimately related. 
A full-fledged quantum theory of gravity and, consequently, 
a more profound understanding of how nature works at fundamental level, 
cannot disregard a proper analysis
of these aspects.

This was indeed the intent of the present work:
following a well-established line of research, we have addressed
the delicate issue of translating thermal features of QFT 
in terms of geometric properties of spacetime. Precious insights 
along this direction have been offered by the introduction of the 
 \EX spacetime, a complex manifold
with a non-trivial horizon structure in which the above merger does arise in 
a very natural way. 
In addition, it has been shown how the different formalisms of field theories 
at finite temperature and density, such as the Path Ordered Method, 
the Closed Time Path formalism,
the Matsubara approach and the Thermo Field Dynamics, can be straightforwardly unified in this background, which thus provides  the ideal ``theater'' for  
analyzing thermal effects and features of QFT.

The picture so outlined has served 
as a framework for the analysis of both standard and
unconventional aspects of the Unruh effect: specifically, 
we have reviewed how the geometry of Rindler space 
induced by hyperbolic (uniformly accelerated) motion plays a crucial r\^ole in the
calculation of the Unruh temperature, and
to what extent deviations from the canonical setting 
do affect such a result. In this connection, 
substantial effects have been derived in the context of gravity theories  
with a minimal fundamental length, where deformations
of the standard uncertainty relation at Planck-scale (Generalized Uncertainty Principle)
give rise to a shift 
of the Unruh temperature depending on the deforming parameter. Besides its intrinsic interest,
we have emphasized how this analysis, extended to the Hawking scenario, 
could clarify whether the
QFT on curved background is  sensitive to Planck-scale corrections, both 
at a theoretical level (via gedanken experiments on the radiation emitted by large
black holes and the formation of micro black holes) and 
from the phenomenological point of view (through the observation of 
induced effects on analogue gravity experiments, such as the analogue 
Unruh radiation in fluids, 
in BEC, in graphene~\cite{silke,Iorio:2017vtw}, 
or in the advanced LIGO experiment~\cite{Bosso:2018ckz}).  
Additionally, it may have non-trivial implications 
on the issue of the information loss paradox in the black hole evaporation theory.
As shown in Refs.~\cite{Nirouei,Nirouei:2011zz}, 
indeed, the embedding of GUP-effects in the quantum tunneling process 
would be responsible for correlations between the tunneling probability of different modes 
in the black hole radiation spectrum, in such a way that
the quantum information is encrypted in the Hawking radiation, and 
recovered as non-thermal GUP correlations between tunneling probabilities of different modes.

Along this line, an unexpected exotic behavior for the Unruh effect
has been derived within the framework of flavor mixing in QFT too.
Starting from previous results in this context, indeed, 
we have found that the thermality of the Unruh spectrum turns out to be non-trivially spoilt
when flavor fields are involved, due to the combination of the Bogoliubov  
transformation arising from mixing and the one related with the causal structure of  
Rindler spacetime. The symmetric way in which such an interplay occurs 
has been interpreted in terms of a possible
geometric origin of mixing, which stems from the association
between the Unruh radiation as an horizon effect and the superposition of fields
with different masses
as a phenomenon
induced by the geometry of the spacetime where we ``live''. 
Although it is still in its infancy, 
we have remarked how this analysis    
might be relevant for better understanding the intimate features of neutrino mixing:
the  dependence of the modified Unruh spectrum on the mixing characteristic 
parameters (i.e., the squared mass difference of mixed fields and the mixing angle), 
combined with future promising experiments on the Unruh radiation and analog phenomena in condensed matter, 
can  be exploited to fix more stringent bounds on these 
unpredicted quantities in the Standard Model, 
providing insights towards the investigation of neutrino physics beyond the 
current knowledge.

In connection with the treatment of vacuum effects in the presence of field mixing, 
the intricate issue of the inverse $\beta$-decay with flavor neutrinos has been later addressed.
Following a pre-existing debate in literature, we have clarified the origin of 
an alleged inconsistency between the decay rates of accelerated protons  in the
inertial and comoving frames, showing the ambiguity 
to be closely related with the problem of neutrino asymptotic nature.
The restored agreement, at least within the (leading order) approximation  of 
small neutrino mass difference, explains how the standard Unruh
effect is mandatory to maintain the general covariance of 
QFT (against the skepticism of part of the community!),
even when mixed fields are considered. More work is inevitably needed 
to extend calculations beyond the adopted framework, 
particularly in view of the highlighted 
non-thermal corrections to the Unruh spectrum appearing in that context.

\medskip
In passing, we have mentioned that the idea of searching for ``new''
physics by looking at violations of the thermality of the Unruh effect also
embraces a variety of other contexts than the ones 
considered in this work; in Ref.~\cite{Marino:2014rfa}, for example, it has been 
shown that the
Casimir-Polder force between two relativistic uniformly accelerated atoms 
exhibits a transition from the short distance 
thermal-like profile predicted by the Unruh effect to a long distance 
non-thermal character, associated with the breakdown of a local inertial description 
of the system. Similarly, deviations from a purely thermal profile
have been highlighted within the framework of the polymer (loop) 
quantization applied to a Rindler trajectory 
in low-energy regime~\cite{Hossainviolat}, and in the evaluation of 
the response function of a detector with 
non-eternal acceleration~\cite{Ahmadzadegan:2018bqz}.
In the latter case, in particular, it has been emphasized that 
the distinctive properties the spectrum  acquires
as a result of a time-dependent acceleration 
allow for the reconstruction of the detector's trajectory, 
in the same way as modulations of the Hawking radiation 
due to an increase in size of the emitting black hole carry information
about the infalling matter responsible for the 
growth itself~\cite{Carney:2017jut}.
The question may naturally arise as to how these exotic behaviors 
match with the findings of our analysis.

To conclude, we stress the present thesis work only provides
a preliminary investigation of non-standard aspects of QFT 
in the context of the Unruh effect, 
together with the proposal
of a unifying geometric perspective for their origin. As already mentioned, 
on the one hand the immediate next step is to generalize
the whole framework to the 
\emph{Hawking black hole theory}, also in light of possible violations of the equivalence principle
arising in that case
(see, for example, Ref.~\cite{Singleton:2011vh} for a more detailed discussion).
On the other hand, a more ambitious prospect is
to figure out how the outlined unconventional features of the Unruh effect 
combine with the \emph{entanglement} properties for accelerated observers;
as highlighted in Ref.~\cite{FuentesSchuller:2004xp}, indeed, 
due to the Unruh radiation, entanglement turns out to be
an observer-dependent quantity in non-inertial frames, as it 
exhibits a decreasing behavior with increasing the acceleration.
Additionally, in Refs.~\cite{Hwang:2010ib}
it has been shown that, unlike the bipartite entanglement,
the tripartite entanglement does not completely vanish in the infinite acceleration 
limit, at least for fermions, which indicates that this kind of quantum information
processing may also be possible if one of the parties approaches to the Rindler horizon (in this sense,
by exploiting similarities between Rindler and Schwarzschild spacetime structures, 
such information processing as quantum tripartite teleportation could be performed even between inside and outside the black hole). 
This begs the question of how the above framework
gets modified when non-thermal corrections to the Unruh distribution are taken into account: 
work is still in progress to clarify this and many other aspects concerning
the rich scenario  of QFT in accelerated frames.

\newpage\null\thispagestyle{empty}\newpage

\appendix 
\addcontentsline{toc}{chapter}{Appendices}
\chapter{Klein-Gordon equation in Minkowski and Rindler coordinates}
\label{plane wave}

\begin{flushright}
\emph{``You cannot teach a man anything; \\
you can only help him discover it in himself.''}\\[1mm]
-  Galileo Galilei -\\[6mm]
\end{flushright}

In this Appendix we review the standard plane wave quantization of a free scalar field in $n$-dimensional Minkowski spacetime. To begin with, let us expand the field in the familiar form
\begin{equation}
\phi(x)\,=\,\int d^{n-1}{k}\, \Big\{a_{{k}}\, \uuu_{{k}}(x)\,+\, {\bar a_{k}}^\dagger\, \uuu_{k}^{\hspace{0.3mm}*}(x) \Big\},
\label{eqn:expans}
\end{equation}
where
\begin{equation}
U_k(x)\,=\,{\big[2\omega_{k}{(2\pi)}^{n-1}\big]}^{-\frac{1}{2}}\, e^{i\left(k.x-\omega_{k} t\right)}
\label{eqn:modes}
\end{equation}
are the plane waves with frequency $\omega_{k}={\sqrt{m^2+{k}^2}}$, and $x\equiv\{t,x^1,x^2,...,x^{n-1}\}$ is the $n$-tuple of Minkowski coordinates. We have denoted by the dot $.$ the $n-1$-dimensional scalar product, i.e. $k.x\equiv\sum_{j=1}^{n-1}k_j\,x_j$.

The modes $U_k$ in Eq.~(\ref{eqn:modes}) are solutions of the Klein-Gordon equation,
\begin{equation}
\bigg\{{\frac{\partial^2}{\partial t^2}\,-\,\sum_{j=1}^{n-1}{\frac{\partial^2}{\partial x_j^2}}}\,+\,m^2 \bigg\}\,\phi(x)\,=\,0\,.
\label{eqn:esplicitKleinGordon}
\end{equation}
They are normalized with respect to the Klein-Gordon (KG) inner product:
\begin{equation}
\Big(\phi_1,\phi_2\Big)\,=\,i\int d^{n-1}{x}\, \Big[\phi_2^*(x)\stackrel{\leftrightarrow}{\partial_{t}}\phi_1(x)\Big]\,,
 \label{eqn:prodscal}
\end{equation}
where the integration is assumed to be performed on a hypersurface of constant $t$. Indeed, we have
\bea
\label{eqn:minkowskinorm}
\Big(U_k,U_{k'}\Big)\,=\,-\Big(U^*_k,U^*_{k'}\Big)\,=\,\delta^{n-1}(k-k')\,, \quad\,\, \Big(U_k,U^*_{k'}\Big)\,=\,0\,.
\eea

The operators $a_k$ and $\bar a_k$ in the field expansion Eq.~\eqref{eqn:expans} are assumed to obey the canonical commutation relations,
\bea
\label{eqn:commutrelationster}
&\Big[a_k, a^\dagger_{k'}\Big]\,=\,\Big[\bar a_k,\bar a^\dagger_{k'}\Big]\,=\,\delta^{n-1}(k-k')\,,
\eea
with all other commutators vanishing. As well known, they can be interpreted as annihilation operators of Minkowski particles and antiparticles, respectively. The Minkowski vacuum $|0\rangle_{\mathrm{M}}$ is accordingly defined by
\begin{equation}
a_k|0\rangle_{\mathrm{M}}\,=\,\bar a_k|0\rangle_{\mathrm{M}}\,=\,0\,,\qquad \forall k\,.
\label{eqn:Minkvac}
\end{equation}


\medskip

Now, in order to extend the quantization formalism to uniformly accelerated observers, let us rewrite  Eq.~(\ref{eqn:esplicitKleinGordon}) in terms of Rindler coordinates $x\equiv\{\eta,\xi,x^2,...,x^{n-1}\}$ defined in Eq.~(\ref{eqn:rindlercoordinatesNONTHERMAL}). A straightforward calculation leads to
\begin{eqnarray}
\non
&&\bigg\{{\frac{\partial^2}{\partial t^2}\,-\,\sum_{j=1}^{n-1}{\frac{\partial^2}{\partial x_j^2}}}\,+\,m^2 \bigg\}\,\phi(x)\,=\,0\,
\\[2mm]
&&\underset{Rindler\;coord.}{\longrightarrow}
\bigg\{\frac{1}{\xi^2}\frac{\partial^2}{\partial\eta^2}\,-\,\frac{\partial^2}{\partial\xi^2}\,-\,\frac{1}{\xi}\frac{\partial}{\partial\xi}\,-\,\sum_{j=2}^{n-1}{\frac{\partial^2}{\partial x_j^2}}\,+\,m^2 \bigg\}\hspace{0.1mm}\,\phi(x)\,=\,0\,.
\label{eqn:esplicitKleinGordonrindler}
\end{eqnarray}

Solutions of positive frequency $\Omega$ with respect to the Rindler time $\eta$ can be written in the form (see Ref.~\cite{Takagi:1986kn})
\begin{equation}
u_\kappa^{\,(\sigma)}(x)\,=\,\theta(\sigma\xi)\!\ {\left[2\Omega{(2\pi)^{n-2}}\right]}^{-\frac{1}{2}}\!\ h_\kappa^{\,(\sigma)}(\xi)\!\ e^{i\left({k}\cdot{x}-\sigma\Omega\eta\right)}\,,
\label{eqn:rindlermodes}
\end{equation}
where $\sigma=+$ refers to the right wedge  $R_+=\{x|x^1>|t|\}$, while $\sigma=-$ to the left wedge $R_-=\{x|x^1<-|t|\}$ (see Fig.~\ref{figure:RindlerNONTHERMAL} in Section~\ref{Unruh}), and we have denoted by the dot $\cdot$ the $n-2$-dimensional scalar product, i.e. $k\cdot x\equiv\sum_{j=2}^{n-1}k_j\,x_j$. The subscript $\kappa$ stands for $\kappa\equiv(\Omega, k^2,...,k^{n-1})$.

 We will refer to the solutions Eq.~(\ref{eqn:rindlermodes}) as Rindler modes. Note that the Heaviside step function $\theta(\sigma\xi)$ has been inserted in order to constrain these modes to only one of the two causally disjoint wedges $R_\pm$. The time dependence
\begin{equation}
u_\kappa^{\,(\sigma)}\propto e^{-i\sigma\Omega\eta}\,,
\end{equation}
reflects the fact that the boost Killing vector $B=\frac{\partial}{\partial\eta}$ is future oriented in $R_+$, while it is past oriented in $R_-$.

The explicit expression of $h_\kappa^{\,(\sigma)}$ in Eq~\eqref{eqn:rindlermodes} can be obtained by requiring the modes $u_\kappa^{\,(\sigma)}$ to be solution of the Klein-Gordon equation in Rindler coordinates, Eq.~\eqref{eqn:esplicitKleinGordonrindler}. This leads to
\begin{equation}
\bigg\{\frac{d^2}{d\xi^2}\,+\,\frac{1}{\xi}\frac{d}{d\xi}\,+\,\frac{\Omega^2}{\xi^2}\,-\,\mu_k^2 \bigg\}\,h_\kappa^{\,(\sigma)}(\xi)\,=\,0\,,
\label{eqn:sostitu}
\end{equation}
which is solved by the modified Bessel functions of second kind. In detail, by imposing that these functions are delta-normalized, one gets
\begin{equation}
h_\kappa^{\,(\sigma)}\,=\,{\left(2/\pi\right)}^{\frac{1}{2}}A_\kappa^{\,(\sigma)}{\left(\alpha\mu_{k}/2\right)}^{i\Omega}\!\ \Gamma(i\Omega)^{-1}K_{i\Omega}(\mu_{k}\xi)\,,
\label{eqn:besselmodified}
\end{equation}
where
\begin{equation}
\label{eqn:fattoredifase}
\quad A_{\kappa}^{\,(\sigma)}\,=\,\left\{ \begin{array}{l}
R^{\, *}_{\kappa}{\left(\alpha\mu_k/2\right)}^{-i\Omega}\, \Gamma(i\Omega)/|\Gamma(i\Omega)|, \qquad \mathrm{for}\, \,  \sigma=+\, ,
\\[2mm]
R_{\kappa}\,{\left(\alpha\mu_k/2\right)}^{i\Omega}\, \Gamma(-i\Omega)/|\Gamma(i\Omega)|, \quad\;\;\;\hspace{-0.3mm}\mathrm{for}\, \,  \sigma=-\, , \\
\end{array}\right.
\end{equation}
and $R_{\kappa}={\left[{\left(\alpha\mu_k/2\right)}^{-i\Omega}\, \Gamma(i\Omega)/|\Gamma(i\Omega)| \right]}^2$. Here $\alpha$ is an arbitrary postive constant of dimension of length and $\Gamma(i\Omega)$ is the Euler Gamma function.

Now, exploiting Eqs.~(\ref{eqn:rindlermodes}), (\ref{eqn:besselmodified}), one can prove that the Rindler modes above defined form a complete and orthonormal set with respect to the KG inner product in Rindler coordinates:
\begin{equation}
\left(\phi_1,\phi_2\right)\,=\,i \int^{+\infty}_{-\infty}\frac{d\xi}{|\xi|}\int d^{n-2}{x}\, \phi_2^*\!\ \overset{\leftrightarrow}\partial_ \eta \!\ \phi_1\,,
\label{eqn:proscalrindlercoord}
\end{equation}
where we have implicitly assumed that the integration is performed on a hypersurface of constant $\eta$. Indeed, 
\begin{equation}
\left(u_\kappa^{\,(\sigma)},\,u_{\kappa'}^{\,(\sigma')}\right)\,=\,-\left(u_\kappa^{\,(\sigma)*},\, u_{\kappa'}^{\,(\sigma)*}\right)\,=\,\delta_{\sigma\sigma'}\!\ \delta^{n-1}(\kappa-\kappa')\hspace{0.2mm},\qquad
\left(u_\kappa^{\,(\sigma)},\,u_{\kappa'}^{\,(\sigma)*}\right)\,=\,0\,.
\label{eqn:rindlernorm1}
\end{equation}

\chapter{Quantum Field Theory of flavor mixing}
\label{QFT of fm}
\begin{flushright}
\emph{``We especially need imagination in science.\\
It is not all mathematics, nor all logic,\\
but it is somewhat beauty and poetry.''}\\[1mm]
-  Maria Mithcell -\\[6mm]
\end{flushright}

\emph{Flavor mixing} is a truly intriguing yet puzzling feature in the Standard Model. Although the idea that particles with different masses can mix is widely accepted in a variety of contexts, ranging from gauge bosons, to quarks and kaons, its origin is still rather elusive, becoming even obscure when referring to neutrinos.
 
The mechanism of mixing has been theoretically addressed in great detail since Pontecorvo's groundbreaking work in 1978~\cite{Pontecorvo}; from then on, experimental developments~\cite{Fukuda:1998mi, Kajita:2016cak, McDonald:2016ixn} have gradually and successfully confirmed the original proposal, thus opening new scenarios beyond
the standard physics of elementary particles. 

The original -- and, for many years, unique -- framework within witch flavor mixing was investigated was the ordinary Quantum Mechanics (see Ref.~\cite{Giunti:2007ry} and references therein). Notwithstanding the efficacy of such an approach at a phenomenological level, however,
it is actually not rigorous from the theoretical point of view, being the coherent superposition of states with different mass forbidden by Bargmann's superselection rule in a non-relativistic theory~\cite{Bargmann:1954gh}. It was not until 1995  that a full QFT formalism was developed, first for Dirac fermions~\cite{Blasone:1995zc} and bosons~\cite{bosonmix} in a simplified two-flavor description, then for three generations with CP violation~\cite{Blasone:2002jv} and for neutral fields (including Majorana neutrinos)~\cite{neutral}, ending with a full comprehensive treatment~\cite{Ji:2002}. Along this line, an interesting achievement was the subsequent discovery that neutrino mixing represents a possible source for the cosmological constant~\cite{Blasone:2004yh}, thus providing  further link between Cosmology and neutrino physics.

In this Appendix, following the above \emph{historical iter}, we review the QFT of flavor mixing in Minkowski spacetime, devoting the final part to generalize the formalism to the Rindler (uniformly accelerated) background. To make our analysis as transparent as possible, we focus on a model involving only two flavors. In spite of such a minimal setting, we show that non-trivial results are obtained, due to a Bogoliubov transformation connecting  definite mass and flavor fields at the level of ladder operators. Furthermore, we investigate the rich condensate structure acquired by the vacuum for flavor fields, paying particular attention to the related problem of the unitary inequivalence
between the flavor and mass representations in the infinite volume limit. We also show that the standard QM formalism is approximately recovered in the relativistic limit of large momenta.

\section{Mixing transformations for fermions}
\label{sec:trasmixferm}
Let us start by reviewing the QFT of flavor mixing for Dirac fermions\footnote{For this purpose, we closely follow Ref.~\cite{Blasone:1995zc}.} (we shall continuously refer to neutrinos, but our results clearly have a wider validity). To this aim, we recall that, for a minimal two-flavor model, Pontecorvo mixing transformations are written as rotations of the particle states
with definite mass $|\nu_i\rangle$ ($i=1,2$) into those with definite flavor $|\nu_\chi\rangle$ ($\chi=e,\mu$) according to Eqs.~\eqref{eqn:U}, \eqref{PMMmatrix},
\begin{equation}
\begin{array}{l}
|\nu_e\rangle\,=\,\cos\theta\,|\nu_1\rangle \,+\,\sin\theta\, \lnu_2\rangle\,,\\[3mm]
|\nu_\mu\rangle\,=\,-\sin\theta\,|\nu_1\rangle \,+\,\cos\theta\, \lnu_2\rangle\,,
\end{array}
\end{equation}
where $\theta$ is the mixing angle. In terms of fields, these relations can be recast into the form
\begin{equation}
\label{eqn:Pontecorvo}
\begin{array}{l}
\lnu_e(x) \,=\, \lnu_1(x)\cos\theta \,+\, \lnu_2(x)\sin\theta\,,\\[3mm]
\lnu_{\mu}(x) \,=\, -\lnu_1(x)\sin\theta \,+\, \lnu_2(x)\cos\theta\, , 
\end{array}
\end{equation}
 being  $\nu_\chi(x)$ and  $\nu_i(x)$ the fields with definite flavor $\chi$ and mass $m_i$, respectively. For the latter, we adopt the standard Dirac expansion:
\begin{equation}
\lnu_i(x)\,=\,\frac{1}{\sqrt{V}}\sum_{\bk,r}e^{i\bk\cdot \bx}\left[\alpha_{\mathbf{k},i}^r\,\lu_{\mathbf{k},i}^r(t)\,+\,
\beta_{-\mathbf{k},i}^{r\dagger}\,\lv_{-\mathbf{k},i}^r(t)\right],\qquad
i=1,2\,,
\label{eqn:expressionexpl}
\end{equation}
where $V$ is the volume of the system under consideration and $\alpha_{\mathbf{k},i}^r$, $\beta_{\mathbf{k},i}^r$ ($i,r\,=\,1,2$) are the annihilators of the vacuum for the fields with definite mass, $|0\rangle_{1,2}\equiv|0\rangle_1\otimes|0\rangle_2$:\footnote{For simplicity, from now on we shall refer to this vacuum as \emph{mass vacuum}.}
\begin{eqnarray}
\alpha^{r}_{{\bk},i}|0\rangle_{1,2}= \beta ^{r }_{{\bk},i}|0\rangle_{1,2}=0\,.
\end{eqnarray} 
To be consitent with the original paper~\cite{Blasone:1995zc}, in what follows we will perform
all computations at finite volume $V$, setting only at the end $V\rightarrow\infty$.

The spinors $\lu_{\mathbf{k},i}^r(t)$, $\lv_{-\mathbf{k},i}^r(t)$ in Eq.~\eqref{eqn:expressionexpl} are given by
\begin{equation}
\lu_{\mathbf{k},i}^r(t)=e^{-i\omega_{\bk,i}t}\lu_{\mathbf{k},i}^r\,,\quad\quad\quad \lv_{-\mathbf{k},i}^r(t)=e^{i\omega_{\bk,i}t}\lv_{-\mathbf{k},i}^r\, , 
\label{eqn:spinoriforma}
\end{equation}
where $\omega_{\bk,i}=\sqrt{m_{i}^{2}+{|\bk|}^{2}}$ and 
\begin{eqnarray}
\label{eqn:spino}
\hspace{-12mm}&\lu^{1}_{{\bk},i}&\hspace{-2mm}=\left( \frac{\omega _{\bk,i}+m_{i}}{2\omega_{\bk,i}} \right) ^{\frac{1}{2}}\left(
\begin{array}{l}
\quad\, 1 \\
\quad\, 0 \\
\frac{k_{3}}{\omega _{\bk,i}+m_{i}} \\
\frac{k_{1}+ik_{2}}{\omega _{\bk,i}+m_{i}}
\end{array}
\right)\hspace{-1mm},\qquad \lu_{{\bk},i}^{2}=\left( \frac{\omega _{\bk,i}+m_{i}
}{2\omega _{\bk,i}}\right) ^{\frac{1}{2}}\left(
\begin{array}{l}
\quad\, 0 \\
\quad\, 1 \\
\frac{k_{1}-ik_{2}}{\omega _{\bk,i}+m_{i}} \\
\frac{-k_{3}}{\omega _{\bk,i}+m_{i}}
\end{array}
\right)\hspace{-0.5mm}, \\[4mm]
\hspace{-12mm}&\lv_{-{\bk},i}^{1}&\hspace{-2mm}=\left( \frac{\omega _{\bk,i}+m_{i}}{2\omega _{\bk,i}}
\right) ^{\frac{1}{2}}\left(
\begin{array}{l}
\frac{-k_{3}}{\omega _{\bk,i}+m_{i}} \\
\frac{-k_{1}-ik_{2}}{\omega _{\bk,i}+m_{i}} \\
\quad\, 1 \\
\quad\, 0
\end{array}
\right)\hspace{-0.5mm}, \qquad\hspace{-1mm} \lv_{-{\bk} ,i}^{2}=\left( \frac{\omega_{\bk,i}+m_{i}}{2\omega _{\bk,i}}\right) ^{\frac{1}{2}}\left(
\begin{array}{l}
\frac{-k_{1}+ik_{2}}{\omega _{\bk,i}+m_{i}} \\
\frac{k_{3}}{\omega _{\bk,i}+m_{i}} \\
\quad\, 0 \\
\quad\, 1
\end{array}
\right)\hspace{-1mm}.
\label{eqn:spinoriespliciti}
\end{eqnarray}
The anti-commutators for the fields are assumed to be canonical, 
\begin{equation}
\left\{ \lnu _{i}^{\alpha }(x),\lnu _{j}^{\beta
\dagger }(x')\right\} _{t=t^{\prime }}=\delta^{3} ({x-x'})\, \delta
_{\alpha \beta }\, \delta _{ij}, \qquad \alpha ,\beta = 1,...4, 
\label{eqn:commutatoricanonicirelazioni}
\end{equation}
that is, at the level of ladder operators:
\begin{equation}
\left\{ \alpha _{{\bk},i}^{r},\alpha _{{\bk'},j}^{s\dagger
}\right\} =\delta _{{\bk\bk'}}\, \delta _{rs}\, \delta _{ij}, \qquad
\left\{ \beta _{{\bk},i}^{r},\beta _{{\bk',}j}^{s\dagger
}\right\} =\delta _{{\bk\bk'}}\, \delta _{rs}\, \delta _{ij},  \qquad
i,j=1,2\,,
\label{eqn:commutatiooperatoricreann}
\end{equation}
with all other anti-commutators vanishing. The orthonormality and completeness relations
read
\begin{eqnarray}
&\lu_{{\bk},i}^{r\dagger }\lu_{{\bk},i}^{s} \,=\, \lv_{{\bk},i}^{r\dagger }\lv_{{\bk},i}^{s}= \delta _{rs}\,,\qquad\,\lu_{{\bk},i}^{r\dagger }\lv_{-{\bk},i}^{s} \,=\, \lv_{-{\bk}
,i}^{r\dagger }\lu_{{\bk},i}^{s}=0\,,  \\[2mm]
& \sum_{r}\left(\lu_{{\bk},i}^{r}\lu_{{\bk},i}^{r\dagger }+\lv_{-{\bk},i}^{r}\lv_{-{\bk},i}^{r\dagger }\right) \,=\, 1\,.
\label{eqn:ortonormalitecompletezzafunzioni}
\end{eqnarray}
Next, we emphasize that the transformations Eqs.~(\ref{eqn:Pontecorvo}) (or their inverse) connect the ``free'' and ``interaction'' Hamiltonians, $H_{1,2}$ and $H_{e,\mu}$, respectively:\footnote{Since we want to analyze flavor oscillations in vacuum, we consider only the mass terms in the two Hamiltonians, neglecting any external interaction.}
\begin{eqnarray}
\label{eqn:Hamiltonianapontec}
&H_{1,2}\,=\,m_1\bar\lnu_1\lnu_1 \,+\, m_1\bar\lnu_2\lnu_2\,,&\\[2mm]
&H_{e,\mu}\,=\,m_{ee}\bar{\lnu}_e\lnu_e \,+\,
m_{\mu\mu}\bar{\lnu}_{\mu}\lnu_{\mu} \,+\,
m_{e\mu}\left(\bar{\lnu}_ e\lnu_{\mu} \,+\, \bar{\lnu}_{\mu}\lnu_e\right)\, ,&
\label{eqn:Hamiltonianapontec2}
\end{eqnarray}
where
\begin{equation}
\label{eqn:masseemu}
\begin{array}{l}
m_{ee}\,=\, m_{1}\cos^{2}\theta \,+\, m_{2} \sin^{2}\theta\,,\\[3mm]
m_{\mu\mu}\,=\, m_{1}\sin^{2}\theta \,+\, m_{2} \cos^{2} \theta\,,\\[3mm]
m_{e\mu}\,=\,\left(m_{2}-m_{1}\right)\sin\theta \cos \theta\,.
\end{array}
\end{equation} 
Equations~(\ref{eqn:Hamiltonianapontec2}), (\ref{eqn:masseemu}) are particularly meaningful, since they show that fields with definite flavor, unlike the ones with definite mass, are interacting. This is somehow at the root of the unitarily inequivalence between flavor and mass representations in the QFT limit, as explained belo
As well known, QFT is built on two levels: on the one hand, the Lagrangian and the ensuing field equations are expressed in terms of Heisenberg (or interacting) fields; on the other, the physical observables are given in terms of asymptotic (in- or out-) fields, also dubbed as physical or free fields. In the LSZ formalism of QFT~\cite{Itzykson}, the latter fields are obtained by the \emph{weak limit} for $t\rightarrow -\infty$ (in- fields) or $t\rightarrow +\infty$ (out-fields) of the former fields. The limit is weak in the sense that the realization of dynamic equations in terms of the
in/out fields is not unique, but it does indeed depend on the adopted representation for the Heisenberg field operators, namely, on the Hilbert space on which they are defined\footnote{This is what happens, for example, 
in the case of spontaneous symmetry breaking, where the same set of Heisenberg field equations describes both the normal and symmetry-broken phases.}. Such a dependence is rooted in the existence of infinite unitarily inequivalent representations of the canonical (anti-)commutation
relations in QFT (for a more detailed discussion, see Refs.~\cite{UM01,UM1,Bogobogo}), a feature which is absent in QM due to the von Neumann theorem~\cite{Stone}.  It goes without saying that, since 
observables are expressed in terms of asymptotic fields, unitarily inequivalent representations describe different, i.e. physically inequivalent, phases of a given system. In order to achieve physically relevant results, we are thus faced with the problem of investigating the \emph{dynamical map} among Heisenberg and free fields,  Not least, it is clear that the rather common approximation of identifying the vacua (and the Fock spaces built on them) for the two sets of fields is very naive and unsuitable for addressing problems of exquisitely theoretical nature, such as the dynamical origin of flavor mixing in the Standard Model.

Aware of these warnings, let us then analyze the mapping Eqs.~\eqref{eqn:Pontecorvo} among flavor and mass fields, and, thus, among the respective Fock spaces $\mathcal{H}_{e,\mu}$ and $\mathcal{H}_{1,2}$. To this aim, let us rewrite Eqs.~\eqref{eqn:Pontecorvo} in terms of the algebraic generator $G_{\theta}(t)$ of flavor mixing,
\begin{equation}
\label{eqn:mixrelaferm}
\begin{array}{l}
\lnu^{\alpha}_e(x)\,=\,
G_{\theta}^{-1}(t)\, \lnu^{\alpha}_1(x)\, G_{\theta}(t)\,,\\[3mm]
\lnu^{\alpha}_{\mu}(x)\,=\, G_{\theta}^{-1}(t)\, \lnu^{\alpha}_2(x)\, 
G_{\theta}(t)\, ,
\end{array}
\end{equation}
where 
\begin{equation}
G_{\theta}(t)\,=\,\exp\left[\theta\int
d^3x\left(\lnu_1^{\dagger}(x)\lnu_2(x)\,-\,
\lnu_2^{\dagger}(x)\lnu_1(x)\right)\right].
\label{eqn:generatoremix}
\end{equation}
It is not difficult to show that, at finite volume $V$, $G_{\theta}(t)$ is a unitary operator that preserves the canonical anti-commutators Eq.~(\ref{eqn:commutatoricanonicirelazioni}). 
More subtle, on the other hand, is the issue of the time dependence of $G_{\theta}(t)$, which we will return to in the following (from now on, however, to streamline the notation, we shall omit such a dependence when no misunderstanding arises).

Now, by introducing the operators
\begin{equation}
S_{+}\, \equiv\,  \int d^{3}{\bx} \; \lnu_{1}^{\dagger}(x)
\lnu_{2}(x) \, , \qquad S_{-}\, \equiv\,  \int d^{3}{\bx} \;
\lnu_{2}^{\dagger}(x) \lnu_{1}(x)\,=\, \left( S_{+}\right)^{\dagger}\, ,
\label{eqn:operatoriS+S-}
\end{equation}
the mixing generator  $G_{\theta}$ can be cast into the form
\begin{equation}
G_{\theta} \,=\, \exp\left[\theta\left(S_{+} \,-\, S_{-}\right)\right].
\label{eqn:newformG}
\end{equation}
The operators $S_{+}$ and $S_{-}$, together with $S_3$,
\begin{equation}
S_{3}\, \equiv\, \frac{1}{2} \int d^{3}{\bx}
\left(\lnu_{1}^{\dagger}(x)\lnu_{1}(x) \,-\,
\lnu_{2}^{\dagger}(x)\lnu_{2}(x)\right),
\label{eqn:ulterioreintrod}
\end{equation}
and the Casimir $S_{0}$,
\begin{equation}
 S_{0}\, \equiv\, \frac{1}{2} \int d^{3}{\bx}
\left(\lnu_{1}^{\dagger}(x)\lnu_{1}(x) \,+\,
\lnu_{2}^{\dagger}(x)\lnu_{2}(x)\right),
\label{eqn:Casimir}
\end{equation}
close the $SU(2)$ algebra, 
\begin{equation}
\left[S_{+} , S_{-}\right]\,=\,2S_{3} \, ,\quad
\left[S_{3} , S_{\pm} \right] \,=\, \pm S_{\pm} \, ,\quad
\left[S_{0} ,S_{3}\right]\,=\, \left[S_{0} , S_{\pm} \right] \,=\, 0\;.
\label{eqn:SU(2)}
\end{equation}
Exploiting Eq.~(\ref{eqn:expressionexpl}), $S_+$, $S_-$, $S_3$ and $S_0$ can be expanded as follows:
\begin{eqnarray}
\label{eqn:s+} 
\non
\hspace{-5mm}S_{+}(t) \,\equiv\, {\sum_{{\bk}} }S_{+}^{{\bk}}(t)&\hspace{-1mm}=\hspace{-1mm}&
{\sum_{{\bk}} }{\sum_{r,s}}\left(\lu_{{\bk},1}^{r\dagger
}(t)\,\lu_{{\bk},2}^{s}(t)\, \alpha _{{\bk},1}^{r\dagger }\alpha
_{{\bk},2}^{s}\,+\,\lv_{-{\bk},1}^{r\dagger}(t)\,\lu_{{\bk},
2}^{s}(t)\, \beta _{-{\bk},1}^{r}\alpha _{{\bk},2}^{s}\right.
\nonumber \\[2mm]
&&\left. \hspace{-6mm}+\,\lu_{{\bk},1}^{r\dagger}(t)\,\lv_{-{\bk},2}^{s}(t)\, \alpha
_{{\bk},1}^{r\dagger }\beta_{-{\bk},2}^{s\dagger}\,+\,\lv_{-{\bk},1}^{r\dagger } (t)\,\lv_{-{\bk},2}^{s}(t)\, \beta _{-{\bk},1}^{r}\beta _{-{\bk},2}^{s\dagger }\right)\hspace{-1mm},\\[6mm]
\label{eqn:s-} 
\hspace{-5mm}S_{-}(t) \,\equiv \,{\sum_{{\bk}} }S_{-}^{{\bk}}(t)&\hspace{-1mm}=\hspace{-1mm}&
{\sum_{{\bk}}}{\sum_{r,s} }\left(\lu_{{\bk}, 2}^{r\dagger
}(t)\,\lu_{{\bk},1}^{s}(t)\, \alpha _{{\bk},2}^{r\dagger }\alpha
_{{\bk},1}^{s}\,+\,\lv_{-{\bk},2}^{r\dagger }(t)\,\lu_{{\bk},
1}^{s}(t)\, \beta _{-{\bk},2}^{r}\alpha _{{\bk},1}^{s}\right.
\nonumber \\[2mm]
&&\left. \hspace{-6mm}+\,\lu_{{\bk},2}^{r\dagger }(t)\,\lv_{-{\bk},1}^{s}(t)\, \alpha
_{{\bk}, 2}^{r\dagger }\beta_{-{\bk},1}^{s\dagger }\,+\,\lv_{-{\bk},2}^{r\dagger }(t)\,\lv_{-{\bk},1}^{s}(t)\, \beta _{-{\bk},2}^{r}\beta _{-{\bk},1}^{s\dagger }\right)\hspace{-1mm},
\end{eqnarray}
and
\begin{eqnarray}
\hspace{-6mm}S_{3}&\equiv& {\sum_{{\bk}} }S_{3}^{{\bk}}\, =\, \frac{1}{2}
{\sum_{{\bk},r} }\left( \alpha _{{\bk},1}^{r\dagger }\alpha
_{{\bk},1}^{r}\,-\,\beta _{-{\bk},1}^{r\dagger }\beta _{-{\bk},
1}^{r}\,-\,\alpha _{{\bk},2}^{r\dagger }\alpha _{{\bk},
2}^{r}\,+\,\beta _{-{\bk},2}^{r\dagger }\beta _{-{\bk},2}^{r}\right),\\[3mm]
\label{eqn:s3} 
\hspace{-6mm}S_{0}& \equiv& {\sum_{{\bk}} }S_{0}^{{\bk}}\, =\, \frac{1}{2}
{\sum_{{\bk},r} }\left( \alpha _{{\bk},1}^{r\dagger }\alpha
_{{\bk},1}^{r}-\beta _{-{\bk},1}^{r\dagger }\beta _{-{\bk},1}^{r}+\alpha _{{\bk},2}^{r\dagger }\alpha _{{\bk},
2}^{r}-\beta _{-{\bk},2}^{r\dagger }\beta _{-{\bk},2}^{r}\right) .
\label{eqn:s0} 
\end{eqnarray}
Note that the operators in Eqs.~(\ref{eqn:s+}), (\ref{eqn:s-}) have the structure
of a rotation combined with a Bogoliubov transformation\footnote{By comparison, mixing transformations in QM are purely rotation.}. Using the above expansions, it is a trivial matter to check that
the original $SU(2)$ algebra Eq.~\eqref{eqn:SU(2)} splits into $k$ disjoint $SU_{{\bk}}(2)$ algebras, 
\begin{equation}
\quad\left[ S_{+}^{\mathbf{k}},S_{-}^{\mathbf{k}}\right] \,=\,2S_{3}^{\mathbf{k}%
},\text{ \quad }\left[ S_{3}^{\mathbf{k}},S_{\pm }^{\mathbf{k}%
}\right] \,=\,\pm S_{\pm }^{\mathbf{k}},\text{ \quad }\left[ S_{0}^{%
\mathbf{k}},S_{3}^{\mathbf{k}}\right]\, =\,\left[ S_{0}^{\mathbf{k}},S_{\pm }^{%
\mathbf{k}}\right] \,=\,0,
\label{eqn:ciascunacomponente}
\end{equation}
\begin{equation} 
\left[ S_{\pm }^{\mathbf{k}},S_{\pm
}^{\mathbf{p}}\right] \,=\,\left[
S_{3}^{\mathbf{k}},S_{\pm }^{\mathbf{p}}\right] \,=\,\left[ S_{3}^{\mathbf{%
k}},S_{3}^{\mathbf{p}}\right] \,=\,0\text{ \qquad }\mathbf{k\,\neq\,
p}\,,
\label{eqn:su(2)k2}
\end{equation}
i.e., we have the group structure $\bigotimes_{{\bk}}SU_{{\bk}}(2)$.

In order to find the connection between the Fock spaces $\mathcal{H}_{e,\mu}$ and $\mathcal{H}_{1,2}$, let us now consider the generic matrix element $_{1,2}\langle a|\lnu^{\alpha}_{1}(x)|b\rangle_{1,2}$ (similarly for $\lnu^{\alpha}_{2}(x)$), where $|a \rangle_{1,2}$ is the generic element of ${\mathcal H}_{1,2}$. Inverting the first of Eqs.~(\ref{eqn:mixrelaferm}), we get
\begin{equation}
_{1,2}\langle a|G_{\theta}\;
\lnu^{\alpha}_{e}(x)\; G^{-1}_{\theta}|b \rangle_{1,2}\,=\,_{1,2}\langle a|\lnu^{\alpha}_{1}(x)|b \rangle_{1,2}\,,
\label{eqn:inversequation}
\end{equation}
which shows that $G_{\theta}^{-1}|a \rangle_{1,2}$ is a vector of $\mathcal{H}_{e,\mu}$, being the field
operator $\lnu_e$ defined on such a space. It thus arises that $G_{\theta}^{-1}$ provides the mapping between ${\mathcal H}_{1,2}$ in ${\mathcal H}_{e,\mu}$, namely
\begin{equation}
G^{-1}_{\theta}: {\mathcal H}_{1,2} \mapsto {\mathcal H}_{e,\mu}\,.
\label{eqn:mappa}
\end{equation}
In particular, for the vacuum $|0 \rangle_{1,2}$ (at finite volume $V$) the following relation holds:
\begin{equation}
|0(\theta,t) \rangle_{e,\mu} \,=\, G^{-1}_{\theta}(t)\;
|0 \rangle_{1,2}\,,
\label{eqn:vuotosapore}
\end{equation}
which means that $|0(\theta,t) \rangle_{e,\mu}$ represents the vacuum for the flavor Fock space ${\mathcal H}_{e,\mu}$.\footnote{In analogy with the mass vacuum, from now on we shall refer to this vacuum as \emph{flavor vacuum}.} In fact, by exploiting Eqs.~\eqref{eqn:Pontecorvo} and using the linearity of $G_{\theta}$, we obtain the annihilators for the fields $\nu_\chi$ $(\chi=e,\mu)$ as
\be
\label{dynamap}
\lu_{{\bk},e}^{r\,\alpha}\;\widetilde\alpha_{{\bk},e}^{\,r}\,=\,G^{-1}_{\theta}\,\lu_{{\bk},1}^{r\,\alpha}\,\alpha_{{\bk},1}^{\,r}\,G_{\theta}\,, \qquad \alpha=1,...,4\,,
\ee
and similarly for the others. It is straightforward to prove that these operators do indeed annihilate the vacuum for definite flavor fields $|0(\theta,t) \rangle_{e,\mu}$. For the sake of completeness, we remark that,  from now on, the following shorthand notation shall be used for this vacuum: 
\be
|0(t)\rangle_{e,\mu}\,\equiv\,|0(\theta,t)\rangle_{e,\mu}\,,\qquad |0\rangle_{e,\mu}\,\equiv\,|0(\theta,t=0)\rangle_{e,\mu}\,.
\ee
Strictly speaking, Eq.~(\ref{eqn:vuotosapore}) has a purely formal meaning, being valid only at finite volume $V$. For $V\rightarrow\infty$, by contrast, one can check that the two vacua become orthogonal, i.e.~\cite{Blasone:1995zc}
\begin{equation}
\qquad\lim_{V \rightarrow \infty}\; _{1,2}\langle0|0(t)\rangle_{e,\mu}\,=\,0, \quad\forall t\,.
\label{eqn:volumeinfinito}
\end{equation}
Of course, such an orthogonality disappears for $\theta=0$ and/or for $m_1=m_2$, consistently with the disappearance of mixing in the Pontecorvo theory.

Equation~(\ref{eqn:volumeinfinito}) is the \emph{heart} of the QFT of flavor mixing; indeed, it expresses  the unitary inequivalence of the flavor and mass representations in the infinite volume limit, showing the absolutely non-trivial nature of the transformations Eqs.~(\ref{eqn:Pontecorvo}) in this regime. Furthermore, it emphasizes the limit of validity of the approximation usually adopted when these two representations are identified.

\medskip
Next, in order to realize how the above technicalities affect the structure of the flavor vacuum, let us  explicitly evaluate the  dynamical map, Eq.~\eqref{dynamap}; to this aim, it is worth redefining the operatorial parts of the fields $\nu_\chi(x)$ $(\chi=e,\mu)$ as\, $\lu_{{\bk},1}^{r\,\alpha}\,\alpha_{{\bk},e}^{\,r}\equiv\lu_{{\bk},e}^{r\,\alpha}\;\widetilde\alpha_{{\bk},e}^{\,r}\,$. The ensuing relation takes the form
\begin{equation}
\alpha _{{\bk},e}^{r}(t) \,\equiv \,G^{-1}_{ \theta}(t)\;\alpha
_{{\bk},1}^{r}\;G_{\theta}(t)\,, 
\label{eqn:dynamicalmap}
\end{equation}
and similarly for the other operators. Using Eq.~\eqref{eqn:generatoremix} and expanding the free fields $\nu_{i}(x)$ $(i=1,2)$ as in Eq.~\eqref{eqn:expressionexpl}, we finally get
\begin{equation}
\alpha^{r}_{{\bk},e}(t)\,=\,\cos\theta\;\alpha^{r}_{{\bk},1}\,+\,\sin\theta\;\sum_{s}\left[\lu^{r\dagger}_{{\bk},1}(t)\,
\lu^{s}_{{\bk},2}(t)\, \alpha^{s}_{{\bk},2}\,+\, \lu^{r\dag}_{{\bk},1}(t)\, \lv^{s}_{-{\bk},2}(t)\, \beta^{s\dagger}_{-{\bk},2}\right],
\label{dynamicalmapexplicit}
\end{equation}
Equation~\eqref{dynamicalmapexplicit} can be further simplified by decoupling the modes with different spins in the sum over $s$: for this purpose, without loss of generality, we can choose the reference frame such that $\bk=\left(0,0,k\right)$, thus obtaining
\begin{equation}
\label{eqn:annihilator} \alpha^{r}_{{k},e}(t)\,=\,\cos\theta\;\alpha^{r}_{{
k},1}\,+\,\sin\theta\,\left( \rho^{{k}\, *}_{12}(t)\; \alpha^{r}_{{
k},2}\,+\,\varepsilon^{r}\; \lambda^{{k}}_{12}(t)\; \beta^{r\dagger}_{-{
k},2}\right),
\ee
and similarly for the other operators, where $\varepsilon^{r}=(-1)^{r}$ and\footnote{In order to avoid confusion among spinors and Bogoliubov coefficients, we have slightly changed the original notation~\cite{Blasone:1995zc}, relabelling these latter as $U_{{k}}\rightarrow\rho^{{k}}_{12}$ and $V_{{k}}\rightarrow\lambda^{{k}}_{12}$.}
\begin{eqnarray}
\label{eqn:Vk2}  \rho^{{k}}_{12}(t)&\equiv &u^{r\dagger}_{{
k},2}(t)\,u^{r}_{{k},1}(t)\,=\, v^{r\dagger}_{-{
k},1}(t)\,v^{r}_{-{ k},2}(t)\,,
\\[4mm]
\label{eqn:uk2}
\lambda^{{k}}_{12}(t)&\equiv & \varepsilon^{r}\, u^{r\dagger}_{{
k},1}(t)\,v^{r}_{-{ k},2}(t)\,=\, -\varepsilon^{r}\, u^{r\dagger}_{{
k},2}(t)\,v^{r}_{-{ k},1}(t)\,,
\end{eqnarray}
By explicit calculations, one has
\begin{equation}
\label{Vk1}  \rho^{{k}}_{12}(t)\,=\,|\rho^{{k}}_{12}|\,e^{i(\omega_{{k},2}-\omega_{{k},1})t}\,,\qquad \lambda^{{k}}_{12}(t)\,=\,|\lambda^{{k}}_{12}|\,e^{i(\omega_{{k},2}+\omega_{{k},1})t},
\end{equation}
where
\begin{eqnarray} 
\label{eqn:Uk}
\hspace{-6mm}|\rho^{{k}}_{12}|&=&\left(\frac{\omega_{{k},1}+m_{1}}{2\omega_{{k},1}}\right)^{\frac{1}{2}}
\left(\frac{\omega_{{k},2}+m_{2}}{2\omega_{{k},2}}\right)^{\frac{1}{2}}
\left(1\,+\,\frac{{k}^{2}}{(\omega_{{k},1}+m_{1})(\omega_{{k},2}+m_{2})}\right),
\\[3mm]
\label{eqn:Vk}
\hspace{-6mm}|\lambda^{{k}}_{12}|&=&\left(\frac{\omega_{{k},1}+m_{1}}{2\omega_{{k},1}}\right)^{\frac{1}{2}}
\left(\frac{\omega_{{k},2}+m_{2}}{2\omega_{{k},2}}\right)^{\frac{1}{2}}
\left(\frac{k}{(\omega_{{k},2}+m_{2})}\,-\,\frac{k}{(\omega_{{k},1}+m_{1})}\right).
\end{eqnarray}
It is a trivial matter  to show that 
\begin{equation}
|\rho^{{k}}_{12}|^{2}\,+\,|\lambda^{{k}}_{12}|^{2}\,=\,1.
\label{eqn:circo}
\end{equation}
The above equation guarantees that flavor operators still obey equal-time canonical anti-commutators.

Now, Eqs.~\eqref{eqn:mixrelaferm}, \eqref{eqn:dynamicalmap}  allow  to adopt the following free field-like expansions for flavor fields:
\begin{equation}
\label{flfieldexpans}
\lnu _{\ell}({ x},t) \,=\,\frac{1}{\sqrt{V}}{\sum_{{ k},r} }
e^{i{ k\,x}}\left[\alpha _{{
k},\ell}^{r}(t)\, \lu_{{ k},j}^{r}(t)\,+\,\beta _{-{ k},\ell}^{r\dagger
}(t)\,\lv_{-{ k},j}^{r}(t)\right]\,,
 \end{equation}
 where $(\ell,j)=(e,1), (\mu,2)$ and the flavor ladder operators are 
 given in Eq.~\eqref{eqn:annihilator} and similar.
 
As previously remarked, in the framework of QFT, flavor mixing arises from a non-trivial interplay between the standard Pontecorvo $\theta$-rotation and an internal Bogoliubov transformation (this is even more evident from Eq.~\eqref{eqn:annihilator}, where the Bogoliubov transformation of coefficients $\rho^{{k}}_{12}$ and $\lambda^{{k}}_{12}$ has been emphasized by enclosing it within round brackets).  As a consequence, the
flavor vacuum acquires a condensate structure of particle/antiparticle pairs with both the same and different masses, as it can be seen by explicitly exhibiting its full expression (for simplicity, we consider again the reference frame such that $\bk=\left(0,0,k\right)$)\footnote{As a remark, note that, due to the existence of various types of particle/antiparticle pairs, flavor vacuum turns out to be different from the ground state of the BCS theory of superconductivity~\cite{Bardeen:1957kj}, where only one type of pair is involved.}:
\begin{eqnarray}
\label{vacstr}
\nonumber |0\rangle_{e,\mu}&\hspace{-1mm}=\hspace{-1mm}& \prod_{{k}}\prod_{r}
\Big[\left(1-\sin^{2}\theta\;|\lambda^{{k}}_{12}|^{2}\right)
\,-\,\varepsilon^{r}\sin\theta\,\cos\theta\, |\lambda^{{k}}_{12}|
\left(\alpha^{r\dagger}_{{\bk},1}\,\beta^{r\dagger}_{-{\bk},2}\,+\,
\alpha^{r\dagger}_{{k},2}\,\beta^{r\dagger}_{-{ k},1}\right)  \\[2mm] \non
&&\hspace{-18mm}+\,\varepsilon^{r}\sin^{2}\theta\,|\lambda^{{k}}_{12}|\,|\rho^{{k}}_{12}|\left(\alpha^{r\dagger}_{{k},1}\,\beta^{r\dagger}_{-{ k},1}\,-\,
\alpha^{r\dagger}_{{k},2}\,\beta^{r\dagger}_{-{ k},2}\right)
\,+\,\sin^{2}\theta\,|\lambda^{{k}}_{12}|^{2}\,\alpha^{r\dagger}_{{
k},1}\,\beta^{r\dagger}_{-{ k},2} \alpha^{r\dagger}_{{
k},2}\,\beta^{r\dagger}_{-{ k},1}
\Big]|0\rangle_{1,2}.\\ \label{eqn:0emu} 
\end{eqnarray}
Using Eqs.~(\ref{eqn:annihilator}), (\ref{eqn:0emu}),  we can finally calculate the condensation density of the flavor vacuum,
\begin{equation} 
\label{eqn:denscondferm}
{}_{e,\mu}\langle 0| \alpha_{{ k},i}^{r \dagger}\, \alpha^r_{{ k},i}
|0\rangle_{e,\mu}\,=\, \sin^{2}\theta\, |\lambda^{{k}}_{12}|^{2}\,, \qquad i=1,2\,,
\end{equation}
with the same result for antiparticles. Clearly, it vanishes for $\theta=0$ and/or for $m_1=m_2$, as it should be in the absence of mixing. Furthermore, using Eq.~\eqref{eqn:Vk}, one can easily show that it  also goes to zero in the relativistic limit of large momenta, $k\gg\sqrt{m_1\,m_2}$, thus recovering the standard QM results.

\section{Flavor charges and states for mixed fields: neutrino oscillations in QFT}
\label{sec:Cariche e stati di sapore dei campi mixati: oscillazioni di neutrino in QFT}
After discussing the vacuum structure, let us investigate the definition and properties of the single-particle flavor states\footnote{In what follows, we shall basically refer to Ref.~\cite{Blasone:2001qa}.}. For the sake of comparison, we first review the case where no mixing occurs, devoting the subsequent analysis to the extension to mixed fields. 

It is well known that the Lagrangian density of two free (non-interacting) Dirac fields $\nu_i$ $(i=1,2)$ with mass $m_i$ reads 
\begin{equation}
\label{eqn:lagrmas} 
{\mathcal L}_{\nu}(x)\,=\, {\bar \lnu_m}(x) \left( i\gamma^{\mu}\partial_{\mu}\, -\, M^{d}_{\nu} \right) \lnu_m(x)\,, 
\end{equation}
where $\lnu_m^T=(\lnu_1,\lnu_2)$ and $M^{d}_{\nu}$ is the diagonal mass operator,  $M^{d}_{\nu}= \mathrm{diag}\, (m_{1},m_{2})$. With a straightforward calculation, it is possible to show that ${\cal L}_{\nu}$  has the global $U(1)$ invariance. This implies the conservation of the Noether's charge
\be
Q_{\nu}=\int d^{3}\, {\bx}\, I^0(x)\,,\qquad I^{\mu}(x)=\bar{\lnu}_{m}(x)\gamma^\mu\lnu_{m}(x))\,,
\ee 
which indeed represents the total lepton number of the system.

Let us now consider the global $SU(2)$ transformation,
\begin{equation}
\label{eqn:SU(2)tran}
\lnu_{m}^{ \prime }(x)\,=\,e^{i\alpha_{j}\cdot \tau_{j}}\lnu
_{m}(x)\,,\qquad  j=1,2,3\,,
 \end{equation}
with $\alpha_{j}$ $(j=1,2,3)$ real constants, ${\tau_{j}} =\sigma_{j}/2$ and $\sigma_j$ being the Pauli matrices. Since $m_1\neq m_2$, the Lagrangian density Eq.~(\ref{eqn:lagrmas}) is no longer invariant. In particular, by use of the
equations of motion, one can derive the following expression for the $SU(2)$ Noether's currents:
 \begin{equation}
 \label{Jmj}
J_{m,j}^{\mu
}(x)\,=\,\bar{\lnu}_{m}(x)\,\gamma ^{\mu }\,\tau _{j}\,\lnu_{m}(x)\,,\qquad j=1,2,3\,.
 \end{equation}
The corresponding Noether's charges, $Q_{m,j}(t)=\int d^{3}{\bx}\, J_{m,j}^{0}(x)$, which in general are not conserved, satisfy the $SU(2)$ algebra (at any time $t$). 

We now point out that the Casimir operator, $C_m\equiv{[\sum_{j=1}^3Q^2_{m,j}(t)]}^{1/2}$, is proportional to the total  (conserved) charge $Q_{\nu}$, since we have 
\begin{equation}
C_m\,=\,\frac{1}{2}Q_{\nu}\,. 
\end{equation}
In the same way, from Eq.~\eqref{Jmj}, we can see that also the charge $Q_{m,3}$ 
is conserved. Thus, we can define the two combinations:
\begin{equation}
\label{eqn:su2noether} 
Q_{\nu_{1}}\,\equiv\,\frac{1}{2}Q_{\nu}
\,+ \,Q_{m,3}\,,\qquad Q_{\nu_2} \,\equiv\,
\frac{1}{2}Q_{\nu} \,- \,Q_{m,3}\,,  
\end{equation}
where 
\be 
Q_{\nu_i}=\int d^{3}{\bx} \,
\lnu_{i}^{\dagger}(x)\hspace{0.3mm}\lnu_{i}(x)\,,\qquad i=1,2\,.
\ee 
These are nothing but the Noether charges associated with 
the non-interacting fields $\lnu_{i}$: in the absence of mixing, they represent the flavor charges.
As expected, they are separately conserved for each generation.


The question now arises as to how the above formalism gets modified when taking into account the  mixing of the two fields. To clear the head, let us then rewrite the Lagrangian density Eq.~(\ref{eqn:lagrmas}) in the flavor basis, 
\begin{equation}
\label{eqn:lagrflav} 
{\cal L}_{\nu}(x)\,=\,  {\bar \lnu_f}(x) \left( \gamma^{\mu}\partial_{\mu}\,-\, M_{\nu} \right) \lnu_f(x)\, , 
\end{equation}
where $\lnu_{f}^{T}=(\lnu _{e},$ $\lnu _{\mu })$ and $M_\nu=\left(
\begin{matrix}
m_e&m_{e\mu}\\
m_{e\mu}&m_\mu
\end{matrix}
\right)$. As in the previous case, one can explicitly calculate the Noether's currents associated with $SU(2)$ transformations acting on the field doublet, obtaining
\begin{equation}
\qquad\qquad\qquad J_{f,j}^{\mu }(x)\,=\,\bar{\lnu}_{f}(x)\,\gamma ^{\mu }\,\tau_{j}\,
\lnu _{f}(x)\,, \qquad j=1,2,3\,.
\end{equation}

The related charges, $Q_{f,j}(t)=\int d^{3}{\bx} \,J_{f,j}^{0}(x)$, still close the $SU(2)$ algebra; in this case, however, due to the \emph{off-diagonal} terms in $M_\nu$, $Q_{f,3}(t)$ turns out to be time-dependent. In other terms, an exchange of charge between the mixed fields $\nu_e$ and $\nu_{\mu}$  is involved, resulting in the phenomenon of flavor oscillations.

In line with Eq.~(\ref{eqn:su2noether}), flavor charges for mixed fields can be now expressed in the form~\cite{Blasone:2001qa}
\begin{equation}
\label{eqn:flavch}
Q_{\nu_e}(t)\,\equiv\,\frac{1}{2}Q_{\nu}\,+\,
Q_{f,3}(t)\,,\qquad Q_{\nu_\mu}(t)\,\equiv\,\frac{1}{2}Q_{\nu} \,-\,
Q_{f,3}(t)\,,
\end{equation}
where
\begin{equation}
\label{B55}
Q_{\nu_\sigma}(t) = \int d^{3}{\bx}\,
 \lnu_{\sigma}^{\dagger}(x)\,\lnu_{\sigma}(x)\,,\qquad \sigma=e,\mu\,.
\end{equation} 
Of course, one has $Q_{\nu_e}(t)\,+\,Q_{\nu_\mu}(t)=Q, \forall t$, consistently with the conservation of the total charge even in the presence of mixing. 

In order to define the single-particle state for mixed fields, let us now introduce the normal ordered charges with respect to the flavor vacuum $|0(t)\rangle_{e,\mu}$:
\begin{equation} 
\hspace{-1mm}::Q_{\nu_{\sigma}}(t)::\,\, \equiv \int d^{3}{\bx} \, ::
\nu_{\sigma}^{\dagger}(x)\,\nu_{\sigma}(x)::\,\, =  \sum_{r}
\int d^3 k \left( \alpha^{r\dagger}_{{\bk},\nu_{\sigma}}(t) \alpha^{r}_{{\bf
k},\nu_{\sigma}}(t) - \beta^{r\dagger}_{-{\bk},\nu_{\sigma}}(t)\beta^{r}_{-{\bk},\nu_{\sigma}}(t)\right).
\end{equation}
Flavor states are thus defined as eigenstates of these charges at a given time $t$ (\emph{e.g.}, $t=0$):
\begin{equation}
\label{eqn:flavstate}
|\nu^{r}_{\;{\bk},\sigma}\rangle \,\equiv\,
\alpha^{r\dagger}_{{\bk},{{\sigma}}}(0) |0(0)\rangle_{{e,\mu}}\,\equiv\,\alpha^{r\dagger}_{{\bk},{{\sigma}}}(0) |0\rangle_{{e,\mu}},
\qquad \sigma \,=\, e,\mu.
\end{equation}
with $\alpha^{r\dagger}_{{\bk},{{\sigma}}}$ given in Eq.~(\ref{eqn:annihilator}) (similarly for the antiparticle states). With such a definition, we have 
\begin{eqnarray}
\label{eqn:su2flavstates} 
&&\hspace{-15mm}::Q_{\nu_e}(0)::|\lnu^{r}_{\;{\bf
k},e}\rangle\,=\,|\lnu^{r}_{\;{\bk},e}\rangle,\hspace{35mm} ::Q_{\nu_\mu}(0)::|\lnu^{r}_{\;{\bf
k},\mu}\rangle\,=\,|\lnu^{r}_{\;{\bk},\mu}\rangle, \\[3mm]
&&\hspace{-15mm}::Q_{\nu_e}(0)::\,|\lnu^{r}_{\;{\bf
k},\mu}\rangle\,=\; ::Q_{\nu_\mu}(0)::\,|\lnu^{r}_{\;{\bk},e}\rangle=0,\qquad\hspace{-2.1mm}::Q_{\nu_\sigma}(0)::\;|0\rangle_{{e,\mu}}\,=\,0.
\end{eqnarray}
In closing, we present the exact QFT formula for neutrino oscillations~\cite{Blasone:1998hf}, which can be derived by computing the expectation value of the flavor charges Eq.~(\ref{B55}) on the states Eq.~(\ref{eqn:flavstate}). A straightforward calculation leads to
\begin{eqnarray}
\label{eqn:oscillfor1}
\hspace{-2mm}{\cal Q}^{\bk}_{{\lnu_e}\rightarrow\lnu_e}(t)
&\hspace{-2.5mm}=\hspace{-2.5mm}& \langle \nu_{{\bf
k},e}^{r}|::Q_{\nu_e}(t)::|\nu_{{\bk},e}^{r}  \rangle
 \\ [2mm]
\nonumber &\hspace{-2.5mm}=\hspace{-2.5mm}& 1\,-\,\sin ^{2}(2\theta )\left[ \left|\bogu\right|
^{2}\sin ^{2}\left( \frac{\omega _{\bk,2}-\omega _{\bk,1}}{2}\,t\right)
\,+\,\left| \bogv\right| ^{2}\sin ^{2}\left( \frac{\omega _{\bk,2}+\omega
_{k,1}}{2}\,t\right) \right],\nonumber 
\end{eqnarray}
and
\begin{eqnarray}
\label{eqn:oscillfor2} 
{\cal Q}^{\bf
k}_{{\lnu_e}\rightarrow\nu_\mu}(t)&\hspace{-2.5mm}=\hspace{-2.5mm}&\langle \lnu_{{\bf
k},e}^{r}|::Q_{\nu_\mu }(t)::|\nu_{{\bk},e}^{r} \rangle
\\ [2mm]
\nonumber
&\hspace{-2.5mm}=\hspace{-2.5mm}& \sin ^{2}(2\theta )\left[ \left|\bogu\right|
^{2}\sin ^{2}\left( \frac{\omega _{\bk,2}-\omega _{\bk,1}}{2}\,t\right)
\,+\,\left| \bogv\right| ^{2}\sin ^{2}\left( \frac{\omega _{\bk,2}+\omega
_{k,1}}{2}\,t\right) \right].\nonumber
\end{eqnarray}
Charge conservation  is guaranteed (at any time) by the condition
\begin{eqnarray} 
{\cal Q}^{\bk}_{{\nu_e}\rightarrow\nu_e}(t) + {\cal Q}^{\bf
k}_{{\nu_e}\rightarrow\nu_\mu}(t) = 1\,,\qquad \forall t\,.
\end{eqnarray}
By comparison with the QM Pontecorvo formulae, 
\begin{eqnarray}
\label{eqn:Pontecorvoee} 
\hspace{-3mm}P_{\nu _{e}\rightarrow \nu e}(t)
\,=\,1\,-\,\sin ^{2}2\theta \text{ }\sin ^{2}\left( \frac{\Delta \omega }{2}%
\,t\right)\hspace{-0.5mm}, \quad  P_{\nu _{e}\rightarrow \nu
_{\mu }}(t) \,=\,\sin ^{2}2\theta \text{ }\sin ^{2}\left( \frac{\Delta \omega }{2}%
\,t\right)\hspace{-1mm}, 
\end{eqnarray} 
two main differences catcth the eye. First, the QFT oscillation amplitudes explicitly 
depend on the momentum via the Bogoliubov coefficients $\bogu$ and $\bogv$; 
second, in Eqs.~\eqref{eqn:oscillfor1}, \eqref{eqn:oscillfor2}  there is an additional 
oscillating term depending on the sum of the frequencies. We also observe that,
for large momenta $|\bk|\gg\sqrt{m_1\,m_2}$, the traditional formulae in Eq.~\eqref{eqn:Pontecorvoee} are recovered, since $\left| \bogu\right| ^{2}\longrightarrow 1$ and $\left| \bogv\right| ^{2}\longrightarrow 0$.

\section{Mixing transformations and flavor states for bosons}
\label{sec:Trasformazioni di mixing di campi bosonici}
The non-trivial nature of mixing transformations in QFT also appears  in the case of bosons\footnote{To make our analysis as clear as possible, we shall apply our considerations to scalar fields.}~\cite{bosonmix}. In this context, however, what oscillate are some other quantum numbers (\emph{e.g.}, the strangeness or the isospin). In spite of this, in the following we shall still refer to such intrinsic properties as ``flavor'' and, thus, to the interacting fields as ``flavor fields''. Additionally, since the skeleton of the formalism is basically the same as for fermions, we shall focus on the aspects that are not in common with the previous analysis, redirecting the reader to Ref.~\cite{bosonmix} for a complete treatment. 
 
Let us start by considering two free charged scalar fields $\phi_i(x)$ ($i=1,2$) with mass $m_i$. As well known, they can be Fourier expanded as follows
\begin{equation}
\label{eqn:Fouscal}
\phi_i(x)\,=\,\int \frac{d^3k}{{\left(2\pi\right)}^{3/2}}\, \frac{e^{i\bk\cdot\bx}}{\sqrt{2\omega_{\bk,i}}}\left(a_{\bk,i}e^{-i\omega_{\bk,i} t}\,+\,\bar a^\dagger_{-\bk,i} e^{i\omega_{\bk,i}t } \right),\qquad i=1,2\,,
\end{equation}
being $a_{\bk,i}$ $(a^\dagger_{\bk,i})$ the annihilators (creators) for the field $\phi_i$. We assume these operators to be canonical.
Following on from what discussed for fermions, mixing transformations relating flavor fields $\phi_\chi$ $(\chi=A,B)$ and free fields  $\phi_i$ $(i=1,2)$ take the form
\begin{equation}
\label{eqn:Pontecorvobis}
\begin{array}{l}
\phi_{A}(x) \,=\, \phi_{1}(x)\, \cos\theta  \,+\, \phi_{2}(x)\, \sin\theta\, ,
\\[3mm]
\phi_{B}(x)\, =\,-  \phi_{1}(x)\, \sin\theta \,  +\,  \phi_{2}(x)\, \cos\theta\, , 
\end{array}
\end{equation}
which can be expressed in terms of the mixing generator
\begin{equation}
G_\theta(t)\,=\,\exp\left[-i\theta\int d^3x\left(\pi_1(x)\phi_2(x)\,-\,\phi^\dagger_1(x)\pi^\dagger_2(x)\,-\,\pi_2(x)\phi_1(x)\,+\,\phi_2^\dagger(x)\pi_1^\dagger(x)\right) \right],
\label{eqn:genmix2}
\end{equation}
 as in Eqs.~(\ref{eqn:mixrelaferm}). Again,  this is a unitary operator (at finite volume $V$) exhibiting the structure of a rotation combined with a Bogoliubov transformation of coefficients
\be
\label{bogcoeffplaw}
 \bogub(t)=|\bogub|\, e^{i\left(\omega_{\bk,2}-\omega_{\bk,1}\right) t}\,,\qquad  \bogvb(t)=|\bogvb|\, \ebogo\,,
 \ee 
 with
\begin{equation}
\label{eqn:coeffbogorholambda}
|\bogub|\equiv\frac{1}{2}\left( \sqrt{\frac{\omega_{\bk,1}}{\omega_{\bk,2}}}\,+\,\sqrt{\frac{\omega_{\bk,2}}{\omega_{\bk,1}}}\right),\qquad |\bogvb|\equiv\frac{1}{2}\left( \sqrt{\frac{\omega_{\bk,1}}{\omega_{\bk,2}}}\,-\,\sqrt{\frac{\omega_{\bk,2}}{\omega_{\bk,1}}}\right). 
\end{equation}
Unlike fermions, however, these coefficients satisfy the ``hyperbolic'' (rather than ``circular'') relation
\begin{equation}
{|\bogub|}^2\,-\,{|\bogvb|}^2\,=\,1 \,,
\label{eqn:iper}
\end{equation}
consistently with the different statistical nature of bosons (compare with Eq.~\eqref{eqn:circo}). This  guarantees that ladder operators for flavor fields are still canonical (at equal times).

Note that the Bogoliubov transformation hiding in $G_\theta(t)$ still induces a non-trivial condensate structure  into the flavor vacuum $|0(\theta,t)\rangle_{A,B}=G^{-1}_\theta(t)|0\rangle_{1,2}$, which is thus a $SU(2)$ coherent state. In analogy with fermions, flavor and mass vacua become orthogonal to each other in the infinite volume limit, giving rise to inequivalent Fock space representations ${\mathcal H}_{1,2}$ and ${\mathcal H}_{e,\mu}$. Of course, such an orthogonality disappears  for $\theta=0$ and/or $m_1=m_2$, according to what discussed in the previous Section.

Now, using the Fourier expansion Eq.~\eqref{eqn:Fouscal} and the mixing generator Eq.~\eqref{eqn:genmix2}, flavor annihilators for the vacuum  $|0(\theta,t)\rangle_{A,B}$ take the form
\begin{equation}
\label{eqn:esbos0}
a _{{\bk},A}(t)\,=\,\cos\theta\, a _{{\bk},1}\,+\,\sin\theta\, \left(\bogubst(t)\, a _{{\bk},2}\, + \, \bogvb(t)\, \bar a^\dagger _{-{\bk},2}\right),
\end{equation}
and similarly for the other annihilators. In terms of these operators, flavor fields can be expanded as
\be
\label{eqn:dynamicalmap2}
\phi_\ell(x)\,=\,\int \frac{d^3k}{{\left(2\pi\right)}^{3/2}}\frac{e^{i\bk\cdot{\bx}}}{\sqrt{2\omega_{\bk,j}}}\left(a_{\bk,\ell}(t)\,e^{-i\omega_{\bk,j}t}\,+\,\bar a^\dagger_{-\bk,\ell}(t)\,e^{i\omega_{\bk,j}t} \right),
\ee
where $(\ell,j)=(A,1),(B,2)$.

It is now interesting to describe an alternative, but equivalent, procedure to derive the formula Eq.~(\ref{eqn:esbos0}).\footnote{Although it has been illustrated only for bosons, such an approach also applies  to the case of fermions.} To this aim, by working backwards, we can start from the expansions of flavor fields Eq.~(\ref{eqn:dynamicalmap2}); exploiting the orthonormality and completeness properties of plane waves, we can immediately write
\begin{equation}
a_{\bk,A}(t)\,\equiv\,(\phi_A,\uuu_{\bk,1})\,,
\label{eqn:scalproda}
\end{equation}
where, to simplify the notation, we have indicated with $U_{\bk,i}$ $(i=1,2)$ the plane wave mode of mass $m_i$. This can be easily checked by using  the mixing transformation Eqs.~\eqref{eqn:Pontecorvobis}:
\begin{eqnarray}
\nonumber
a_{\bk,A}(t)&=&\cos\theta\int{d^3k'}\left[a_{\bk',1}\,(\uuu_{\bk',1},\uuu_{\bk,1})\,+\,\bar a^{\dagger}_{-\bk',1}\,(\uuu^*_{-\bk',1},\uuu_{\bk,1})\right]\\[1.5mm]\nonumber
&&+\,\sin\theta\int{d^3k'}\left[a_{\bk',2}\,(\uuu_{\bk',2},\uuu_{\bk,1})\,+\,\bar a^{\dagger}_{-\bk',2}\,(\uuu^*_{-\bk',2},\uuu_{\bk,1})\right]\\[2.5mm]
&=&\cos\theta\, a_{\bk,1}\,+\,\sin\theta\left[\tilde{\rho}^{\bk\, *}_{12}(t)\,a_{k,2}\,+\,\tilde{\lambda}_{12}(t)\,\bar a^{\dagger}_{-\bk,2}\right],
\end{eqnarray}
that is indeed the same expression obtained in Eq.~(\ref{eqn:esbos0}).

By comparison with Eq.~(\ref{eqn:denscondferm}), we also illustrate the expression of the condensation density of the flavor vacuum,
\begin{equation}
\label{eqn:denscondbosoni2} 
{}_{A,B}\langle 0(t)| a_{{\bk},i}^{\dagger}\, a_{{\bf
k},i} |0(t)\rangle_{A,B}\,=\, \sin^{2}\theta\; {|\bogvb|}^2, \qquad i=1,2\,,
\end{equation}
with the same result for antiparticles. As for the case of  fermions, it is a trivial matter to show that it vanishes  for $\theta=0$ and/or for $m_1=m_2$, as it should be in the absence of mixing. Additionally,  it  goes to zero in the relativistic limit ${|\bk|}^2\gg\frac{{m_1}^2+{m_2}^2}{2}$, correctly reproducing the standard Pontecorvo results.

\bigskip

The structure of currents and charges for mixed scalar fields can be  defined  consistently  with what discussed in Section~\ref{sec:Cariche e stati di sapore dei campi mixati: oscillazioni di neutrino in QFT}. Here we only report the QFT generalization of the standard Pontecorvo formulae, 
\begin{eqnarray} 
\non
{\cal Q}^{\bk}_{A\rightarrow A}(t)&\hspace{-1.5mm}=\hspace{-1.5mm}& 1\hspace{0.5mm}-\hspace{0.5mm}  \sin^2( 2
\theta)  \left[ |\bogub|^2 \, \sin^2 \left( \frac{\omega_{\bk,2} -
\omega_{\bk,1}}{2}\, t \right) - |\bogvb|^2  \sin^2 \left( \frac{\omega_{\bk,2} + \omega_{\bk,1}}{2}\,t
\right)\right]
\label{eqn:bososc1}\\
\\[4mm]
\non
\hspace{-10mm}{\cal Q}^{\bk}_{A\rightarrow B}(t) &\hspace{-1.5mm}=\hspace{-1.5mm}&
 \sin^2( 2 \theta)
\left[\, |\bogub|^2 \, \sin^2 \left( \frac{\omega_{\bk,2} -
\omega_{\bk,1}}{2}\, t \right) - |\bogvb|^2  \sin^2 \left( \frac{\omega_{\bk,2} + \omega_{\bk,1}}{2}\, t
\right)\right], \\
\label{eqn:bososc2}
\end{eqnarray}
to be compared with the corresponding expressions for fermions, Eqs.~\eqref{eqn:oscillfor1}, \eqref{eqn:oscillfor2}. Note the negative sign in front of the $|\bogvb|^2$ terms, in contrast with Eqs.~\eqref{eqn:oscillfor1}, \eqref{eqn:oscillfor2}: the
boson flavor charge thus can assume also negative values (see Ref.~\cite{bosonmix} for the physical interpretation). Clearly, the total charge
is conserved at any time $t$, being  ${\cal Q}^{\bk}_{A\rightarrow A}(t)+{\cal Q}^{\bk}_{A\rightarrow B}(t)=1, \forall t$.

As seen for fermions, Eqs.~\eqref{eqn:bososc1}-\eqref{eqn:bososc2} do not match with the usual QM oscillation formulae, to which, however, reduce in the relativistic limit of large momenta. Apart from the extra oscillating term depending on the sum of the frequencies and the $\bk$-dependent amplitudes, 
 Eqs.~\eqref{eqn:bososc1}, \eqref{eqn:bososc2} contain the information about the statistics of mixed fields: for bosons and fermions the Bogoliubov coefficients appearing in the amplitudes are indeed significantly different from each other, the former satisfying the ``hyperbolic'' condition Eq.~\eqref{eqn:iper}, the latter the ``circular'' one Eq.~\eqref{eqn:circo}.

\chapter{Entanglement in neutrino mixing and oscillations}
\label{Entmeas}
\begin{flushright}
\emph{``Guilt is an indulgence, \\
it entangles you in the past.
''}\\[1mm]
-  Gregg Hurwitz -\\[6mm]
\end{flushright}
In this Appendix we illustrate how the issue of flavor mixing 
and oscillations is intimately related to the \emph{entanglement} 
phenomenon. In particular, with explicit reference to neutrinos, 
it is shown  that oscillations can be equivalently described in terms
of dynamical entanglement of flavor modes. In order to set the stage, 
we first review the quantum mechanical treatment of the problem, relying on
the use of Pontecorvo states (for a more thorough discussion of this, 
see Ref.~\cite{Blasone:2007vw}); the obtained results are then extended
to the quantum field theory framework, where a richer structure
of quantum correlations appears~\cite{Blasone:2013zaa}, 
due to the non-trivial condensation of particle-antiparticle pairs in flavor vacuum (see Eq.~\eqref{vacstr}).

To begin with, let us recast Pontecorvo transformations Eq.~(\ref{eqn:U}) into the form
\be
\label{Pontransf}
|\underline{\nu}^{(f)}\rangle\,=\,U(\theta)|\underline{\nu}^{(m)}\rangle\,,
\ee
where $U(\theta)$ is the Pontecorvo matrix in Eq.~(\ref{PMMmatrix})
and we have introduced the shorthand notation
$|\underline{\nu}^{(f)}\rangle={\left(|\nu_e\rangle,|\nu_\mu\rangle\right)}^{T}$, 
$|\underline{\nu}^{(m)}\rangle={\left(|\nu_1\rangle,|\nu_2\rangle\right)}^{T}$ for
states with definite flavor and mass, respectively.
Flavor states evolve in time according to
\be
\label{flaccto}
|\underline{\nu}^{(f)}(t)\rangle\,=\,U(t)|\underline{\nu}^{(f)}\rangle\,
\equiv\,U(\theta)\hspace{0.2mm}U_0(t)\hspace{0.2mm}U^{-1}(\theta)|\underline{\nu}^{(f)}\rangle\,,
\ee
where $|\underline{\nu}^{(f)}\rangle$ are the flavor states at $t=0$, 
$U_0(t)\,=\,\mathrm{diag}(e^{-i\hspace{0.1mm}\omega_1\hspace{0.1mm}t},e^{-i\hspace{0.1mm}\omega_2\hspace{0.1mm}t})$, and $\omega_i$ $(i=1,2)$ is the 
energy eigenvalue associated with $|\nu_i\rangle$.

If we now consider the neutrino occupation number
for a given flavor as reference quantum number, the
following correspondence with two-qubit states naturally arises:
\be
|\nu_e\rangle\,\equiv\,|1\rangle_{\nu_e}|0\rangle_{\nu_\mu},\qquad 
|\nu_\mu\rangle\,\equiv\,|0\rangle_{\nu_e}|1\rangle_{\nu_\mu}\,,
\ee
with $|j\rangle_{\nu_\alpha}$ $(\alpha=e,\mu)$ standing for a $j$-occupation number state
of a neutrino in mode $\alpha$. Therefore, as a result of mixing, 
entanglement is established among flavor
modes, in a single-particle setting. Eq.~(\ref{flaccto}) can then be
recast as
\be
\label{receqas}
|\nu_\alpha(t)\rangle\,=\,U_{\alpha e}(t)|10\rangle\,+\,U_{\alpha\mu}(t)|01\rangle\,, 
\ee
where $|ij\rangle$ denotes the two-qubit vector $|i\rangle_{\nu_e}|j\rangle_{\nu_\mu}$, and
${|U_{\alpha\beta}(t)|}^2={|\langle\nu_\beta|\nu_\alpha(t)\rangle|}^2$ $(\beta=e,\mu)$ is the transition
probability between $|\nu_\alpha\rangle$ and $|\nu_\beta\rangle$ at time $t$, with
${|U_{ee}(t)|}^2+{|U_{e\mu}(t)|}^2=1$. Thus, the time-evolved states
$|\underline{\nu}^{(f)}(t)\rangle$ appear as \emph{entangled Bell-like
superpositions} of the two flavor eigenstates with time-dependent
coefficients.

As well-known, entanglement  can be quantified via several different
measures. According to  
Ref.~\cite{Blasone:2007vw}, here we use the \emph{linear entropy} 
(for a review of the most common entanglement quantifiers and their significance for various
topics in quantum information, see Section~\ref{enme}). To this aim,  
let us consider the density matrix 
$\rho^{(\alpha)}=|\nu_\alpha(t)\rangle\langle\nu_\alpha(t)|$ describing the
two-qubit Bell state $|\nu_\alpha(t)\rangle$ in Eq.~\eqref{receqas}. 
Exploiting the definition given in Eq.~\eqref{entr}, 
the linear entropy $S_{L\alpha}^{(\nu_e;\nu_\mu)}\equiv
S_{L}^{(\nu_e;\nu_\mu)}(\rho^\alpha)$ associated
to the reduced state after tracing over one (flavor) mode
takes the form
\begin{eqnarray}
\label{linentr}
\non
S_{L\alpha}^{(\nu_e;\nu_\mu)}\,=\,S_{L\alpha}^{(\nu_\mu;\nu_e)}&=& 4 |U_{\alpha e}(t)|^2\hspace{0.2mm}|U_{\alpha \mu}(t)|^2\\[2mm]
\non
&=&4|U_{\alpha e}(t)|^2\hspace{0.2mm}(1-{|U_{\alpha e}(t)|}^2)\\[2mm]
&=&4|U_{\alpha \mu}(t)|^2\hspace{0.2mm}(1-{|U_{\alpha \mu}(t)|}^2)\,.
\end{eqnarray}
This expression establishes that the linear entropy of the reduced state
is equal to the product of the two-flavor
transition probabilities. On the one hand, for $t=0$, that is, when flavors are not mixed, 
the entanglement is zero, and the global state is factorized.  
On the other hand, for $t>0$ flavor oscillations occur and entanglement of flavors
develops, reaching its maximal value  at largest mixing, 
i.e. for ${|U_{ee}(t)|}^2={|U_{\mu e}(t)|}^2=1/2$. We emphasize that a 
formally similar result has been obtained 
in Ref.~\cite{Alok:2014gya}, where it is also shown that
all the well-known quantum correlations, such as the Bell's inequality, are directly related to the
neutrino oscillation probabilities. Furthermore, in Ref.~\cite{Bittencourt:2014pda}, 
quantum entanglement and oscillation damping in the context of flavor
quantum transition are found to have the same origin owing to localization and decoherence
effects brought up by an external wave packet picture.

Let us now extend the above results to the QFT framework.
To this aim, consider field mixing transformations in Eq.~\eqref{eqn:Pontecorvo}
and the definition of neutrino flavor state (at $t=0$) in Eq.~\eqref{eqn:flavstate}, here 
rewritten as
\be
\label{flst}
|\nu_{\beta}\rangle\, \equiv\,\alpha^{\dagger}_{{{\beta}}}(0) |0\rangle_{{e,\mu}},\qquad (\beta=e,\mu)\,,
\ee
where we have omitted for simplicity the spin and momentum indices of the state.
At time $t$, for example, the electron neutrino state takes the form
\be
\label{flaventstate}
|\nu_{e}(t)\rangle\,=\,e^{-iHt}|\nu_e\rangle\,,
\ee
(and similarly for the muon neutrino) where $H$ is the QFT free Hamiltonian. 
Working in the flavor Hilbert space $\mathcal{H}_{e,\mu}$, we can rewrite 
Eq.~\eqref{flaventstate} as follows:
\be
\label{bass}
|\nu_{e}(t)\,=\,\Big[U_{ee}(t)\hspace{0.3mm}\alpha^\dagger_e\,+\,U_{e\mu}(t)\hspace{0.3mm}\alpha^\dagger_\mu\,+\,
U^{e{\bar{e}}}_{e\mu}(t)\hspace{0.3mm}\alpha^\dagger_e\hspace{0.3mm}\alpha^\dagger_\mu\hspace{0.3mm}\beta^\dagger_e\,+\,U^{\mu{\bar{\mu}}}_{ee}(t)\hspace{0.3mm}\alpha^\dagger_e\hspace{0.3mm}\alpha^\dagger_\mu\hspace{0.3mm}\beta^\dagger_\mu\Big]|0\rangle_{e,\mu},
\ee
where the (time-dependent) coefficients are given by
\begin{eqnarray}
\hspace{-5mm}U_{ee}(t)&=&e^{-i\omega_1t}\Big[\cos^2\theta\,+\,\sin^2\theta\left(e^{-i(\omega_2-\omega_1)t}{|\rho_{12}|}^2\,+\,e^{-i(\omega_2+\omega_1)t}{|\lambda_{12}|}^2\right)\Big]\hspace{-0.3mm},\\[2mm]
\hspace{-5mm}U_{e\mu}(t)&=&e^{-i\omega_1t}\hspace{0.2mm}\rho_{12}\hspace{0.2mm}\cos\theta\sin\theta\left(e^{-i(\omega_2-\omega_1)t}\,-\,1\right),\\[2mm]
\hspace{-5mm}U_{e\mu}^{e\bar{e}}(t)&=&e^{-i\omega_1t}\hspace{0.2mm}\lambda_{12}\hspace{0.2mm}\cos\theta\sin\theta\left(1\,-\,e^{-i(\omega_2+\omega_1)t}\right),\\[2mm]
\hspace{-5mm}U_{ee}^{\mu\bar{\mu}}(t)&=&e^{-i\omega_1t}\hspace{0.2mm}\rho_{12}\hspace{0.4mm}\lambda_{12}\hspace{0.2mm}\sin^2\theta\left(e^{-i(\omega_2+\omega_1)t}\,-\,e^{-i(\omega_2-\omega_1)t}\right)\,
\end{eqnarray}
with $\rho_{12}$ and $\lambda_{12}$ defined in Eq.~\eqref{Vk1} and
\be
{|U_{ee}(t)|}^2\,+\,{|U_{e\mu}(t)|}^2\,+\,{|U_{e\mu}^{e\bar{e}}(t)|}^2\,+\,{|U^{\mu\bar{\mu}}_{ee}(t)|}^2\,=\,1\,.
\ee
Unlike the previous treatment, in the QFT framework the time-evolved state Eq.~\eqref{bass}
is a \emph{multi-particle} entangled state, due to the richer structure
of flavor vacuum with respect to the corresponding quantum mechanical state.
As a consequence, we have a multipartite entanglement in a
four-qubit state of the form:\footnote{
In the same way as for the Pontecorvo states Eq.~\eqref{receqas}, 
we consider as reference quantum number the neutrino occupation
number.}
\be
|\nu_e(t)\rangle\,=\,U_{ee}(t)\hspace{0.2mm}|1000\rangle\,+\,U_{e\mu}(t)\hspace{0.2mm}|0100\rangle\,+\,U_{e\mu}^{e\bar{e}}(t)\hspace{0.2mm}|1110\rangle\,+\,U_{ee}^{\mu\bar{\mu}}(t)\hspace{0.2mm}|1101\rangle\,,
\ee
where $|ijkl\rangle$ denotes the four-qubit vector $|i\rangle_{\nu_e}\hspace{0.2mm}|j\rangle_{\nu_\mu}\hspace{0.2mm}|k\rangle_{\bar\nu_e}\hspace{0.2mm}|h\rangle_{\bar\nu_\mu}$, with $i,j,k,h=0,1$. Following the derivation of Eq.~\eqref{linentr},  
the linear entropies $S_L^{(a;b,c,d)}$ associated with $|\nu_e(t)\rangle$
are simply given by
\begin{eqnarray}
\label{entaga}
S_L^{(\nu_e;\nu_\mu,\bar\nu_e,\bar\nu_\mu)}&=&4\hspace{0.2mm}{|U_{e\mu}(t)|}^2\left(1\,-\,{|U_{e\mu}(t)|}^2\right),\\[2mm]
\label{entagb}
S_L^{(\nu_\mu;\nu_e,\bar\nu_e,\bar\nu_\mu)}&=&4\hspace{0.2mm}{|U_{ee}(t)|}^2\left(1\,-\,{|U_{ee}(t)|}^2\right),\\[2mm]
\label{entagc}
S_L^{(\bar\nu_e;\nu_e,\nu_\mu,\bar\nu_\mu)}&=&4\hspace{0.2mm}{|U^{e\bar e}_{e\mu}(t)|}^2\left(1\,-\,{|U^{e\bar e}_{e\mu}(t)|}^2\right),\\[2mm]
\label{entagd}
S_L^{(\bar\nu_\mu;\nu_e,\nu_\mu,\bar\nu_e)}&=&4\hspace{0.2mm}{|U^{\mu\bar \mu}_{ee}(t)|}^2\left(1\,-\,{|U^{\mu\bar \mu}_{ee}(t)|}^2\right).
\end{eqnarray}
It is worth noting that, in the QM limit, Eqs.~\eqref{entaga} and (\ref{entagb}) reduce
to the Pontecorvo analogs Eq.~\eqref{linentr}, while Eqs..~\eqref{entagc} and (\ref{entagd}) 
go to zero, as it could be expected.

\section{Quantum entanglement measures}
\label{enme}
In the previous Section, it has been shown 
that flavor oscillations admit an equivalent description 
in terms of entanglement of flavor modes. In that case, 
specifically, following Ref.~\cite{Blasone:2007vw},
we have used the \emph{linear entropy} to quantify the entanglement
of neutrino states (see Eq.~\eqref{linentr}
and the corresponding QFT relations, Eqs.~\eqref{entaga}-\eqref{entagd}). 
In order to define such a measure\footnote{This Section
is not intended to be an exhaustive review of all possible entanglement measures. For
a more detailed analysis of this, we remand the 
reader to the appropriate literature (see, for instance, Refs.~\cite{Plenio:2007zz,Horodecki:2009zz,tzc,Blasone:2008sh} and therein).}, consider 
a quantum state described by the density matrix $\rho$ 
in a $D$-dimensional Hilbert space $\mathcal{H}$.
Denoting by $\mu=\mathrm{Tr}\rho^2$ the purity of the state, 
the linear entropy $S_L(\rho)$ is defined as
\be
\label{entr}
S_L(\rho)\,=\,\frac{D}{D-1}[1-\mathrm{Tr}\rho^2]\,=\,\frac{D}{D-1}[1-\mu(\rho)]\,.
\ee
For pure states $\rho=|\psi\rangle\langle\psi|$ and $\mu=1$;
by contrast, for mixed states $\mu<1$, and it acquires its minimum value
$1/D$ on the maximally mixed state $\rho=\mathbb{1}_D/D$. 
The normalization in 
Eq.~\eqref{entr} ensures that the entropy $S_L$
ranges from 0 (pure states) to 1 (maximally mixed states). 

Linear entropy is widely employed to quantify the entanglement
of quantum systems, even though it is not the only possible measure. 
Bipartite entanglement of pure states, for instance, can be unambiguously
quantified by the \emph{von Neumann entropy} 
\be
\label{vnent}
S_V(\rho)\,=\,-\mathrm{Tr}[\rho\hspace{0.1mm}\log_D\rho]\,,
\ee
or by any other
monotonic function of the former (among these, the linear entropy itself 
has a special physical significance, since it is linked to the purity of the
reduced states, and enters in the fundamental monogamy
inequalities for distributed entanglement in the multipartite
setting). 

For bipartite mixed states, several entanglement measures
have been proposed~\cite{Bennett:1996gf,Vedple,Vidal:2002zz}.
Although providing interesting operative definitions, the \emph{entanglement of formation}
and of \emph{distillation} are very hard to compute. 
A renowned result in this context is the Wootters formula for the entanglement of
formation for two-qubit mixed states~\cite{Wootters:1997id}. 
An alternative measure, which
is closely related to the entanglement
of formation, is the \emph{concurrence} (the entanglement of formation is a monotonically increasing function of the concurrence)~\cite{Coffman:1999jd}:
\be
\label{conc}
C(\rho^{(\alpha,\beta)})\,=\,\mathrm{max}\{0,\lambda_1\,-\,\lambda_2\,-\,\lambda_3\,-\,\lambda_4\},
\ee
where $\{\lambda_i\}_{i=1}^{4}$ are the square roots of the eigenvalues, 
in decreasing order, of the non-Hermitian matrix $\rho^{(\alpha,\beta)}\tilde\rho^{(\alpha,\beta)}$, 
with $\rho^{(\alpha,\beta)}\,=\,\mathrm{Tr}_{\gamma\neq \alpha,\beta}[\rho]$ being the reduced
density operator and $\tilde\rho^{(\alpha,\beta)}\,=\,(\sigma_y\otimes\sigma_y)\rho^{(\alpha,\beta)*}(\sigma_y\otimes\sigma_y)$ the spin-flipped state. With reference to the entanglement of flavor modes, for example, 
the concurrence can be used to measure the entanglement between two
flavors in the neutrino three flavor state, after tracing
the third flavor. The quantities to compute in that case 
are $C\left(\rho_\alpha^{(\beta,\gamma)}(x)\right)\equiv C_\alpha^{(\beta,\gamma)}$,
with $\alpha, \beta, \gamma, \eta=e,\mu,\tau$.

Computational difficulties in the quantification of bipartite entanglement for mixed states
are also encountered with the 
\emph{relative entropy}~\cite{Vedple}. On the other hand, a more straightforwardly 
computable entanglement monotone is the \emph{logarithmic negativity} $E_{\mathcal{N}}$,
based on the requirement of positivity of the density
operator under partial transposition 
$E_{\mathcal{N}}=\log_2 ||\,\tilde\rho_{12}\,||_1$, where $||\,\cdot\,||_1$
denotes the trace norm, i.e. $||\,\mathcal{O}\,||=\mathrm{Tr}[\sqrt{\mathcal{O}^\dagger\,\mathcal{O}}]$ 
for any Hermitian operator $\mathcal{O}$~\cite{Vidal:2002zz}.
The so-called \emph{bona fide} density matrix $\tilde\rho_{12}$
is obtained through the partial transposition
with respect to one mode, say mode $2$, of $\rho_{12}$, i.e.
$\tilde\rho_{12}\equiv\rho_{12}^{PT\hspace{0.4mm}2}$.
Given an arbitrary orthonormal product basis $|i_1,j_2\rangle$, 
the matrix elements of $\tilde\rho_{12}$ are determined by
$\langle i_1,j_2|\tilde\rho_{12}|k_1,l_2\rangle=\langle i_1,j_2|\rho_{12}|k_1,j_2\rangle$.
Cleary, for pure states such a measure provides the same results as the von Neumann entropy
defined above.

In the instance of multipartite states, 
the quantification of entanglement
becomes much more convoluted, as the
various measures are not equivalent and sometimes ill-defined.
Important achievements have been reached in understanding the different ways in which multipartite systems can be entangled. The intrinsic
nonlocal character of entanglement imposes its invariance under any local quantum operations; therefore, equivalence
classes of entangled states can be defined through the group of reversible stochastic local quantum operations assisted
by classical communication (SLOCCs)~\cite{Pope}. 
Such an approach allows to demonstrate that three and four qubits can
be entangled, respectively, in two and nine inequivalent ways~\cite{Dur:2000zz}
(we remand to Ref.~\cite{Blasone:2007wp} for a more detailed discussion on this).

Starting from the above classification, various attempts 
to introduce efficient entanglement measures for multipartite systems 
have been done. The
characterization of the quantum correlations through a measure embodying a collective property of the system, should be based on the introduction of quantities invariant 
under local transformations. A successful step in this
direction has been put forward by Coffman, Kundu, and Wootters in Ref.~\cite{Coffman:1999jd}, 
where, with reference to the entanglement in three qubits systems, 
the so-called \emph{residual, genuine tripartite entanglement}, or 3-tangle,
has been constructed in terms of the squared concurrences associated with 
the global three qubit state and the reduced two-qubit states. 
Several generalizations of this quantifier have been later proposed: among these, 
we mention the \emph{Schmidt measure}, defined as the minimum of $\log_2 r$,  
with $r$ being the minimum of the number of terms in an expansion of the state in product basis~\cite{eiser}, and the \emph{global entanglement} of Meyer
and Wallach~\cite{Meyer}, which is given by the sum of concurrences between one qubit and all others,
and can be expressed as
the average subsystem linear entropy~\cite{Brennenqinf}.
A further measure for (pure) multipartite states
has been considered in Ref.~\cite{Blasone:2007wp} 
in terms of the possible bipartitions of the system:  the \emph{average von Neumann entropy}, as
its name suggests, is indeed a functional of the von Neumann entropy averaged on a given bipartition
of the system. In order to define it, let  $\rho$ be the density operator for 
a pure state $|\psi\rangle$ describing the $N$-partite system $S=\{S_1,S_2,\dots,S_N\}$.
Consider the bipartition of such a system in two 
subsystems $S_{A_n}=\{S_{i_1},S_{i_2},\dots,S_{i_n}\}$, with
$1\leq i_1< i_2< .\,.\,. <i_n\leq N$ $(1\leq n<N)$, and 
$S_{B_{N-n}}=\{S_{j_1},S_{j_2},\dots,S_{j_{N-n}}\}$, with
$1\leq j_1< j_2< .\,.\,. <j_{N-n}\leq N$, and $i_q\neq j_p$. 
Denoting by $\rho_{A_n}$ the density matrix reduced 
with respect to the subsystem $S_{B_{N-n}}$,
\be
\rho_{A_n}\,\equiv\,\rho_{i_1,i_2,\dots,i_n}=\mathrm{Tr}_{B_{N-n}}[\rho]\,=\,\mathrm{Tr}_{j_1,j_2,\dots,j_{N-n}}[\rho]\,,
\ee
and using Eq.~(\ref{vnent}) for the von Neumann entropy $S_V(\rho_{A_n})$
associated with such a bipartition, the average von Neuman entropy is 
defined as
\be
\langle S^{(n;N-n)}_{V}\rangle\,=\,{N\choose n}^{-1}\sum_{A_n}S_V(\rho_{A_n})
\ee
where the sum is intended over all the possible bipartitions of the system in two subsystems each with 
$n$ and $N-n$ elements $(1 \leq n < N)$.

Entropic measures, however, cannot be used to quantify the entanglement of mixed states.
In this context, it is useful to introduce a generalized version of the logarithmic negativity
defined above. Again, let $\rho$ be the density matrix for a multipartite 
mixed state associated with a $N$-partite system $S$, and consider
the bipartition into two subsystems $S_{A_n}$ and $S_{B_{N-n}}$.
Denoting by 
\be
\tilde{\rho}_{A_n}\,=\,\rho^{PT\,B_{N-n}}\,=\,\rho^{PT\,j_1,j_2,\dots,j_{N-n}}
\ee
the \emph{bona fide} density matrix obtained by the 
partial transposition of $\rho$ with respect to the parties 
belonging to $S_{B_{N-n}}$, the logarithmic negativity 
associated with the fixed bipartition is given by $E_{\mathcal{N}}^{(A_n;B_{N-n})}=\log_2||\,\rho_{A_n}\,\log_2\tilde\rho_{A_n}\,||_1$. Following the same procedure as for 
the average von Neumann entropy, one can define the \emph{average logarithmic negativity} as~\cite{Blasone:2007wp}
\be
\langle E^{n;N_n}_\mathcal{N}\rangle\,=\,{N\choose n}^{-1}\sum_{A_n}E_{\mathcal{N}}^{(A_n;B_{N-n})}\,,
\ee
where the sum is intended over all the possible bipartitions of the system.

We finally remark that a completely different characterization of multipartite entanglement can be defined
in terms of purely geometric measures~\cite{tzc,Blasone:2008sh}. The relative entropy 
of entanglement (generalized for multipartite settings) and
the \emph{geometric entanglement} are among the most common
quantifiers belonging to this class. 
On the one hand, the relative entropy is defined
as the distance of a given state from the set of
fully separated states, quantified in terms of the quantum
relative entropy~\cite{Vedralpl}. On the other hand,
the geometric entanglement is the
Euclidean distance of a given
multipartite state to the nearest fully separable state~\cite{tzc,Shi}. One of
the benefits of working with such a measure is the interesting connection that it exhibits with
other quantifiers~\cite{tzc}. Moreover, it can be efficiently estimated
by quantitative entanglement witnesses amenable of experimental
verification~\cite{Reimpell}. Given an $N$-partite pure
state $|\psi\rangle$, the geometric measure of entanglement 
introduced by Wei and Goldbart is defined as~\cite{tzc}:
\be
E_G(|\psi\rangle)\,=\,1\,-\,\underset{{|\Phi\rangle}}{\mathrm{max}}{\big|\langle\phi|\psi\rangle\big|}^2\,,
\ee
where the maximum is taken with respect to all pure
states that are fully factorized, i.e. the $N$-separable
states $\phi\rangle$ such that
\be
\label{factorization}
|\phi\rangle\,=\,\bigotimes_{s=1}^{N}|\phi_s\rangle\,,
\ee
where $|\phi_s\rangle$ being single-qubit pure states.
As it can be shown, this is an intrinsically geometric measure because it coincides
with the distance (in the Hilbert-Schmidt norm) between
a given pure state and the set of fully separable (i.e. fully
product) pure states.

The geometric measure can be extended by the convex
roof procedure to the case of mixed states; additionally, it is a
proper multipartite entanglement monotone. In spite of its
very appealing properties and the large number of cases where it can
be usefully exploited~\cite{altepeter}, however,
the global nature of the Wei-Goldbart geometric
entanglement does not allow to discriminate among
the different bipartite and multipartite contributions to
the overall entanglement, to determine their properties,
and to establish a systematic hierarchy among them.
In this connection, in Ref.~\cite{Blasone:2008sh} it has been defined
a natural and powerful multipartite generalization of such a quantifier 
for pure states of many-qubit
systems. Consider a $N$-qubit system, corresponding to a tensor-product
state space $\mathcal{H}^{d_{N}}$ of dimension $d_{N} = 2^{N}$.
Let us introduce an integer $K$, $2 \leq K \leq N$, and the
ordered sequence of integers $\{ M_1,M_2,\ldots,M_K \}$, where
$M_1 \leq M_2 \leq \ldots \leq M_K$, and $\sum_{s=1}^{K} M_s = N$.
For a $K$-partition of the system in $K$
subsystems described by the sets $\{Q_{s}\}_{s=1}^{K}$, 
each set $Q_{s}$ will be composed of $M_{s}$ elementary parties, i.e.
$Q_{s}=\{ i_{1}^{(s)},i_{2}^{(s)},\ldots, i_{M_{s}}^{(s)} \}$,
where $i_{j}^{(s)} \in \{1,\ldots,N\}$ is a discrete index labeling the $N$
elementary parties, and $Q_{s}\bigcap Q_{s'} = \emptyset$ for $s \neq s'$.
Given a generic $K$-partition $Q_1|Q_2|\ldots|Q_K$ of the $N$-qubit system,
any $K$-separable state associated to such a partition is defined as
the tensor product of $K$ $M_{s}$-qubit
pure states $|\Phi_{s}^{(Q_{s})}\rangle$. 
Each state $|\Phi^{(Q_{s})}\rangle$
belongs to the Hilbert space $\mathcal{H}^{d_{Q_{s}}}$ of dimension
$d_{Q_{s}} = 2^{M_{s}}$.
Starting from Eq.~\eqref{factorization}, a $K$-separable state 
can then be written as
\begin{equation}
\bigotimes_{s=1}^{K} \,
|\Phi_{s}^{(Q_{s})}\rangle \,,
\label{KseparablePhi}
\end{equation}
and the Hilbert space $\mathcal{H}^{d_{N}}$ is accordingly decomposed 
as $\bigotimes_{s=1}^{K}\mathcal{H}^{d_{Q_{s}}}$.
Varying the integers $M_s$,
one obtains different $K$-partitions $Q_1|Q_2|\ldots|Q_K$ and,
correspondingly, different possible $K$-separable states.
Let us now denote by ${\mathcal{\mathbf{S}}}_K$ the set of all $K$-separable states,
i.e.
\begin{equation}
{\mathcal{\mathbf{S}}}_K = \bigcup_{\{Q_1,\ldots,Q_K\}} S_K 
(Q_1|\ldots|Q_K) \,,
\label{CheClass}
\end{equation}
with $S_K (Q_1|Q_2|\ldots|Q_K)$ being the set of all the $K$-separable
states associated to a fixed $K$-partition.
The geometric measures of entanglement
with respect to $K$-separable pure states for
an arbitrary $N$-qubit pure state $|\Psi^{(N)}\rangle$ can be then
respectively defined as:
\begin{equation}
E_{G}^{(K)}(Q_1 |\ldots |Q_K ) \,=\, 1\,-\,\Lambda^2_{K}(Q_1 |\ldots |Q_K ) \,,
\label{GMErel}
\end{equation}
where 
\be
\label{GMErel2}
\Lambda_{K}^{2}(Q_1 |\ldots |Q_K ) \,=\, \max_{|\varphi\rangle \in
S_{K}(Q_1 |\ldots |Q_K ) } \,
\Big|\langle \varphi| \Psi^{(N)}\rangle\Big|^{2}\,,
\ee 
and
\be
\label{GMErel3}
E_{G}^{(K)}(|\psi^{(N)}\rangle) \,=\, 1\,-\,\Lambda^2_{K}(|\psi^{(N)}\rangle)\,,
\ee
with
\be
\label{GMErel4}
\Lambda_{K}^{2}(|\psi^{(N)}\rangle)\,=\,\max_{|\Phi\rangle \in
\mathcal{\mathbf{S}}_K}  \,
\Big|\langle \Phi| \Psi^{(N)}\rangle\Big|^{2}\,.
\ee
From Eqs.~\eqref{GMErel}, \eqref{GMErel2}, \eqref{GMErel4}, 
it thus arises that $E_{G}^{(K)}(|\psi^{(N)}\rangle)$ in Eq.~\eqref{GMErel3} 
provides a measure of the absolute minimum distance of a state from the set of
all $K$-separable states. Trivially, for any $N$-partition of the system (i.e. $K=N$), 
one has $M_1=M_2=\dots M_N=1$ and $N-$separability 
coincides with full separability,
while $1$-separability is a common feature of any state,
i.e. $E_G^{(1)}=0$ for all states $\{|\psi^{(N)}\rangle\}$.

\smallskip

\chapter{Quantization of mixed boson fields in hyperbolic modes}
\label{QMBFHM}
\begin{flushright}
\emph{``The beginning of knowledge\\
is the discovery of something\\ we do not understand.''}\\[1mm]
-  Frank Herbert -\\[6mm]
\end{flushright}


In the original approach~\cite{Blasone:1995zc, bosonmix}, the QFT
of flavor mixing  has been addressed by quantizing the fields in the conventional plane wave basis.
By virtue of Lorentz covariance, however, such a formalism can be equivalently analyzed within a representation diagonalizing any other Lorentz group generator. In particular,  one may wonder  how the mixing transformations Eq.~(\ref{eqn:esbos0}) would appear in the hyperbolic ``boost'' scheme, that is, the scheme which diagonalizes the boost momentum operator (spatio-temporal rotation).
In what follows, we try to answer this question. Specifically, by working in the usual framework of  Minkowski spacetime, we show how the results and definitions reported here provide the basis for extending the field theoretic treatment of flavor mixing to the Rindler frame\footnote{We shall focus on the case of scalar fields. Similar considerations can be readily extended to fermions.}.

In line with what we discussed  in Section~\ref{sec:Trasformazioni di mixing di campi bosonici} of Appendix~\ref{QFT of fm}, let us consider two scalar fields $\phi_i$ $(i=1,2)$ with mass $m_i$. Exploiting  
the completeness and orthonormality properties of boost modes $\Big\{\widetilde{U}_{\kappa,i}^{\,(\sigma)},\,\widetilde{U}_{\kappa,i}^{\,(\sigma)*} \Big\}$, Eq.~(\ref{eqn:Uwidetilde}), we can expand  the mixed fields $\phi_{\chi}$ $(\chi=A,B)$ in Eqs.~\eqref{eqn:Pontecorvobis} as follows (to be compared with the expansion Eq.~(\ref{eqn:expansionfieldutilde}) for a free field):
\begin{equation}
\label{eqn:expancampmix}
\phi_{\ell}(x)\,=\, \sum_{\sigma,\hspace{0.4mm}\Omega}\,\int d^{2} {k}\,\Big\{d_{\kappa,\ell}^{\,(\sigma)}\,\widetilde{U}_{\kappa,j}^{\,(\sigma)}(x)\,+\,\bar d_{\kappa,\ell}^{\,(\sigma)\dagger}\,\widetilde{U}_{\kappa,j}^{\,(\sigma)*}(x)\Big\}\,,
\end{equation}
where $(\ell,j)=(A,1),(B,2)$ and we have used the shorthand notation introduced in Eq.~(\ref{eqn:simpnot}). Consistently with Eq.~\eqref{eqn:scalproda}, the flavor annihilation operators in the hyperbolic representation are given by
\begin{equation}
\label{eqn:daoperator}
d_{\kappa,A}^{\,(\sigma)}\,=\,
\Big(\phi_A,\,\widetilde{U}_{\kappa,1}^{\,(\sigma)}\Big)
\,=\,\cos\theta\, d_{\kappa,1}^{\,(\sigma)}
\,+\,\sin\theta \sum_{\sigma',\hspace{0.4mm}\Omega'}\,\Big(d_{(\Omega',\vec{k}),2}^{\,(\sigma')}\;\bogocoeffAlpha\, +\,\bar d_{(\Omega',-\vec{k}),2}^{\,(\sigma')\dagger}\;\bogocoeffBeta\,\Big)\,,\\
\end{equation}
and similar for the other operators.  As in the framework of plane waves, mixing transformations exhibit the structure of a rotation combined with an internal Bogoliubov transformation. The Bogoliubov coefficients $\bogocoeffAlpha$ and $\bogocoeffBeta$ now read
\begin{equation}
\bogocoeffAlpha \,=\,\int^{+\infty}_{-\infty}\frac{dk_1}{4\pi}\lf(\frac{1}{\omega_{\bk,1}}+\frac{1}{\omega_{\bk,2}}\ri){\lf(\frac{\omega_{\bk,1}+k_1}{\omega_{\bk,1}-k_1}\ri)}^{i\sigma\Omega/2} {\lf(\frac{\omega_{\bk,2}+k_1}{\omega_{\bk,2}-k_1}\ri)}^{-i\sigma'\Omega'/2}
e^{i(\omega_{\bk,1}-\,\omega_{\bk,2})t}\,,
\label{eqn:coefficientuno}
\end{equation}
\vspace{2mm}
\begin{equation}
\bogocoeffBeta \,=\,\int^{+\infty}_{-\infty}\frac{dk_1}{4\pi}\lf(\frac{1}{\omega_{\bk,2}}
-\frac{1}{\omega_{\bk,1}}\ri){\lf(\frac{\omega_{\bk,1}+
k_1}{\omega_{\bk,1}-k_1}\ri)}^{i\sigma\Omega/2} {\lf(\frac{\omega_{\bk,2}+k_1}{\omega_{\bk,2}-k_1}\ri)}^{-i\sigma'\Omega'/2}
e^{i(\omega_{\bk,1}+\,\omega_{\bk,2})t}\,,\\
\label{eqn:coefficientdue}
\end{equation}
to be compared with the corresponding relations in the plane wave representation, Eq.~(\ref{bogcoeffplaw}). For $m_1\rightarrow m_2$, it is a trivial matter to check that $\bogocoeffBeta\rightarrow 0$ and $\bogocoeffAlpha\rightarrow \delta(\sigma-\sigma')\,\delta(\Omega-\Omega')$, as it should be in the absence of mixing.

The analytic resolution of the integrals Eqs.~(\ref{eqn:coefficientuno}), (\ref{eqn:coefficientdue}) is absolutely non-trivial. In spite of these technicalities, however, a useful approximation can be obtained for $t=\eta=0$ and $\frac{\Delta m^2}{m_i^2}\equiv \frac{m_2^2-m_1^2}{m_i^2}\ll 1$, where one has (to the leading order)~\cite{Blasone:2017nbf}
\bea
\bogocoeffAlpha\Big|_{t=0}&=&\delta_{\sigma\sigma'}\,\delta{(\kappa-\kappa')}\,-\,\frac{\Delta m^2}{8\mu_{k,1}^{{\hspace{0.3mm}}2}}\;\frac{\sigma\,\Omega-\sigma'\Omega'}{\sinh[\frac{\pi}{2}(\sigma\,\Omega-\sigma'\Omega')]}\,,
\label{eqn:coefficientunoapprox}
\\ [4mm]
\bogocoeffBeta\Big|_{t=0}&=&-\,\frac{\Delta m^2}{8\mu_{k,1}^{{\hspace{0.2mm}}2}}\;\frac{\sigma\,\Omega-\sigma'\Omega'}{\sinh[\frac{\pi}{2}(\sigma\,\Omega-\sigma'\Omega')]}\,.
\label{eqn:coefficientdueapprox}
\eea
In closing, let us remark that  the flavor operators Eq.~\eqref{eqn:daoperator} obey the canonical commutation relations (at equal times), 
\begin{equation}
\label{eqn:commutatorflavor}
\left[d_{\kappa,\chi}^{\,(\sigma)}(t)\,,\, d_{\kappa',\chi'}^{ ( \sigma')\dagger}\,(t')\right]\bigg|_{t=t'}=\left[\bar d_{\kappa,A}^{\,(\sigma)}\,(t)\,,\, {\bar d_{\kappa',A}}^{\,(\sigma')\dagger}\,(t')\right]\bigg|_{t=t'}\,=\,\delta_{\chi\chi'}\,\delta_{\sigma\sigma'}\,\delta^3(\kappa-\kappa'),
\end{equation}
with $\chi,\chi'=A,B$ and all other commutators vanishing. This allows for the derivation of  the following conditions for the Bogoliubov coefficients $\bogocoeffAlpha$ and $\bogocoeffBeta$:
\begin{eqnarray}
\label{eqn:primprop}
\sum_{\sigma'',\hspace{0.4mm}\Omega''}\left({\cal A}_{(\Om',\Om''),\,\vec{k}'}^{(\si',-\si'')}\;\,{\cal B}_{(\Om,\Om''),\,\vec{k}}^{(\si,\si'')\,*}\,-\,{\cal A}_{(\Om,\Om''),\,\vec{k}}^{(\si,\si'')\,*}\;\,{\cal B}_{(\Om',\Om''),\,\vec{k}'}^{(\si',-\si'')}\right)&=&0\,,\\[3mm]
\label{eqn:secprop}
\sum_{\sigma'',\hspace{0.4mm}\Omega''}\left(\bogocoeffAlphater\;\,\bogocoeffAlphaquintst\,-\,\bogocoeffBetater\;\,\bogocoeffBetaquintst\right)&=&\delta_{\sigma\sigma'}\delta^3(\kappa-\kappa')\,.
\end{eqnarray}
The latter, in particular, is the hyperbolic representation equivalent of Eq.~(\ref{eqn:iper}).

\chapter{Inverse $\beta$-decay with neutrino mass eigenstates: conceptual and computational flaws of the treatment}
\label{Cozzella}
\begin{flushright}
\emph{``Our responsibility\\ is to do what we can, \\
lear what we can,\\
improve the solutions, \\
and pass them on.
''}\\[1mm]
-  Richard P. Feynman -\\[6mm]
\end{flushright}

In this Appendix we briefly comment on some  results  found by Cozzella et al.
in the context of the inverse $\beta$-decay with neutrino mixing~\cite{Cozzella:2018qew}. 
Before going into details, we emphasize that the aforementioned work~\cite{Cozzella:2018qew}  has been 
submitted on arXiv just one day apart from ours~\cite{Blasone:2018czm}.
In spite of addressing the
same issue, however, it sounds bizarre that the 
arguments put forward by these two papers about the real nature of vacuum for mixed fields are
diametrically opposed, thus making the
debate on the treatment of flavor mixing in QFT even more lively (a similar analysis has
been recently performed within the framework of the Casimir effect in Ref.~\cite{Casimirmix}, 
where it has been shown that different choices of vacuum for mixed fields enclosed in a
finite region inevitably 
lead to different expressions of the Casimir force between the confining plates). 
Here we explain why the analysis of Ref.~\cite{Cozzella:2018qew} 
is missing of some fundamental points, 
focusing on both conceptual and computational flaws there reported.

To begin with, let us observe that in Ref.~\cite{Cozzella:2018qew} it is claimed that no problem arises at all when involving neutrino mixing in the decay of accelerated protons, and that the transition rates in the laboratory and comoving frames exactly match. Sifting through the analysis of Ref.~\cite{Cozzella:2018qew}, however, we conclude that there are no grounds at all for supporting these strong statements, also in light of the results obtained in Section~\ref{prdec}. To argue this, it is useful to extract from the  technical review of Appendix~\ref{QFT of fm} the essential ingredients for quantizing mixed neutrino fields. In Section~\ref{Comments on the work by Cozzella} we expose our criticism on the paper~\cite{Cozzella:2018qew}. Setion~\ref{Conclusionscomment} is devoted to a brief discussion.

\section{Neutrino mixing in QFT}
\label{Neutrino mixing in QFT}

Neutrino mixing~\cite{Pontecorvo} is embedded in the Standard Model in analogy with quark mixing, at least for the case of Dirac neutrinos. Considering for simplicity the case of two flavors, we have that the Lagrangian associated to neutrino fields with definite flavor $\nu_e$, $\nu_\mu$ acquires off-diagonal terms in the mass matrix, which can be  diagonalized by means of the mixing transformations, Eqs.~\eqref{eqn:Pontec1}: 
\begin{equation}
\label{fermion}
\begin{array}{l}
\nu_e(x)\,=\,\cos\theta\,\nu_1(x)+\sin\theta\,\nu_2(x)\,,\\[4mm] 
\nu_\mu(x)\,=\,-\sin\theta\,\nu_1(x)+\cos\theta\,\nu_2(x)\,, 
\end{array}
\end{equation}
where $\nu_1$, $\nu_2$ are fields with definite mass.

Quantization of flavor fields is achieved by identifying  the algebraic generator of the rotation Eq.~\eqref{fermion} (see Appendix~\ref{QFT of fm})
\begin{equation}
\label{bogolgene}
\begin{array}{l}
{\nu}_e(x)\,=\,{G}^{-1}\,{\nu}_1(x)\,{G}\,,\\[4mm]
{\nu}_\mu(x)\,=\,{G}^{-1}\,{\nu}_2(x)\,{G}\,,
\end{array}
\end{equation}
with $ G$ given in Eq.~\eqref{eqn:generatoremix},
\be\label{Genegenegene}
{G}\,=\,\exp\lf\{\theta\int d^3x\,\lf[{\nu}^\dagger_1(x){\nu}_2(x)-{\nu}^\dagger_2(x){\nu}_1(x)\ri]\ri\},
\ee
here reported for convenience.
The expansions for flavor fields are obtained by inserting into Eq.~\eqref{bogolgene}  the corresponding expansions for free fields. A straightforward calculation leads to Eq.~\eqref{flfieldexpans}
\be
\label{eqn:planewavemix}
\nu_\ell(x)\,=\,\sum_{r} \int d^{\hspace{0.2mm}3}\hspace{0.1mm}k\hspace{0.3mm} e^{ik\cdot x}\left[\hspace{0.2mm} a^{r}_{k,\ell}\hspace{0.7mm}u^{r}_{k,j}(t)\,+\, b^{r\hspace{0.2mm}\dagger}_{-k,\ell}\hspace{0.7mm}v^r_{-k,j}(t)\hspace{0.2mm}\right],
\ee
where $\quad (\ell,j)=(e,1), (\mu,2)$ and $r$ is the spinor polarization. 
We remark that the continuous
limit relation $\sum_\bk\rightarrow \frac{V}{{(2\pi)}^3}\int d^3k$ has been used to derive Eq.~\eqref{eqn:planewavemix};
in this way, flavor ladder operators are consistently defined, 
as explicitly shown in Appendix~\ref{QFT of fm}.


A subtle point in the above quantization procedure arises when considering the action of the mixing generator ${G}$ on the free field vacuum $|0\rangle_{\mathrm{M}}$ (mass vacuum). This indeed generates a non-trivial state $|0\rangle_{\mathrm{f}}$ (flavor vacuum), defined as in Eq.~\eqref{eqn:vuotosapore}:\footnote{Consistently with the original paper, in this Section we have slightly changed the notation of flavor and mass vacua. See by comparison Eq~\eqref{eqn:vuotosapore} of Appendix~\ref{QFT of fm}.}
\be\label{flavac}
|0\rangle_{\mathrm{f}} \,\equiv\,{G}^{-1}\,|0\rangle_{\mathrm{M}}\,.
\ee
In Appendix~\ref{QFT of fm} it has been argued that the Fock spaces for neutrinos with definite mass and definite flavor ($\mathcal{H}_m$ and $\mathcal{H}_f$, respectively),
which are connected by a unitary transformation at
finite volume, become
unitarily \emph{inequivalent} in the infinite volume limit~\cite{Blasone:1995zc}, being
\be\label{limit}
\lim_{V\rightarrow\infty} {}_{\mathrm{m}}\langle 0|0\rangle_{\mathrm{f}}\,=\,0\,.
\ee
Furthermore, the following properties for the flavor vacuum have been highlighted:
\bit
\item
it is an eigenstate of the flavor charge operators ${Q}_{\nu_\alpha}$, obtained by means of Noether's theorem;
\item
it is a generalized Gilmore-Perelomov coherent state~\cite{Gilmore};
\item 
it is a condensate of neutrino-antineutrino pairs with both same and different mass, as it 
is evident from Eq.~(\ref{vacstr}).
\eit
Flavor neutrino states are then consistently defined as eigenstates of flavor charges 
\be
\label{Blastates}
{Q}_{\nu_\alpha}|\nu_{\alpha}\rangle\,=\,|\nu_{\alpha}\rangle, \quad \alpha=e,\mu\,.
\ee
In the relativistic limit, these states  reduce to the usual Pontecorvo flavor states~\cite{Pontecorvo},
\be\label{Pontecstates}
|\nu_\alpha\rangle_P\,=\,\sum_i U_{\alpha,i}|\nu_i\rangle\,,
\ee
where $U_{\alpha,i}$ is the generic element of the mixing matrix, Eq.~\eqref{PMMmatrix}. This has been largely discussed in Appendix~\ref{QFT of fm}.

\section{Remarks}
\label{Comments on the work by Cozzella}
In this Section, we list our criticisms to some of the findings of Ref.~\cite{Cozzella:2018qew}.
\medskip

\emph{a.}\hspace{1cm}In the Introduction, those authors consider the recent investigation of Ref.~\cite{Ahluwalia:2016wmf} about the inequality of the inertial and the comoving decay rates for the process $p\rightarrow n\,e^+\,\nu_e$ with mixed neutrinos, asserting that in principle no discrepancy should exist between the calculations done in the two reference frames. On the basis of this, they then affirm that  ``\emph{either the Unruh effect is wrong (contradicting
several previous results~[7], including what we consider to be a virtual observation of it~[8])  or some mistake was
made in the previously mentioned analysis}''.

We basically agree with the authors of Ref.~\cite{Cozzella:2018qew} on the fact that general covariance should be taken as a guiding principle in the analyzed problem. However, a (possible) mismatch of the above decay rates in the presence of neutrino mixing is not in contradiction with Refs.~[7, 8] of their work, where neutrino mixing is not taken into account at all. 
Indeed, it is perfectly possible that corrections to the Unruh effect (in the form of nonthermal contributions) may arise in connection with neutrino mixing. This has been explicitly shown  in Chapter~\ref{Non-thermal signature}, where the quantization of both bosonic and fermionic fields in the presence of mixing has been performed in accelerated frames.

On the other hand, the analysis of the decay of accelerated protons with mixed neutrinos has been developed in Chapter~\ref{The necessity of the Unruh effect in QFT: the proton}, where it has been shown that general covariance does hold in the approximation in which Pontecorvo states can be used to represent flavor neutrinos. Beyond such an approximation, general covariance must hold, but Unruh effect may be modified.

\medskip
\emph{b.}\hspace{8mm} In Section III of Ref.~\cite{Cozzella:2018qew}, the claim ``\emph{Attempts to canonically quantize the ${\nu}_\alpha$ fields} [...] \emph{give annihilation (and creation) operators whose physical meaning is unclear~[20]}'' is not really motivated, and it is argued only in a note (Ref.~[20] of that paper), which is flawed. Indeed, as shown in Section~\ref{Neutrino mixing in QFT}, the flavor annihilation operators are self-consistently defined with respect to the flavor vacuum and the flavor neutrino state has a neat physical interpretation as eigenstate of flavor charge.

\medskip
\emph{c.}\hspace{8mm} 
Above we have remarked that the exact flavor states in Eq.~\eqref{Blastates} reduce to the ordinary Pontecorvo states Eq.~\eqref{Pontec} in the relativistic limit. This is indeed in agreement with the statement at the beginning of Appendix B of Ref.~\cite{Cozzella:2018qew}. We stress that, in Chapter~\ref{The necessity of the Unruh effect in QFT: the proton}, we only deal with Pontecorvo states, which are valid within the  approximation there considered\footnote{Note that such states can be efficiently generalized also to the case of curved backgrounds, and in particular of accelerated frames~\cite{Cardall}.}. The use of Pontecorvo states in the calculations of Chapter~\ref{The necessity of the Unruh effect in QFT: the proton} produces ``off-diagonal'' terms (namely, contributions proportional to $\cos^2\theta\sin^2\theta$) in the expression of decay rates. These terms are completely absent in the corresponding treatment of Ref.~\cite{Cozzella:2018qew}, thus giving rise to a fundamental contradiction.

The origin of this discrepancy can be readily understood: in Ref.~\cite{Cozzella:2018qew}, the authors consider the decay process
\be\label{process}
p\,\rightarrow\, n\,\bar{l}_{\alpha}\,\nu_i,
\ee
with $l_\alpha=\{e,\mu,\tau\}$ and $\nu_i=\{\nu_1,\nu_2,\nu_3\}$.

It is evident that, with such a choice, the aforementioned off-diagonal terms can never appear in any calculation of the decay rates. Thus, the outcome of Ref.~\cite{Cozzella:2018qew} - which is claimed to be exact - cannot reproduce the one of Chapter~\ref{The necessity of the Unruh effect in QFT: the proton} within the relativistic approximation, where Pontecorvo states are generally recognized to be well-defined.

\medskip
\emph{d.}\hspace{8mm} 
We further point out that in Ref.~\cite{Cozzella:2018qew} mixing is correctly taken into account at the level of fields in the weak interaction action
\be\label{action}
{S}_I\,=\,\int d^4x\,\sqrt{-g}\,\left(\sum_\alpha{\bar{\nu}}_\alpha\,\gamma^\mu\,{P}_L\,{l}_\alpha\,{j}_\mu\,+\,\mathrm{h.\,c.}\right),
\ee
but not at the level of the states\footnote{This is the reason why the total decay rate Eq.~(48) in Ref.~\cite{Cozzella:2018qew} contains second powers of the mixing matrix elements, while in Eq.~(38) of Chapter~\ref{The necessity of the Unruh effect in QFT: the proton} they appear with the fourth power.}, not even in the particular situations in which Pontecorvo states are known to be valid (see Appendix B of Ref.~\cite{Cozzella:2018qew}). 

It must be emphasized that the combined use of the action Eq.~\eqref{action} with the states for the process Eq.~\eqref{process} in Ref.~\cite{Cozzella:2018qew} does not satisfy normalization conditions. Fields and states adopted in the evaluation of transition amplitudes should indeed be ``compatible'' in the sense of normalization. Namely, if the action is written in terms of a generic field ${\phi}$ as\footnote{To avoid technicalities, in what follows we refer to scalar fields. Analogous considerations hold for  fermionic fields.}
\be
{S}\,=\,\int d^4x\,\mathcal{L}\left({\phi},\partial\,{\phi}\right),
\ee
then one expects to work with states $|\phi\rangle$ that 
satisfy the relation~\cite{qftsm,Cheng:1985bj}
\be\label{norm}
\langle 0|\,{\phi}(0)\,|\phi(p)\rangle\,=\,1\,.
\ee
As an example, suppose to consider a scalar field ${\phi}$; adopting the standard plane wave expansion,
\be\label{bfieldexp}
{\phi}(x)\,=\,\int\frac{d^3k}{(2\pi)^32\,\omega_{\bk}}\left[{a}_{\bk}\,e^{-ik.x}\,+\,{b}^\dagger_{\bk}\,e^{ik.x}\right],
\ee
then the commutation relations for the ladder operators read~\cite{Greiner:1996zu,qftsm}
\be\label{comm}
\left[{a}_{\bk},{a}^\dagger_{\bp}\right]\,=\,\left[{b}_{\bk},{b}^\dagger_{\bp}\right]\,=\,(2\pi)^3\,2\,\omega_{\bk}\,\delta^3(k-p),
\ee
with all other commutators vanishing, which validate the use of Eq.~(\ref{norm}).
 
If now we consider the case involving mixing, we have
\begin{equation}
\label{boson}
\begin{array}{l}
|\phi_A\rangle\,=\,\cos\theta\,|\phi_1\rangle\,+\,\sin\theta\,|\phi_2\rangle,\\[4mm]
|\phi_B\rangle\,=\,-\sin\theta\,|\phi_1\rangle\,+\,\cos\theta\,|\phi_2\rangle,
\end{array}
\end{equation}
with a similar relation holding for fields, in analogy with Eq.~\eqref{fermion}. Then, in the expansion Eq.~\eqref{bfieldexp}, for each  field ${\phi}_i\,$ ($i=1,2$), we have $\omega_{\bk}\rightarrow\omega_{\bk,i}\,$, ${a}_{k}\rightarrow{a}_{k,i}$ and ${b}^\dagger_{k}\rightarrow{b}^\dagger_{k,i}$. 

Assuming for the mixed fields $\phi_a$ $(a=A,B)$ an action analogous to the one in Eq.~(\ref{action}) and considering a decay process involving $|\phi_i\rangle$ similar to the one in Eq.~(\ref{process}), we get
\be\label{ineq}
\sum_a\,\langle 0|\,{\phi}_a\,|\phi_i\rangle\,=\,\sum_{a,i} U_{a,i}\,\neq\, 1\,.
\ee
thus showing that the states $|\phi_i\rangle$ are at odds with the fields $\phi_a$ in the sense of the normalization Eq.~(\ref{norm}).

A proper choice for the process Eq.~(\ref{process}) should instead be
\be
p\,\rightarrow \,n\,\bar{l}_\alpha\,\nu_\alpha\,,
\ee
with
\be
\langle 0|\,{\phi}_\alpha\,|\phi_\alpha\rangle\,=\,\cos^2\theta\,+\,\sin^2\theta\,=\,1\,,
\ee
which is the result one would expect consistently with Eq.~(\ref{norm}). 

It must be noted that, with the normalization Eq.~(\ref{norm}), the transition probabilities between the states $|\phi_A\rangle$ and $|\phi_B\rangle$ are slightly different from the usual ones. Indeed, it is a trivial matter to show that
\begin{equation}
\left|{\langle\phi_A(t)|\phi_A\rangle}\right|^2\,=\,4(2\pi)^6\omega^2_A\left[1\,-\,\frac{\omega_1\omega_2}{\omega^2_A}\sin^22\theta\sin^2\left(\frac{\Delta\omega\,t}{2}\right)\right]\,,
\label{prob}
\end{equation}
with $\omega_A=\omega_1\cos^2\theta+\omega_2\sin^2\theta$ and $\Delta\omega=\omega_1-\omega_2$. 
 By defining
\be
|\tilde{\phi}(p)\rangle\,=\,\frac{|\phi(p)\rangle}{(2\pi)^3\,2\,\omega_p}\,,
\ee
Eq.~(\ref{prob}) becomes
\be\label{nprob}
\left|\langle\tilde{\phi}_A(t)|\tilde{\phi}_A\rangle\right|^2\,=\,1\,-\,\frac{\omega_1\omega_2}{\omega^2_A}\,\sin^22\theta\,\sin^2\left(\frac{\Delta\omega\,t}{2}\right)\,,
\ee
which, in the relativistic limit, reduces to the usual Pontecorvo oscillation formula.  

\medskip
\emph{e.}\hspace{8mm} 
Finally, we comment on Section II.C of Ref.~\cite{Cozzella:2018qew}. There the authors aim to prove that mixed neutrinos do not spoil the thermality of the Unruh effect. Similarly to what done in Ref.~\cite{Ahluwalia:2016wmf}, they assume a thermal bath of neutrinos with definite mass. The probability of detecting a neutrino with flavor $\alpha$ per unit proper time is then given by Eq.~(34) of Ref.~\cite{Cozzella:2018qew}: 
\begin{equation}
\frac{dP_{\alpha}}{d\tau}\Big|_{\Delta m^2_{ij}\sim 0}\,=\,\sum_i|U_{\alpha,i}|^2\,\frac{dP_{exc,i}}{d\tau}\Big|_{\Delta m^2_{ij}\sim 0}\,\approx\,\frac{dP_{exc,i}}{d\tau}\Big|_{m_i\approx const}\,,
\label{mat}
\end{equation}
 with $\Delta m^2_{ij}=|m^2_i-m^2_j|$.

The last step of Eq.~(\ref{mat}) practically erases mixing, which is not equivalent to consider it in the relativistic regime. Hence, their claim that the thermality of the Unruh effect is also preserved for flavor neutrino is unfounded. A rigorous analysis of field mixing in accelerated frame  has been given in Chapter~\ref{Non-thermal signature}, where non-thermal contributions have been found, which correctly vanish when the relativistic limit is taken into account. Of course, such contributions also vanish in the trivial equal-mass limit considered by Cozzella \emph{et al.}

We remark once again that the assumption of Ref.~\cite{Cozzella:2018qew} about thermal states for mixed neutrinos is the same as the one adopted in Ref.~\cite{Ahluwalia:2016wmf} (where it is motivated by the requirement of KMS condition). Such an observation has been shown  in Chapter~\ref{The necessity of the Unruh effect in QFT: the proton} to violate the general covariance of QFT.

\smallskip

\section{Comparison with results of Chapter~\ref{The necessity of the Unruh effect in QFT: the proton}}
\label{Conclusionscomment}
In this Appendix we have discussed some claims of Ref.~\cite{Cozzella:2018qew}, where it is stated that  the standard Unruh effect is perfectly valid for mixed neutrinos and that the decay rates in the inertial and comoving frames do indeed match also in the presence of mixing. 

In line with the results found in Chapter~\ref{The necessity of the Unruh effect in QFT: the proton}, we share with the authors of Ref.~\cite{Cozzella:2018qew} the belief that general covariance should act as a guiding principle in the comparison of the two decay rates. However, we have shown that the analysis performed by Cozzella \emph{et al.} fails to pinpoint the peculiar aspects of mixing involved in the problem under consideration. In particular, this last statement is supported by the fact that calculations of Ref.~\cite{Cozzella:2018qew} do not recover the outcome of Chapter~\ref{The necessity of the Unruh effect in QFT: the proton}
(where Pontecorvo  states are employed to investigate the inverse $\beta$-decay) in the limit there considered. Indeed, there is no approximation that can provide the decay rates of Ref.~\cite{Cozzella:2018qew} with the off-diagonal contributions arising from the study conducted in Chapter~\ref{The necessity of the Unruh effect in QFT: the proton}.

As a final remark, we would like to stress that the problem considered in Refs.~\cite{Blasone:2018czm,BlasonePOS,Cozzella:2018qew} has to do with the internal consistency of QFT, and thus it is necessary to proceed as far as possible with an exact approach without assumptions based on phenomenological considerations.  In our analysis, by resorting to the concept of flavor neutrino states defined as exact eigenstates of flavor charges obtained via Noether's theorem, we define an approximation in which usual Pontecorvo states for flavor neutrinos can be safely used. In Chapter~\ref{The necessity of the Unruh effect in QFT: the proton}, we have shown that, in such an approximation, the decay rates in the two frames do indeed agree. It remains an open question if corrections to the Unruh effect are necessary in the more general case of exact flavor states. In any case, corrections will be small and this is to be regarded as a virtue of the formalism which allows to highlight novel, unpredicted, effects. Further investigation is required along this line.

%
%

\cleardoublepage

\end{document}